\documentclass[fleqn,10pt,twoside]{report}
\usepackage[square,comma]{natbib}

\bibpunct{[}{]}{,}{n}{}{;}
\usepackage{graphicx}
\usepackage[english,dutch]{babel}
\usepackage[ps2pdf,bookmarks,breaklinks,pdfborder={0 0 0},pdfpagescrop={92 75 546 750}]{hyperref}

\usepackage{placeins} 
\usepackage{color}
\usepackage{booktabs}
\usepackage{textcomp} 
\usepackage{pifont}   
\usepackage{textpos}
\usepackage{subfigure}
\definecolor{gray}{rgb}{0.600,0.600,0.600}

\usepackage{microtype}
\usepackage{lmodern}
\usepackage{sectsty}
\allsectionsfont{\sectiontypeset\selectfont}
\partfont{\sffamily\fontsize{32}{50}\selectfont}

\newcommand{\chaptertypeset}{\sectiontypeset\selectfont}
\newcommand{\sectiontypeset}{\sffamily}
\DeclareMathAlphabet{\mathpzc}{OT1}{pzc}{m}{it}
\usepackage[times]{quotchap_Bert}

\usepackage[titles]{tocloft}
\renewcommand{\cftpartfont}{\large\sectiontypeset\bfseries\selectfont}
\renewcommand{\cftchapfont}{\sectiontypeset\bfseries\selectfont}
\renewcommand{\cftchappagefont}{\sectiontypeset\bfseries\selectfont}
\def\startchapter{
  \ifodd\count0
    \clearpage\thispagestyle{myheadings}\vspace*{\fill}\clearpage
  \else
    \vspace*{\fill}\clearpage\fi}

\makeatletter
\def\cleardoublepage{\clearpage\if@twoside \ifodd\c@page\else%
\hbox{}%
\thispagestyle{empty}%
\newpage%
\if@twocolumn\hbox{}\newpage\fi\fi\fi}
\makeatother

\def\clap#1{\hbox to 0pt{\hss#1\hss}}

\def\mathrlap{\mathpalette\mathrlapinternal}

\def\mathrlapinternal#1#2{\rlap{$\mathsurround=0pt#1{#2}$}}

\def\dtilde#1{\mathrlap{\rlap{\raise1.2pt\hbox{$\mathsurround=0pt%
\tilde{\phantom{\hbox{$#1$}}}$}}\tilde{\phantom{\hbox{$#1$}}}}#1}

\usepackage{slashed}

\usepackage{color}		
\usepackage{epsfig}
\usepackage{metre}

\usepackage{curves}
\usepackage{bbding}
\usepackage{pifont}
\usepackage{subfigure}
\usepackage[small]{caption}
\newcommand{\captionfonts}{\small \em}

\usepackage{multirow}

\makeatletter  
\long\def\@makecaption#1#2{%
  \vskip\abovecaptionskip
  \sbox\@tempboxa{{\captionfonts #1: #2}}%
  \ifdim \wd\@tempboxa >\hsize
    {\captionfonts #1: #2\par}
  \else
    \hbox to\hsize{\hfil\box\@tempboxa\hfil}%
  \fi
  \vskip\belowcaptionskip}
\makeatother 

\usepackage{fancyhdr}
\usepackage{enumerate}
\usepackage{amssymb}
\usepackage[mathscr]{eucal}
\usepackage[errorshow]{tracefnt}
\usepackage{makeidx}
\usepackage{amsmath}
\usepackage{amssymb}
\usepackage{array}
\usepackage{amsthm}
\usepackage{eucal}
\usepackage{setspace}
\usepackage{tikz}

\usepackage{epsfig}
\usepackage{pgf}

\newcommand{\outlook}[2]
{
\begin{itemize}
  \item[\rh] \underline{\emph{#1}} #2
\end{itemize}
}
\newcommand{\punchline}[1]
{
\begin{center}
\parbox{11cm}{\vspace{.3cm}{{\sectiontypeset\selectfont Summary{\hrule}\vspace{.3cm}}{\footnotesize \it
 #1}}\vspace{.3cm}}
\end{center}
}

\newcommand{\outline}[1]
{
\subsubsection{Outline}
{#1}
}
\def\RN{Reissner-Nordstr\"om }

\def\C{{C}}

\def\B{{B}}

\def\unit{\hbox{$1\hskip -1.2pt\vrule depth 0pt height 1.6ex width 0.7pt
\vrule depth 0pt height 0.3pt width 0.12em$}}

\newcommand{\SU}{\mathop{\rm SU}}
\newcommand{\SO}{\mathop{\rm SO}}

\newcommand{\U}{\mathop{\rm {}U}}

\newcommand{\Sp}{\mathop{\rm {}Sp}}

\DeclareMathOperator{\tr}{Tr}
\DeclareMathOperator{\Tr}{Tr}

\def\ansatz{{Ansatz}}
\def\ansatze{{Ans\"atze}}
\def\worldvolume{{world volume{}}}
\def\spacetime{{space-time{}}}
\def\spacetimes{{space-times{}}}
\def\worldvolumes{{world volumes{}}}

\newcommand{\AdS}[1]{{\rm AdS}_{#1}}
\newcommand{\CY}[1]{{\rm CY}_{#1}}
\newcommand{\sS}[1]{{\rm S}^{#1}}

\renewcommand{\d}{\textrm{d}}
\newcommand{\ft}[2]{{\textstyle\frac{#1}{#2}}}

\newcommand{\Real}{\mathbb{R}}
\newcommand{\Db}[1]{\overline{D}#1}

\def\e{{\mathrm e}}

\def\veps{{\varepsilon}}

\def\rmi{{\rm i}}
\def\rmd{{\rm d}}
\def\Re{\mathop{\rm Re}\nolimits}
\def\Im{\mathop{\rm Im}\nolimits}

\newcommand{\de}{\textrm{d}}

\newcommand{\SL}{\mathop{\mathrm{SL}}}
\newcommand{\GL}{\mathop{\mathrm{GL}}}
\newcommand{\cN}{\mathcal{N}}
\newcommand{\QE}{q_e}
\newcommand{\QM}{q_m}
\newcommand{\QT}{Q_\mathrm{T}}
\newcommand{\Qc}{{\cal Q}}

\newcommand{\im}{\mathrm{i}}

\newcommand{\ra}{\ding{242}}
\newcommand{\rh}{\HandRight}

\newcommand{\rab}{{\ArrowBoldDownRight}}

\newcommand{\tickb}{\ding{52}}

\def\ca{{\cal{A}}}

\def\cg{{\cal{G}}}

\def\ch{{\cal{H}}}

\def\cl{{\cal{L}}}
\def\cm{{\cal{M}}}
\def\cn{{\cal{N}}}
\def\Cn{{\cal{N}}}
\def\co{{\cal{O}}}
\def\ck{{\cal{K}}}
\def\cw{{\cal{W}}}
\def\cv{{\cal{V}}}

\def\Mg{{\cal M}}

\def\Conv{{\gamma}}
\def\En{{E}}

\makeindex

\newcommand{\pf}[2]{\frac{\partial {#1} }{\partial {#2}}}
\newcommand{\ta}[2]{\frac{\de {#1} }{\de {#2}}}

\newcommand{\bSL}{\mathop{\rm SL}}

\newcommand{\f}[3]{f_{#1#2}{}^{#3}}

\newcommand{\bSLn}[2]{\bSL({#1},\mathbb{#2})}

\newcommand{\bU}{\mathop{\rm {}U}}
\newcommand{\bSU}{\mathop{\rm {}SU}}

\newcommand{\diag}{{	\rm diag}}

\newcounter{exercise}[section]
\renewcommand{\theexercise}{\thesection.\arabic{exercise}}
\newcommand{\bexer}{
   \begin{list}{\bf Exercise \theexercise : }{\refstepcounter{exercise}
           \setlength {\rightmargin}{\leftmargin}}\item }
\newcommand{\eexer}{\end{list}}
\newcounter{example}[section]
\renewcommand{\theexample}{\thesection.\arabic{example}}
\newcommand{\bexam}{
   \begin{list} {\noindent \emph{Example \theexample: }}{\refstepcounter{example}
           \setlength {\rightmargin}{\leftmargin}}\item }
\newcommand{\eexam}{\end{list}}

\RequirePackage{hyperref}

\numberwithin{equation}{section}

\usepackage{multicol}
\makeatletter
\renewcommand\@idxitem{\footnotesize\par\hangindent 20\p@}
\renewcommand\subitem{\@idxitem \hspace*{10\p@}}
\renewcommand\subsubitem{\@idxitem \hspace*{15\p@}}
\renewcommand\indexspace{\par \vskip 15\p@ \@plus2\p@ \@minus1.5\p@\relax}
  {\end{multicols}\if@restonecol\onecolumn\elseclearpage\fi}
\makeatother

\usepackage{appendix}

\usepackage{threeparttable}
\usepackage{ctable2}

\usepackage{wrapfig}

\def\pa{\partial}

\def\to{\rightarrow}
\def\be{\begin{equation}}
\def\ee{\end{equation}}
\def\bea{\begin{eqnarray}}
\def\eea{\end{eqnarray}}
\def\nonu{\nonumber \\{}}
\def\half{{1 \over 2}}
\def\ca{{\cal{A}}}

\def\cg{{\cal{G}}}

\def\ch{{\cal{H}}}

\def\cl{{\cal{L}}}
\def\cm{{\cal{M}}}
\def\cn{{\cal{N}}}

\def\cw{{\cal{W}}}

\def\sF{{{\rm F}\!\!\!\!\hskip.8pt\hbox{\raise1pt\hbox{/}}\,}}

\def\a{\alpha}
\def\b{\beta}
\def\d{\delta}

\def\f{\phi}

\def\m{\mu}
\def\n{\nu}
\def\o{\omega}
\def\p{\pi}
\def\q{\theta}
\def\r{{\rho}}
\def\s{\sigma}

\def\G{\Gamma}

\def\L{\Lambda}

\def\S{\Sigma}

\def\norm{\Conv}

\usepackage{epsfig}
\usepackage{latexsym}
\usepackage{amsfonts}
\usepackage{amsmath}
\usepackage{amsthm}
\usepackage{amssymb}
\usepackage{amsbsy}
\usepackage{multirow}
\usepackage{slashed}

\newcommand{\figref}[1]{Fig. \ref{#1}}

\newcommand{\tableref}[1]{Table \ref{#1}}

\def\rmi{{\rm i}}
\def\rmd{{\rm d}}

\def\({\left (}
\def\){\right )}

\def\cald         {{\cal D}}

\def\calm         {{\cal M}}

\newsavebox{\uuunit}
\sbox{\uuunit}
    {\setlength{\unitlength}{0.825em}
     \begin{picture}(0.6,0.7)
        \thinlines
        \put(0,0){\line(1,0){0.5}}
        \put(0.15,0){\line(0,1){0.7}}
        \put(0.35,0){\line(0,1){0.8}}
       \multiput(0.3,0.8)(-0.04,-0.02){12}{\rule{0.5pt}{0.5pt}}
     \end {picture}}

\def\be{\begin{equation}}
\def\ee{\end{equation}}
\def\bea{\begin{eqnarray}}
\def\eea{\end{eqnarray}}

\newcommand{\beq}{\begin{eqnarray}}
\newcommand{\eeq}{\end{eqnarray}}

\def\a{\alpha}
\def\b{\beta}

\def\G{\Gamma}
\def\d{\delta}

\def\L{\Lambda}

\def\f{\phi}

\def\m{\mu}
\def\n{\nu}
\def\o{\omega}

\def\p{\pi}
\def\r{{\rho}}

\def\s{\sigma}
\def\S{\Sigma}

\newcommand{\R}{\ensuremath{\mathbb{R}}}
\newcommand{\Z}{\ensuremath{\mathbb{Z}}}

\def\pa{\partial}
\def\T{\tau}

\def\to{\rightarrow}
\def\nonu{\nonumber \\{}}
\def\half{{1 \over 2}}
\def\sF{{{ F}\!\!\!\!\hskip.8pt\hbox{\raise1pt\hbox{/}}\,}}
\def\som{{{ \omega}\!\!\!\!\hskip.8pt\hbox{\raise1pt\hbox{/}}\,}}
\def\sJ{{{\rm J}\!\!\!\!\hskip.8pt\hbox{\raise1pt\hbox{/}}\,}}

\begin{document}

 \pagestyle{empty}
\vskip-10pt

\begin{center}


\noindent
{\huge\sffamily\textcolor{black}{\textbf{Hidden Structures of Black Holes}}}
\vskip 1.5truecm


\begin{center}
{\large \textbf{\textsf{Bert Vercnocke}}} \
\vskip .3 cm
\it 
CEA Saclay - DSM/IPhT\\
B\^at. 774\\
91191 Gif sur Yvette Cedex, France\\[3mm]{
e-mail:} {\tt Bert.Vercnocke@cea.fr} \\
\end{center}
\vskip .2 cm

\small{Dissertation presented in partial fulfillment of the requirements\\ for the degree of Doctor of Science at K.U.Leuven, Belgium}

\vskip .7 cm
\end{center}
{\centering
\centerline{\sffamily\bfseries Abstract}
}

\noindent 

This thesis investigates two main topics concerning black holes in extensions of general relativity that can arise as low energy limits of string theory.

First, the structure of the equations of motion underlying black hole solutions is considered, in theories of $D$-dimensional gravity coupled to scalars and vectors. The presence of scalar fields complicates the analysis of black hole solutions. However, for solutions preserving supersymmetry, the equations of motion have a dramatic simplification: they become first-order instead of the second-order equations one would expect. Recently, it was found that this is a feature some non-supersymmetric black hole solutions exhibit as well. We investigate if this holds more generally, by examining what the conditions are to have first-order equations for the scalar fields of non-supersymmetric black holes, that mimic the form of their supersymmetric counterparts. This is illustrated in examples.

Second, the structure of black holes themselves is investigated. String theory has been successful in explaining the Bekenstein-Hawking entropy for (mainly supersymmetric) black holes from a microscopic perspective. However, it is not fully established what the interpretation of the corresponding  `microstates' should be in the gravitational description where the black hole picture is valid. There have been recent advances to understand the nature of black hole microstates in the gravity regime, such as the fuzzball proposal. A related idea says that black hole configurations with multiple centers are related to microstates of single-centered black holes. We report on work relating both pictures. As an aside, through a connection with the black hole deconstruction proposal, a relation between violations of causality for certain spacetimes (presence of closed timelike curves in the geometry) and a breakdown of unitarity in the dual conformal field  theory is given.

\newpage
\thispagestyle{empty} \mbox{}
\newpage

\thispagestyle{empty} \mbox{}
%


%
\begin{textblock}{160}(-17.5,-32.5)
\vspace{-\parskip}
\begin{center}
\includegraphics[width=8mm]{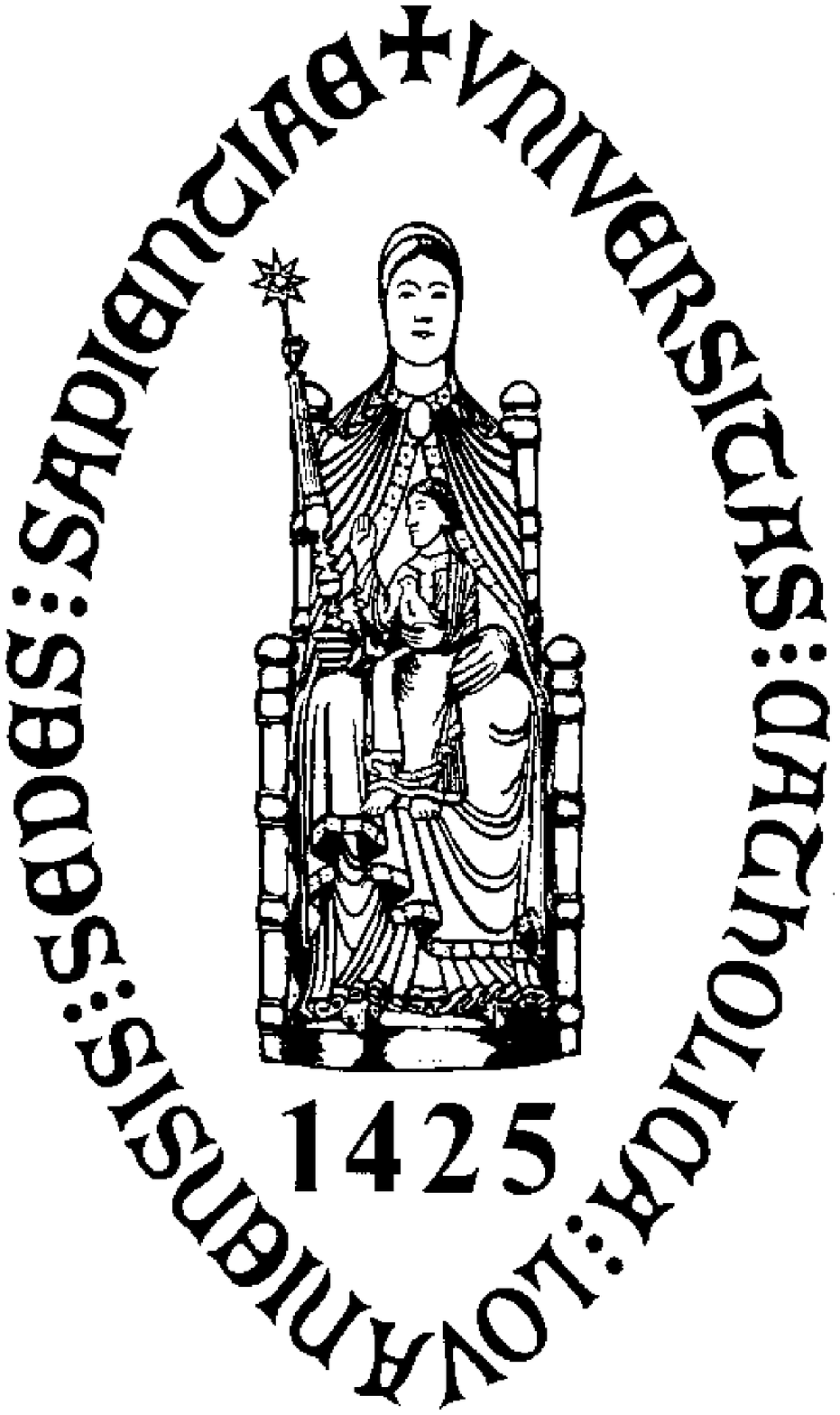}
\end{center}
\end{textblock}
\begin{textblock}{100}(42.5,-12.5)
\vspace{-\parskip}
\colorbox{gray}{\parbox[c][18mm]{98mm}{
{\bf \footnotesize \textcolor{white} {Arenberg Doctoral School of Science, Engineering \& Technology}}\\
{\footnotesize \textcolor{white} {Faculty of Science}}\\
{\footnotesize \textcolor{white} {Department of Physics and Astronomy}}
}}
\end{textblock}
\begin{textblock}{60}(-18.35,-12.55)
\textblockcolour{}
\vspace{-\parskip}
\includegraphics*[width=60.8mm]{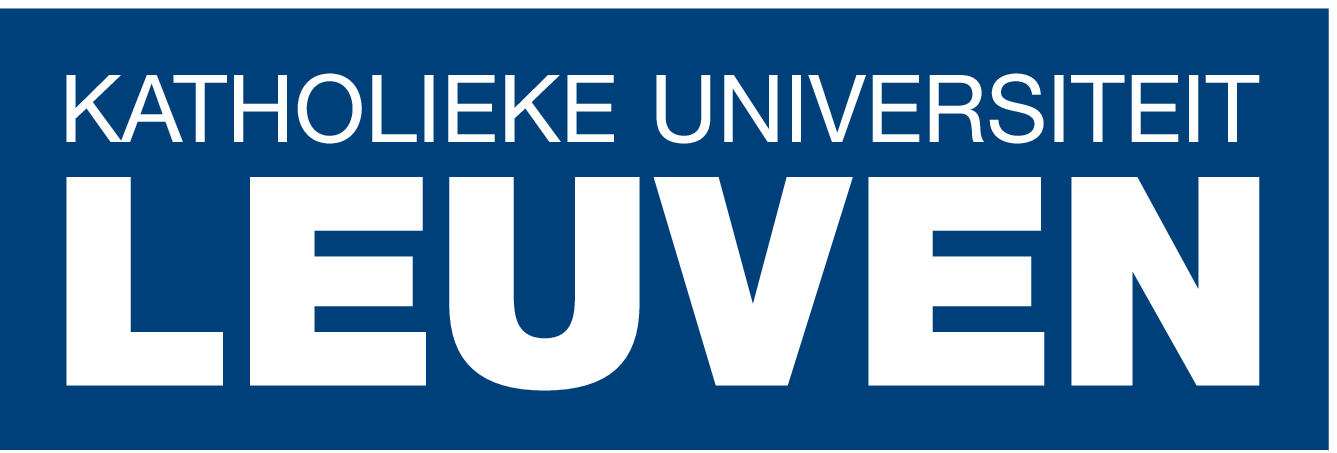}
\end{textblock}

\begin{textblock}{125}(0,15)
\textblockcolour{}
\centering
\huge\sffamily\textcolor{black}{\textbf{Hidden Structures of Black Holes}}\\[5mm]
\includegraphics*[width=.8\textwidth]{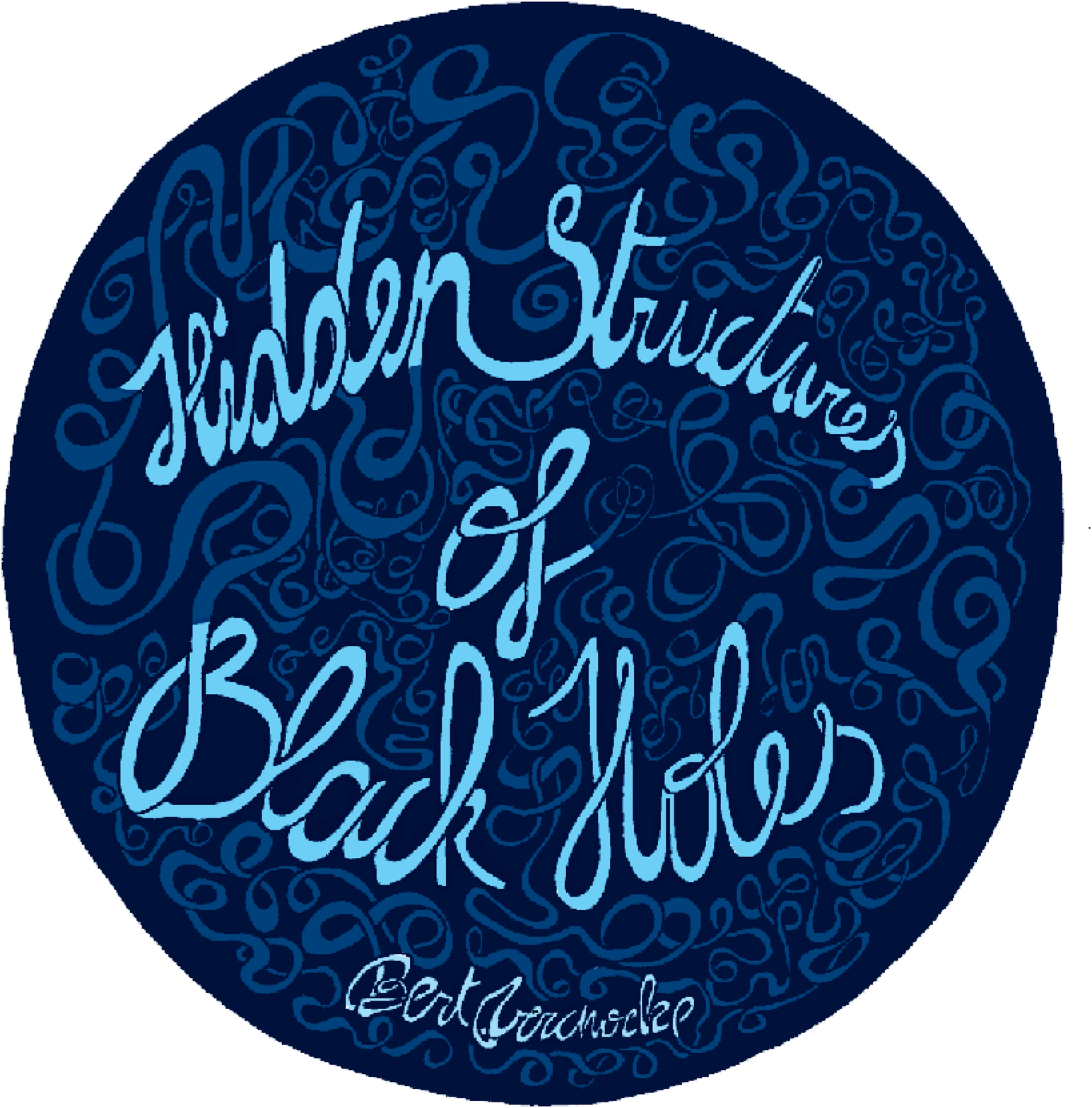}
\end{textblock}

%
\begin{textblock}{120}(0,57.5)
\end{textblock}
\begin{textblock}{120}(0,130.5)
\textblockcolour{}
\vspace{-\parskip}
\centering
\large\sffamily\textbf{Bert Vercnocke}
\end{textblock}
\begin{textblock}{120}(0,145)

\textblockcolour{}
\vspace{-\parskip}
%

\begin{tabular}{p{6.5cm}l}
Supervisor:&\\
Prof.\ Dr.\ A.\ Van Proeyen&\\[\parskip]
Examination Board:                                                   & \\
 Prof.\ Dr.\ D.\ Boll\'e, president           & \\
 Prof.\ Dr.\ J.\ Indekeu, secretary            & \\
 Prof.\ Dr.\ W.\ Troost                             & \\
 Dr.\ R.\ I.\ Bena                             & \\
\rule{0.25cm}{0pt}(\emph{CEA Saclay, France})  & Dissertation presented in partial\\
Prof.\ Dr.\ J.\ M.\ Figueroa-O'Farrill             & fulfillment of the requirements for\\
\rule{0.25cm}{0pt}(\emph{University of Edinburgh, UK})   & the degree of Doctor of Science\\
\end{tabular}

\end{textblock}
\vfill
\vspace{2cm}

\begin{textblock}{120}(0,200)
\centering
{August 2010} 
\end{textblock}

\newpage
\mbox{}
\vfill

Cover illustration by Wide Vercnocke

%
%
%
%
%
%
%


\cleardoublepage
\selectlanguage{dutch}
\pagenumbering{roman}
\cleardoublepage
\selectlanguage{english}

\mbox{}
\vspace{.3\textheight}
\begin{center}
\emph{My old man told me one time\\
you never get wise, you only get older\\
and most things, you never know why, but that's fine
}
\\[12pt]
\raggedleft
The Dandy Warhols
\end{center}
\mbox{}
\vfill

\cleardoublepage
\mbox{}
\vspace{.3\textheight}
\begin{center}
{\it Voor mijn meisjes} 
\end{center}
\mbox{}
\vfill

\cleardoublepage
\chapter*{
Reading guide}
This thesis gives an overview of the work performed during my doctorate. The aim is to gain a better understanding of various aspects in the theoretical study of black holes, in the wider context of string theory and supergravity.


It consists of a general introduction, followed by four parts. 

The introduction is aimed at readers with some notion of (or interest in) physics, but not necessarily of this particular field. Part \ref{pt:intro} gives the general context of the research, aimed at readers with a knowledge of physics at the undergraduate level (especially basic knowledge of general relativity). Then follow two parts describing most of the research done during the PhD. Each describes different aspects in the study of black holes and starts with its own introduction. In particular, part \ref{pt:FirstOrder} discusses the work on first-order equations for non-supersymmetric black holes. Section \ref{s:BJ-Introduction} to \ref{s:BJ-History_FoForm_BHs} constitute an introduction to the ideas presented in this part of the thesis. Part \ref{pt:Entropy} contains the research on black hole `microstates' in the (super)gravity regime.  The reader is advised to go to chapter \ref{c:FB} for an introduction to this field. Finally, part \ref{pt:Conc} gives a conclusion and a technical appendix. 

This version of my PhD thesis differs slightly from the printed version. The Dutch summary is left out, as it contains a very non-technical account of the work performed. For an online version of the printed PhD thesis including the Dutch summary, see \href{http://itf.fys.kuleuven.be/hep/phd/ThesisBertVercnocke_Cover.pdf}{\textcolor{blue}{http://itf.fys.kuleuven.be/hep/phd/ThesisBertVercnocke\_Cover.pdf}}.


\selectlanguage{english}
\cleardoublepage


\cleardoublepage
\phantomsection
\markboth{{Contents}}{{\chaptertypeset Contents}}
\tableofcontents

%

\cleardoublepage
\pagenumbering{arabic}

\pagestyle{fancy} 
\renewcommand{\chaptermark}[1]{\uppercase{\markboth{#1}{#1}}}
\renewcommand{\sectionmark}[1]{\uppercase{\markright{#1}}}
\fancyhf{}
\fancyhead[RO]{\scriptsize\sf\rightmark\ \hrulefill\ \thepage} 
\fancyhead[RE]{\scriptsize\sf\thepage\ \hrulefill\ \leftmark}
\renewcommand{\headrulewidth}{0pt}
\fancypagestyle{plain}{
\fancyhf{}
\fancyfoot[C]{\scriptsize\sf\thepage}
\renewcommand{\headrulewidth}{0pt}
\renewcommand{\footrulewidth}{0pt}
}

\cleardoublepage

\chapter{
Introduction}\label{c:Intro}


\section{The large and the small}

Remember the last time you were walking in the outdoors. Imagine again how you looked around and noticed your surroundings, felt the wind in your face, smelled the fresh air. Your senses certainly gave (and continuously \textit{give}) you the impression that there `is' something outside of your own consciousness. We can call that something `our world' or `the universe'. Truth is that, whatever we called it, we humans have always tried to understand it, tried to answer the questions that pop into our heads when we look at the world. Who or what controls the weather elements that keep us dry or surprise us with a  thunderstorm? How do things move? What makes the world go round?  

Luckily for us, we do not need all details of the world around us to understand parts of it: for example, we do not have to know the smallest structure of metal (atoms, molecules\ldots), to mend a bike or to build a car. We say that science, and physics in particular, is organized in scales. Science proposes models to describe the world at a given scale. The scales we are mostly interested in, and most comfortable with, are the scales set by our everyday lives. Distances of meters and kilometers; time laps of seconds, hours, years; the speed of pedestrians, bikes, cars. In these circumstances, a successful (and accurate) way to model physics is through classical mechanics: the movement of objects is given through their acceleration, which in turn is given by the forces that act on them. In this picture, forces can be of many kinds: frictional forces, like the headwind plaguing a cyclist, the normal forces that keep a bottle of wine on the table, the gravitational force that keeps us from flying off to touch the stars. What happens if we broaden our scope and go to smaller, or larger scales? Then the classical mechanics model loses accuracy. 

When speeds grow large and/or masses of objects are far beyond what we encounter in our daily lives (think of planets, stars, galaxies), we enter into the realm of Einstein's \textit{relativity} theory. In the beginning of the last century, Einstein reformulated the laws of mechanics under two principles. First, he stated that the laws of physics should be independent of the observer (principle of relativity), second, that the speed of light in vacuum is the same for every observer.\footnote{{}For \textit{special} relativity, this is more correctly stated by saying that the laws of physics apply in any \textit{inertial} system and that the speed of light in vacuum is the same in any \textit{inertial} system. \textit{General} relativity further eploits the principle of covariance to handle accelerated frames.}  Following these principles, the generalization of the laws of classical mechanics goes in two steps: first, one has \textit{special relativity} (S.R.), which is the extension to speeds that are comparable to the speed of light. Second, with the inclusion of gravity, the observer-independence is assured through the extension to  \textit{general relativity} (G.R.): dynamics of particles is encoded in the curvature of \spacetime{}. 

When we go down to model the very small -- sub-atomic scales, elementary particles -- quantum mechanics and \textit{quantum field theory} enter the picture. In quantum field theory, elementary particles are represented as points,  moving around in  \spacetime{}. Point particles interact with each other through the exchange (emission and absorption) of other particles. (For example, the electromagnetic forces between electrons is realized through the exchange of photons.) The theoretical model of quantum field theory that describes to extremely high accuracy all elementary particles observed so far and the forces between them, is called the \textit{standard model} of elementary particle physics.

We are able to model the world extremely well in the respective domains of these theories: \textit{general relativity} (`the large') and the \textit{standard model} (`the small'). Through them, we are able to describe phenomena from sub-atomic length-scales ($10^{-15}$ m) up to astronomical scales (light years, $10^{16}$ m). At small length scales, probed by present-day particle accelerators, the standard model describes the sub-atomic particles that are observed and the three forces between them relevant at those scales (electromagnetic force, the weak and strong force), while gravity is negligible at these distances (see below). In the intermediate range, at everyday scales, general relativity reduces to the gravitational force we experience in everyday life and the fundamental three forces of the standard model make up the remaining forces we experience. At cosmological scales, general relativity models accurately observed reality, for instance by explaining the observed expansion of the universe, or through the confirmed prediction of black holes. Maybe this is reason enough for you to say `stop, we are done': if these theories work so well, where is the point in pursuing theoretical physics? 
 
Actually, there are enough reasons to believe that this is not the end of the story. Most of these reasons are not passed on to us by experiment, but are rather thorns in the theorist's eye, pointing to corners of the fundamental theories above that are not (fully) understood. 
Maybe the most compelling are the \textit{problems with gravity}. 
First, relativity only works under the assumption that the world is purely classical and does not use the principles of quantum mechanics. But, as a classical theory, general relativity breaks down at singularities, points where the gravitational field becomes infinite. These occur for instance in the big bang model and in the descriptions of black holes (see below).  Second, the standard model is only a good description when gravity is weak, and only works well when we exclude gravity from the picture. Even worse: trying to commonly describe quantum field theory with gravity results in severe problems. Said in another way, we do not have an accurate \textit{quantum} theory describing gravity at the smallest scales. 

In the rest of this introduction, we treat these two problems. First, we go into the problems with general relativity related to the existence of black holes in the next section. We expect these problems to be addressed in a consistent quantum theory of gravity. This is the subject of the section thereafter, where we propose string theory as a good quantum gravity theory that reconciles the standard model and gravity in one common picture, and we discuss how this theory can explain black hole physics. Finally, the content of this thesis is placed against this background.



\section{Black holes can guide us\ldots}

We need to go to general relativity, to really make the concept of a black hole clear. It is understood through the nature of the curvature of \spacetime{} induced by a massive object. When a mass is put in a region of space, it curves \spacetime{} around it.
By stacking more and more mass in a given \spacetime{} volume, eventually the curvature will become so large, that particle trajectories and even light rays are so extremely bent around the mass, they cannot escape from its pull and fall back in: nothing can escape from the region around the mass. Since even light does not come out, we call this a black hole.

The boundary between the black hole region, from which nothing can escape, and the rest of \spacetime{} is called the event horizon. Due to gravitational attraction, all matter inside the event horizon will eventually collapse into a single point. All mass of the black hole is then stacked into this point, making it singular: the mass density is infinite. We can thus picture the black hole as a singularity, hidden behind an event horizon (the singularity is hidden, because no information about it can pass through the event horizon). 
\begin{figure}[ht!]
\centering
 \includegraphics[width=.27\textwidth]{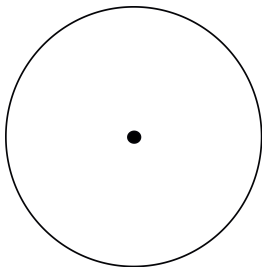}
\label{fig:Intro_BH_Sing}
\caption[Black hole as a singularity surrounded by an event horizon.]{Cartoon of a black hole as a singularity surrounded by an event horizon.}
\end{figure}

\paragraph{Black holes exist.}
This may seem like a lot of theoretical mumbo-jumbo, but black holes are really `out there'. 
A black hole can form when objects collapse due to their own gravitational pull. In particular, this happens at the end of the life of giant stars. During its lifetime, a star is a huge fusion reactor. In the incredibly hot star interior, light atom nuclei form heavier nuclei through fusion reactions, releasing energy.  This energy is associated with an outward pressure, opposing the gravitational force that wants to make the star collapse. Near the end of the star's lifetime, when the amount of fuel inside the star becomes insufficient, the pressure, too, will drop and gravity starts winning. When the star is massive enough (mass of more than about 4 times that of the sun), it will eventually collapse into a black hole. However, since black holes are `black', we cannot detect them directly. By using indirect methods (influence of the black hole on its surroundings), certain systems have been identified as black holes and it is commonly believed that entire galaxies (including our own, the Milky Way) revolve around supermassive black holes in their midst.

\paragraph{Black holes radiate.}

In a sense, black holes represent gravity at its strongest. Therefore they form a natural way to try to probe the possible simultaneous description of gravity and quantum physics. 
A first step towards a quantum mechanical understanding of gravity, was ignited by Stephen Hawking in the seventies \cite{Hawking:1974sw}. He considered quantum field theory in the fixed gravitational background of a black hole. Such a setup is called a semi-classical treatment of gravity. Hawking showed that, semiclassically, black holes are not really black: they emit radiation! 
Intuitively, this is understood from particle creation from the vacuum near the black hole horizon: one particle of the pair falls into the black hole, while the other one escapes to infinity. 
The main result is that the black hole emits particles with the spectrum of a perfect black body, attributing thermal properties to the black hole: we can associate a temperature and an entropy to the black hole. These results came as a real surprise. The temperature and entropy are defined through geometric quantities of the black hole \spacetime{}. 
In particular, the entropy is proportional to the area $A_H$  of the event horizon. This entropy goes by the name of Bekenstein-Hawking entropy (Bekenstein was the first to advocate that the black hole horizon area should be seen as an entropy):
\begin{equation}
 S_{BH} =  \frac14 \frac{k_B c^3}{\hbar G_N} A_H\,,
\end{equation}
where the proportionality constants are Boltzmann's constant $k_B$ and the combination  $\hbar G_N/c^3$, which has dimensions of length squared, made up out of Planck's constant $\hbar$, the speed of light $c$ and Newton'sconstant $G_N$. For instance, for a black hole with mass $M$, the entropy scales as $S_{BH} \sim M^2$. The temperature typically goes as $T\sim 1/M$, i.e.~grows smaller for larger black holes. (More information on black hole thermodynamics is given in the next chapter.)

\paragraph{Black holes have problems.}
Through the semiclassical treatment of black holes, we can make an analogy with equilibrium thermodynamics.  In thermodynamics, we do not know, or use, the microscopic degrees of freedom of the system. Instead, the system is described by a set of macroscopic properties: energy $E$, entropy $S$, temperature $T$\ldots~Likewise, in semiclassical gravity, a black hole is described by its mass $M$ (or energy $E=M c^2$), its Bekenstein-Hawking entropy $S_{BH}$,  its (Hawking) temperature $T_H$. However, there is a difference. In the case of thermodynamics, we \textit{have} a microscopic description at hand, which can be related to the macroscopic description through statistical mechanics. In particular, we can relate the entropy of a system with a certain value of the macroscopic parameter(s) (e.g.~fixed energy), to a number of microstates $\Omega$ that give rise to the same value of the macroscopic parameter(s) through
\begin{equation}
 S = k_B \ln (\Omega)\,.\label{eq:Intro-Entropy}
\end{equation}
For black holes, a microscopic description and a `statistical mechanics' is not at hand, at least not in the context of (semiclassical) general relativity: there is only one black hole for a given mass $M$, we do not have access to a number of `microstates'. An interpretation of the entropy as in eq.~\eqref{eq:Intro-Entropy} is not possible.

We expect a (or the?) quantum theory of gravity to provide us with an answer to the microscopic nature of black holes: how can we account for the entropy? Can we construct `microstates' for a black hole? What do they look like?  There are also other interesting problems a quantum theory should resolve. Think about the singularity, which we expect to be resolved somehow through quantum effects (in quantum theory, physics at small length scales can differ significantly from the classical theory). Or the information paradox: as a black hole radiates, it loses mass and will eventually evaporate. In the end, we would be left with a universe filled with thermal radiation that has a very high entropy: this radiation reveals no information about the initial state, leading to possible unitarity violation in quantum mechanics. More detailed information about black holes and their problems is given in the next chapter.

We thus have a motivation for the study of black holes in the context of a quantum gravity theory. In particular, this thesis studies aspects of string theory, a promising quantum gravity theory, through black hole solutions.

\section{\ldots to test string theory}

\paragraph{String theory.}

At the moment, there is a promising candidate that unites both the standard model and gravity in a consistent quantum theory, called \textit{string theory}.\footnote{Although there are other attempts at quantum theories of gravity, as loop quantum gravity, they do not consider the inclusion of standard model physics.} In string theory, the idea of a point particle is abandoned. Instead, particles are pictured as strings, moving around in \spacetime{}. The typical length $l_s$ of a string is much smaller than the smallest distance scales probed in experiments so far, such that in particle accelerators, strings effectively look like point particles. 

\paragraph{Why strings are so attractive.} However simple the onset may be, any consistent quantum theory built out of the interactions of strings, instead of point particles, is very rich and has a whole range of beautiful properties, answering the difficulties raised in the first section. First, a consistent string theory includes \textit{gravity}, since it always contains a state, called graviton,  with the right properties to mediate the gravitational force (as for instance the photon mediates electrodynamic forces). Moreover, it includes a whole range of other particles and forces, among which those of the standard model. (More particles and forces can arise, but there is not necessarily a clash with present-day experiments.) We thus get a natural inclusion of gravity and standard model physics in one picture, a \textit{unification} of all fundamental forces. Furthermore, the fact that particles are no longer points, but are extended in space as strings, has two far-reaching consequences: on the one hand,
it follows that the structure of the interactions is uniquely fixed by the free theory and there are \textit{no free dimensionless parameters}: there are no arbitrary interactions to be chosen, as in the standard model (which has about 20 free parameters: particles masses, values of various coupling parameters determining the strength of the forces). E.g.~string theory predicts the existence of a scalar field, the dilaton $\phi$. The vacuum expectation value of its exponential plays the role of an effective coupling constant $g_s = \langle \e^\phi \rangle$. 
In principle this would suggest that there \textit{is} a free parameter $g_s$. However, as it is related to the dynamical scalar field $\phi$, its value is supposedly fixed by the string dynamics and is not put in arbitrarily. On the other hand, there are no short-distance singularities, and in this way one finds that string theory provides a consistent formulation of quantum gravity, at least in string perturbation theory (in powers of $g_s$), because perturbative string theory is finite order by order. 
Finally, string theory is essentially \textit{unique}. There are in principle several consistent string theories that can be constructed, but they are all related by dualities (see below) and should be seen as different aspects of one underlying theory.

There are more features a string theory automatically includes. First, string theory needs \textit{extra dimensions}.\footnote{This is for critical string theory.} String theory lives in ten dimensional \spacetime{} (nine spatial directions, one time direction). 
Second, string theory requires \textit{supersymmetry}, a symmetry relating bosons (such as the particles that are responsible for forces) and fermions (such as the particles that build up matter). 
The prediction of supersymmetry is maybe the best candidate to be confirmed experimentally, thus supporting the ideas behind string theory. With the start of the operation of the Large Hadron Collider in CERN, Geneva, a new range of energy scales (up to $1.4 \times 10^4$ GeV) opens up and may lead to the detection of supersymmetric partners of known particles, since there are arguments (not relying on string theory, by the way) relating the scale of the weak interactions (around 100 GeV), to the scale of supersymmetry breaking and the latter scale in turn  determines the masses of the hypothetical supersymmetric partner particles.

\paragraph{String theory is patchwork.}


It turns out that one can construct five consistent string theories, all having the properties above. These go by the name of type I, type IIA, type IIB and heterotic $\SO(32)$ and $E_8\times E_8$ superstring theory. In none of the string theories, we have a complete handle on the physics: in particular, we typically have an idea about perturbative string theory, but non-perturbative aspects are hazy. Things changed drastically when in the mid-1990s, it became clear that all five consistent superstring theories that had been constructed, are related through a web of \textit{dualities}, i.e.~they are physically equivalent. We recognize T-dualities, which relate different string theories on different \spacetime{} geometries, and S-dualities, which map a weakly coupled string theory into another strongly coupled one. We thus have a handle on strongly coupled string theories through perturbative calculations in a dual, weakly coupled string theory. These dualities suggest that all five string theories are aspects of one underlying theory, dubbed M-theory, a theory in eleven \spacetime{} dimensions. Through the web of dualities, we have several viewpoints to study this underlying theory. However, a complete view is lacking, since we can only study corners of the underlying theory, by considering perturbative calculations in each dual string theory. In particular, we do not have a handle on solutions (vacua) where the coupling is in an intermediate range, being neither small (amenable to perturbation theory) nor large (amenable to perturbation theory in a dual string theory picture). Rather, we have only caught glimpses of what M-theory should look like.

Another breakthrough from the same period gave more information about the non-perturbative formulation of string theories. Even though string theory was introduced as a theory of strings, which extend along one spatial dimension, it became clear that there are also objects extending in more dimensions. These objects are called $p$-branes, where $p$ stands for the number of spatial dimensions of the objects (the word `$p$-brane' is a  generalization of the two-dimensional membrane). All $p$-branes (except for the string itself) become infinitely massive as $g_s\to0$, explaining why they did not turn up in string perturbation theory. An interesting subset of these $p$-branes are D$p$-branes, or D-branes for short.   At $g_s=0$, these D-branes are described as rigid surfaces in \spacetime{} on which open strings can end. However, when the string coupling is non-zero, these D-branes have a dynamics of their own. They are characterized by masses that go as  $1/g_s$: in strongly coupled string theory, these D-branes become light and should be seen as the fundamental degrees of freedom of the theory. 

We see that we know several patches of the fundamental theory, but that the story is far from being finished. In a sense, even though the theory is older than the author of this doctoral thesis,  string theory is still in its childhood years.

\paragraph{String theory and the real world.}
%
%
String theory has suffered from a lot of criticism, because it has not made any falsifiable predictions at the moment. How can we understand this? Before we noted that \textit{`string theory has no adjustable parameters'} and \textit{`string theory is unique'}. If the theory is uniquely fixed, then where is the trouble with providing quantitative predictions?

The answer lies in the fact that a unique theory need not have a unique solution.  In case of string theory, there seems to be an enormous amount of solutions, which are (meta)stable, and many of which have properties that are similar to our world. These solutions have been referred to as the landscape of string vacua. If the number of solutions would be small, say ten, then we would be happy still, since we could check them one by one and compare them to the observed facts. However, the estimated number associated to these so-called `string vacua' lies around $10^{500}$. Moreover, this is just an estimate: we are not able to construct all these solutions.

To understand the appearance of this huge amount of possible solutions, consider how we can link string theory to the real world. All different string theories live in ten dimensions and the underlying M-theory even has eleven dimensions: this seems to be far off from the observed four-dimensional world (3 dimensions of space, one of time). To make contact with four-dimensional physics, one assumes that the other six (or seven) dimensions are sufficiently small, such that they escaped detection. In terms of the scales mentioned in the beginning of this introduction, this means that the length scale of these extra dimensions should be smaller than the length scales probed in particle accelerators. A compactification of string theory obtained in this way still is rich enough to account for our realistic world: it contains gravity and also fields of the correct form to describe the standard model (i.e.~gauge fields and fermions). However, there are also many other (scalar) fields that appear, describing the geometrical details of the curled up extra dimensions. In total, there is a huge freedom in choosing the details of the compact space  and this leads to a large number of solutions in four dimensions.

Is the story of string theory over, if we cannot obtain real-life information?  We cannot count on accelerator experiments to give us detailed information about string theory. Unless some of the internal dimensions are unnaturally large, the energy scale at which we would see string physics, is way beyond what we observe today.  Therefore, we must go back to theory. Maybe the flaw is in the estimates of the number of vacua, as these use perturbative string theory, and a full non-perturbative formulation singles out a few, or maybe even one unique vacuum? At present, this is not really expected. The most conservative approach would be to give up the hope to find an exact `our world'-scenario from string theory, at least for the time being. This means we should focus on general results (as opposed to very concrete predictions) we can extract from string theory. We can divide this in two broad ideas. On the one hand, we can still try to use string theory still as a `theory of everything' and look for general features the theory can teach us, e.g.~by making statistical predictions from the string landscape. The other idea is to use string theory as a tool, for instance to study totally different systems, related through to string theory through dualities as the AdS/CFT correspondence. This correspondence relates string theory on certain geometries (Anti-de Sitter or AdS spaces) to quantum field theories with a scaling symmetry (conformal field theories of CFTs). The AdS/CFT correspondence has been used to obtain qualitative results for non-perturbative (and in conventional calculations inaccessible) results for quantum field theories describing for instance the strong interaction (QCD) or condensed matter systems (e.g.~superconducting materials), through the study of string theory or its gravity limit in AdS geometries. 

The idea that is followed in this work is to see if string theory can teach us about microscopic features of black holes: we consider the study of black holes as a `theoretical test' for string theory as a quantum gravity theory.

\paragraph{String theory and black holes.}
As string theory provides a window on quantum gravity, this opens perspectives to study the origin of the Bekenstein-Hawking entropy. In fact, it has been explained for the special class of \textit{supersymmetric} black holes and forms one of the major successes of string theory.

The first microscopic derivation of black hole entropy from string theory is due to Strominger and Vafa \cite{Strominger:1996sh}. They considered a certain black hole in a string theory compactification, made out of a set of D-branes (the D-branes are wrapped on the internal directions, such that in four-dimensions we see a point particle). By counting the number $N$ of quantum mechanical states associated to the D-brane system in string theory, they reproduced the Bekenstein-Hawking entropy as $S_{BH} = k_B \ln N$. 

There are two caveats. First, the calculation of the entropy is done in a regime where gravity can be neglected (formally, this is done by taking the string coupling $g_s$ to zero), such that the system under study is a gas of weakly interacting D-branes. A priori, it is unclear that the quantum states of this system should correspond to the number of states associated to description as a black hole, where gravity is no longer neglected. It can be shown that only for \textit{supersymmetric black holes}, the number of states is invariant under variation of the coupling $g_s$. However, supersymmetric black holes carry charge and are extremal (they do not emit radiation and have zero Hawking temperature), which means their charge is maximal for a given mass. They are highly unrealistic, as a typical astrophysical black hole carries no (or very little)  charge.

Second, this type of calculation does not give an understanding of what the structure of the black hole microstates is. As we only have an idea of these states in a dual regime, where gravity is weak, we do not know how to interpret a `black hole microstate' in the gravity description. 

\outlook{}{The above leads to two interesting research questions: what about \textit{non-supersymmetric} black holes? What \textit{is} a black hole microstate? These are (partially) addressed in this thesis, see below.}

%


\section{Overview of the research in this thesis}

In this thesis, we do not consider the full scope of string theory. Instead, we focus on classical extensions of general relativity that are inspired by string theory, so-called supergravity theories. Supergravity is the effective description of string theory on length scales much larger than the string scale: the only relevant excitations of the string are massless point particles, coupled to the metric describing the \spacetime{}. We always consider classical supergravity, such that quantum effects are suppressed. 

It is in this regime, the supergravity description, that the gravitational attraction is strong and very massive states of string theory can form black holes. We study the two questions raised above in the context of supergravity:
\begin{enumerate}
\item Properties of non-supersymmetric black holes in supergravity
\item Interpretation of a microscopic black hole state in the regime where gravity effects are important, i.e.~in supergravity
\end{enumerate}
We first discuss these subjects and then give an overview of the chapters to come.


\subsection*{Topic 1: first-order formalisms for non-supersymmetric black holes}
String theory gives rise to many scalar fields. These are unobserved in nature and their presence complicates the connection of string theory to the real world. 
Also in the supergravity description, the low energy gravity description of string theory, scalar fields are omnipresent. Of special interest is the study of solutions to supergravity,  think of black hole solutions in four dimensions, but also black $p$-branes in ten dimensions, which are essential in the formulation of string theory. Typically, these solutions are characterized by the presence of at least one (string theory dilaton) to many (compactification moduli) scalar fields. 

The first broad  research topic treated in this thesis, concerns the structure of the equations of motion that govern the dynamics of these scalar fields for \textit{non-supersymmetric} black holes. In general, in supergravity a black hole solution is characterized by the metric (describing the black hole \spacetime{}) and a plethora of other fields (gauge fields generalizing the electromagnetic field and scalar fields). When we demand that the black hole solution is spherically symmetric, the form of the metric is very much constrained and the gauge fields are fixed by symmetry. However, the many scalar fields that are present have a non-trivial dynamics. The dynamics of these scalar fields is governed by equations of motion that are in general second-order differential equations. 

As we noted before, for black holes that preserve some of the supersymmetry, we have met with success in discussing questions as the entropy problem. The underlying reason is that the constraint of supersymmetry makes the black hole solution `simple': it is more constrained because of the requirement of supersymmetry. Also on the level of the equations of motion this simplicity is pertained: supersymmetric solutions obey first-order equations of motion, instead of second-order ones.

It has been noticed in the literature that for some non-supersymmetric black holes, a similar simplification of the equations of motion takes place. Where does this extra structure in the field equations come from? Part of the work presented in this thesis is concerned with that question. \textit{In this thesis, we discuss generalizations of previous work on first-order field equations for non-supersymmetric black holes and find a condition for the existence of the first-order formulation.}

\subsection*{Topic 2: Interpretation of black hole `microstates' in supergravity.}
The entropy of supersymmetric black holes has been explained by counting states in  a dual regime of string theory, where gravity is negligible. However, this does not shed light on the nature of the microstates in the regime where we have an interpretation as a black hole, since this requires that the gravitational interaction cannot be neglected. 

A fruitful approach is the so-called fuzzball proposal, initiated by Mathur and collaborators in the early 2000s. This states that the correct interpretation of a black hole microstate is a state where the matter is spread out in a sort of `fuzzy' ball, that is of a size comparable to the size of the black hole horizon. Each individual state has no horizon and no singularity. The black hole should be seen as an artefact of an averaging procedure over all these microstate geometries. This view is comparable to describing the gas in a room through a set of macroscopic parameters (volume, temperature, pressure). However, the correct state of the gas is one of very many possible microstates, possible configurations the gas molecules can have. Similarly, a black hole with Hawking temperature and a certain mass, is a thermodynamic description of very many possible microstates, i.e.~smooth fuzzball geometries. This picture is supported by calculations in the context of string theory and supergravity. In particular,  one can often construct a large set of classical solutions that have the same macroscopic parameters (charges, mass) as the black hole and have the same form at large distance, but have no horizon. These solutions typically differ from the black hole on scales of horizon-size. It is hoped that a proper quantization of these solutions can explain the black hole entropy. There has been partial success in this direction, mainly for supersymmetric black holes in five dimensions. For these cases, one can often show that the supergravity state corresponds to one of the dual D-brane states that are used in deriving the entropy of the black hole.\footnote{Note that the existence of such fuzzball microstates only became clear after the advent of string theory -- in general relativity alone (or even in Einstein-Maxwell theory), such microstates cannot be constructed. The fuzzball proposal can only hold in richer theories, as string theory.}

The fuzzball programme is most successful for solutions in five non-compact \spacetime{} di\-men\-sions.
It is of interest to understand black hole mi\-cro\-states in \textit{four} dimensions.  In this context, multi-center black holes have been shown to play an important role. Multi-center black holes arise naturally in four-dimensional supergravity as bound states of several black holes, when the gravitational attraction is exactly cancelled by the repulsion due to (generalized) electromagnetic forces. First, the five-dimensional analogs of these four-dimensional multi-center solutions play a role in the fuzzball proposal, see for instance the work summarized in the review \cite{Bena:2007kg}. Second, there are arguments that multi-center configurations in four dimensions can explain the entropy of ordinary, single-center black holes. \textit{The related research in this thesis follows this line and investigates the role of multi-center solutions in understanding the nature of black hole entropy and the role of multi-center configurations as microstates  of black holes in the supergravity regime.}


\subsection*{Chapter overview}

This thesis is divided into four parts. Part \ref{pt:intro} gives more background on black holes in general relativity and string theory. The two middle parts (part \ref{pt:FirstOrder} and part \ref{pt:Entropy}) discuss the original work of this doctorate and the specific background material needed to understand it. Finally, part \ref{pt:Conc} contains the final conclusions and appendices, including the Dutch summary. In detail, we have:
\begin{itemize}
\item[\ra] \textbf{Part \ref{pt:intro}}: \emph{\underline{Black holes as a playground}}\\
This part is aimed at readers with no detailed knowledge of general relativity, nor of string theory. Researchers in these fields can skip this part.

Chapter \ref{c:BH_Playground} treats black holes in general relativity and the appearance of Hawking radiation, black hole entropy and the issues it causes in some more detail. Special emphasis is laid on the concept of extremal black holes (which are stable, have vanishing Hawking temperature and do not radiatiate) and non-extremal black holes (which have a finite temperature and radiate). 

The purpose of chapter \ref{c:IS} is to show how black holes in four-dimensional (super)gravity fit into the framework of string theory. 
The microscopic derivation of the entropy for supersymmetric black holes is discussed and the relation between extremal and supersymmetric solutions is clarified.

\item[\ra]\textbf{Part \ref{pt:FirstOrder}}: \emph{\underline{A simplified description for non-supersymmetric black holes}}\\
This part contains contributions in the research field of first-order formalisms for black holes and other (super)gravity solutions. The systems under study are classical extensions of general relativity inspired by string theory, containing gauge fields and scalar fields. We look at the properties of the equations of motion for the scalar fields.

Chapter \ref{c:BJ} is built around the work \cite{Janssen:2007rc}.\footnote{Work I contributed to that has led to publications, will be referred as \cite{VanProeyen:2007pe}--\cite{Raeymaekers:2009ij}, these citations correspond to the publication list that is given after the bibliography.} It starts with a review of the appearance of first-order equations for supersymmetric solutions, in particular for black holes. 
Then follows a discussion of our work  \cite{Janssen:2007rc}. 
We investigate a general class of theories  including gravity, a Maxwell field and a dilaton in arbitrary dimension greater than
three. We  show that timelike and spacelike $p$-brane solutions (not only black holes), can be derived from first-order equations, as
opposed to the second-order differential equations one would normally expect. The novelty here
is that the rewriting in terms of first-order differential equations is not restricted to supersymmetry or extremality,
but applies to a generic $p$-brane \ansatz{}.

Chapter \ref{c:JP-Gradient_Flow} discusses the work in \cite{Perz:2008kh}. It gives a systematic study of first-order equations for non-supersymmetric solutions.  In particular, we study first-order flows for the scalar fields of black hole solutions and concentrate on the possibility of finding an existence criterion for a so-called `fake superpotential', a function of the scalars in the theory, which generalizes the role the central charge plays for supersymmetric solutions. The gradient of the fake superpotential determines the radial evolution of the scalars and the metric warp factor. We  illustrate the criterion in several examples. For computational simplicity, we focus on supergravities where the scalars parameterize a symmetric space. 

\item[\ra] \textbf{Part \ref{pt:Entropy}}: \emph{\underline{Entropy in supergravity: a search for microstates}}\\
In this part of the thesis, the construction of supergravity solutions that are the classical counterparts of black hole microstates is considered. The focus is on using multi-centered configurations to construct such microstates.

Chapter \ref{c:FB} gives an overview of the fuzzball proposal, an approach which has had most success in five dimensions, and the use of multi-center configurations to explain black hole microstates in four dimensions.

In chapter \ref{c:JR}, the work of \cite{Raeymaekers:2008gk} is summarized, which relates certain fuzzball solutions in five dimensions to configurations of multi-center black holes in four dimensions. 

Chapter \ref{c:TL} reviews the work of \cite{Levi:2009az}, which tried to understand the entropy of a four-dimensional black hole made from D-branes (namely the D0-D4 black hole). In earlier work, a deconstruction of the D0-D4 black hole was proposed in terms of a certain multi-center system. 
Each center of the multi-center configuration has zero entropy and therefore the resulting configuration is one entropyless `microstate' for the D0-D4 black hole. Two different ways of counting the microscopic entropy using this multi-center realization disagree, however (see \cite{Denef:2007yt} and \cite{deBoer:2009un}). 
The original aim of the work \cite{Levi:2009az} was to settle the issue. 
The result was not conclusive, it could not answer the entropy question. However, it was still interesting in its own right, showing that M-theory admits a compactification to the three-dimensional G\"odel universe.

Chapter \ref{c:TL} discusses the related work of \cite{Raeymaekers:2009ij}, where we reconsidered the three-dimensional G\"odel universe. The G\"odel universe is a \spacetime{} with closed timelike curves. However, the presence of closed timelike curves leads to problems as causality violation. 
By embedding G\"odel space into an asymptotically Anti-de Sitter \spacetime{}, we use the AdS/CFT correspondence to show that when closed timelike curves are present in  G\"odel space, unitarity is violated in the dual CFT and vice versa. This leads to a quantum mechanical argument for causality protection.
\item[\ra] \textbf{Part \ref{pt:Conc}}: \emph{\underline{Conclusions and Appendices}}\\
We end the thesis with a conclusion in chapter \ref{c:Conclusions}. Appendix \ref{app:SS} gives some technical background to clarify calculations in chapters \ref{c:JP-Gradient_Flow}.
\end{itemize}

%
%

\cleardoublepage

\part{Black holes as a playground}
\label{pt:intro}
\cleardoublepage
\chapter{
Introducing black holes\label{c:BH_Playground}}

\punchline{
This chapter gives a short review of black holes in general relativity, for readers who have some idea about general relativity, but lack a  detailed knowledge of the theory. The aim is to give a quick refreshing of the most important ideas and to introduce the concepts of black holes and the Hawking temperature and Bekenstein-Hawking entropy that can be associated to them. Section \ref{s:BHin-GR} gives a very brief sketch of general relativity, section \ref{s:BHin-BH} treats the Schwarzschild and \RN black holes. In section \ref{s:BHsemiclass}, the relation between black hole mechanics and thermodynamics is discussed and section \ref{s:BH-infoparadox} gives a conclusion.
}

\section{General relativity}\label{s:BHin-GR}%
In general relativity, the concepts of absolute space and time are abandoned. Instead, space and time are intricately related to the matter distribution in the universe. We can break this up into two steps. First, time and space are treated on an equal footing in a covariant framework. This is opposed to Newtonian mechanics, where time has a privileged role as being an outside parameter measuring the evolution of the system under consideration. 
Space-time is described as a four-dimensional manifold endowed with a metric, measuring distances between points in four-dimensional \spacetime{}. The peculiarity is that this metric is not just the usual one describing Euclidean geometry: we are not dealing with a four-dimensional Euclidean space! On the contrary, the metric is not even positive definite and has Lorentzian  signature $(-,+,+,+)$. Choosing coordinates $x^\mu,\mu=0\ldots3$, the metric can be written symbolically as
\begin{equation}
\de s^2 = g_{\mu\nu}(x)\de x^\mu \de x^\nu
\end{equation}
The \textit{second step} concerns the question:
``What happens if we were to consider matter in an otherwise empty \spacetime{}?'' The metric, with components $g_{\mu\nu}$, is central in this discussion. First, since it governs the geometry of \spacetime{}, it tells matter how to move. This is not really shocking. But second, the theory of general relativity also shows us that matter tells \spacetime{} how to curve, which may come as a surprise. We conclude that the details of a \spacetime{} (encoded in the metric $g$) are determined by the type of matter content and vice versa. 


In general relativity, the relation between the geometry of \spacetime{} and its matter content is given by the \textit{Einstein equations}. These are tensorial equations relating the curvature of the four-dimensional \spacetime{} manifold, to the matter content under consideration.
The Einstein equations are written down as:
\begin{equation}
R_{\mu\nu} - \frac12 R g_{\mu\nu} = 8\pi G_4 T_{\mu\nu}\,.
\end{equation}
The left-hand side is given in terms of the metric, and contains the Ricci tensor $R_{\mu\nu}$ and Ricci scalar $R$, who are built up out of contraction of the Riemann curvature tensor with the metric. On the right-hand side, we recognize the energy-momentum tensor $T_{\mu\nu}$, which is in general a two-derivative expression containing the matter fields of the configuration one is studying. Finally, $G_4$ is Newton's constant in four dimensions, it determines the strength of the gravitational coupling. The Einstein equations are second-order partial differential equations and can be used to find the solutions for the metric corresponding to a given energy-momentum tensor. For instance, in case of the vacuum ($T_{\mu\nu} = 0$), we  find flat space as a solution. Strangely enough, as Schwarzschild showed in 1915, there is also a spherically symmetric black hole solution, describing empty space outside a concentration of mass located in a point.

We continue with a discussion of black hole solutions to Einstein's equations, first in vacuum $T_{\mu\nu}=0$, and then in general relativity coupled to the electromagnetic field ($T_{\mu\nu}$ is then the energy-momentum tensor of electrodynamics). These black holes serve as prototypes for the black holes we study in string theory.



\section{Black holes in general relativity}\label{s:BHin-BH}

In this section we give an overview of black hole solutions in general relativity. We pick out two cases as guiding examples for the future, namely the Schwarzschild solution and the \RN solution. The first one is the unique static spherically symmetric black hole solution in vacuum (i.e.\ no angular momentum) and will be used to obtain some intuition about black holes. The second one is the unique static solution with electric or magnetic charge and forms a useful toy model for more complicated charged black holes in string theory.  We will not pay as much detail to the concept of rotating black holes.

\subsection{Vacuum solution\label{ss:BHvacuum}: Schwarzschild}
A few months after the publication of Einstein's theory of general relativity, German physicist Karl Schwarzschild surprised the physics community with a publication of an exact non-trivial (i.e.~not Minkowski) solution to Einstein's equation in vacuum. The Schwarzschild metric describes the metric outside of a spherically symmetric body with a total mass $M$. It is a solution to the Einstein equations in vacuum, $R_{\mu\nu} - \frac12 \sqrt{-g} R=0$, since there is no energy-momentum outside of the body. If we denote the radial coordinate as $r$, 
 the Schwarzschild metric is
\begin{equation}
\de s^2 = -(1-\frac{r_s}r)\de t^2 + (1-\frac{r_s}r)^{-1}\de r^2 + r^2 (\de \theta^2 + \sin^2 \theta )\de \phi^2\,,\label{eq:BHgen-Schwarzschild_Metric}
\end{equation}
The time coordinate is denoted $t$, while the spatial part of the metric is sliced in spheres of radius $r$, with a standard metric on the two-sphere in terms of angular coordinates $\theta$ and $\phi$. The Schwarzschild radius $r_s$ is determined by the mass of the spherically symmetric object and is given by:
\begin{equation}
r_s=\frac{2 G_4 M}{c^2}\,,
\end{equation}
with $M$ the total mass (we put in the factor of $c^2$ for completeness).
To understand the relation with gravitational physics in classical mechanics, take a look at large distance $r\to \infty$. First, the Schwarzschild metric approaches that of flat Minkowski space arbitrarily closely: $g_{\mu\nu}\sim \diag (-1,1,1,1)$. Following a standard prescription, e.g.~by examining the motion of a test particle moving around in the Schwarzschild metric at large $r$, at a speed that is low compared to the speed of light $c$, one finds the particle behaves as in classical mechanics, under influence of the Newtonian gravitational potential $V(r)$, given through the asymptotic time-time component of the metric as $g_{tt}|_{r\to \infty} = -(1 + 2 V(r)/c^2)$. One has
$V(r) = -\frac{G_4 M}r\,,$
which justifies the picture of the Schwarzschild metric as the relativistic description of a spherically symmetric body with mass $M$.

The Schwarzschild metric has two apparent singularities. At $r = r_s$ and $r=0$, components of the metric blow up. However, when describing the metric outside of a spherically symmetric body, one does not need to consider these cases. The Schwarzschild metric is an accurate description \textit{outside} of the body and for real-life systems, like stars, the outer radius of the body is larger than the Schwarzschild radius. Take for example our sun: the radius of the sun measures $7\times 10^{8}\,m$, while the Schwarzschild radius is about $r_s = 3\times 10^{3}\,m$!

However, we could imagine taking the Schwarzschild metric describing a geometry in the entire region $r\geq0$. What does the Schwarzschild metric then describe? First, we need to consider the radii $r=0,~r=r_s$: 
\begin{itemize}
\item[\ra] At $r=0$ real trouble appears. This can be seen by calculating scalar invariants from the (curvature of the) metric. For instance, the square of the Riemann tensor blows up at $r\to0$ as:
\begin{equation}
 R_{\mu\nu\rho\sigma}R^{\mu\nu\rho\sigma} = 12 \frac {r_s^2}{r^6}\,.
\end{equation}
The point $r=0$ corresponds to a physical singularity, the curvature of \spacetime{} becomes infinite.
\item[\ra] It turns out that $r=r_s$, is just a singularity of the coordinate system we use to describe the solution. This can be seen for instance by calculating the curvature invariants and noticing that all of these are well-behaved at the Schwarzschild radius.
\end{itemize}
The solution \eqref{eq:BHgen-Schwarzschild_Metric} is called a black hole and describes a singularity in \spacetime{} where the curvature blows up, which is not so surprising, as we have packed all mass into one point. Any celestial body (e.g.~a very massive star) that shrinks (due to gravitational attraction) to a radius $r<r_s$, will eventually collapse into such a black hole. The locus  $r=r_s$ is an \textit{event horizon}: all particles, including light, falling into the region behind the event horizon, can never re-emerge (hence the name `black' hole).  

\subsection{Solution in electromagnetism: \RN }\label{ss:BHelectromagn}
We now consider black holes with electric and magnetic charge. They are found by solving the Einstein equations with an electromagnetic radiation source:
\begin{equation}
 R_{\mu\nu} - \frac12\sqrt{-g} R =  8\pi G_4 T_{\mu\nu}^{(em)}\,.
\end{equation}
Writing electromagnetic fields through a four-vector potential $A_\mu$, we introduce the field strength $F_{\mu\nu} = \partial_\mu A_\nu - \partial_\nu A_\mu$.\footnote{The components of the electric and magnetic field $\vec E, \vec B$ are related to the field strength as
$F_{0i} = -E_i\,, F_{ij} = \sqrt{-g} \epsilon_{ijk} B_k$, where $i,j,k$ are indices ranging over the spatial components and $\epsilon$ is the antisymmetric symbol with $\epsilon_{123}=1$.
} We then have  $T_{\mu\nu}^{(em)}= F_{\mu\rho}F_\nu{}^{\rho} - \frac14 F_{\rho\sigma}F^{\rho\sigma}$. The Einstein equations follow as Euler-Lagrange equations for the metric, from the \textit{Einstein-Maxwell action}\index{Einstein-Maxwell|defn}:
\begin{equation}
S = \int\de^4 x \sqrt{-g}\,\left(\frac1{16\pi G_4} R - \frac 14 F_{\mu\nu}F^{\mu\nu}\right)\,,\label{eq:BH-EM_action}
\end{equation}
where $g$ is the determinant of the metric $g_{\mu\nu}$.

\subsubsection{\RN solution}
Analogously to the Schwarzschild metric, we can write down a solution to the Einstein-Maxwell system, describing the metric of a spherically symmetric mass distribution with total mass $M$, electric charge $\QE$ and magnetic charge $\QM$. This solution is named after the two people who were first in writing it down as the \RN solution. One finds the metric is given by:
\begin{equation}
 \de s^2 = -\Delta(r)\de t^2 + \Delta(r)^{-1}\de r^2 + r^2 \de \Omega\,,\quad \Delta(r) = 1 - \frac {2G_4 M}r + \frac {G_4 Q^2}{r^2}\,,\label{eq:BJ-RN_metric}
\end{equation}
where $Q$ is the total charge of the solution, given in terms of the electric and magnetic charges as $Q^2 = \QE^2 + \QM^2$. Spherical symmetry puts severe restrictions on the field strength as well. Only the radial components $E_r,B_r$ of the electric and magnetic fields are non-zero and give the following non-vanishing components of the field strength:
\begin{align} 
F_{rt} = \frac \QE{r^2}\,,\qquad F_{\theta\phi} = \QM \sin \theta\,.
\end{align}
{}For vanishing charges ($Q=\QE=\QM = 0$), we recover the Schwarzschild solution. 

As in the Schwarzschild solution, the point $r = 0$ describes a physical singularity, where the curvature of \spacetime{} and the field strength blow up.\footnote{This is understood through the Einstein equations and equations of motion for the gauge field. The source is a massive, charged point: a delta-function source for the gravitational field and a delta-function charge density sourcing the gauge field.} Again, there can be additional metric singularities of the metric, when $g_{tt} = \Delta(r)$ is zero. These additional singularities are no physical singularities, but denote shortcomings of the coordinate system. This happens for $r=r_\pm$, with:
\begin{equation}
 r_\pm = G_4 M \pm\sqrt{(G_4 M)^2 - G_4 Q^2}\,.
\end{equation}
Again, $r=r_\pm$ correspond to horizons. We have three regimes:
\begin{itemize}
\item $G_4 M^2 < Q^2$. Now $\Delta(r)$ has no real zeroes ($r_\pm$ become imaginary). No horizons form and the singularity at $r=0$ is 
unshielded from the rest of \spacetime{}. The singularity is called a `naked singularity'. On physical grounds, naked singularities are rejected by most physicists. In order to be able to exclude these solutions, Penrose stated the \emph{cosmic censorship conjecture} \cite{Penrose:1969pc}, saying that naked singularities are never formed. In the next chapter, we see how cosmic censorship is often guaranteed in extensions of general relativity with supersymmetry.
\item $G_4 M^2 > Q^2$: In this case $\Delta(r)$ has two real zeroes $r_+>r_- >0$, corresponding to an inner and an outer horizon. The outer horizon $r=r_+$ is an event horizon. Nothing can escape from the region of \spacetime{} hidden by the event horizon. 
\item $G_4 M^2 = Q^2$: $\Delta(r)$ has only one zero and the two horizons coincide. The black hole has the maximal amount of charge for its mass $M$ that is allowed by the cosmic censorship conjecture. The black hole is therefore called \textit{extremal}.  Note that this solution will not occur in nature, as this requires the black hole to have an enormous amount of charge. For instance, a one-solar mass extremal black hole would have a charge of more than $10^{20}$ Coulomb.
\end{itemize}

In conclusion, under the cosmic censorship conjecture, we consider two kinds of solutions: the \textit{extremal} \RN solution ($\sqrt{G_4} M = Q$, which is hard to obtain in nature) and the \textit{non-extremal} \RN solution ($\sqrt{G_4} M > Q$, which covers the class of stationary, non-rotating astrophysical black holes). 

In string theory applications, one has mostly studied extremal black holes, as these are easiest to study. For later reference, we write down the extremal \RN  solution. In the coordinate system as above, it reads:
\begin{equation}
\de s^2 = -(1-\frac{\sqrt{G_4}Q}r)^2\de t^2 + (1-\frac{\sqrt{G_4}Q}r)^{-2}\de r^2 + r^2 \de \Omega\,.
\end{equation}
In following chapters, we always discuss spherically symmetric, charged black holes in another coordinate system, that is widely used in supergravity literature. Perform first the coordinate transformation $\tilde r = r- r_0$, where $r_0 = \sqrt{G_4} Q$ denotes the location of the horizon. In terms of this new radial coordinate, and dropping again the tilde, the metric can be written as
\begin{equation}
 \de s^2 = -\e^{2U(r)} \de t^2 + \e^{-2U(r)} \left(\de r^2+ r^2(\de \theta^2 + \sin^2 \de \phi^2)\right)\,, \label{eq:BHgr-RN_Extremal}
\end{equation}
where the positive factor $\e^{2U}$ is 
\begin{equation}
 \e^{-2U} = (1 + \frac {\sqrt{G_4} Q}r)^2\,.
\end{equation}

\subsection{Most general solution and black hole uniqueness}
Above we have presented spherically symmetric solutions. We can also consider a system that rotates, or think about what the most general stationary black hole solution is. 
This is a justified question, as we expect spherical collapse of an object to result in a black hole, which should reach an equilibrium state asymptotically in  time. Properties of equilibrium states were intensely studied in the 60s, 70s and also 80s of the last century. For a (partial) overview, see the book by Hawking and Ellis \cite{Hawking:1973uf}. In that period, it was proven that the most general, stationary (equilibrium) solution to general relativity coupled to an electromagnetic field is the Kerr-Newman solution. It is a generalization of the \RN solution, depending on one extra parameter, the momentum $J$, describing a stationary rotating black hole. We do not give its form  here, as we are almost exclusively interested in non-rotating, spherically symmetric black holes in this thesis.

We see that a stationary black hole is fully determined by three parameters: the mass $M$, the total charge $Q$ and the angular momentum $J$. This result is the product of a series of black hole uniqueness theorems, initiated by Israel in 1967 \cite{Israel:1967za,Israel:1967wq} and completed by many others. In the years 1967-1975 black hole uniqueness was investigated for `vacuum' configurations ($Q=0$), building on \cite{Israel:1967za},  by Carter, Hawking and Robinson in independent works \cite{Carter:1971zc, Hawking:1971vc,Robinson:1975bv} (the only pure vacuum equilibrium states are those of the Kerr family, see table \ref{tab:BHs_Charges}). Later it was shown that the most general stationary black hole solutions with charge are those of the Kerr-Newman family, by Mazur \cite{Mazur:1982db} and independently by Bunting \cite{Bunting:1983}, see also \cite{Carter:1985}. (For a more extensive list of references regarding black hole uniqueness, see \cite{Mazur:2000pn}.)

Black hole uniqueness for stationary solutions is linked to  the `no-hair theorem', first put forth as a conjecture by Wheeler in 1971 \cite{Ruffini:1971}: a black hole carries no hair, in the sense that there are no extra parameters (for instance, dipole, quadrupole moments of the charge distribution) determining the solution. Moreover, the no hair theorem states that gravitational collapse of a celestial body should result in a black hole with exactly the three parameters $M,Q,J$ (all other properties of the collapsing body cannot be extracted from the resulting black hole). 

We give an overview in table \ref{tab:BHs_Charges}.
\begin{table}[ht!]
\centering{
\begin{tabular}{|l|l|l|l|}
\hline
&$M$ & $J$ & $Q$\\
\hline
Schwarzschild& \tickb{} &&\\
Kerr &\tickb & \tickb & \\
Reissner-Nordstrom&\tickb&&\tickb\\
Kerr-Newman&\tickb&\tickb&\tickb\\
\hline
\end{tabular}
}
\caption[Black holes in Einstein-Maxwell in terms of $(M,Q,J)$.]{Overview of black hole solutions to general relativity (3+1 dimensions) with an electromagnetic field, in terms of $(M,Q,J)$. The several black holes are named after their discoverers. A tick \tickb 
denotes that the corresponding parameter is non-zero.}
\label{tab:BHs_Charges}
\end{table}

\section{Black holes at the semiclassical level\label{s:BHsemiclass}}

\subsection{A suggestive parallel}
In the beginning of the seventies of the last century, Hawking proved that, at least in classical general relativity, the total area of the black hole horizon(s) can never decrease. We write $\Delta A \geq 0$.
For instance, an evolving black hole's horizon never shrinks, or the total area of the event horizons in a process of two black holes colliding grows larger. At that time, authors (including Hawking) noticed the parallel with the second law of thermodynamics, stating that entropy of a closed system never decreases $(\Delta S \geq 0)$. This led Bekenstein \cite{Bekenstein:1972tm} to conjecture through a series of thought experiments, that the area of the event horizon really \textit{is} an entropy, for some proportionality factor $\alpha$:\footnote{Intuitively, one can understand that for an outside observer, all classical information of collapsing matter is lost. One only has access to the black hole horizon, whose area gives an idea about the lack of information, measured through entropy.}
\begin{equation}
 S_{BH} = \alpha \frac{k_B}{l_P^2} A = \alpha \frac{k_B c^3}{G_4 \hbar} A
\end{equation}
The subscript `BH' stands for `Bekenstein-Hawking' (and not `black hole'), the Boltzmann constant $k_B$ and the Planck length $l_P$ are included to obtain the correct dimensions. In the following, we will put $c=\hbar=k_B=1$, so that $S_{BH} = \alpha \frac{A}{G_4}$.
\ctable[
cap = The four laws of thermodynamics and black hole mechanics.,
caption = The parallel between the laws of thermodynamics and the laws of black hole mechanics for the Schwarzschild black hole.\label{tab:BHgen-BH_Thermo}
,
label = width,
width = 120mm,
pos = ht!,
center
]{ c>{\small}X>{\small}X}{
}{ \FL
\multicolumn{3}{c}{4 Laws in Comparison} \ML
\textbf{Law}& \textbf{Thermodynamics}& \textbf{Black hole mechanics (G.R.)} \NN
\cmidrule(r){1-1}\cmidrule(rl){2-2}\cmidrule(l){3-3}
0$^{th}$ law&For a system in equilibrium, temperature $T$ is constant&On the black hole  horizon, the \emph{surface gravity} $\kappa$ is constant\NN[2mm]
\cmidrule(rl){2-2}\cmidrule(l){3-3}
1$^{st}$ law&$\boxed{\de E = T \de S}$&$\boxed{\de M = \frac1{8\pi}\kappa \de A}$\NN
&(energy is conserved)&(mass is conserved)\NN[2mm]
\cmidrule(rl){2-2}\cmidrule(l){3-3}
2$^{nd}$ law&$\boxed{\de S \geq0}$&$\boxed{\de A\geq 0}$\NN
&(The entropy of a closed system never decreases)
& (The area of the horizon a black hole system never decreases) \NN[2mm]
\cmidrule(rl){2-2}\cmidrule(l){3-3}
3$^{rd}$ law&Zero temperature $T=0$ is never reached in a physical process&It is not possible to form a black hole with vanishing surface gravity ($\kappa=0$)\LL
}

Hawking and others did not agree with this identification. After all, systems that have an entropy should have a temperature and associated (thermal) radiation. But nothing comes out of a black hole in general relativity: a black hole is black, right? According to Hawking, the area-entropy identification had to be seen as merely an analogy. Soon after, Bardeen, Carter and Hawking \cite{Bardeen:1973gs} put the analogy between black hole mechanics in general relativity and thermodynamics on a firmer footing, by proving the so-called `\textit{four laws of black hole mechanics}' which hold for any black hole \spacetime{}. 
This suggests that not only the horizon area behaves as an entropy, $S_{BH} = \alpha A/G_4$, but that also the surface gravity $\kappa$ of the black hole at the horizon behaves as a temperature $T_H = \frac{1}{8\pi \alpha}\kappa$, called Hawking temperature.\footnote{The surface gravity at the horizon is defined as the acceleration needed to keep a point particle at the horizon (thus opposing the gravitational attraction), when exerted from infinity. (Imagine the thought experiment of lowering the particle from spatial infinity on an infinitely long, massless wire up to the black hole horizon and pulling on the wire exactly hard enough to keep it in place).} See table \ref{tab:BHgen-BH_Thermo}.

At this point, the parallel with thermodynamics is merely an analogy, since black holes do not exhibit any `real' thermodynamic properties, as radiation. 

\subsection{More than an analogy}
Hawking lifted the analogy with thermodynamics to a real identification, by considering a semiclassical treatment of black holes in general relativity. In a semiclassical treatment of gravity, one treats matter quantum mechanically, but the gravitational field (the metric) is treated classically on the level of general relativity. In particular, matter is represented through a quantum field theory in a curved background (as opposed to ordinary quantum field theory, which considers quantum fields in a Minkowski background). A good analogy to keep in mind is that of electrodynamics where one does not consider quantization of the electromagnetic field $A_\mu$, such that `quantum' electrons move around in a `classical' background.

In the semiclassical analysis, one finds that black holes emit radiation. Intuitively, this can be understood through quantum mechanical pair creation of the vacuum just outside the black hole horizon: one virtual particle of the pair falls into the black hole, while the other escapes to infinity. One finds the black hole acts as a perfect black body, radiating  with temperature 
\begin{equation}
 T_H  = \frac1 {2\pi}\kappa\,,
\end{equation}
where $\kappa$ is the surface gravity at the horizon. This fixes the constant introduced above to $\alpha=1/4$. The Bekenstein-Hawking entropy of the black hole is then:
\begin{equation}
 S_{BH} = \frac{A}{4 G_4}\,.
\end{equation}
For the Schwarzschild black hole, we have, reinstating all constants, $S_{BH} =  2\pi k_B M^2/M_P^2 = 2\pi k_B \frac{G_4}{\hbar c} M^2$ (where $M_P$ is the Planck mass) and $T_H = \frac{\hbar c^3}{k_B} \frac 1{8 \pi G_4 M}$, see also the example below.
The expelled particles making up the radiation carry energy and one sees the black hole gradually loses mass. For astrophysical black holes, this effect is negligible, but it is most certainly there. The attentive reader may have noticed an inconsistency with the interpretation of the horizon area as an entropy: as the black hole loses mass, the horizon area will shrink, leading to a seeming violation of the second law of thermodynamics. The solution is that one has to consider the total change of entropy as the sum of the Bekenstein-Hawking entropy of the black hole and the entropy of the world outside of the black hole, in a \textit{generalized second law of thermodynamics}\index{second law of thermodynamics!generalized}:
\begin{equation}
 \Delta S_{tot} = \Delta S_{BH} + \Delta S_{outside}\,.
\end{equation}
In particular, the thermal radiation carries an entropy, such that the sum of the black hole and radiation entropy is never decreasing.

\paragraph{Example: the \RN black hole.}
For the \RN black hole, one finds that the Hawking temperature and entropy are given as (we put again $k_B = c = 1$):
\begin{equation}
S_{BH} = \frac{\pi r_+^2}{G_4}\,,\qquad T = \frac{r_+ - r_-}{4\pi r_+^2}\,.
\end{equation}
For an \emph{extremal} \RN black hole ($r_\pm= G_4^2M^2 = G_4 Q^2$), this is
\begin{equation}
S_{BH} = \pi G_4 M^2 = \pi Q^2\,,\qquad T=0\,.\label{eq:BH-RN_SBH_extremal}
\end{equation}
Note that the entropy scales as charge (or mass) squared, something we will encounter in following chapters for more intricate extremal charged black holes as well. An extremal black hole thus has zero temperature, and is unobtainable in a physical process. This links to the observation before, that extremal black holes are not realized in nature, as they carry an enormous amount of charge. All real-life black holes are non-extremal.

\section{Lessons and open questions\label{s:BH-infoparadox}}
\paragraph{Issues.} The identification of a black hole as a perfect black body with a temperature and entropy leads to several questions and problems.
\begin{itemize}
\item[\ra] \emph{\underline{Nature of Bekenstein-Hawking entropy:}} What is the meaning of the entropy? In thermodynamics, the macroscopic Clausius entropy $S$ can be related to microscopic degrees of freedom through statistical mechanics as:
\begin{equation}
 S = k_B \ln \Omega\,,
\end{equation}
where $\Omega$ is the number of microstates corresponding to the observed thermodynamic microstate. Is there a similar meaning one can associate to the `macroscopic' Bekenstein-Hawking entropy? In view of general relativity, or even its semiclassical treatment, such an interpretation is impossible: for any given value of the parameters $M,Q,J$, there is exactly \emph{one} black hole. (In the context of Einstein Maxwell theory, we cannot associate a `geometric' entropy to the black hole, that is found by having a certain number of geometries with the same macroscopic parameters, i.e.\ asymptotic charges, as the black hole that can account for the entropy. This is due to the black hole uniqueness theorems, see earlier. In extensions of general relativity with more electromagnetic fields, such as string theory and supergravity, such geometries can be constructed and can (hopefully) account for the entropy of the black hole. This is the subject of the research presented in Part \ref{pt:Entropy}.)
\item[\ra] \emph{\underline{Information paradox:}} (Discussion based on \cite{Giddings:1995gd,Harvey:1992xk}) . The Hawking radiation process seems to allow for the possibility that quantum mechanical pure states evolve into mixed states. This would violate unitarity, since in a unitary quantum theory, pure states evolve into pure states.\footnote{ In density matrix language, a pure state $|\psi\rangle$ is described as $\rho = |\psi\rangle \langle \psi|$. Under unitary evolution, $|\psi(t)\rangle = U |\psi\rangle  = \e^{i H t/\hbar}|\psi\rangle$ and $\rho$ remains pure, it can never be in a mixed form $\rho = \sum_i|\psi_i\rangle\langle\psi_i|$.} Say we start with a pure state of infalling matter $|\psi\rangle$ which collapses and forms a black hole.  In density matrix language, we have $\rho = |\psi\rangle\langle\psi|$.
Once the black hole forms, it emits Hawking radiation, through a pair production process. One member of the pair flies off and is correlated with the other member that is drawn into the black hole. This process continues until the black hole has completely evaporated. 
%
However,  in Hawking's original interpretation of the evaporation process, the infalling particles of each pair are destroyed in the singularity. This means that, if the black hole evaporates away, we are left with only radiation quanta. They are in a highly entangled state, but we no longer see any other states they are entangled with: the density matrix of the radiation is in a mixed state of the form $\rho = \sum_i p_i |\psi_i\rangle\langle\psi_i|$. Thus the most direct interpretation of black hole evaporation through Hawking radiation leads to the violation of unitarity in quantum mechanics, as pure states can evolve into mixed states.
The \textit{name} information paradox derives from the consequential information loss, measured through an entropy difference of $\Delta S \sim M^2$, where $S = \tr(-\rho \ln \rho)$ is the (Shannon) entropy of the state. Before the gravitational collapse, $S = -\tr(\rho \ln \rho) = 0$, while the entropy of the end state is of the order of the Bekenstein-Hawking entropy ($S\sim S_{BH}\sim M^2$). Alternatively, this entropy $S\sim M^2$ is obtained by noting that the Hawking radiation one is left with is \textit{thermal}. 

\end{itemize}
One expects these issues to be dealt with in a consistent theory of quantum gravity. Remember that Hawking's calculations revealing black body radiation were performed in a semi-classical treatment of the black hole. Because the black hole horizon region is a region of low curvature (at least for large black holes), this approach is a good approximation of a full quantum gravity theory, but still an approximation. Any quantum gravity theory should provide an answer to the two questions raised above: we can see those questions	 as a theoretical `test' a quantum theory should pass.

The research presented in this thesis mostly deals with black holes in the context of superstring theory, as a quantum theory of gravity. In the `low energy regime' (energies low compared to the energy scale set by the string length), superstring theories are well described by (super)gravity theories. These gravity theories are extensions of Einstein-Maxwell theory. When we study black holes in the supergravity context we see that all properties and issues raised in this chapter remain: black holes in supergravity have a Hawking temperature, a Bekenstein-Hawking entropy, etc. Because of the relation to superstring theory, however, those issues are now amenable to study in the regime where the full string theory is valid and we can understand black holes from a quantum mechanical viewpoint. 

Some literature on this topic is reviewed in the next chapter, where it is also shown how the black hole entropy can be derived microscopically in string theory. Our work on black holes in supergravity theories is presented in part \ref{pt:FirstOrder} and our work on the comparison of classical black holes in supergravity and their quantum mechanical origin in string theory is given in part \ref{pt:Entropy}.

%

%
\cleardoublepage
\chapter{
Black holes in supergravity and string theory}\label{c:IS}

\punchline{This chapter is aimed at readers with no detailed knowledge of string theory. Building on the concepts introduced in the previous chapter, the main goal is to give an idea of what a typical black hole in a compactification of string theory looks like and how its Bekenstein-Hawking entropy is explained. To elucidate the main ideas, an example of a black hole made from string theory objects called D-branes is presented. The focus is on the special class of supersymmetric black holes, for which the entropy can be explained from string theory. We end with an important discussion on the difference with the more realistic (and much larger) class of non-supersymmetric black holes.}


\section{Supergravity in four dimensions}\label{s:IS-sugra4d}
Supergravity theories are supersymmetric extensions of general relativity that follow naturally from string theory. We  consider the embedding of supergravity in string theory in the next section.  In this section, we focus on black hole solutions in supergravity. 


In four dimensions, we can have supergravity theories with up to 8 types of supersymmetry. We speak of $\cN$-extended supergravity, where $\cn=1$ up to 8, depending on the theory. The more supersymmetry is present (the higher $\cn$), the more constrained the theory becomes. In general, we like theories with a low amount of supersymmetry, as there is more freedom in the possible couplings between the fields and it becomes possible to describe physics of the real world. On the other hand, when we want to discuss black hole solutions, we find that especially black hole solutions that preserve some of the supersymmetry are interesting, as these allow for a statistical derivation of the Bekenstein-Hawking entropy. We thus have to find a balance between the amount of supersymmetries $\cn$ of the theory and the possible amount of supersymmetry a solution preserves. 
Therefore, we shift our interest to $\cn=2$ supergravity in four dimensions -- the theory with the lowest amount of supersymmetry that can  have supersymmetric black hole solutions.

\subsection{\texorpdfstring{$\cn=2$} {} supergravity}
We consider the $\cN=2$ supersymmetric extension of the Einstein-Maxwell Lagrangian and show how to construct charged black holes to this extension. As a minimal input, we take the metric $g_{\mu\nu}$ and $n_V$ vector fields $A^\alpha_\mu$ with $\alpha=1\ldots n_V$. Due to the requirement of  supersymmetry, each of these fields is accompanied by fermionic and bosonic superpartners. The collection of the metric and its superpartners is called the \textit{gravity multiplet}, while each vector and its superpartners make up one \textit{vector multiplet}. For each, we have:
\begin{itemize}
\item[\ra] \textit{\underline{Gravity multiplet}}: Contains the metric $g_{\mu\nu}$ (describing excitations of a spin 2 field called `graviton'), 2 spin 3/2 fermions  (`gravitinos') and a spin 1 vector field $A_\mu^0$, called `graviphoton'.
\item[\ra] \textit{\underline{Vector multiplet}}: Each consists of 1 of the vector fields $A_\mu^\alpha$, 2 spin 1/2 fermions (`gauginos') and a complex scalar $z^\alpha$.
\end{itemize}
We only consider the bosonic content of the theory, since we are interested in \textit{classical black hole solutions}. In a classical solution, all fermions are put to zero. The bosonic part of the $\cN=2$ supergravity Lagrangian contains $n_V+1$ vectors $A^I_\mu = (A^0_\mu,A^\alpha_\mu)$   and $n_V$ complex scalars $z^\alpha,\alpha=1\ldots n_V$, coupled to the metric $g_{\mu\nu}$. Arranging the vector fields in field strengths $F_{\mu\nu}^I = \partial_\mu A_\nu^I - \partial_\nu A_\mu^I$ we have:
\begin{align}
S =  \int \rmd^4 x \sqrt{-g}&\Big{(}\frac1{16\pi G_4}R +\frac12(\Im
\mathcal{N}_{IJ} )F_{\mu\nu}^I F^{\mu\nu J}\label{eq:N-=2SUGRA}\\
&-\frac1{4\sqrt{-g}}(\Re \mathcal
N_{IJ})\veps^{\mu\nu\rho\sigma} F_{\mu\nu}^I
F_{\rho\sigma}^J - 2 g^{\mu \nu}G_{\alpha\bar
\beta}\partial_\mu z^\alpha \partial_\nu \bar z^{\bar \beta}\Big{)}\,. \nonumber
\end{align}
The first two terms on the first line are the straightforward generalization of the Einstein-Maxwell action \eqref{eq:BH-EM_action} to $n_V+1$ vector fields $A^I_\mu$ (As compared to the Einstein-Maxwell action before, we rescaled the gauge fields $A^I\to \sqrt 2 A^I$, to make sure the black hole charges are integers, see below).  The terms on the second line are required by supersymmetry and give an extra kinetic term for the vector fields and a term describing the scalars. The couplings in front of the vector field are set by a scalar dependent matrix $\cn (z,\bar z) = \Re \cn + \rmi \Im \cn$. The scalars can be seen as coordinates on a manifold with metric $G_{\alpha\bar \beta}(z,\bar z)$. Due to the constraints set by supersymmetry, the form of $\cN$ and the details of this manifold are of a special type (special K\"ahler geometry, for reviews, see e.g.~\cite{Fre:1995dw,Craps:1997gp}).  The bosonic action for the supergravity multiplet only ($n_V=0$, $\cN=-\rmi$), is the Einstein-Maxwell action \eqref{eq:BH-EM_action}, while $n_V\geq1$ gives other interesting extensions of Einstein-Maxwell theory. 

We set out for black hole solutions to this system. If we leave out the scalars from the discussion, an asymptotically flat and spherically symmetric black hole is the straightforward generalization of the \RN black hole: it would be determined by a mass $M$ and a set of $n_V+1$ electric charges $q_I$ and $n_V+1$ magnetic charges $p^I$, one for each gauge field $A^I_\mu$. However, due to the requirement of $\cN=2$ supersymmetry, the scalar fields also couple to the vector fields and metric and we have to take their effect into account. We discuss this a bit below, a rigorous discussion of the effects of the scalar fields $z^\alpha$ is the subject of part \ref{pt:FirstOrder}. 


\subsection{Supersymmetric black holes} 
We derive the form of a spherically symmetric black hole solution and more specifically its Bekenstein-Hawking entropy for \textit{supersymmetric} black holes. A detailed derivation of the equations of motion is found in chapter \ref{c:BJ}, here we concentrate on the form of the solution. 

A supersymmetric black hole is invariant under some of the supersymmetry transformations that relate bosons and fermions. It can be shown that such a black hole is also \textit{extremal} (see section \ref{s:IS-nonsusyBH} below). A spherically symmetric, extremal black hole has a metric of the form:
\begin{equation}
 -\e^{2U(r)} \de t^2 + \e^{-2U(r)}\left(\de r^2 + r^2\de \Omega^2\right)\,.
\end{equation}
(Compare to the extremal \RN solution \eqref{eq:BHgr-RN_Extremal}, for which $\e^{2U} = \left(1 - Q/r\right)^2$.) By spherical symmetry, also the  gauge fields are fully determined by the electric and magnetic charges. We do not discuss their form here, but refer to chapter \ref{c:BJ} for an in-depth treatment.
We are thus left with finding the dynamics of $U,z^\alpha$. Demanding that the black hole is left invariant by some of the supersymmetry transformations, gives the following first-order equations for the redshift factor $U(r)$ and the scalars $z^\alpha(r)$, which are most easily written in terms of the inverse radial coordinate $\tau = 1/r$:
\begin{align}
\ta U \tau &=-\e^{U} |Z(q,p;z,\bar z)|\,,\qquad \quad
\ta {z^\alpha}\tau=-\frac12\e^U G^{\alpha\bar\beta}\pf {|Z(q,p;z,\bar z)|}{z^{\bar \beta}}\,.\label{eq:IS-attr_eqs}
\end{align}
The function $Z(q,p;z,\bar z)$ is complex and is often called central charge. It is a function of on the scalar fields and the conserved vector charges. Its exact form depends on the details of the supergravity model (how many vector multiplets $n_V$, details of the geometry of the scalars).

The solution to these first-order equations can be inferred from the function $Z$. One can show that the radial dependence of the scalar fields is such that the solution for the scalars flows to definite values at the horizon (in terms of $p,q$), regardless of the values $z^\alpha = z_\infty^\alpha$ at spatial infinity. This behaviour goes by the name of attractor mechanism and the solution is called an attractor. It makes sure that the area of the event horizon depends on the charges only. In particular, one finds the area of the event horizon of the black hole solution is given in terms of the minimum of $|Z|$ defined as $Z_\star(p,q) = \min_z Z(p,q;z,\bar z)$:
\begin{equation}
 A_H = 4 \pi G_N |Z_\star(p,q)|^2\,.\label{eq:IS-Area_EventHor}
\end{equation}
We conclude the black hole has a Bekenstein-Hawking entropy:
\begin{equation}
 S_{BH}(p,q) = \pi |Z_\star(p,q)|^2
\end{equation}
In general, the function $|Z_\star(p,q)|$ is linear in the charges, such that this is the straightforward generalization of the entropy for the \RN black hole ($S_{BH}=\pi Q^2(\QE,\QM) = \pi(\QE^2 + \QM^2)$, see \eqref{eq:BH-RN_SBH_extremal}). A detailed discussion of the attractor mechanism and the Bekenstein-Hawking entropy is given in chapter \ref{c:BJ}.
%
\paragraph{Example: D0-D4 black hole.}

To make things less abstract, we give the explicit metric for one specific solution to the particular case of the so-called STU-model. This model describes $\cN=2$ supergravity with three vector multiplets ($n_V=3$). Its full action is given in chapter \ref{c:JR}, for our purposes just note that there are in total four vector fields $(A^I_\mu = A^0_\mu,A^1_\mu,A^2_\mu,A^3_\mu)$ and a black hole can thus carry four electric charges $q_0,q_1,q_2,q_3$ and four magnetic charges $p^0,p^1,p^2,p^3$. We consider the black hole with only the following four charges $q_0,p^1,p^2,p^3$, called the  D0-D4 black hole. Its name is clarified below. The metric is given as:
\begin{equation}
 \de s^2 = -(H_0 H^1 H^2 H^3)^{-1/2} \de t^2 + (H_0 H^1H^2H^3)^{1/2}\left(\de r^2 + r^2\de \Omega^2_2\right)\,,\label{eq:IS-metric-D0D4}
\end{equation}
where $\de \Omega^2_2 = \de \theta^2 + \sin^2\theta \de \phi^2$ is the metric on the unit two-sphere and $H_0,H^\alpha$ are harmonic functions 
\begin{equation}
 H_0 = 1 + \frac{\sqrt{2G_N}q_0}{r}\,,\qquad H^\alpha = 1 + \frac{\sqrt{2G_N}p^\alpha}r\,, (\alpha=1\ldots3)\,. 
\end{equation}
When we take all charges equal to $Q$, this describes the metric of the extremal \RN black hole \eqref{eq:BH-RN_SBH_extremal}. (The factors of $\sqrt 2$ are due to the specific normalization of the vector kinetic terms ($A^I\to \sqrt{2}A^I)$). The mass of the solution is read off from the asymptotic behaviour of the metric component $g_{tt}$:
\begin{equation}
 M = \frac{\sqrt{2}}{4}\left(q_0 + p^1 + p^2 + p^3\right)\,.
\end{equation}
Remember that the event horizon in these coordinates is located at $r=0$, from which we find the area of the event horizon is $A_H = 8\pi G_N \sqrt{q_0p^1p^2p^3}$, leading to a Bekenstein-Hawking entropy:
\begin{equation}
S_{BH} = 2\pi \sqrt{q_0p^1p^2p^3}\,.\label{eq:IS-SBH_D0D4}
\end{equation}
In other words, for this black hole the central charge  has the attractor value $|Z|^2_\star = 2\sqrt{q_0p^1p^2p^3}$.
We show below how this black hole is found from string theory.

\section{\texorpdfstring{$\cN=2$}{} supergravity from IIA string theory}\label{s:IS-ST}
Since all consistent string theories are related by dualities, we can concentrate on one of them to show how to obtain black hole solutions from string theory. We give the low energy description of type IIA string theory and then we show how this relates to the four-dimensional action given above.\footnote{For more information on string theory see the classic textbooks \cite{Green:1987sp,Green:1987mn,Polchinski:1998rr,Polchinski:1998rq,Becker:2007zj}.}

\paragraph{IIA supergravity action.}

We give the low energy effective action of type IIA string theory. (`Low energy ' is relative and means energies $E$ well below the scale set by the string length $E \ll 1/l_s$. The energies reached in present-day accelerators are `low' in this terminology.) In this limit, the only vibration modes of the string that are of relevance are the massless modes and are described by type IIA supergravity.
Type IIA supergravity is one of the two possible ten-dimensional supergravity theories invariant under $\cN=2$ supersymmetry, namely the one for which the two supersymmetry generators (spinors) have opposite chirality.\footnote{The other $\cN=2$ supergravity in ten dimensions is the type IIB supergravity, the low-energy limit of IIB superstring theory, which has two supersymmetry generators with the same chirality.} 
We are only concerned with the bosonic content of the theory, given by the following fields (note that we use indices $M,N\ldots$ in ten dimensional space-time):
\begin{itemize}
\item The ten-dimensional \textit{metric}, with components $G_{MN}$.
\item The \textit{dilaton} $\phi$. This is a scalar field, its vacuum expectation value sets the value of the string coupling as $g_s = \langle \e^\phi\rangle$.
\item An \textit{antisymmetric two-form} field with components $B_{MN}$.
\item \textit{Higher-form gauge fields}. These are generalizations of the Maxwell field $A_\mu$ of four dimensions. We have a one-form potential with components $C_{M}$ and a three-form with components $C_{MNP}$. 
\end{itemize}
{}From now on, we use differential form notation and write $B_{(2)}, C_{(1)}, C_{(3)}$ (for instance $C_{(1)} = C_M \de x^M$ and $B_{(2)} = \frac12 B_{MN}\de x^M \wedge \de x^N$) with associated field strengths $H_{(3)} = \de B_{(2)}$,  $F_{(2)} = \de C_{(1)}$, $F_{(4)} = \de C_{(3)}$.  All the fields above form the bosonic content of the type IIA supergravity action, which is, up to two derivatives, completely determined by supersymmetry to have the form
\begin{align}
S = \frac1{16\pi G_{10}}\int  &\de^{10} x\,\e^{-2\phi}\sqrt{-G}\Big{(} R -\frac12 |H_{(3)}|^2 - \frac12 \partial_\mu \phi \partial^\mu \phi
 \\
&- \frac 12 |F_{(2)})|^2 - \frac12 |\tilde F_{(4)}|^2\Big{)}
-\frac1{16\pi G_{10}}\int\frac12 B_{(2)}\wedge F_{(4)}\wedge F_{(4)}\,,\nonumber
\end{align}
where $G_{10}$ is Newton's constant in ten dimensions, we introduced $\tilde F_{(4)} = F_{(4)} - C_{(1)}\wedge H_{(3)}$ and we have the notation $|F_{(n)}|^2 = \frac1 {n!} F_{\mu_1\ldots \mu_n}F^{\mu_1\ldots \mu_n}$ and likewise for $|H_{(3)}|^2$. We do not give the contribution of the fermionic fields, as we are always concerned with classical solutions to the equations of motion (e.g.~black hole solutions), for which all fermions are identically zero. 


\paragraph{Compactification to four dimensions.}
To relate ten-dimensional string physics to the four-dimensional world, we assume that ten-dimensional \spacetime{} with Min\-kow\-ski signature is a direct product of four-dimensional space-time (Minkowski signature) and a six-dimensional (Euclidean) compact space:
\begin{equation}
 M_{1,9} = M_{1,3} \times M_6\,.
\end{equation}
Furthermore, we assume the six-dimensional manifold $M_6$ to have a very small volume, such that to very good approximation the world is four-dimensional. We say we have a four-dimensional compactification of string theory.

The ten-dimensional IIA theory has 32 real supersymmetries. Depending on the structure and symmetries of the internal manifold, the resulting four-dimensional theory is invariant under all, some, or none of these supersymmetries. We are intrested in four-dimensional theories with $\cN=2$ supersymmetry in four dimensions which have 8 real supercharges (see previous section). It turns out that to break 3/4 of the supersymmetry, we need to compactify on a Calabi-Yau manifold, a Ricci-flat complex manifold with $\SU (3)$ holonomy. 

\paragraph{Four-dimensional action. }The low energy effective action of a Calabi-Yau com\-pac\-tification of type IIA string theory,  gives $\cN=2$ supergravity in four dimensions. The ten-dimensional fields arrange in $\cN=2$ multiplets (see table \ref{tab:IS-Comp_Results}):
\begin{itemize}
\item \textit{Gravity multiplet}: The four-dimensional components of the metric $G_{MN}$ give rise to the four-dimensional metric $g_{\mu\nu}=G_{\mu\nu}$. The components along the four-dimensional space of $C_{(1)}$, give rise to a graviphoton $A^0_\mu=C_\mu$.
\item \textit{Vector multiplets}: These contain a set of gauge fields from $C_{(3)}$: $A^\alpha_\mu = C_{\mu i\bar j}$, where $i,\bar j$ represent directions along the Calabi-Yau and $\alpha$ numbers the $n_V$ possibilities, which depend on the topological properties of the Calabi-Yau manifold. Furthermore, the components of the metric and the antisymmetric two-tensor along the internal manifold are scalars from the four-dimensional point of view. They arrange in complex scalars $z^\alpha$.
\item \textit{Hypermultiplets}: All other field components arrange in hypermultiplets, containing scalar fields and fermions. We do not consider those here.
\end{itemize}
\begin{table}[ht!]
\centering
 \begin{tabular}{|c|ccc|}
\hline
{\bf Gravity multiplet}&$g_{\mu\nu}$&$\leftrightarrow$&$G_{\mu\nu}$\\
&$A^0_\mu$&$\leftrightarrow$&$C_\mu$\\
\hline
{\bf Vector multiplets}& $A^\alpha_\mu$&$\leftrightarrow$&$C_{\mu i\bar j}$\\
&$z^\alpha$&$\leftrightarrow$&$g_{i\bar j},B_{i\bar j}$\\
\hline
 \end{tabular}
\caption[10d origin of 4d $\cN=2$ bosonic fields.]{Ten-dimensional origin of the bosonic content of the four-dimensional $\cN=2$ action \eqref{eq:N-=2SUGRA}.\label{tab:IS-Comp_Results}}
\end{table}

In conclusion, we have a four-dimensional theory of gravity, scalars and vectors, described by four-dimensional $\cN=2$ supergravity given earlier, see \eqref{eq:N-=2SUGRA}.

\section{Four-dimensional black holes from \texorpdfstring{$p$}{}-branes}\label{s:IS-Dbranes}
Even though superstring theory was originally conceived as a theory of strings, also objects of higher dimensionality appear. We discuss these objects from the point of view of ten-dimensional IIA supergravity and show how they can form a black hole in four dimensions through compactification. 

\paragraph{Black $p$-branes.} 
Consider a charged point particle. It can act as a source for an electromagnetic potential $A_\mu$. Thus the one-form $A_{(1)} = A_\mu \de x^\mu$ couples naturally to a point particle. In the same way, a ($p+1)$-form  potential $A_{(p+1)}$ couples to an object that is stretched out in $p$ (spatial) dimensions. Such an object is called a $p$-brane. 
When the gravitational interaction is taken into account, the charged point particle can form a charged black hole and similarly we can have black $p$-branes in supergravity when $(p+1)$ forms are present.

Type IIA supergravity has several $(p+1)$-forms: $C_{(1)}$ and $C_{(3)}$.\footnote{There is also the two-form $B_{(2)}$. The 1-branes it couples to are just the fundamental strings from which the theory is built.} This suggests the existence of 0-branes (point particles) and 2-branes (surfaces). But there are more. One can show that in ten dimensions a $(p+1)$-form potential $C_{(p+1)}$ not only couples naturally to $p$-branes (through an `electric' coupling), but also to $(6-p)$-branes (through a `magnetic' coupling).  In conclusion, in the supergravity approximation, the type IIA theory has black $p$-brane solutions: electric 0-branes and 2-branes, and magnetic 4-branes and 6-branes. For reasons to become clear below, these are referred to as D0, D2, D4 and D6 branes, see table \ref{tab:IS-branes}.
\begin{table}[ht!]
\centering
 \begin{tabular}{|c|c|c|c|}
  \hline
$p$&\textbf{Source for}& \textbf{Type of coupling}&\textbf{4D Charge}\\
\hline
0&$C_{(1)}$&electric&$q_0$\\
2&$C_{(3)}$&electric&$q_\alpha$\\
4&$C_{(3)}$&magnetic&$p^\alpha$\\
6&$C_{(1)}$&magnetic&$p^0$\\
\hline 
\end{tabular}
\caption{Various $p$-branes in IIA that source $C_{(1)}, C_{(3)}$.\label{tab:IS-branes}}
\end{table}


\paragraph{Ten-dimensional origin of four-dimensional black hole.}
In the four-dimensional $\cN=2$ supergravity description, the black hole acts as a pointlike source for the gauge fields $A^I_\mu$. The ten-dimensional origin of the gauge fields $A^I_\mu$ lies in the forms $C_{(1)},C_{(3)}$,  which can have legs on the internal manifold. The configurations of $C_{(1)},C_{(3)}$ that give  rise to the gauge field $A^I_\mu$ in four dimensions are exactly of the form that they couple to $p$-branes extending along the Calabi-Yau (see table \ref{tab:IS-Comp_Results}). We conclude a four-dimensional black hole is made from black $p$-branes that extend along the internal directions.
One finds the electric and magnetic charges $(q_0,q_\alpha;p^\alpha,p^0)$ of the four-dimensional black hole are due to branes of different dimensionality. Electric charges arise through placing $q_0$ units of 0-brane charge and $q_\alpha$ units of 2-brane charge along the Calabi-Yau, while the magnetic charges are due to $p^\alpha$ units of 4 brane charge  and $p^0$ units of 6-brane charge, see table
 \ref{tab:IS-branes}.

\section{Explaining the entropy from D-branes}\label{s:IS-entropy}
The revolutionary insight by Polchinski in the mid ninetees \cite{Polchinski:1995mt}, was that the \mbox{$p$-branes} that act as (electric or magnetic) sources for the potentials $C_{(p+1)}$, are accurately described at small string coupling  ($g_s\to0$) as Dirichlet $p$-branes, or D-branes for short. These D-branes are confined on submanifolds of ten-dimensional \spacetime{}, open strings can end on them.\footnote{These D-branes provide Dirichlet boundary conditions for the open strings, hence their name.} In particular, D-branes of different dimensionality can form bound states and from the open string perturbation theory, it is in certain cases possible to count the quantum mechanical degeneracy of a state with a given fixed number of D-brane charges (fixed charges $q_I$ and $p^I$).

By quantizing the open strings, one can determine the dynamics of the D-branes. From the discussion above, we find that type IIA string theory has D0 branes, D2 branes, D4 branes and D6 branes. 
The identification as D$p$-branes, suggests  we can find an estimate for the black hole entropy by finding the number of quantum mechanical ground states that correspond to the given total D-brane charge of the four-dimensional black hole in the supergravity approximation.




However, there is an important caveat: string perturbation theory and the supergravity description as a black hole are two different pictures. In terms of a typical number of D-branes $Q$, one finds that perturbation theory is valid (and we can easily study D-brane bound states) when $g_s Q \ll 1$, while supergravity (and the black hole picture) is a good approximation only when $g_s Q\gg 1$.

If we were to study the number $N$ of quantum mechanical ground states of the D-brane system in the `microscopic' regime $g_s Q\ll 1$, there is no reason to believe this number gives the Bekenstein-Hawking entropy of the black holes as $S_{BH} = \ln N$ in the `macroscopic' supergravity regime. However, there is a class of solutions for which the result is independent of the value of the coupling $g_s Q$, namely for 	supersymmetric black solutions.\footnote{Actually, the number of supersymmetric (BPS) states with a given charge is only independent of $g_s$ when we consider large charges $q,p$. In that regime, the degeneracy of states is well approximated by an index that is guaranteed to be invariant under continuous deformations. Such an index counts the difference between bosonic and fermionic states. For an exponentially large number of states, one expects that the ground state consists mainly of either bosonic or fermionic states, such that the index is a good approximation to the total number of states.} Following that idea, Strominger and Vafa  where the first to successfully explain the entropy from a microscopic D-brane picture for five-dimensional supersymmetric black holes in  \cite{Strominger:1996kf}. Many successful identifications followed, in particular the result of Maldacena, Strominger and Witten is of prime relevance \cite{Maldacena:1997de}.  It explained  the entropy of supersymmetric black holes with D4, D2 and D0 charges in Calabi-Yau compactifications of IIA string theory, from a microscopic point of view, through the connection with eleven-dimensional M-theory. Later, the microscopic entropy counting has been further understood and put on a firmer putting from the AdS/CFT correspondence, see for instance \cite{Peet:2000hn} for a review.
So far, no successful counting has been done for general non-supersymmetric black holes. 

\paragraph{Example: D0-D4 black hole revisited.}
We show how the black hole with metric \eqref{eq:IS-metric-D0D4} of $\cN=2$ supergravity is obtained from string theory. Consider type IIA supergravity compactified on a six-torus, which we take to be a direct product of three two-tori $T_1\times T_2\times T_3$. Wrap $p^1$ D4 branes on $T_2\times T_3$, $p^2$ D4-branes on $T_1\times T_3$ and $p^3$ D4 branes on $T_1\times T_2$ and consider $q_0$ 0-branes (points) in the $T^6$. The reduction of the resulting supergravity configuration to four dimensions exactly gives rise to the black hole discussed in \eqref{eq:IS-metric-D0D4}.
Strictly speaking, a torus reduction gives a four-dimensional $\cN=8$ supergravity theory. By counting the number $N$ of quantum mechanical states of the D-brane system with total charges $(q_0,p^1,p^2,p^3)$, one finds for large charges the leading order result \cite{Maldacena:1997de}
\begin{equation}
 N = \e^{2\pi\sqrt{q_0p^1p^2p^3}}\,,
\end{equation}
which agrees with the Bekenstein-Hawking entropy \eqref{eq:IS-SBH_D0D4} as $S_{BH} = \ln N$.

\section{Non-supersymmetric black holes?}\label{s:IS-nonsusyBH}
 

Above we discussed only supersymmetric black holes, which preserve some of the supersymmetry of the $\cN=2$ supergravity theory. In this section, we comment on non-supersymmetric black holes. The main point is to clarify the relation between the condition for supersymmetry and the condition for extremality. In short, all supersymmetric black holes are extremal, while non-supersymmetric black holes can be either extremal or non-extremal, see table \ref{tab:Susy}. This distinction is important in light of the research in part \ref{pt:FirstOrder}.

\ctable[
cap = {Overview of supersymmetric, extremal and non-extremal black holes.},
caption = {Overview of general black hole solutions. Most are non-extremal, only a subset are extremal. When embedded in a supersymmetric theory, all supersymmetric black hole solutions are extremal as well. Note: there are extremal, non-supersymmetric solutions.},
label = tab:Susy,
width = 120mm,
pos = ht!,
center
]{ c>{\small}X>{\small}Y>{\small}X}{
}{ \FL
\multicolumn{4}{c}{\textit{Extremality and supersymmetry properties}} \ML
\multicolumn{3}{c}{\bf EXTREMAL}& \multicolumn{1}{c}{\textbf{\mbox{NON-EXTREMAL}} }\NN
\multicolumn{3}{c}{\textit{$T_H=0$, no radiation}}&\multicolumn{1}{c}{\textit{$T_H\neq0$, emits radiation}}\NN[2mm]
\cmidrule(rl){1-3}\cmidrule(l){4-4}
\multicolumn{1}{c}{Supersymmetric} &OR&\multicolumn{1}{c}{Non-supersymmetric}&\multicolumn{1}{c}{\mbox{Non-supersymmetric}}\LL
}
\vspace{-.03\textheight}


\paragraph{\texorpdfstring{$\cN=2$}{} supergravity and (non-)supersymmetric solutions.} 
A solution of $\cN=2$ supergravity does not have to be invariant under all $\cN=2$ supersymmetry. In terms of the amount of supersymmetry that is preserved, we distinguish three types of solutions: solutions invariant under all supersymmetry, half  or none of the supersymmetry. It turns out that a black hole necessarily breaks at least half of the supersymmetry. We thus have two types of black holes:
\begin{itemize}
\item \textit{supersymmetric} black holes preserving 1/2 of the supersymmetries 
\item \textit{non-supersymmetric} black holes, breaking all supersymmetry.
\end{itemize}
 
\paragraph{Black holes and the BPS bound.}
For black hole solutions in $\cN=2$ supergravity theories carrying electric and magnetic charges, one finds a relation between the mass $M$ and the charges $q_I,p^I$: 
\begin{equation}
M\geq \sqrt{G_N} |Z(p,q;z_\infty,\bar z_\infty)|\,.\label{eq:IS-BPS_bound}
\end{equation}
where $Z(p,q;z_\infty,\bar z_\infty)$ is the evaluation at spatial infinity of the  complex function $Z$ that appears in the attractor equations \eqref{eq:IS-attr_eqs}. In section \ref{ss:BJ-BPS_SUSY}, this relation is explained as a consequence of the superalgebra underlying the supergravity theory. In this context, this relation is known as the BPS bound.
Solutions that saturate this bound ($M= \sqrt{G_N}|Z|_\infty$), are called BPS states and leave half of the supersymmetry unbroken, states with $M> \sqrt{G_N}|Z|_\infty$ break all supersymmetry. (We use $\infty$ solely to refer to spatial infinity in this section.) We summarize in table \ref{tab:IS-Susy_BHs}.
\begin{table}[ht!]
\centering 
\begin{tabular}{|c|c|}
\hline
\multirow{2}{*}{
\bf Supersymmetric (`BPS')}&\multirow{2}{*}{$M=\sqrt{G_N} |Z(p,q;z_\infty,\bar z_\infty|$}\\[-1mm]&\\
\hline
\multirow{2}{*}{\bf Non-supersymmetric} &\multirow{2}{*}{$M>\sqrt{G_N} |Z(p,q;z_\infty,\bar z_\infty|$}\\[-1mm]
&\\
\hline
 \end{tabular}
\caption{Black hole solutions in $\cN=2$ supergravity and the BPS bound.\label{tab:IS-Susy_BHs}}
\end{table}
The appearance of the BPS bound is strikingly similar to the extremality bound of the \RN black hole. Remember from the previous chapter that, to exclude naked singularities (cosmic censorship), one has to impose 
\begin{equation}
 M \geq \sqrt{G_N} Q(\QE,\QM)\,,
\end{equation}
where $\QE$ and $\QM$ are the electric and magnetic charges of the \RN black hole and $Q = \sqrt{\QE^2 +\QM^2}$. A \RN black hole with $M = \sqrt{G_N} Q$ is extremal (Hawking temperature $T_H=0$, no radiation) and with $M > \sqrt{G_N} Q$ is non-extremal ($T_H>0$, radiates). This suggests that for  charged black hole solutions in supergravity, supersymmetric black holes are extremal and vice versa. In many instances this is true: the BPS bound often acts as a `cosmic censor', ensuring that naked singularities cannot appear. This was first shown for dilatonic black holes in $\cN=4$ supergravity in \cite{Kallosh:1992ii}, where it was detailed how the extremal dilatonic black hole is supersymmetric. However, \textit{in general, all supersymmetric solutions are extremal, but not necessarily vice versa}, see below.



\paragraph{Black holes and the extremality bound.}
We follow the discussion of \cite{Kallosh:2006bt}. A black hole solution to $\cN=2$ supergravity with mass $M$, temperature $T$ and Bekenstein-Hawking entropy $S_{BH}$, can be shown to obey the constraint:\footnote{To understand where the relation \eqref{eq:IS-ExtremalityBound2} originates, we have to jump ahead to chapter \ref{c:BJ}. There it is shown that a spherically symmetric black hole solution to $\cN=2$ supergravity obeys the constraint \eqref{eq:HamConstraintN=2}. Evaluation of this constraint at spatial infinity gives \eqref{eq:IS-ExtremalityBound2}. (To compare to the quantities $U$, $V_{BH},$ and $c^2$ in eq.~\eqref{eq:HamConstraintN=2},  note that $V_{BH} = |Z(p,q;z,\bar z)|^2 + \frac14 \left|\pf {Z(p,q;z,\bar z)}{z^\alpha}\right|^2$, $\left.\dot U\right|_\infty = -M$ and $c^2 = T_HS_{BH}$.)}
\begin{align}
 M^2 &= 2 T_HS_{BH} +  G_N |Z(p,q;z_\infty,\bar z_\infty)|^2 + \frac14 \left|\pf Z {z^\alpha}(q,p;z_\infty,\bar z_\infty)\right|^2 - \left|\ta {z^\alpha} \tau\right|^2_\infty\,.\nonumber\\
&\geq G_N |Z(p,q;z_\infty,\bar z_\infty)|^2 + \frac14 \left|\pf Z {z^\alpha}(q,p;z_\infty,\bar z_\infty)\right|^2 - \left|\ta {z^\alpha} \tau\right|^2_\infty\,.\label{eq:IS-ExtremalityBound2}
\end{align}
The two last squares on the right-hand sides are formed with the scalar metric $G_{\alpha\bar\beta}$ and the inequality follows because we have that $2T_HS_{BH}\geq0$.
We have equality only for extremal black holes $(T_H=0)$. 

When the scalars are constant, the extremality bound \eqref{eq:IS-ExtremalityBound2} coincides with the BPS bound \eqref{eq:IS-BPS_bound} (e.g.~for the \RN black hole) and supersymmetric black holes are extremal ($T_H=0$) and vice versa. However, for non-trivial scalars $z_\alpha$, the extremality bound \eqref{eq:IS-ExtremalityBound2} and the BPS bound \eqref{eq:IS-BPS_bound} can differ. 

We can classify supersymmetric and non-supersymmetric solutions as:
\begin{itemize} 
 \item[\ra] \textit{\underline{Supersymmetric black holes}} saturate the BPS bound \eqref{eq:IS-BPS_bound} 
and the scalars obey the first-order equations \eqref{eq:IS-attr_eqs} (which read at spatial infinity  $\left.\ta {z^\alpha}\tau\right|_{\infty}=-\frac12 G^{\alpha\bar\beta}\pf {|Z(q,p;z_\infty,\bar z_\infty)|}{z^{\bar \beta}}$, since $U_\infty =0$), such that we conclude from \eqref{eq:IS-ExtremalityBound2} that $T_H=0$. \textit{A supersymmetric black hole is extremal}.
\item[\ra] \textit{\underline{Non-supersymmetric black holes}} do not saturate the BPS bound and do not obey the first-order equations \eqref{eq:IS-attr_eqs}: $\left.\ta {z^\alpha}\tau\right|_{\infty}\neq-\frac12 G^{\alpha\bar\beta}\pf {|Z(q,p;z_\infty,\bar z_\infty)|}{z^{\bar \beta}}$.  We have two possibilities
\begin{enumerate}
\item The case $T_HS_{BH}>0$ describes a \textit{non-extremal} black hole with non-vanishing temperature, with two non-coincident horizons. It evaporates quantum mechanically until $T_H=0$.
\item The second possibility describes the endpoint of the evaporation process above. The black hole has vanishing temperature $T_H=0$ and is \textit{extremal}. But since still $\ta {z^\alpha}\tau \neq -\frac12 G^{\alpha\bar\beta}\partial {|Z|}/\partial z^{\bar\beta}$, it has no unbroken supersymmetry.
\end{enumerate}
\end{itemize}
In summary, we have the setup of table \ref{tab:IS-Susy_BHs}.

\section{Looking forward}\label{s:IS-conclusions}

We have discussed an embedding and entropy counting for supersymmetric black holes in string theory.

In the following, we look at two types of extensions. Part \ref{pt:FirstOrder} is concerned with the field equations for the scalars of extremal and non-extremal black holes solutions in extensions of general relativity inspired by string theory (such as various supergravity theories). The aim is to extend the property of supersymmetric solutions, that obey first-order equations, to both extremal and non-extremal non-supersymmetric solutions. Part \ref{pt:Entropy} is concerned with finding the interpretation of the microstates one counts for the entropy of supersymmetric black hole solutions in the regime where supergravity is valid, instead of in the perturbative string theory regime.

\pagestyle{fancy} 
\renewcommand{\chaptermark}[1]{\uppercase{\markboth{#1}{#1}}}
\renewcommand{\sectionmark}[1]{\uppercase{\markright{#1}}}
\fancyhf{}
\fancyhead[RO]{\scriptsize\sf\rightmark\ \hrulefill\ \thepage} 
\fancyhead[RE]{\scriptsize\sf\thepage\ \hrulefill\ \leftmark}
\renewcommand{\headrulewidth}{0pt}
\fancypagestyle{plain}{
\fancyhf{}
\fancyfoot[C]{\scriptsize\sf\thepage}
\renewcommand{\headrulewidth}{0pt}
\renewcommand{\footrulewidth}{0pt}
}

\cleardoublepage
\part{A simplified description for non-supersymmetric black holes}
\label{pt:FirstOrder}
\cleardoublepage
\chapter{
A first look at first-order formalisms in supergravity}
\label{c:BJ}\label{C:BJ}
\punchline{
{}For a generic solution to a gravitational system, coupled second-order equations have to be solved. However, first-order equations that follow from writing the action as a sum of squares have been found for supersymmetric and extremal black holes for some time now. Recently, Miller\index{Miller}, Schalm\index{Schalm} and Weinberg\index{Weinberg} \cite{Miller:2006ay} have shown that it is even possible to construct first-order equations for non-extremal \RN{} black holes in this way as well. In this chapter, we show that this idea can be extended to branes of arbitrary dimension and, more importantly, to time-dependent solutions. We present the first-order equations of this type for all stationary branes (Lorentzian \worldvolume{}) and all time-dependent branes (Euclidean \worldvolume{}) of an Einstein-dilaton-$p$-form system in arbitrary dimensions. The account is mainly based on our work work with Bert Janssen, Paul Smyth and Thomas Van Riet, see ref. \cite{Janssen:2007rc}.}
\section{Introduction and overview\label{s:BJ-Introduction}}
In this chapter, we wish to investigate the structure of the equations of motion underlying many interesting solutions to gravity and supergravity. As was stressed in the introductory chapters, we always keep  black hole solutions, be it in four dimensions or not, in the back of our heads. Here, also specific $p$-brane solutions are discussed, with both radial and time-dependence.\footnote{Such $p$-branes have a $(p+1)$-dimensional \worldvolume{}. In this terminology, a black hole is a 0-brane and has a one-dimensional worldline, much like a relativistic particle.} In general, one has to solve coupled second-order PDEs in order to find a typical $p$-brane solution to a gravitational theory. Often, symmetry arguments have been used to show that the equations of motion simplify. {}For instance for supersymmetric solutions, the equations of motion reduce to a set of first-order equations. In this chapter, we begin an investigation to see if such a simplified structure can exist for solutions that are not supersymmetry, or exhibit a similar property as extremality\footnote{Extremal black holes and branes are those that are stable, in the sense that they do not emit Hawking radiation. They are seen as the endpoint of the radiation process of a general, non-extremal black hole/brane.}. We  use such a simplifying principle to shed light on the structure of black hole and $p$-brane solutions, but also to see if we can learn more about time-dependent solutions of supergravity and string theory, since especially these latter ones are ill understood in the context of string theory.

It has been known for a long time that particular stationary $p$-brane solutions of supergravity preserve some fraction of supersymmetry. Typically, one considers solutions where only bosonic modes (such as the metric and scalar fields) are excited and the fermionic fields are put to zero. Practically this means that the solutions fulfill some \emph{first-order} differential equations that arise from demanding the supersymmetry transformations to be consistently satisfied for vanishing fermions. E.g.~for a supersymmetric black hole \ansatz{} in $D$ dimensions, such first-order equations for the scalar fields $\phi^a$, provide an integrated form of the second-order equations of motion and are of the type
\begin{equation}
\dot{\phi}^a=\pm G^{ab}\partial_b|Z|\,,\label{Floweq}
\end{equation}
where $Z=Z(\phi)$ is a complex function of the scalars (more on this function below).

Such first-order equations have become known as Bogomol'nyi or BPS equations\index{Bogomol'nyi!equations}, after Bogomol'nyi's\index{Bogomol'nyi} \cite{Bogomolny:1975de}, and Prasad and Sommerfield's \cite{Prasad:1975kr} work on first-order equations and exact solutions for magnetic monopoles in Yang--Mills--Higgs theory. In those works, and generalizations thereof, first-order equations were obtained for certain solutions to non-gravitational field theories by rewriting the energy (a functional of the fields) as a sum of squares plus a boundary term. Letting $\phi$ denote the collection of all fields, one finds:
\begin{equation}
 E = \int \de^3 x \sum_i(G_i(\phi,\nabla\phi) )^2  +\int_{S^2_{\infty}} H(\phi,\nabla\phi)\label{eq:JP-intro_BPS_SumOfSquares}\,,
\end{equation}
where $G_i$ are functions of the fields $\phi$ and their (first-order) derivatives, and the second integral on the right-hand side is a boundary term, evaluated at spatial infinity. This expression shows that all solutions to this system must obey the so-called BPS bound $E\geq \int_{S^2_{\infty}} H$. Moreover, the solutions that saturate this bound are obtained by putting each individual term $G_i$ to zero, thereby giving rise to first-order equations the fields have to obey. Note that these solutions are stable, since they cannot decay to configurations with less energy. From now on we exclusively use the term \textit{Bogomol'nyi equations}\index{Bogomol'nyi!equations} as a shorthand for equations of motion\index{equations of motion} that are inferred by rewriting the energy (and later also the action) as a sum of squares\index{sum of squares} as in eq.~\eqref{eq:JP-intro_BPS_SumOfSquares}.\footnote{Note that in the literature, the term `BPS equations'\index{BPS!equation|fn} is often  used for equations of this type, but this does not necessarily mean the solutions are supersymmetric. 
On the other hand, 
the term `BPS solution'\index{BPS!solution|fn} is often used to express that a solution preserves a certain amount of supersymmetry\index{supersymmetry!fn}. To avoid possible confusion, we  refer to equations of motion that are found  by rewriting the action as a sum of squares as `equations of Bogomol'nyi type' or as `Bogomol'nyi equations'. We come back to this below.} It was later shown that the saturation of the BPS bound is intimately linked to the preserved supersymmetry of solitons in supersymmetric theories by Witten and Olive \cite{Witten:1978mh}. They showed that when embedded in a supersymmetric theory, the BPS bound can be derived from the associated supersymmetry algebra and it is only saturated for a supersymmetric solution. In addition, the first-order Bogomol'nyi equations correspond to the ones obtained by putting the supersymmetry variations to zero.

So much for the discussion about non-gravitational field theories. What can we say when gravity is included? Consider for example stationary non-extremal or time-dependent solutions. Such solutions cannot preserve supersymmetry in ordinary supergravity theories. {}{}From the field theory intuition, one therefore expects that such solutions cannot be found from Bogomol'nyi equations, but rather by solving the full second-order equations of motion.
Luckily, there are at least three instances in the literature where it has been shown this view has been too pessimistic:
\begin{enumerate}
 \item Not all extremal black hole solutions of supergravity have to
be supersymmetric. It turns out that many non-supersymmetric but
extremal solutions fulfill first-order equations in a given
supergravity theory (see for instance \cite{Ceresole:2007wx,
Andrianopoli:2007gt, LopesCardoso:2007ky}). More surprisingly,
Miller\index{Miller} et.~al.~showed that the \emph{non-extremal}
\RN\index{\RN} black hole solution of Einstein--Maxwell
theory can be found from first-order equations \cite{Miller:2006ay}
by a clever rewriting of the action as a sum of squares \`a la
Bogomol'nyi. The method of \cite{Miller:2006ay} is the main tool for
the results presented in this chapter.
\item Many stationary domain wall\index{domain wall} solutions that do not preserve any
supersymmetry have been shown to allow for first-order-equations
\cite{Freedman:2003ax, Celi:2004st, Papadimitriou:2006dr}. The resulting solutions are called fake supersymmetric, and the formalism has been dubbed \textit{fake supergravity}\index{fake supergravity}.
\item FLRW-cosmologies\index{cosmology!FLRW--} are very similar to domain walls
\cite{Cvetic:1994ya, Skenderis:2006fb, Skenderis:2006jq}. The
difference in the metrics of both kinds of solutions are given by a few signs. Up to these sign differences, the same first-order equations for domain walls exist for cosmologies. These relations have become known as {pseudo-BPS} conditions \cite{Skenderis:2006fb,Skenderis:2006jq} (see also \cite{Salopek:1990jq, Bazeia:2005tj}  for the first-order framework in  cosmology), the formalism in this context is often referred to as \textit{pseudo-supersymmetry}\index{pseudo-supersymmetry}. As for domain walls one readily checks that these
first-order equations arise from the fact that the action can be
written as a sum of squares \cite{Chemissany:2007fg}. The structure
underlying the existence of these first-order equations can be
understood from Hamilton--Jacobi theory \cite{Skenderis:2006rr,
Townsend:2007aw,Townsend:2007nm}.\footnote{In ordinary
supergravity theories the pseudo-BPS relations cannot be related to
supersymmetry preservation. However, in the case of supergravity
theories with `wrong sign' kinetic terms the pseudo-BPS relations
are related to true supersymmetry \cite{deWit:1987sn,
Behrndt:2003cx, Bergshoeff:2007cg, Skenderis:2007sm,Vaula:2007jk}. In this thesis
we consider ordinary supergravity theories and therefore
pseudo-BPS conditions are not related to supersymmetry. Practically
this means that we have first-order equations which can be
understood to originate from a Bogomol'nyi rewriting of the
action.}
\end{enumerate}

Especially the first point merits further attention. It is clear that there is a big difference between properties of solutions to field theories without and with the inclusion of gravity. As the authors of \cite{Miller:2006ay} point out, for stationary solutions to gravitational field theories, the action functional, rather than the energy, is crucial in deriving  Bogomol'nyi equations\index{Bogomol'nyi!equations}. A relation of the form \eqref{eq:JP-intro_BPS_SumOfSquares} can  be written down, but for the action instead of the energy, with the crucial difference that the different terms $G_i^2$ can appear with relative minus signs. This allows for a broader class of solutions that can be found by putting each individual term $G_i$ to zero, but such solutions do not have to saturate a bound any more. We conclude \textit{there exist solutions that follow from Bogomol'nyi equations (by writing the action as a sum and difference of squares), but that do not saturate a BPS-like bound.} In this chapter, we show that also many $p$-brane solutions to supergravity are of this form.

Note that the examples given above (1--3) are only a subset of the different $p$-branes\index{$p$-brane} that exist, namely timelike $0$-branes in $D=4$ (the \RN\index{\RN} black holes) and $(D-2)$-branes (domain walls and FLRW-cosmologies). It is the aim of this chapter to understand in general when (non-extremal) stationary and time-dependent $p$-branes in arbitrary dimensions can be found from first-order equations that follow by writing the \textit{action} as a sum of squares. We  generalize the method \cite{Miller:2006ay} to such $p$-brane solutions. A broader understanding of the underlying principles that guarantee first-order equations\index{first-order!equation} of the form presented below, is investigated in chapter \ref{c:JP-Gradient_Flow} for black hole\index{black hole} solutions in $D$ dimensions.\\

The rest of this chapter is organized as follows. In section \ref{s:BJ-Historical_Overview}, we give an overview of the use of first-order formalisms\index{first-order!formalism} in the literature. We start with non-gravitational field theory and work our way up to the appearance of first-order equations for solutions to (super)gravity. The focus is on the underlying principle of supersymmetry. In section \ref{s:BJ-History_FoForm_BHs}, we give a detailed account of the appearance of first-order equations of motion for black hole solutions that can be derived by rewriting the action as a sum of squares\index{sum of squares}. We stick to $\Cn=2$ supergravity in four dimensions and discuss supersymmetric and extremal\index{extremal} solutions, to end with some comments on non-extremal\index{non-extremal} solutions.

After the introductory part, we delve into the possibilities of extending the first-order properties of the equations of motion to non-supersymmetric solutions. In section
\ref{s:BJ-MainResults} we consider Einstein--Maxwell theory and
repeat the construction of \cite{Miller:2006ay} for the first-order equations\index{first-order!equations} for the
non-extremal Reissner-Nordstr\"om black hole. Also in section \ref{s:BJ-MainResults}, we discuss the generalization of the argument of \cite{Miller:2006ay} to stationary and time-dependent $p$-brane solutions to Einstein-Maxwell theory coupled to a dilaton in an arbitrary number of $D$ \spacetime{} dimensions. We explain how the BPS equations for general $p$-branes follow from for the BPS equations  of $(-1)$-branes via an uplifting procedure. We finish with conclusions in section \ref{s:BJ-Discussion_Outlook}, where we lay the basis for the more systematic investigation of chapter \ref{c:JP-Gradient_Flow}  of first-order formalisms for static\index{static}, spherically symmetric \index{spherically symmetric} black holes.

%
%

\section{Historical overview of first-order formalisms\label{s:BJ-Historical_Overview}}

\subsection{BPS-Monopoles  and supersymmetry\label{ss:BJ-BPS_SUSY}}
We review Bogomol'ny, Prasad and Sommerfield's (BPS) description of 't Hooft and
Polyakov's magnetic monopole as an energy-minimizing object \cite{Bogomolny:1975de,Prasad:1975kr} in a form that is applicable in other non-gravitational field theories. We show the relation between the so-called BPS-bound  for the energy and supersymmetry that was first explained by Witten and Olive \cite{Witten:1978mh} and repeat the argument showing that solutions that preserve some supersymmetry always saturates a BPS-like bound on its energy. Often, such solutions can be found by solving first-order equations.
\subsubsection{BPS-bound}
Consider Yang-Mills theory coupled to a Higgs field (i.e.\ a scalar with a potential that allows for
spontaneous symmetry breaking).
't Hooft and Polyakov \cite{'tHooft:1974qc,Polyakov:1974ek} showed that there exist magnetic monopole solutions to this theory of a specific nature.  They considered the gauge group $\SU(2)$. Through the Brout-Englert-Higgs mechanism, the Higgs scalar acquires a vacuum expectation value and the gauge symmetry is spontaneously broken to $\U(1)$. The electric and magnetic fields $\vec E,\vec B$ associated with the unbroken $\U(1)$ symmetry, obey Maxwell's equations with sources given in terms of the scalar field, allowing for monopole-like solutions carrying magnetic charge.

Bogomol'nyi \cite{Bogomolny:1975de}, and independently Prasad and Sommerfield \cite{Prasad:1975kr} showed that the 't Hooft-Polyakov monopole solution has minimal energy, ensuring stability, and can be obtained by solving first-order equations. This is realized by rewriting the energy, a functional of the fields describing the solution, as a positive integral and a boundary term. The positive integrand is a sum of squared expressions. Such a rewriting has later also been achieved for many other solutions to non-gravitational field theories and can schematically be written as in equation \eqref{eq:JP-intro_BPS_SumOfSquares},
\begin{equation}
 E = \int \de^3 x \sum_i(G_i(\phi,\nabla\phi) )^2  +\int_{S^2_\infty} H(\phi,\nabla\phi)\,,\label{eq:BJ-Energy_BPS}
\end{equation}
where the the scalar(s) and gauge fields are collected into one notation $\phi$ and the  second term is a boundary term, obtained by evaluated an integral over a function $H$ of the fields at a 2-sphere at spatial infinity. {}For specific examples, the functions $G_i$ have explicit and often simple forms. {}For instance in the special case of the 't Hooft-Polyakov monopole, there is only one quadratic term in the action of the form $G_i^2 = (\vec \nabla \varphi \pm \vec\B)^2$, $\varphi$ being the Higgs scalar field and $\vec B$ the magnetic field.

The equation \eqref{eq:BJ-Energy_BPS} shows two things:
\begin{enumerate}
\item The existence of a lower bound on the energy:
\begin{equation}
 E \geq \int_{S^2_{\infty}} H\,.\label{eq:BJ-BPS_Bound_Monopole}
\end{equation}
The integral on the right hand side gives a contribution that is solely a function of the topological charges of the solution. In the
case of the Polyakov-'t Hooft monopole, this would be the magnetic charge. {}For obvious reasons, we refer to the bound \eqref{eq:BJ-BPS_Bound_Monopole} as a BPS bound.
\item Solutions that minimize the energy are such that the squared terms between brackets in eq.~\eqref{eq:BJ-Energy_BPS} are identically zero and obey
first-order equations:
\begin{equation}
 G_i(\phi,\nabla\phi)=0\,. \label{eq:BJ-Bogomolnyi_Equations}
\end{equation}
\end{enumerate}
The BPS authors showed that the 't Hooft-Polyakov monopole, and its generalization to solutions including both magnetic and electric charges, are exactly those solution that saturate the BPS bound.

Note that in the literature, equations of the type \eqref{eq:BJ-Bogomolnyi_Equations} are often referred to as BPS equations.
However, we  refer to such equations as
\textit{Bogomol'nyi-type} equations and  refer to a rewriting of an energy or action functional in terms of
a sum of squares as in \eqref{eq:BJ-Energy_BPS} as a rewriting \emph{\`a la Bogomol'nyi}. We  reserve the epithet \emph{BPS}
for solutions that preserve part of the supersymmetry, as we explain now.

\subsubsection{BPS bound and supersymmetry}
Around the time of the description of the monopoles above, also supersymmetry entered the game. Soon it was discovered by Olive and Witten \cite{Witten:1978mh} that the BPS bound \eqref{eq:BJ-BPS_Bound_Monopole} has a deeper meaning, when the system described above, or any other system with topological charges, can be embedded in a supersymmetric theory. They make use of the fact that the supersymmetry algebra contains central charges, which are non-vanishing for solutions with topological charges. The supersymmetry algebra for $N$-extended supersymmetry is (we suppress spinor indices):
\begin{equation}
 \{Q_i,Q_j\} = \left(\delta_{ij}\gamma^\mu P_\mu + \tilde Z_{ij}\right) {\cal C}^{-1}\,, \qquad i = 1\ldots N\,,
\end{equation}
where $\cal C$ is the charge conjugation matrix. The first term  expresses the fact that two subsequent supersymmetry transformations act as a \spacetime{} translation $P_\mu$. The second term was introduced in \cite{Haag:1974qh}, where $\tilde Z_{ij}$ is an antisymmetric complex matrix that plays the role of a central charge in the supersymmetry algebra: it commutes with all other elements of the algebra.

{}For concreteness and later reference, we focus on the case of $\Cn=2$ supersymmetry. In that case, since any  two-dimensional antisymmetric matrix is proportional to $\epsilon_{ij} = \begin{pmatrix}0&-1\\1&0      \end{pmatrix}$,  the matrix $ \tilde Z_{ij}$ in the central charge term is determined by one complex number $\tilde Z$.\footnote{It is common in the literature to denote the central charge as $Z$. We include a tilde to differ between the central charge $\tilde Z$ of the superalgebra and the function $Z$ appearing in the discussion of the attractor mechanism, see section \ref{s:BJ-History_FoForm_BHs}.} We have $\tilde Z_{ij} =\epsilon_{ij}(\Re \tilde Z \unit + \rmi \Im \tilde Z \gamma_5 )$ and we  henceforth use the term `central charge' exclusively for the complex number $\tilde Z$.

One can then construct combinations $\tilde Q$ of the supersymmetry generators $Q$
 and their complex conjugates $Q^*$, such that one obtains $\{\tilde Q,\tilde Q^*\}$ as an expression in terms of the mass $M$ and central charge $\tilde Z$.\footnote{Remember that for a massive state, we can always do a Lorentz transformation to the rest frame, such that the only non-vanishing component of the four-momentum is $P^0 = E = M$ is the rest energy, which is equal to the mass $M$ of the state, since we work in units where $c = 1$).} It can be shown that the requirement that $\{\tilde Q,\tilde Q^*\} \geq 0$, leads to a bound, relating the mass and the modulus of the complex central charge:
\begin{equation}
 M\geq |\tilde Z|\,.\label{eq:BJ-BPS_bound_susy}
\end{equation}
There is even more that one can say. {}From the expression $\{\tilde Q,\tilde Q^*\}$, it follows that when a solution is supersymmetric, such that at least one of the supersymmetry generators $Q$ gives zero when evaluated on the solution, the BPS bound is saturated. The reverse argument also holds: whenever the BPS bound \eqref{eq:BJ-BPS_bound_susy} is saturated, a solution preserves part of the supersymmetry.

One can compute the central charge for a certain theory in terms of the topological charges. In the case of the BPS monopole, the
central charge is just the expression $\tilde Z = \int_{S^2_\infty}  H$ above. In this way, Olive and Witten showed that for the BPS-monopole
(and its generalizations carrying both electric and magnetic charges)  the BPS bound on the energy \eqref{eq:BJ-BPS_Bound_Monopole} follows from supersymmetry. Of course, the existence of a BPS bound and its relation to supersymmetry preservation applies to all extended supersymmetry
theories with topological charges.

\outlook{Upshot:}{We highlight two main points for later reference. The BPS-bound, \eqref{eq:BJ-BPS_bound_susy} in its general form, is saturated for a solution to a certain supersymmetric theory \textit{if and only if} the solution preserves part of the supersymmetry. Moreover, such solutions can often be found by a Bogomol'nyi trick of completing the squares.}


\subsection{Supersymmetric solutions in supergravity?}
So far, we have concentrated on the use of first-order equations and their relation to supersymmetry, in theories without gravity. Our ultimate goal is learning more about black holes, so the question naturally rises what the equivalence is of the first-order formalisms discussed above in supergravity and more specifically what we can learn from applying it to black hole solutions. In the rest of this historical overview, we restrict to $\Cn=2$ supergravity in four dimensions for concreteness.

Two of the central issues raised above have found their place in the framework of supergravity. On the one hand, we have the existence of a BPS bound from the supersymmetry algebra. In supergravity, as in any other theory of supersymmetry, a representation of the symmetry algebra satisfies the BPS bound. In the case of $\Cn=2$ supergravity, this bound again takes on the form
\begin{equation}
 M \geq |\tilde Z|\,,
\end{equation}
where $Z$ is the (complex) central charge of the supersymmetry algebra. A supersymmetric solution to the supergravity equations of motion (forming a representation of the local $\Cn=2$ supersymmetry algebra) minimizes this bound, being a stable solution of course. On the other hand, supersymmetric solutions to supergravity satisfy first-order equations as well. This can be heuristically understood as follows. When studying black hole solutions to supergravity, people mostly consider bosonic backgrounds, in which the bosonic degrees of freedom are excited and have a non-trivial profile, while all fermions are zero. In order to see the appearance of first-order equations such solutions have to satisfy, one takes the realization of the supersymmetry variations on the fields into consideration, with supersymmetry parameter $\epsilon$:
\begin{align}
 \delta_{\epsilon} |{\rm bosons}\rangle &\sim | {\rm fermions} \rangle\,, \nonumber\\
\delta_{\epsilon} | {\rm fermions} \rangle&\sim|{\rm bosons}\rangle\,.\label{eq:SUSYvar}
\end{align}
The supersymmetry variation on the bosons corresponds to an operator  acting on the fermions, while the supersymmetry variation of the fermions is some other operator acting on the bosons. We can now ask how we can find a solution that is invariant under (part of) the supersymmetry transformations. When one chooses a bosonic background solution, where all fermions are put to zero, the first of the equations \eqref{eq:SUSYvar}  implies that the supersymmetry transformations for the bosons are zero ($\delta_{\epsilon} |{\rm bosons}\rangle=0)$. The only non-trivial information follows from putting the second equation in \eqref{eq:SUSYvar} to zero. The key thing to note is that in supergravity theories, the operator appearing in the supersymmetry variation of the fermions, gives rise to a set of first-order differential equations on the supersymmetry parameter $\epsilon$ and on the bosonic fields:
\begin{equation}
 \delta_{\epsilon} | {\rm fermions} \rangle = 0\quad \Leftrightarrow \quad \left\{  \begin{array}{l} \slashed\partial\epsilon(x) + \cdots = 0\\ A({\rm bosons})\cdot \epsilon (x) = 0 
\,. \end{array} \right.\label{eq:KSEgen}
\end{equation}
where $A$ is a field-dependent matrix, containing first-order derivatives of the bosons and acting on the spinor $\epsilon$. The first equation of \eqref{eq:KSEgen} comes from demanding the supersymmetry variation of the gravitino (the superpartner of the graviton) to vanish, while the other equations follow from vanishing supersymmetry variations of the other fermion fields. One sees these so-called \emph{Killing spinor equations} give rise to a set of first-order equations for the bosonic fields and a set of differential equations and possibly projection conditions on the spinors $\epsilon$. To solve for the Killing spinor $\epsilon$, one makes an \ansatz{} $\epsilon = B(x)\cdot \epsilon_0$, where $B$ is a \spacetime-dependent matrix and $\epsilon_0$ is constant. If we denote the (real) dimensionality of the space of spinors $\epsilon_0$ as $N$, than a solution to the equations \eqref{eq:KSEgen} is characterized by the number $n$ of independent real spinors $\epsilon_0$ that can be found, that satisfy the Killing spinor equations. It follows that the solution is invariant under $n/N$ of the total symmetry. One says the solution is $n/N$ BPS. In particular, we  concentrate on 1/2 BPS solutions of $\cN=2$ supergravity.

\outlook{Summary:}{A supersymmetric \textit{bosonic} background (like a black hole solution), can often be obtained by solving the Killing spinor equations \eqref{eq:KSEgen}. When a solution for the spinor parameter $\epsilon$ is found, this comes down to solving first-order equations for the bosonic fields under consideration.}

\subsection{First order formalisms for specific black hole solutions}
Now we discuss how supersymmetric black hole solutions were found from first-order equations. A lot of attention in the literature has gone to supersymmetric black hole solutions to $\Cn=2$ supergravity and higher dimensional analogs, so we  focus on those. The Killing spinor equations can be solved to find supersymmetric black hole solutions to $\Cn=2$ supergravity coupled to vector multiplets (containing vectors, scalars and their fermionic superpartners).\footnote{Already earlier, the most general form of a supersymmetric solution had been discussed in \cite{Tod:1983cq,Tod:1995cq} and supersymmetric solutions were found from the supersymmetry transformations in other theories, as $\cn = 4$ supergravity in four dimensions, see for instance \cite{Kallosh:1992ii}.} After one makes the correct \ansatz{} for the vector gauge fields, one obtains a set of first-order equations determining the radial flow of the scalars and warp factor of the metric. The main structure leading to solving these equations and giving one the details of the near-horizon geometry, lies in the \emph{attractor mechanism} \cite{Ferrara:1995ih,Strominger:1996kf,Ferrara:1996dd}: whatever the initial values of the scalars at radial infinity, they take on universal values at the black hole horizon. These values are determined by black hole charges only, assuring the horizon area depends on the electric and magnetic charges, but are insensitive to the initial conditions of the scalar fields. This is particularly useful to obtain the (macroscopic) Bekenstein-Hawking entropy of those solutions, since the black hole geometry (and the horizon area determining the Bekenstein-Hawking entropy) is fully determined by the black hole charges.  {}For generic non-supersymmetric black holes  this is not true -- for such black hole solutions, the horizon area and other near-horizon quantities will depend on the values of the scalars at radial infinity. We go into the attractor mechanism in more detail in section \ref{ss:BJ-AM}.

It was later noticed that the attractor behaviour of the scalars and the associated first-order equations, are not just a matter of supersymmetry. In \cite{Ferrara:1997tw} already it was noted that the attractor mechanism\index{attractor mechanism} is in fact due to the extremality of the solution. Indeed, one can obtain the attractor equations by reducing the second-order equations of motion to a set of first-order equations without reference to the Killing spinor equations. This can be done on the level of the equations of motion, but also by considering an adapted \emph{Bogomol'nyi trick}\index{Bogomol'nyi!trick} on the effective Lagrangian of the system. Again, we expand on this in section \ref{ss:BJ-AM}. It was only a few years ago that the full potential of the attractor mechanism\index{attractor mechanism} for extremal solutions was realized, see references \cite{Ferrara:1997tw,Goldstein:2005hq,Goldstein:2005rr,Kallosh:2006bt}, whence several groups started investigating the structure of the attractor equations for non-supersymmetric extremal black holes \cite{Andrianopoli:2006ub,Ceresole:2007wx,Cardoso:2007ky}. References on later work are given below.

We now go back to the two leading ideas we posed at the end of section \ref{ss:BJ-BPS_SUSY}. We saw how supersymmetric solutions saturate the BPS bound and that the fields of the solution satisfy specific first-order equations of motion. What can we say when a solution is no longer supersymmetric? Obviously, the BPS bound is longer satisfied by definition. But does this also hold for the existence of first-order equations? {}For a long time, the existence of first-order equations has been thought to be tantamount to supersymmetry. But the attractor behaviour of extremal non-supersymmetric solutions contradicts this. Moreover, in 2006 Miller\index{Miller}, Schalm\index{Schalm} and Weinberg\index{Weinberg} \cite{Miller:2006ay} showed that for the \RN\index{\RN} black hole, even its non-extremal (and thus non-supersymmetric) solutions can even be found by solving a set of first-order equations. The article \cite{Miller:2006ay} formed the starting point of the our work on first-order formalisms for non-supersymmetric black holes and branes published in \cite{Janssen:2007rc}. Both articles are the subject of section \ref{s:BJ-MainResults}.

\section{Attractor mechanism and first-order equations for black holes\label{s:BJ-History_FoForm_BHs}}
We now discuss the status of the application of first-order formalisms for black hole solutions prior to the result of Miller\index{Miller}, Schalm\index{Schalm} and Weinberg\index{Weinberg} \cite{Miller:2006ay}, or in other words, a state of affairs at the end of 2006. First we treat the supersymmetric case, then the non-supersymmetric extremal case. The focus is on $\Cn=2$ supergravity in four dimensions. A clear account can for instance also be found in \cite{Denef:1999,Denef:2000nb} for supersymmetric solutions and \cite{Andrianopoli:2006ub} for extremal ones. In \cite{Ferrara:2008hw} some more physical thoughts can be found. We follow a similar line of argument as in those references.

\subsection{Attractor mechanism for supersymmetric black holes\label{ss:BJ-AM}}
As we have seen in the previous chapter, supergravity coupled to a number of vector multiplets is a natural (supersymmetric) extension of Einstein-Maxwell theory. Again, this supergravity theory has black hole solutions, but generically they exhibit a richer structure than solutions to Einstein-Maxwell theory. This is because the supergravity setup also includes scalar fields, which in general have a non-trivial behaviour in a black hole solution. {}For supersymmetric, static, spherically symmetric \index{spherically symmetric} black hole solutions to $\Cn=2$ supergravity coupled to $n_V$ vector multiplets one can show that the black hole metric and the scalar fields can be derived from a set of first-order equations. Moreover, these equations show that the values of the scalars at the horizon on the black hole only depend on (electric and magnetic) charges. As we mentioned above, this is known as the attractor mechanism. Let us consider this mechanism in detail.

We start from the action for $\Cn=2$ supergravity coupled to $n_V$ vector multiplets. The bosonic part of this Lagrangian contains a metric, coupled to $n_V$ scalar fields and $n_V+1$ vectors with scalar-dependent gauge couplings:
\begin{align}
S &= \frac1{16\pi G_4} \int \rmd^4 x \sqrt{-g}\left(R -2 g^{\mu \nu}G_{\alpha\bar
\beta}\partial_\mu z^\alpha \partial_\nu \bar z^{\bar \beta} \right)\label{eq:N-=2SUGRA_2}\\
&+\frac{\Conv^2}{4\pi G_4}\int\rmd^4 x \sqrt{-g}\left(\frac14(\Im
\mathcal{N}_{IJ} )F_{\mu\nu}^I F^{\mu\nu J} -\frac1{8\sqrt{-g}}(\Re \mathcal
N_{IJ})\veps^{\mu\nu\rho\sigma} F_{\mu\nu}^I
F_{\rho\sigma}^J\right).\nonumber
\end{align}
The constant $G_4$ is Newton's gravitational constant in four dimensions, while the constant $\Conv$ is convention dependent. The latter constant is kept as a book-keeping device and it encodes the normalization of the vector fields. {}For example, by taking $\Conv$ to have dimensions of length (or mass$^{-1}$), in natural units where $\hbar = c = 1$, we see that we can describe vector fields and charges which are dimensionless in natural units. We  later put $\Conv = 1$, as in \cite{Denef:2000nb}.  Note that $\Im \cn$ is a negative-definite matrix in order to have a positive energy density of the electromagnetic fields. We show how static, spherically symmetric \index{spherically symmetric}  black hole solutions to this action can be obtained from an effective action depending on one (radial) parameter.
\subsubsection{Symmetries and \ansatz{}}
Starting from symmetry arguments, we can make a very restrictive black hole \ansatz{} for the metric, and even determine the form of the gauge fields completely. This justifies the result advertised before, namely that we can obtain an effective theory for the scalars of the theory. First note that requiring the metric to describe a static, spherically symmetric \index{spherically symmetric} \spacetime{}, dictates it should be of the form:
\begin{equation}
\de s^2 = -\e^{2U(\tau)}\de t^2 + \e^{-2U(\tau)}\left(\frac{c^4}{\sinh^4 c\tau}\rmd\tau^2 +\frac{c^2}{\sinh^2 c\tau}(\rmd\theta ^2 + \sin^2\theta \rmd\varphi ^2)\right)\,,\label{eq:BJ-Ansatz_Metric_N=2_AM}
\end{equation}
in terms of a radial coordinate $\tau$, $c$ is a constant. The function $U$ is often called redshift factor or black hole warp factor. It measures the redshift in the frequency of radiation as it moves away from the black hole.  Both the metric warp factor and the scalars can only depend on the radial parameter, due to spherical symmetry:
\begin{equation}
 U(\tau),\qquad z^\alpha(\tau)\,.
\end{equation}
The constant $c$ determines the curvature of the three-dimensions spatial metric ($R^{(3)}\sim c^2)$. It should be interpreted as a non-extremality parameter: for extremal black holes, which have vanishing Hawking temperature, we have $c=0$, while for non-extremal black holes this is no longer the case.\footnote{The constant $c$ is actually given by $c^2 = 2 S_{BH} T_H$, where $S_{BH}$ is the entropy and $T_H$ is temperature of the black hole,
see for instance \cite{Kallosh:2006bt}.} The interpretation as a non-extremality parameter in our context becomes clear below eq.~\eqref{eq:HamConstraintN=2}. {}For later use, we give the Ricci scalar\index{Ricci!scalar} of the metric \eqref{eq:BJ-Ansatz_Metric_N=2_AM}:
\begin{equation}
R = 2\e^{2U}\frac{\sinh^4 c\tau}{c^4}\left(c^2 - \dot U^2 +\ddot U\right)\,,\label{eq:AM-Ricci_Scalar}
\end{equation}
where a dot means differentiation w.r.t. the radial parameter $\tau$.

To relate the \ansatz{} \eqref{eq:BJ-Ansatz_Metric_N=2_AM} to a more familiar expression for the metric, consider the relation of the coordinate system above to more standard radial coordinates:
\begin{equation}
\frac{c^2}{\sinh^2 (c \tau)} \equiv (r-r_0)^2 - c^2 \equiv (r-r_+)(r-r_-)\,,
\end{equation}
where we defined the constant $r_0$ as a reference point on the $r$-axis and $r_\pm$ are the roots of the quadratic polynomial $(r-r_0)^2 - c^2$. This coordinate identification is done, anticipating the expected near-horizon behaviour. {}For general, non-extremal solutions $(c \neq0)$, the black hole has two horizons, located at $r = r_\pm = r_0 \pm c$. {}For extremal solutions, $c=0$ and the two horizons coincide, at the radius $r = r_0$. Note that for the extremal solutions, the radial coordinate $\tau$ is related to the standard radial coordinate $r$ as $\tau = 1/(r-r_0)$.

Let us discuss the asymptotic form of the metric. As the radial parameters $r\to \infty, \tau \to 0$, we go to spatial infinity. Demanding that we have an  asymptotically flat solution, sets the warp factor $\e^{2U(\tau)}$ to 1 as $\tau$ approaches zero. Moreover, from an asymptotic analysis of the $g_{tt}$ component of the metric, see for example \cite{Misner:1974qy}, the ADM mass $M$ of the solution is given by  $g_{tt}= \left.\e^{2U}\right|_{\tau \to 0} = 1 - 2 G_4 M \tau + {\cal O}(\tau^2)$. There are more boundary conditions one can put on the metric. As $\tau \to \infty$ ($r\to r_+$), we approach the black hole horizon. Therefore, the black hole horizon area $A_H$ is determined by the warp factor $U(\tau = \infty)$. The boundary conditions at spatial infinity and the horizon are summarized in table \ref{tab:BJ-Asymptotic_metric}. We will use the asymptotic expansions of the warp factor $U$ later.

\begin{table}[ht!]
\centering{
\begin{tabular}{|l|l|ll|}
\hline
{\bf REGION}& {\bf RADIAL COORDs.} & &{\bf METRIC}\\
\hline
 &&&\\[-6pt]
Spatial infinity&$\tau \to 0,\quad r\to \infty $ &$U = 0,$ &$\quad \dot U = -G_4 M$\,. \nonumber\\[2mm]
Near-horizon&$\tau \to \infty,\quad r\to r_+$ :& $A_H/4\pi$ &$=\e^{-2U}\frac{c^2}{\sinh^2 c\tau}$
\\[2mm]
&&& $=\e^{-2U}(r-r_+)(r-r_-)$\,.\\[2mm]
\hline
\end{tabular}
}
\caption{Asymptotics for the metric function $U$ in terms of the radius.\label{tab:BJ-Asymptotic_metric}}
\end{table}

Note from table \ref{tab:BJ-Asymptotic_metric} that general \emph{(non-)extremal} solutions have the same, universal near-horizon geometry, as for $\tau\to0, r\to r_+$ the metric can be written as:
\begin{equation}
  \de s^2 = - \frac {(r-r_+)(r-r_-)} {r_H^2} \de t^2 +\frac{ r_H^2}{(r-r_+)(r-r_-)}\de  r^2 + r_H^2(\de \theta^2 + \sin^2 \theta \de \phi^2)\,,\label{eq:BJ-Metric_AM_NearHorizon_NonExtremal}
\end{equation}
where the radius $r_H$ is defined in terms of the area of the event horizon as $A_H = 4\pi r_H^2$.
Notice that this is the near-horizon geometry of a \RN\index{\RN} black hole, see section \ref{ss:BHelectromagn}. {}For \emph{extremal} solutions ($c=0$), $r_0=r_+=r_-$. By a shift of the radial variable $\tilde r = r - r_0$,
the near-horizon region takes on the universal form:
\begin{equation}
  \de s^2 = - \frac {\tilde r^2} {r_H^2} \de t^2 + r_H^2\frac{\de \tilde r^2}{\tilde r^2} + r_H^2(\de \theta^2 + \sin^2 \theta \de \phi^2)\,.\label{eq:BJ-Metric_AM_Near_Horizon}
\end{equation}
This metric describes the direct product space $\AdS 2 \times \sS 2$, which is known as the Bertotti-Robinson metric ($\AdS 2$ is spanned by $(t,\tilde r)$, while the $\sS 2$ is parametrized by $(\theta,\phi)$).  The radius of the two sphere and the Anti-de Sitter space are both equal to $r_H = \sqrt{A_H/4\pi}$. We note that this geometry is the \emph{same near-horizon region as the one for the extremal \RN\index{\RN} metric}, see section \ref{ss:BHelectromagn}. These extremal black holes have Bekenstein-Hawking entropy:
\begin{equation}
  S_{BH} = \frac{A_H}{4 G_4} = \frac{\pi r_H^2}{G_4}\,.
\end{equation}
We will see later how the constant $r_H$ is found in terms of the black hole charges by solving the equations of motion for the metric, gauge field and scalars.

What about the gauge fields? We have $n_V+1$ vectors in the game, representing a global $\left(U(1)\right)^{n+1}$ symmetry. The solution is specified by giving the charges under these symmetries. In terms of the gauge fields, the solution carries electric and magnetic charges. As for the Einstein-Maxwell case, these are determined by Gauss' law: the enclosed charges are given as the surface integrals of the electric and magnetic flux through a closed surface $\Sigma$ surrounding the source. The complication now is that, due to the scalar-dependent gauge couplings, the field strengths have a more intricate relation to the electric and magnetic charges. Denoting the $n_V+1$ electric charges as $q_I$ and magnetic charges as $p^I$, we have
\begin{equation}
p^I=\frac 1 {4\pi }\int_{\Sigma^2} F^I\,,\qquad q_I=\frac 1 {4\pi }\int_{\Sigma^2} G_I\,,
\end{equation}
i.e.~the magnetic charges $p^I$ are defined in the usual fashion in terms of the field strengths $F^I$, while the electric charges are defined in terms of a dual field strength $G_I$. The field strength $G_I$ is defined through the Lagrangian as
\begin{equation}
 G_{I\mu\nu} = \veps_{\mu\nu\rho\sigma} \frac{\partial S}{\partial F_{\rho\sigma}^I} \qquad \Rightarrow \qquad G_I = {\rm Im} \, \cn_{IJ} \star F^J + {\rm Re}  \,\cn_{IJ} \label{eq:DualFieldStrength}
F^J \,,
\end{equation}
where $\star$ denotes the Hodge dual. {}For Einstein-Maxwell theory (one vector, $\cn = -\rmi$) this definition correspond to the usual $G = -\star F$. Note that integer charges are then found up to a factor of $\Conv$: denoting $Q = (q,p)$, we have that \begin{equation}
 \Gamma = \frac{\Conv}{\sqrt{G_4}}Q\,,
\end{equation}
with $\Gamma$ a vector of integer charges. See  \cite{Denef:2000nb}. E.g.\ by putting the normalization constant $\Conv = \sqrt{G_4}$, we would obtain $Q = \Gamma$ and $(p,q)$ are then integers. 

The requirement of staticity\index{static}\footnote{In a static\index{static} \spacetime{}, one demands the fields to have vanishing Lie derivative along the timelike Killing vector, to have a static\index{static} solution. This translates to $\cl_X F^I = \cl_X G_I =0$, where $X=\partial_t$ is the timelike Killing vector of the \spacetime{}. } and spherical symmetry dictates how the angular components of the field strengths are determined by the charges: $F^I{}_{\theta\phi}=p^I\sin \theta,G_{I\theta\phi}=q_I\sin \theta$\,. Using the definition of $G_I$ \eqref{eq:DualFieldStrength}, we can translate the requirement of spherical symmetry\index{spherical symmetry} into an \ansatz{} for both the electric and magnetic\index{electric}\index{magnetic} components of the field strengths\index{field strength} $F^I$ and $G_I$:
\begin{equation}
 \begin{pmatrix}
  F\\ G
 \end{pmatrix}
 = \Omega\cm  \begin{pmatrix}
  p\\ q
 \end{pmatrix}\e^{2U} \de t \wedge \de \tau + \begin{pmatrix}
  p\\ q
 \end{pmatrix} \sin \theta \de \theta \wedge\de \phi\,,
 \label{eq:BJ-Ansatz_GaugeField_AM}
\end{equation}
where $\Omega = \begin{pmatrix}0 &\unit\\-\unit &1 \end{pmatrix}
$ and the $(2n_V +2 )\times (2n_V+2)$ symplectic matrix\index{symplectic matrix} $\cm(z,\bar z)$ is defined 
in terms of the period matrix\index{period matrix} $\cn$ as:
\begin{equation}
 {\cal M}=\begin{pmatrix}-I-RI^{-1}R&RI^{-1}\cr
  I^{-1}R&-I^{-1}\end{pmatrix}\,, \qquad R=\Re\,{\cal N}\,,\quad I=\Im\,{\cal
  N}\,.
\end{equation}
We summarize the vector field content (including charges) and effective potential for $\Cn=2$ supergravity  in table \ref{tab:ReissnerNordstrom}.
\begin{table}[ht!]
\centering
  \begin{tabular}{|l|l|l|}
    \hline
    {\bf Theory/truncation}& {\bf $\Cn=2$ supergravity} & {\bf Einstein-Maxwell}\\
    \hline
    Content& gravity multiplet  &gravity multiplet \\
    &$+ n_V$ vector multiplets&no vector multiplets\\
    Gauge fields&$F^I,I=0\ldots n$&$F \equiv F^0$\\
    Elec.~charges&$q^I,I=0\ldots n$&$\QE\equiv q^0,$ \\
    Magn.~charges&$p_I,I=0\ldots n$&$\QM\equiv p_0$\\
    Gauge couplings&$\cn_{IJ}$&$\cn_{00} = \rmi$\\
    Scalars& $z^\alpha $& $z^\alpha = cst$\\
    Eff.~potential&$V_{BH} = \frac 12\Conv^2 \begin{pmatrix}
      p&q
    \end{pmatrix} \cm \begin{pmatrix}p\\q
    \end{pmatrix}$&$V_{BH} = \frac12\Conv^2(\QE^2 + \QM^2)$\\
\hline
\end{tabular}
\caption[The field content of $\Cn =2$ supergravity coupled to vector multiplets.]{The field content of $\Cn =2$ supergravity coupled to vector multiplets and a truncation to Einstein-Maxwell theory used in this thesis. Only the non-zero fields are given. We also give the notation used throughout for the excited gauge fields and non-zero charges.\label{tab:ReissnerNordstrom}}
\end{table}

\subsubsection{Effective one-dimensional action\index{effective action!one-dimensional}}
The \ansatz{} for the gauge fields\index{gauge field} solves the equation of motion and allows us to eliminate the gauge fields from the Lagrangian \eqref{eq:N-=2SUGRA_2}. One should be careful in doing this: eliminating fields in terms of an \ansatz{} is a matter best dealt with on the level of the equations of motion. It can be shown that the equations of motion for  $U$ and the scalars can be derived from an effective action in one dimension, which should be seen as the equivalent of the action \eqref{eq:N-=2SUGRA_2} per unit time $T$:
\begin{equation}
S_{\rm eff}=S/T = -\frac1{8\pi G_4}\int \de \tau \,\Bigl(\dot{U}^2
+G_{\alpha \beta}(z,\bar z)\dot{z}^\alpha\dot{\bar z}^\beta +
\e^{2U}V_{BH}(z,\bar z)-c^2\Bigr),\label{eq:EffAction}
\end{equation}
where a dot means differentiation w.r.t. the radial parameter $\tau$.\footnote{Remember this action cannot be obtained from direct substitution of the \ansatze{}\index{\ansatz} into the four-di\-men\-sio\-nal action \ref{eq:N-=2SUGRA_2}. One can however write down another, equivalent, action for the vector fields, such that it is possible to plug in the \ansatz{} in the equations of motion. Therefore, one needs to find a form of the action for the vectors which is manifestly duality invariant. Such an action fails to be generally covariant, however. This is discussed for example  in \cite{Henneaux:1988gg,Denef:2000nb} and for general stationary black hole\index{stationary} \ansatze{} (describing rotating black holes and multi-center\index{multi center} configurations) in my work  with Antoine Van Proeyen\index{Van Proeyen} \cite{VanProeyen:2007pe}. Below, we encounter many more effective actions obtained in this way, but we will not keep repeating the subtleties one can encounter in deriving an effective action.} The terms $\dot U^2 - c^2$ in the effective action originate from the Ricci scalar\index{Ricci scalar} \eqref{eq:AM-Ricci_Scalar} (the term $\ddot U$ in the Ricci scalar only contributes a total derivative to the action and is omitted). The black hole potential\index{black hole potential} encompasses the effects of the gauge fields\index{gauge field}, it is given by:
\begin{equation}
V_{\rm BH}=\ft12 \Conv^2\begin{pmatrix} p& q\end{pmatrix} {\cal M}\begin{pmatrix}
  p\\ q
 \end{pmatrix}\geq0\,.\label{eq:BJ-BH_Potential_Defn}
\end{equation}
The inequality follows because $\Im \cn$ is a negative-definite\index{negative-definite} matrix, so $\cm$ is positive-definite\index{positive-definite}. Remember that the kinetic terms of the gauge fields have scalar dependent couplings, hence it is no wonder that eliminating the gauge fields in terms of their charges through the \ansatz{} \eqref{eq:DualFieldStrength}, gives rise to a scalar potential. This scalar potential, a non-negative function, is proportional to the electromagnetic energy density and is invariant under electric-magnetic duality transformations.\footnote{Electric-magnetic duality acts on the vectors $(F^I, G_I)$ and $(p^I,q_I)$ through multiplication with symplectic matrices (elements of $\Sp (n_V,\Real)$). See section \ref{ss:W_dual_invariant} for some more information.} The one-dimensional effective action\index{effective action!one-dimesional} \eqref{eq:EffAction} is that of a particle living on $\mathbb{R} \times M_{\rm scal}$, subject to an external force field given by the effective black hole potential\index{black hole potential} $V_{BH}$, where $\Real$ denotes the black hole warp factor and $M_{\rm scal}$ is the scalar manifold. The radial parameter $\tau$ plays the role of time. 

To make sure that solutions to the effective action\index{effective action} \eqref{eq:EffAction} are also solutions to the full supergravity system \eqref{eq:N-=2SUGRA_2}, we have to check that the Einstein equations stemming from other metric modes than the $U$-mode are also fulfilled. One can verify that the only extra information residing in the Einstein equations is the following equation, which acts as a constraint that solutions to the effective action should obey:
\begin{equation}
\dot{U}^2
+G_{\alpha \beta}(z,\bar z)\dot{z}^\alpha\dot{\bar z}^\beta -
\e^{2U}V_{BH}(z,\bar z) = 2c^2\,. \label{eq:HamConstraintN=2}
\end{equation}
What is the meaning of the constraint equation \eqref{eq:HamConstraintN=2}? First notice that the Lagrangian has the form $T - V$, whereas that the constraint equation takes the form $T + V = 0$. Hence the name `Hamiltonian constraint'\index{Hamiltonian constraint}. More on the interpretation of the effective black hole description as a Hamiltonian system, with $\tau$ playing the role of the time coordinate, can be found in the next chapter, in section \ref{ss:JP-Eff_BH_Hamiltonian}. One can check that for any solution to the action \eqref{eq:EffAction} the left-hand side of \eqref{eq:HamConstraintN=2} is constant. We conclude that the constraint only fixes this integration constant to be  equal to the non-extremality parameter $c^2$. In principle this information should be contained in the boundary conditions: the values of the fields at spatial infinity. An intuitive understanding of the extremality parameter can be obtained by referring for example to an electrically charged \RN\index{\RN} solution. This is a solution to the $\Cn=2$ supergravity Lagrangian, where all scalar\index{scalar} fields are constant and $\Im \cn = -\rmi$, see table \ref{tab:ReissnerNordstrom}. Evaluating the Hamiltonian constraint \eqref{eq:HamConstraintN=2} at spatial infinity then gives $ G_4^2 M^2 = \Conv^2 \QE^2 + c^2$, where $\QE$ is the total electric charge. Only when $c = 0$ do we have an extremal  black hole.

\subsubsection{Deriving a BPS bound, finding first-order equations}
We show how we can easily obtain supersymmetric\index{supersymmetric} solutions to the one-dimensional effective action \eqref{eq:EffAction}, obtaining a relation to the BPS bound and first-order equations along the way.  The key thing to note is that the black hole potential $V_{BH}$ has a very restrictive form. Using the special geometry\index{special geometry} properties of the scalar manifold\index{scalar!manifold}, one can show that it has the quadratic form:
\begin{equation}
V_{BH} (z,\bar z) = \Conv^2(|Z|^ 2 + 4 G^{\alpha \bar \beta}\partial_\alpha |Z|  \bar \partial_{\bar \beta} |Z|) \label{eq:BHpot_DZDZ}
\end{equation}
The complex function $Z(z,\bar z) \in \mathbb{C}$ takes complex values on the scalar manifold and it is defined by:
\begin{equation}
Z = \e^{\ck /2}(X^I q _I - F_I p^I) \label{eq:BJ-Z}
\end{equation}
where $X^I = (X^0,X^\alpha)$ are projective coordinates on the scalar manifold, such that $z^\alpha = X^\alpha/X^0$ and $F_I$ denotes derivatives of a scalar prepotential $F(X)$ as $F_I = \partial F / \partial X^I$. The scalar prepotential is a homogeneous function of degree two in the  $X^I$ that is specific to the $\cN=2$ supergravity Lagrangian under consideration. Finally, $\ck$ is the K\"alher potential that determines the scalar metric as $G_{\alpha\bar\beta} = \partial_\alpha \partial_{\bar\beta}\ck(z,\bar z)$.

Evaluated at spatial infinity, the function $Z$ equals the central charge of the $\Cn=2$ supersymmetry algebra\index{supersymmetry!algebra}. {}For this reason, people refer to the function $Z$ as the central charge function, or even plainly central charge\index{central charge}\index{central charge!function}. Note that this is misleading terminology, since $Z$ is not a constant!

Having obtained the quadratic form \eqref{eq:BHpot_DZDZ} of the black hole potential, we see that the effective action \eqref{eq:EffAction} is a sum of quadratic terms. This begs for an application of the Bogomol'nyi trick\index{Bogomol'ny!trick}. By completing the squares, we can rewrite the action \eqref{eq:EffAction} as a sum of squares\index{sum of squares} plus a boundary term:
\begin{align}
 S_{\rm eff} &= -\frac1{8\pi G_4}\int \de \tau \,\Bigl((\dot{U}
\pm \Conv  |Z|\e^{U})^2 + \big{|}\big{|}\dot z^\alpha \pm 2\Conv\e^U G^{\alpha\bar \beta }\partial_{\beta} |Z|\,\big{|}\big{|}^2 - c^2\Bigr)\nonumber\\
& \pm \left.\frac{\Conv}{4\pi G_4}|Z|\e^{U}\right|^{\tau = \infty}_{\tau=0}\,.\label{eq:Uz_action}
\end{align}
The norm $\big{|}\big{|}\cdot\big{|}\big{|}^2$ is taken w.r.t. the metric $G$ on the scalar manifold.

We discuss two far-reaching consequences of this form of the action, similar to the discussion for the BPS monopole\index{BPS monopole}. 

\begin{enumerate}
\item \underline{\emph{BPS bound.}} Remember that we are discussing static\index{static} solutions, i.e.~they are independent of time $t$ in the coordinates \eqref{eq:BJ-Ansatz_Metric_N=2_AM}. As such, the associated Hamiltonian\index{Hamiltonian} can be written as minus the action\footnote{Remember that for a general mechanical system with generalized coordinate vector $q$ and momenta $p$, the Hamiltonian density $\ch$ and the Lagrangian density\index{Lagrangian} $\cl$ are related by the Legendre transform\index{Legendre transform} $\ch(p,q,t) = p \frac {\partial q}{\partial t} - \cl(q,\frac {\partial q}{\partial t} ,t)$. In our case, the generalized coordinates $q$ are the scalars $z$ and warp factor\index{warp factor} $U$. Since we describe a static\index{static} solution, the velocities $ \frac {\partial q}{\partial t}$ are zero.}:
{}For the solution at hand, we know that the Hamiltonian w.r.t. the time coordinate $t$ corresponds to the ADM mass of the black hole (up to a proportionality factor depending on chosen normalizations). This leads to the BPS bound:
\begin{equation}
 {G_4}M \geq {\Conv} |Z|_{\tau =0}\label{eq:BH_BPSbound}
\end{equation}
We dropped the $\tau = \infty$ boundary term in \eqref{eq:Uz_action}, since we expect the warp factor $\e^U$ to vanish at the horizon. Any solution for which each individual term of the squares in \eqref{eq:Uz_action} is identically zero, satisfies this bound. We show below that these correspond to supersymmetric black hole solutions.

\item \underline{\emph{First order equations.}} We can saturate the BPS bound \eqref{eq:BH_BPSbound} by demanding that all the individual squares in \eqref{eq:Uz_action} are zero. This leads to a system of $n_V+1$ first-order differential equations for the scalars $z$ and the warp factor $U$:
\begin{align}
\dot U &= \mp \Conv\e^{U} |Z|\nonumber\\
\dot z^\alpha &= \mp2\Conv\e^U G^{\alpha\beta}\partial_\beta |Z|\label{eq:AttractorEqs}
\end{align}
These equations fix the radial dependence of the scalars in terms of a gradient flow on the scalar manifold, determined by the scalar function $Z(z,\bar z)$. We can readily throw away the solutions with the lower (plus) signs, as these give rise to solutions that are gravitationally repulsive (negative ADM mass $G_4 M = -\dot U$, as seen from table \ref{tab:BJ-Asymptotic_metric}) and do not correspond to well-behaved black hole solutions.
\end{enumerate}

\subsubsection{Attractor mechanism}
\begin{figure}
\centering
\begin{picture}(150,115)
\put(-10,128){\epsfig{figure=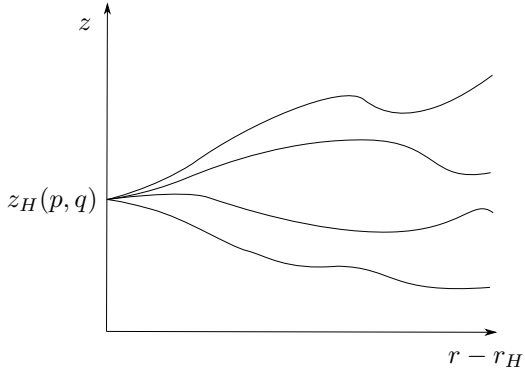,width=.35\textwidth,angle=-90}}
\put(-20,118){$z$}
\put(-47,51){$z_H(p,q)$}
\put(120,-8){$r-r_H$}
\end{picture}
\caption[The attractor mechanism.]{The attractor mechanism for the scalars $z^\alpha$ in terms of the radial distance $r-r_H$ from the horizon. The horizon values $z=z_H(q,p)$ are fixed by the electric and magnetic charges $(q,p)$.\label{fig:JP-AM}}
\end{figure}

The first-order equations contain a lot of physics we can extract without having to solve the system explicitly. In particular, we show that the scalar fields converge to fixed values at the black hole horizon ($\tau \to \infty)$ and these values are given entirely in terms of the conserved electric and magnetic charges, but insensitive to the initital values of the scalars at spatial infinity ($\tau = 0)$. This feature goes by the name of attractor mechanism and is presented pictorially in figure \ref{fig:JP-AM}.

In order to unravel this behaviour, we focus on the second of the equations \eqref{eq:AttractorEqs}. Using this equation, we see that (remember we stick to the upper (minus) signs)
\begin{equation}
\ta {|Z|} {\tau} = -2\e^U G^{\alpha \bar \beta}\partial_\alpha |Z|\partial_{\bar \beta} |Z| \leq 0\,.\label{eq:}
\end{equation}
The inequality on the right-hand side follows because the scalar metric is positive definite. We conclude from this equation that the modulus $|Z|$ of the central charge is a decreasing function of $\tau$. Furthermore, since the modulus $|Z|\geq 0$, it is bounded from below and we conclude it will reach its minimum value $|Z|=|Z|_{\rm min}$ as $\tau \to \infty$ (and we approach the black hole horizon). 

The condition $|Z|=|Z|_{\rm min}$ at the black hole horizon puts severe restrictions on the horizon values of the scalar fields. In particular, the values of the scalars at the black hole horizon are found by solving the minimizing condition:
\begin{equation}
  \pf {|Z|}{z^\alpha} = 0\,.\label{eq:BJ-minimizing_condition}
\end{equation}
The solutions $z^\alpha \equiv z_H^\alpha$ for which \eqref{eq:BJ-minimizing_condition} holds are found without reference to the initial conditions. {}For instance, if we were to continuously vary the values of the scalars at spatial infinity, the values $z_H^\alpha$ for which \eqref{eq:BJ-minimizing_condition} holds would of course not change.\footnote{Barring crossing walls of marginal stability, see for instance \cite{Moore:1998pn,Denef:2007vg}.} We conclude the scalar fields flow to fixed values at the horizon, which only depend on the electric-magnetic charges, but are insensitive to the values of the scalar fields at spatial infinity. This feature is known as the \textit{attractor mechanism}.


{}From the first of the equations \eqref{eq:AttractorEqs}, for $\tau \to \infty$, the warp factor is:
\begin{equation}
\e^{U} = \Conv |Z|_{\rm min} \tau\,, \qquad \text{as $\tau \to \infty$}\label{eq:BJ-eU_AM}
\end{equation}
This behaviour justifies the interpretation of $\tau\to \infty$ as the near horizon region: the timelike Killing vector $\partial/\partial t$, whose norm is given as $\e^{-U}$, becomes null at $\tau \to \infty$. Plugging the near-horizon behaviour \eqref{eq:BJ-eU_AM} into the metric \ansatz{} \eqref{eq:BJ-Ansatz_Metric_N=2_AM}, we see the black hole horizon is a sphere of radius $\pi \Conv^2 |Z|^2_{\rm min}$ and the near horizon geometry ($\tau \to \infty$) has the AdS$_{2}$ $\times$ S$^2$ form \eqref{eq:BJ-Metric_AM_Near_Horizon}, with the radius $r_H = \Conv |Z|_{\rm min}$. The Bekenstein-Hawking entropy of the supersymemtric  black hole is thus:
\begin{equation} 
 S_{BH} = \frac{\pi \Conv^2|Z|_{\rm min}}{G_4}\,,
\end{equation}
and only depends on the black hole charges, not on the values of the scalar fields at infinity. Note that we really need the attractor mechanism in order to have agreement with microstate counting methods to obain the entropy from string theory, as for supersymmetric solutions, one finds the number of ground states only depends on the electric-magnetic charges.

In conclusion, we see that the central charge function is really `central'. It gives the lower bound on the mass of a solution (the value of $|Z|$ at infinity, the true central charge of the supersymmetry algebra, turns up in the BPS bound), giving the ADM mass for supersymmetric solutions, but it also determines the behaviour of the scalars as a gradient flow. The scalars reach fixed values at the black hole horizons, which are found by minimizing the central charge function. We proceed below with a discussion of non-supersymmetric solutions. Some properties are readily extended to extremal solutions, but non-extremal solutions were for a long time an open question.


\subsection{Attractor mechanism for non-supersymmetric black holes\label{ss:BJ-AM_Non-susy_BHs}}
In this section, we discuss some properties of black hole solutions that do not preserve any supersymmetry. The aim is to give a flavour of the literature on the subject, before we start with the discussion of new research in section \ref{s:BJ-MainResults}.

Up till now, we restricted to black hole solutions of $\Cn=2$ supergravity in four dimensions that preserve half of the supersymmetries. We have seen that the scalars for those supersymmetric black hole solutions are driven to fixed values at the black hole horizon by the attractor mechanism and that the attractor flow equations that govern this behaviour are derived from writing the action as a sum of squares.

Following the logic of section \ref{s:BJ-Historical_Overview}, one could be tempted to think that the first-order equations and the attractor behaviour derived from them are solely a manifestation of supersymmetry. Indeed, in the context of the non-gravitational theories discussed before, the saturation of the BPS bound for a supersymmetric solution was related to rewriting the energy functional as a sum of squares from which first-order equations of motion followed.  Since non-supersymmetric solutions do not saturate the BPS bound (${G_4} M\geq {\Conv}|Z|$), we would conclude that the attractor mechanism does not hold for solutions that break supersymmetry.

We explain now that for gravitational theories, this view is too pessimistic. It has been shown in the literature that the attractor mechanism still holds for \textit{extremal} solutions \cite{Gibbons:1996af,Ferrara:1997tw,Goldstein:2005hq,Goldstein:2005rr,Kallosh:2006bt} (including all supersymmetric and some non-supersymmetric solutions). And recently, it has been shown that also non-supersymmetric \textit{extremal} black holes can allow for flow equations analogous to those of the non-supersymmetric case \cite{Andrianopoli:2007gt,Ceresole:2007wx,Cardoso:2007ky}.

\subsubsection{Extremal black holes ($c = 0$) -- attractor mechanism holds}
The flow equations \eqref{eq:AttractorEqs} and the attractor mechanism were first described in \cite{Ferrara:1995ih,Strominger:1996kf}, based on supersymmetry arguments. It was soon noticed  \cite{Gibbons:1996af,Ferrara:1997tw} that the attractor mechanism has a broader validity and is a feature common to \emph{extremal} black hole solutions to $\Cn=2$ supergravity. Examples of non-supersymmetric extremal attractors were worked out in detail from 2005, for instance in refs.~\cite{Sen:2005wa,Goldstein:2005hq,Kallosh:2005ax,Tripathy:2005qp,Giryavets:2005nf}.
We give an argument based on demanding the scalars to be regular (have finite values) at the horizon and follow the clear explanation of \cite{Kallosh:2006bt,Andrianopoli:2006ub}. 

Intuitively, one can understand the attractor behaviour by analogy with systems in classical mechanics. In that context, fixed points of the motion are only reached in the limit $t\to \infty$, where $t$ is the natural evolution parameter (mostly time). {}For extremal black holes, a natural evolution parameter that determines the flow of the sclar fields is the `physical distance' to the black hole horizon. This physical distance is defined as follows. Consider a point $x_0^\mu$ in \spacetime{} outside of the horizon. Choose a path $\gamma$ by keeping the time $t$ and angles $\theta,\phi$ fixed in the black hole metric \eqref{eq:MetricSinh} and varying only the radial parameter from the point $x_0^\mu$ to the black hole horizon. The total metric distance $\rho \equiv \int_\gamma \de s$ measured over this path is called the physical distance (this is the analog of  proper time for spacelike separations). The physical distance can be obtained as a certain function (coordinate redefinition) of the radial coordinate $\tau$. In the near-horizon region of extremal black holes, this redefinition is given as $\rho = r_H\log(\tau/\tau_\star) = r_H\log((r-r_0)/(r_\star-r_0))$, where $\tau_\star$ (or $r_\star$) is the radial position of the reference point $x_0^\mu$ and $r_H = \sqrt{A_H/4 \pi}$, with $A_H$ the area of the event horizon. The near-horizon metric \eqref{eq:BJ-Metric_AM_Near_Horizon} is then written as:
\begin{equation}
\de s^2 = - \frac {\tilde r^2} {r_H^2} \de t^2 + {\de\rho^2} + r_H^2(\de \theta^2 + \sin^2 \theta \de \phi^2)\,,
\end{equation}
It follows that for extremal black holes, the physical distance $\rho$ from any point at $\tau = \tau_\star$ in \spacetime{} to the event horizon is infinite, since $\rho = r_H\log(\tau/\tau_\star)\to \infty$ as $\tau\to \infty$ : extremal black holes develop an infinitely deep `throat', with the horizon at the bottom.  Since the natural evolution parameter $\rho$ goes to infinity as we approach the horizon, one can intuitively understand that the scalars `lose their memory' and attain values at the horizon that are independent of their initial conditions at spatial infinity.



This can be made more precise. {}From the (second-order) equations of motion to the effective action \eqref{eq:EffAction}, it follows that near the horizon, the scalars behave as $z^\alpha \sim G^{\alpha\bar \beta}\pf {V(\phi)}{\phi^{\bar \beta}}\Big{|}_{\phi = \phi_H}\rho^2$. Since $\rho\to \infty$ as we approach the black hole horizon, regularity of the scalar fields demands that the black hole potential reaches an extremum:
\begin{equation}
 \text{$z^\alpha$ regular at horizon} \qquad \Rightarrow \qquad \pf {V_{BH}(z,\bar z)}{z^\alpha}\Big{|}_{z = z_H}=0\,.
\end{equation}
Since the potential $V_{BH}(z,\bar z)$, defined in eq.~\eqref{eq:BJ-BH_Potential_Defn}, depends only on the scalar fields and the electric and magnetic charges ($q_I,p^I$), it follows that the values of the scalars $z = z_H$ for which the potential is minimized are completely determined by the conserved charges ($q_I,p^I$). In particular, the horizon values of the scalars $z_H$ do not depend on the values of the scalars at spatial infinity. We conclude that the attractor mechanism holds for all static, spherically symmetric, \textit{extremal} black hole solutions, supersymmetric or not. Note that the argument is similar to that  for supersymmetric solutions: for supersymmetric solutions, the attractor mechanism follows from minimizing the central charge function $|Z(z,\bar z)|$, for general extremal solutions, one minimizes the effective black hole potential $V_{BH}(z,\bar z)$.

But there is more to say. It took almost ten years after the discovery of the attractor mechanism to realize that not only the attractor behaviour extends to non-supersymmetric solutions, but also the form of the flow equations \eqref{eq:AttractorEqs}. In \cite{Andrianopoli:2007gt,Ceresole:2007wx,Cardoso:2007ky},  it was noted that formally, equations of the form \eqref{eq:AttractorEqs}, where the central charge $|Z|$  is replaced by a scalar-dependent function $W(z,\bar z)$, also give rise to extremal solutions. {}For these solutions, the black hole potential has the form:
\begin{equation}
 V(z,\bar z) = \Conv^2(W^2 +  4 G^{\alpha\bar\beta} \partial_\alpha W \partial_{\bar \beta} W)\,.\label{eq:BJ-BHpot_DWDW}
\end{equation}
When $W=|Z|$, we are describing supersymmetric solutions, but when $W\neq |Z|$ the solutions necessarily breaks all supersymmetry. {}For now we take this novel way of deriving first-order equations for non-supersymmetric, extremal solutions by rewriting the black hole  potential as in equation \eqref{eq:BJ-BHpot_DWDW} as an encouraging result. We thoroughly review the observations of \cite{Andrianopoli:2007gt,Ceresole:2007wx,Cardoso:2007ky} in chapter \ref{c:JP-Gradient_Flow}, when we discuss the form of the flow equations. In particular, we review the literature on this subject in section \ref{ss:FlowEqsLiterature}.


\subsubsection{Non-extremal black holes ($c \neq 0$) -- attractor mechanism cannot hold}
In this case, the attractor mechanism is no longer applicable, see for instance \cite{Garousi:2007zb} for a discussion. One can see this from a similar argumentation as before, by considering the concept of `physical distance' as defined above. Starting at a certain position outside the black hole horizon, it follows that the physical distance to the black hole horizon is finite, for any starting point. {}For the near-horizon region  of non-extremal solutions ($\tau\to\infty$), we have $\sinh(c\tau) \sim \e^{c\tau}$ and we see from table \ref{tab:BJ-Asymptotic_metric} that $\e^{-2U}\sim \e^{2c\tau}$. The coordinate redefinition from $\tau$ to a radial variable $\tilde \rho$ measuring the physical distance between spacelike separated points (still in the near-horizon region), is then given through $\de \tilde \rho = r_H c\, \e^{-c\tau}\de \tau$  and the metric \eqref{eq:BJ-Ansatz_Metric_N=2_AM} takes on the form:
\begin{equation}
  \de s^2 = - \frac {(r-r_+)(r-r_-)} {r_H^2} \de t^2 +\de \tilde \rho^2 + r_H^2(\de \theta^2 + \sin^2 \theta \de \phi^2)\,.
\end{equation}
Integrating the defining condition for $\tilde \rho$, we get $\tilde \rho = r_H (\e^{-c\tau_\star}-\e^{-c \tau})$, where $\tau_\star$ is the radial position of the reference point. We see that as $\tau\to \infty$ (we approach the horizon), $\tilde \rho$ tends to a finite value.

Because the physical distance from a point in \spacetime{} to the horizon is finite, one sees that in general  the scalars do not `lose their memory' and the values of the scalars at the horizon depend on the boundary conditions they obey at spatial infinity. On the possibility of first-order equations for non-extremal solutions, we cannot yet make a conclusion. We immediately proceed with this discussion.

\subsection{Questions about the non-extremal case}
We have studied in some detail supersymmetric and non-supersymmetric \emph{extremal} black hole solutions and we have seen they exhibit a simple structure. The scalars for these solutions reach fixed universal values at the horizon due to the attractor mechanism and explicit solutions for extremal black holes can be obtained through solving first-order differential equations. These equations can be found by writing the action as a sum of squares --- as a result the BPS bound for supersymmetric solutions follows. But what about their \emph{non-extremal} counterparts? Two things we know for sure. We know that non-extremal solutions do not saturate any BPS-type bound: their mass is not minimal for a given amount of charge, these black holes expel the excess mass by emitting Hawking radiation. Furthermore, we know from the argumentation in section \ref{ss:BJ-AM_Non-susy_BHs} that non-extremal black holes do not allow for the attractor mechanism to work: the horizon values of the scalars  vary as one changes the boundary conditions at spatial infinity. Does this imply that no nice features, as finding a set of first-order equations, of the extremal case survive at all? {}For years, physicist have thought, but not proven, the answer to this question to be a convincing `no'. It was commonly believed that non-extremal solutions cannot be obtained from Bogomol'nyi-type\index{Bogomol'nyi!equation} first-order equations\footnote{Remember that we speak of Bogomol'nyi equations whenever a set of first-order equations of motion can be found be rewriting the action as a sum of squares.}. On top of this, people silently assumed that a rewriting of the action \` a la Bogomol'nyi\index{Bogomol'nyi!rewriting of the action} is tantamount to having supersymmetry. We have seen that indeed such a rewriting of the action is possible for spherically symmetric \index{spherically symmetric} black holes which are supersymmetric and even for some non-supersymmetric extremal ones. But what about non-extremal solutions (which necessarily break all supersymmetry)?

Recently, in \cite{Miller:2006ay} Miller\index{Miller}, Schalm\index{Schalm} and Weinberg\index{Weinberg} came up with a surprising result. In the context of a simple theory, Einstein-Maxwell\index{Einstein-Maxwell} theory in four dimensions, they pointed out that also non-extremal \RN\index{\RN} solutions can be obtained from a set of first-order equations that are of Bogomol'nyi-type. Thus first-order equations of motion can be derived for non-extremal solutions, from a clever rewriting of the action as a sum (and difference) of perfect squares\index{sum of squares}.\footnote{A little warning for the reader is appropriate here. In \cite{Miller:2006ay}, the authors refer to first-order equations that can be found by rewriting the action as a sum of squares as `BPS equations'\index{BPS!equation|fn}, even if the solutions to this equations do not preserve supersymmetry. Since this is easily confused with common usage `BPS solutions' for solutions that preserve a certain amount of supersymmetry, we dislike this terminology and opt for the notion `Bogomol'nyi  equations'\index{Bogomol'nyi!equations|fn}.} This contradicts the common lore that such Bogomol'nyi equations would not exist for non-extremal solutions. As a bonus, the authors of \cite{Miller:2006ay} were able to show that, under one mild assumption, these solutions are \emph{not} embeddable as supersymmetric solutions to any supergravity theory, thereby destroying the common belief that the existence of Bogomol'nyi-type first-order equations\index{first-order!equation} are indicative of supersymmetry\index{supersymmetry}.

Given the intriguing results of \cite{Miller:2006ay}, the question that naturally comes to mind is how general their observations really are. Can one extend their results to a more general setup than Einstein-Maxwell theory? Can a \emph{generic non-extremal black hole} solution be derived from first-order equations, either or not of Bogomol'nyi type? Of special interest are the non-extremal black hole solutions to $\Cn=2$ supergravity. And what about objects of different dimensionality: is it possible to reach conclusions for extremal vs. \emph{non-extremal branes}?  We wish to answer these questions in several steps. In the remainder of this chapter, we discuss our work \cite{Janssen:2007rc} with Bert Janssen, Paul Smyth and Thomas Van Riet. We took the results of \cite{Miller:2006ay} at face value and extended them to Einstein-Maxwell\index{Einstein-Maxwell} theory coupled to an extra scalar (dilaton), in arbitrary \spacetime{} dimensions. The Bogolmol'nyi trick can be applied and all non-extremal brane solutions to this setup can be found from first-order equations. The class of branes studied also includes time-dependent\index{time-dependent} solutions and forms a fruitful step to a better understanding of time-dependent solutions in supergravity and even string theory, to date an outstanding issue. {}For the case of generic black hole solutions with more than one scalar field, the question of the possibility of Bogomol'nyi type equations\index{Bogomol'nyi!equations} and a possible richer structure of these equations in terms of a gradient flow\index{gradient flow} reminiscent of the extremal case is deferred to the next chapter.

\section{A first-order formalism for timelike and spacelike brane solutions\label{s:BJ-MainResults}}
In this section, we present the work with Bert Janssen\index{Janssen}, Paul Smyth\index{Smyth} and Thomas Van Riet\index{Van Riet}, reported in \cite{Janssen:2007rc}. First we repeat the arguments of Miller\index{Miller}, Schalm\index{Schalm} and Weinberg\index{Weinberg} \cite{Miller:2006ay} for the appearance of first-order Bogomol'nyi-type  equations\index{Bogomol'nyi!equation} for the non-extremal \RN\index{\RN} black hole, which formed the starting point of our work, in section \ref{ss:BJ-Review_MSW}.\footnote{Note that the Bogomol'nyi trick was first applied in the context of gravitational systems to self-gravitating solutions in the case of cosmic strings \cite{Comtet:1987wi}; see also
\cite{Collinucci:2006sp,Davis:2008ps} for recent discussions. The same procedure can be applied to time-dependent gravitating solutions \cite{Chemissany:2007fg}.}  In section \ref{ss:BJ-WhyDoesItWork} we sketch the condition presented in \cite{Miller:2006ay} for their first-order formalism to be applicable. In section \ref{ss:BJ-PossGener}, we explain that those conditions are met for both radial and time-dependent $p$-brane solutions to Einstein-Maxwell theory coupled to one scalar (a dilaton). In sections \ref{ss:BJtimelike_0} and \ref{ss:BJtimelike_p}, we explicitly perform the Bogomol'nyi trick\index{Bogomol'nyi!trick} for non-extremal\index{non-extremal} and time-dependent solutions\index{time-dependent!solutions} to that system. A conclusion and an extensive discussion can be found in the next section \ref{s:BJ-Discussion_Outlook}.

\subsection{First order equations for non-extremal Reissner-Nordstr\"om}\label{ss:BJ-Review_MSW}\index{\RN{} black hole!first-order eqs.|imp}
We repeat the argument of \cite{Miller:2006ay} for non-extremal black holes in Einstein-Maxwell theory. In the subsequent sections, we discuss our work \cite{Janssen:2007rc} that extends their result to black hole, $p$-brane and cosmological solutions\index{$p$-brane!solution}\index{black hole!solution}\index{cosmological!solution} to Einstein-Maxwell\index{Einstein-Maxwell} theories with an extra scalar (`dilaton') and a scalar-dependent gauge coupling\index{gauge coupling}.

Consider Einstein-Maxwell\index{Einstein-Maxwell} theory in four dimensions, consisting of gravity coupled to an electromagnetic field\index{electromagnetic field}. The Lagrangian\index{Lagrangian!Einstein-Maxwell} for this system is given by:
\begin{equation}\label{eq:Maxwell-Einstein-action-4d}
S=\frac{1}{16\pi G_4}\int \de ^4 x \sqrt{-g} \left(R - \Conv^2 F_{\mu\nu}F^{\mu\nu}\right)\,,
\end{equation}
Again, $\Conv$ is a constant that is convention dependent, cfr.~\eqref{eq:N-=2SUGRA_2}.
Note this is a subsector of the full $\Cn=2$ supergravity Lagrangian\index{Lagrangian!$\Cn=2$ supergravity} \ref{eq:N-=2SUGRA_2}, where only the gravity multiplet is excited. In terms of the $\Cn=2$ supergravity fields, we have put $F_{\mu\nu} \equiv F_{\mu\nu}^0$, and none of the other gauge fields $F^I$ nor the scalars are present, see also table \ref{tab:ReissnerNordstrom}. The \RN\index{\RN} solution to this action was discussed in section \ref{ss:BHelectromagn}. It is a static, spherically symmetric \index{spherically symmetric} solution carrying electric and/or magnetic charges. We wish to see how the \RN\index{\RN} solution is derived from a first-order formalism. {}For the extremal case, the result is already contained in the discussion of the attractor mechanism\index{attractor mechanism}, while the non-extremal case is discussed here.

As in section \ref{s:BJ-History_FoForm_BHs}, we take a static\index{static} spherically symmetric \index{spherically symmetric} \ansatz{}.\index{static}\index{spherically symmetric}\index{\ansatz{}}  This would lead to the metric \eqref{eq:EffAction}, but following \cite{Miller:2006ay}, we make a slight generalization:
\begin{equation}
\de s^2 = -\e^{2U(\tau)}\de t^2 + \e^{-2U(\tau)}\Bigl(\e^{2\C(\tau)}\e^{4\B(\tau)}\de \tau^2 +\e^{2\B(\tau)}\de \Sigma_k^2 \Bigr)\,,\label{eq:MetricSinh}
\end{equation}
again in terms of a radial variable $\tau$. The function $C(\tau)$ expresses the parametrization invariance of the radial direction and corresponds to the freedom in choosing a radial coordinate\index{radial coordinate}. Later, we will see that by fixing the parametrization invariance of the $\tau$ direction by putting $C=0$, the equations of motion dictate $\e^{2\B} = c^2/ \sinh^2(c\tau)$ and we reproduce the earlier metric \ansatz{} \eqref{eq:BJ-Ansatz_Metric_N=2_AM}. We give the Ricci scalar of the metric \eqref{eq:MetricSinh} for later reference:
\begin{equation}
R = 2\, \e^{2(U-\B-\C)}\Bigl(\ddot{U} - \dot{U}^2
-\dot{U}\dot{C} -2\ddot{\B} + 2\dot{\B}\dot{\C} +
\dot{B}^2 \Bigr) + 2 \e^{2(U-\B)}\,.\label{eq:BJ-RiciScalar_MSW}
\end{equation}

{}For the gauge field\index{gauge field}, we only consider turning on electric charge. The earlier gauge-field \ansatz{} \eqref{eq:BJ-Ansatz_GaugeField_AM} for the Einstein-Maxwell truncation, with one electric charge\index{electric charge} $\QE \equiv q^0$ and all other charges put to zero, reads (see table \ref{tab:ReissnerNordstrom})
\begin{equation}\label{eq:BJ-RN_GaugeField_Ansatz}
 F = \QE \e^{2U + \C} \de t \wedge \de \tau\,.\
\end{equation}
As explained in the discussion of section \ref{ss:BJ-AM}, the above \ansatz{} for the gauge field allows us to obtain an effective action for the warp factor $U$.
Plugging in the \ansatz{} into the equations of motion, one sees that they can be derived from the following effective action:
\begin{align}
S_{\rm eff} =&\int \de \tau\,\,\left(e^{-\C}\Bigl(\dot{\B}^2-\dot{U}^2\Bigr)+\e^{\C}\Bigl(\e^{2\B} - \,\Conv^2\QE^2\e^{2U}\Bigr) \right)\\
&+\left[e^{-C}(- 2\dot B+\dot U )\right]_{\tau=0}^{\tau =\infty} \,,\label{eq:Uaction}
\end{align}
where a dot again denotes a derivative w.r.t. $\tau$. In the rest of this section, we drop the prefactor $\frac1{8\pi G_4}$ and we keep the boundary terms  explicit.

We know that the field $C$ does not appear with a derivative in the action and is therefore not a propagating degree of freedom. This was to be expected, since it expresses the
reparameterization invariance of the radial direction. It acts as a Lagrange multiplier, enforcing a Hamiltonian constraint. In the gauge $\C=0$, the constraint becomes:
\begin{equation}
\left(\dot{\B}^2-\e^{2\B}\right)-
\left(\dot{U}^2-\Conv^2 \QE^2\e^{2U}\right)= 0\,.
\label{eq:constraintU}
\end{equation}
Note that we shuffled some terms for later convenience. By using the correspondence in table \ref{tab:ReissnerNordstrom}, we can compare to the discussion of section \ref{ss:BJ-AM} and the effective action \eqref{eq:EffAction} obtained for a more general $\Cn=2$ setup. Remember that the scalars are constants now. The difference in the effective action and the constraint equation lies in the new terms involving $B(\tau)$, whereas earlier we had the non-extremality parameter $c^2$. We see below that plugging in the solution for $B$ exactly reproduces this term.

One can obtain the \RN\index{\RN} solution by writing the effective action as a sum of squares, both for extremal solutions as for non-extremal solutions. Let us discuss these two cases in detail.

\subsubsection{The extremal solution}
Obtaining first-order equations for extremal solutions, follows the arguments of section \ref{ss:BJ-AM}. We go over the procedure quickly, since this is just an extension of the general setup for $\Cn=2$ supergravity with the field $B(\tau)$. Applying the Bogomol'nyi trick as in section \ref{ss:BJ-AM}, gives us:
\begin{equation}
 S_{\rm eff} =\int \de \tau \,\Bigl( \left(\dot{B}\pm \e^{B}
\right)^2-\left(\dot{U}
\pm\Conv \QE\e^{U}\right)^2\Bigr) + Z_{\rm MSW}\,.\label{eq:Uaction_ext}
\end{equation}
The boundary term is given by:
\begin{equation}
Z_{\rm MSW} = \int\de \tau \frac{\de}{\de \tau }\left(- 2\dot B +\dot U \mp 2\e^B \pm 2\Conv \QE \e^U \right)\,.
\end{equation}
The first two terms follow from completing the squares, the last two carry over from \eqref{eq:Uaction}. As before, demanding that the squared terms in the action vanish gives rise to first-order equations of motion.  Again, the solution with the lower signs gives is not a well-behaved black hole solution (negative ADM mass). We restrict to the equations with the upper signs. Solving the first-order equations then gives us:
\begin{equation}
 \e^{-U} = 1 + \Conv \QE \,\tau\,,\qquad \e^{-B} = \tau\,,\label{eq:BJ-FO_EQS_MSW_RN_EXT}
\end{equation}
where we fixed the integration constant by demanding the asymptotics of table \ref{tab:BJ-Asymptotic_metric}. In the last equation, we absorbed the integration constant in a shift of the origin of the $\tau$ axis. If we perform the change of coordinates $r = 1 / \tau$, then we see that this solution is none other than the electrically charged extremal \RN\index{\RN} solution of chapter \ref{c:BH_Playground}:
\begin{align}
 \de s^2 &= -H(r)^2\de t^2 + H(r)^{-2}(\de r^2 + r^2 (\de \theta^2 + \sin^2 \theta \de \phi^2))\,,\nonumber\\
 F& = \de (H(r) \,\de t) = \frac{\Conv \QE} {r^2} \de t \wedge \de r\,,\qquad
H(r) = 1+ \frac {\Conv \QE}{r}\,.
\end{align}
We see that the ADM mass $M$ of the black hole solution is given as $G_4 M = \Conv \QE$.
In other words, the solution saturates the BPS bound, as can also be seen
by evaluating the action \eqref{eq:Uaction_ext} on the solution and remembering the standard relation between (minus) the action and the (potential) energy for static\index{static} solutions (the only non-zero contribution comes from the boundary term and reads $G_4 M \sim Z_{\rm MSW} =  \Conv \QE \e^U|_{\tau = 0} = \Conv \QE$).


\subsubsection{The non-extremal solution}
What about the non-extremal case? The effective action \eqref{eq:Uaction} for the Reissner-Nordstrom solution can be written, up to boundary terms whose form we do not further discuss, by completing the squares in the more general way:
\begin{equation}
S_{\rm eff}=-\int \de \tau\,\,\left(\left(\dot{B}+\sqrt{\e^{2B}
+\beta^2}\right)^2 -\left(\dot{U}+ \sqrt{
\Conv^2 \QE^2\e^{2U}+\beta^2} \right)^2\right)\,. \label{eq:BJ-RN_NonExtr_SumSqs}
\end{equation}
The addition of a constant $\beta\geq0$ under the square roots still allows to rewrite the action as a difference of squares, since the cross-terms are again total derivatives.  {}{}From now on, we only consider the sign possiblity in each squared term that gives a well-behaved black hole solution. Note that this specific rewriting of the effective action as a sum/difference of squares only holds when kinetic and potential terms decouple. {}For a generic effective supergravity action as \eqref{eq:EffAction}, the potential $V_{BH}$ is a complicated function of the scalar fields and a naive sum/difference of orthogonal terms is not possible.
One could also wonder if it is possible to consider two different constants under the square root. However, this is  undesirable, since it would require the addition of an infinite boundary term. Moreover, one can check the Hamiltonian constraint dictates these constants to be the same.

A stationary point of this action is given by a solution to the following first-order differential equations:
\begin{align}
\dot{U}&= -
\sqrt{\Conv^2 \QE^2\e^{2U}+\beta^2}
\,,\nonumber\\
\dot{\B}&=-\sqrt{\e^{2\B} +\beta^2}\,.
\label{eq:BJ-1stOrderEqs_NonExtr_RN}
\end{align}
The solutions of the first-order equations \eqref{eq:BJ-1stOrderEqs_NonExtr_RN} are
\begin{equation}
\e^{-U} = \Conv \QE\frac{\sinh(\beta \tau +
c_{U})}{\beta}\,,\qquad\e^{-\B} = \frac{\sinh (\beta \tau)}\beta \,,
\label{eq:BJ-MSW_RN_FirstOrderEqs}
\end{equation}
where $c_U$ is a constant and we put the integration constant in the solution for $B$ to zero by shifting the origin of the $\tau$-axis. If we compare to the metric \ansatz{} \eqref{eq:BJ-Ansatz_Metric_N=2_AM}, we see the constant $\beta$ is just the non-extremality parameter introduced earlier:
\begin{equation}
 \beta = c\,.
\end{equation}
Substituting the above solution into our \ansatz{} may not yield a metric that is easily recognized. Therefore, we go to standard coordinates:
\begin{equation}\label{transformation}
r(\tau)= \Conv \QE \frac{\sinh(\beta \tau + c_{U})}{\sinh (\beta \tau) }\,,
\hspace{2cm} \tilde t = \frac{1}{\Conv \QE }\frac{\beta}{\sinh c_U} t\,,
\end{equation}
such that the area of the two-sphere slicing in the metric \eqref{eq:MetricSinh} is given by $4\pi r^2$. We can then also use our intuition that $c$ is a non-extremality parameter. Defining a mass $M$ through
\begin{equation}
\beta^2 = c^2 =  G_4^2 M^2 -\Conv^2 \QE^2\label{eq:BJ-MSW_BPS_bound}\,,
\end{equation}
we see that the solution takes the form \eqref{eq:BJ-RN_metric} for the non-extremal
electrically charged \RN\index{\RN} solution
\begin{align}
\de s^2 &= -\Delta(r)\de t^2 + \Delta(r)^{-1} \de r^2 + r^2 \de \Omega^2\,,\\
\Delta(r)&=1 - \frac{2 G_4 M} r +  \frac{\Conv^2\QE^2}{r^2}\,.
\end{align}
We recognize the constant $M$ as the ADM mass, which obeys the BPS bound $G_4 M \geq \Conv^2 \QE$ from equation \eqref{eq:BJ-MSW_BPS_bound}.

\subsection{Mechanism behind first-order equations for non-extremal \RN\label{ss:BJ-WhyDoesItWork}}
We can find first-order equations for the scalars of the non-extremal \RN black hole by starting from the equations for the extremal case through  a constant parameter $\beta$ (eq.\eqref{eq:BJ-FO_EQS_MSW_RN_EXT} vs.~eq.~\eqref{eq:BJ-MSW_RN_FirstOrderEqs}). We investigate how general this observation is. The idea is taken from \cite{Miller:2006ay} and is the following. Say we have an effective action for a set of scalar fields $q^i$, such that we have a solution that obeys first-order equations $\dot q^i = f^i(q)$. When can we find another solution that obeys similar equations $\dot q^i = \tilde f^i(q)$, with $\tilde f^i =\sqrt{(f^i)^2 + (\beta^i)^2}$, with $\beta^i$ a set of constants?



Consider an action with a potential for a collection of scalar fields $q^i$
\begin{equation}
  S = \int \de \tau \left( \cg_{ij}(q) \dot q^i \dot q^j + \cv(q)\right)\,,\label{eq:BJ-Action_q_V}
\end{equation}
where the fields depend on $\tau$ as $q^i(\tau)$. We allow the metric $\cg$ to have indefinite signature. (This is the straightforward generalization of the effective action \eqref{eq:Uaction} for the \RN black hole, which has scalars $q^i = (U,\C)$). Suppose the potential can be written in terms of a vector with components $f^i(q)$ as 
\begin{equation}
\cv(q) = \cg_{ij} f^i f^j\,,\label{eq:BJ_Why_V=Gff}
\end{equation}
such that the functions $f^i$ are determined by the gradient of a function $\cw(q)$:
\begin{equation}
  \cg_{ij} f^j = \pf {\cw (q)}{q^i}\,.\label{eq:BJ-f_W}
\end{equation}
Typically, this rewriting in terms of $\cw (q)$ is possible for a supersymmetric solution (think of the role the central charge $Z$ played before) and also applies to the extremal \RN black hole. The action can then be written in terms of squared expressions\footnote{This is not necessarily a `sum' of squared terms, since we have not specified the signature of the metric $\cg$ yet. We only get a true sum of squares when the metric is positive definite.} and a boundary term:
\begin{equation}
  S = \int \de \tau  \,\cg_{ij}\left( \dot q^i -f^i(q) \right)\left( \dot q^j-f^j(q)\right) - 2 {\cw}\Big{|}_{\tau=0}^{\tau = \infty}\,.\label{eq:BJ-Action_q_W}
\end{equation}

We ask the question if it is possible to write the action (and the equations of motion) in terms of a second set of functions $\tilde f^i(q)\neq f^i(q)$ such that also
\begin{equation}
  \cg_{ij}\tilde f^i = \pf {\tilde \cw(q)}{q^i}\,.\label{eq:BJ-Condition_Vectors_q}
\end{equation}
Clearly, this can be done if  $\cv= \cg_{ij}f^i f^j = \cg_{ij}\tilde f^i \tilde f ^j$.
{}For the rest of the discussion, we choose to work in terms of the one forms $f_i = \cg_{ij} f^j$ and $\tilde f_i = \cg_{ij} f^j$. Then the condition for \eqref{eq:BJ-Condition_Vectors_q} to hold reads:
\begin{equation}
   \cg^{ij}(f_i f_j - \tilde f_i \tilde f_j)=0\,.\label{eq:BJ-Condition_f_ftilde}
\end{equation}
If the (inverse) metric $\cg^{ij}$ is positive definite, we have $f_i = \pm \tilde f_i$, which means no new rewritings would open up. However, in gravity theories, the effective action can have a metric $\cg$ of indefinite signature, such that null vectors $\beta^i$ exist. This opens up new possibilities.

We consider the metric $\cg$ to have indefinite signature and make two simplifying assumptions:
\begin{enumerate}
\item \emph{\underline{The metric $\cg_{ij}$~is diagonal.}}  Then we can solve
\eqref{eq:BJ-Condition_f_ftilde} in terms of a null vector  $\beta_i$ of the inverse metric (i.e.~$\cg^{ij}\beta_i \beta_j =0$) by demanding that
\begin{equation}
\tilde f_i = \pm \sqrt{\left(f_i\right)^2 + \left(\beta_i\right)^2} \qquad\text{(no sum)}\,.\label{eq:BJ-f_ftilde_beta}
\end{equation}
We take the null vector $\beta$ to be constant.
\item \emph{\underline{Every function $f_i$ depends on the corresponding $q^i$ only.}}
It immediately follows from \eqref{eq:BJ-f_ftilde_beta} that $\tilde f_i$ can indeed be written as in equation \eqref{eq:BJ-Condition_Vectors_q}. When the metric $\cg_{ij}$ is constant, this can be rephrased as having a potential $\cv$ which is a sum of terms that each depend on one $q^i$, as follows from eq.\ \eqref{eq:BJ_Why_V=Gff}.
\end{enumerate}

\textit{When these two conditions are met, a solution with equations of motion $\dot q^ i = f^i(q)$ automatically guarantees that we have another solution with equations of motion $\dot q^ i = \tilde f^i(q) = \sqrt{(f^i(q))^2 + (\beta^i)^2}$, by the Bogomol'nyi rewriting \eqref{eq:BJ-Action_q_W} in terms of $f$ and $\tilde f$.}\footnote{The effective action \eqref{eq:Uaction} of the \RN\index{\RN} black hole is a special case of this setup. The coordinates are $q^i= (U,\B)$, the metric is $\cg = {\rm diag} (1,-1)$ and we have $f_i = (\e^\B, \Conv \QE \e^U)$. The two conditions above are satisfied and the null vector $(\beta,\beta)$ gives another Bogomol'nyi rewriting in terms of $\tilde f_i$ as in \eqref{eq:BJ-f_ftilde_beta}, giving the equivalent action \eqref{eq:BJ-RN_NonExtr_SumSqs}.} In chapter \ref{c:JP-Gradient_Flow}, we turn to the question when a rewriting of the action is possible in terms of functions $\tilde f_i = \partial_j \tilde \cw$, without any simplifying assumptions.


\subsection{Possible generalizations\label{ss:BJ-PossGener}}
We now motivate three ways in which we can generalize the derivation of Bogomol'nyi equations for the \RN\index{\RN} example. We consider an extension to Einstein-Maxwell\index{Einstein-Maxwell} theory coupled to a \emph{dilaton}\index{dilaton} in an \emph{arbitrary number of \spacetime{} dimensions}, which allows for \emph{$p$-brane} solutions which can have \textit{radial} or \textit{time} dependence.

\subsubsection{How can we generalize?}
We have discussed Bogomol'nyi equations for the \RN\index{\RN} black hole, the most general static\index{static} solution to Einstein-Maxwell theory (the theory of general relativity coupled to a vector field or `photon'). 

What can we say about the generality of the argument? Does it extend to other theories than Einstein-Maxwell, which are more capable of describing real-world phenomena? Do such theories also hide a first-order structure in their defining equations of motion? And what class(es) of theories can form a starting point for a systematic study of the problems raised by these questions? In \cite{Janssen:2007rc}, we considered these issues and saw three possible ways in which one would naturally like to seek an extension. We restrict to solutions that are determined by one coordinate, much like the radial dependence of the static\index{static} \RN\index{\RN} solution.  Then we see three natural ways in which one can ask if a rewriting of the equations of motion \`a la Bogomol'nyi into first-order equations is possible:
\begin{enumerate}
\item \emph{Generalize the kind of solutions}. This does not (yet) require to go beyond Einstein-Maxwell theory, but is a question of taking other classes of \ansatze{}. If we consider dependence on one coordinate only, why should we stay with radial dependence? We can try to have a time coordinate as the one variable in the problem. Conceptually, this is interesting since finding a simplified description for time-dependent solutions is much sought after in the literature: time-dependent backgrounds are especially ill  understood in string theory. A handle on implementing time-dependence in the formalism of the previous section can be found from knowledge of the domain wall--cosmology correspondence, as was briefly touched upon in the introduction to this chapter, section \ref{s:BJ-Introduction}. Domain walls are codimension-one objects which only depend on the transverse radial coordinate, while cosmologies describe the time evolution of a codimension-one spatial slice. The correspondence between these two dictates that both seemingly different types of solutions have similar properties, as for instance an analogous description in terms of Bogomol'nyi equations. The differences between the forms of these equations lies in but a few minus signs. These similarities and subtle sign differences can be proven by performing a double Wick rotation, mapping the radial variable of the domain wall to the time variable of the corresponding cosmological solution and vice versa. We use this idea and describe time-dependent metrics and gauge fields, related to their static\index{static} counterparts by a Wick rotation of the radial and time coordinates.

In addition, we consider solutions for which the transverse space need not necessarily be sliced by spheres as in the \RN\index{\RN} case, see the metric \eqref{eq:MetricSinh}. We generalize to transverse spaces whose hypersurfaces of constant radius/time coordinate are maximally symmetric spaces (i.e.\ hyperbolic, flat or spherical).
\item \emph{Generalize to more involved theories}. {}From a phenomenological point of view, it is certainly interesting to investigate theories with a richer spectrum than Einstein-Maxwell theory. On the bosonic side, we immediately think of trying to perform the Bogomol'nyi trick for theories including more vector and scalar fields or even for a bosonic truncation of a full-fledged supergravity system.\footnote{Remember that most focus is given to bosonic background, with vanishing fermions. E.g.~the black holes we study in this thesis fall under this class.} This links to our original motivation of understanding black hole solutions in string theory and supergravity, as a starting point for testing string theory as a quantum theory of gravity. {}For simplicity, we initiate the study of more general theories by considering the toy model consisting of Einstein-Maxwell theory with an extra scalar, dubbed dilaton. The dilaton is taken to dictate the value of the electromagnetic coupling through the exponential $1/ g^2 = \e^{a\phi}$, with $a$  a constant and $g$ the electromagnetic coupling. The addition of this type of scalar is well-motivated, since Einstein-Maxwell-dilaton theory can be obtained as a truncation of many supergravity theories.
\item \emph{Generalize to other dimensions}, both of \spacetime{} and of the branes we are considering (remember we were only talking about black holes, or 0-branes, up till now). In the light of string theory and its low energy limit, supergravity, there is no reason we should restrict to four-dimensional \spacetime{} or even zero-dimensional objects (as black holes). Many higher dimensional supergravity theories do exist, going up to eleven dimensions. And while conventional wisdom tells us that general relativity in four dimensions with static\index{static} spherical symmetry (only radial dependence see chapter \ref{c:BH_Playground}) has unique black hole solutions, more possibilities appear in higher dimensions. Especially higher dimensional objects are of prime importance. Think of the D-branes of type II string theory, or strings themselves, for instance. Therefore we  upgrade the number of \spacetime{} dimensions from four to a general number $D\geq4$ and study $p$-brane solutions. These $p$-brane solutions are defined below in detail. {}For now, know that $p$ is an integer and think of $p$-branes as objects tracing out a ($p+1$)-dimensional \worldvolume{} in \spacetime{}. {}From the physical interpretation in terms of the \worldvolume{}, we consider all cases $p\geq -1$.
\end{enumerate}

\subsubsection{Setting the stage}
Taking the above three points together, we have motivated the study of $p$-brane solutions to gravity, coupled to a $(p+1)$-form gauge field and a dilaton in $D$ dimensions. These $p$-branes solutions can be static\index{static} and  have radial dependence, or can have time dependence, according to the first generalization we wish to make. We first  give the action of the theory and then consider the details of the radial- or time-dependent $p$-brane \ansatz{}e.

We start with the action of general relativity in $D$ dimensions, coupled to a ($p+2$)-form field strength and a dilaton:
\begin{equation}\label{eq:BJ-action}
S=\frac{\Conv_D^2}{4\pi G_D}\int \de^D x\sqrt{-g}\Bigl( \tfrac{1}{4\Conv_D^2}R_D-
\tfrac{1}{2}(\partial\phi)^2-\tfrac{1}{2 (p+2)!}\e^{a\phi}F_{(p+2)}^2
\Bigr)\,,
\end{equation}
where $R$ is the Ricci-scalar, $G_D$ is the $D$-dimensional Newton constant, $\phi$ is a scalar field (the dilaton)
and $F_{(p+2)}$ is the field strength of some $(p+1)$-form if $p\geq -1$. Again, $\Conv_D$ is a convention-dependent constant. In order for the gauge field and dilaton to have standard dimensions, one takes $\Conv_D$ to have dimensions of $L^{\frac{D-2}2}$, where $L$ is a length and we set $\hbar = c = 1$. Often, the notation $\kappa_D = \Conv_D/\sqrt{2}$ is used to denote this gravitational coupling.
{}For the special case $p=-2$, the field strength $F_{(p+2)} = F_0$ is a zero-form, in
other words a scalar. In that case we consider $F_{(p+2)}^2$ to
be a cosmological term (giving rise to a scalar potential). The parameter $a$ is
fixed and is called the dilaton coupling. \textit{Note that in the following, we always drop the normalization $\frac{\gamma^2_D}{4\pi G_D}$ in front of the action, as it plays no role in the mathematical results we derive}.
%

The potential $A_{(p+1)}$ is sourced by extended objects. The equations of motion
derived from the action \eqref{eq:BJ-action} allow for electrically charged  $p$-branes
and magnetically charged ($D-p-4$)-branes. We only consider electric charge. By electrically charged $p$-branes, we mean solutions to the action \eqref{eq:BJ-action} which:
\begin{itemize}
 \item depend on one coordinate, which can be either timelike or spacelike.
 \item have a  $(p+1)$-dimensional \worldvolume{}. In this sense we call them $p$-branes. The \worldvolume{} is taken to be flat.
 \item  only carry electric charge under the gauge field. This means the gauge potential lives on the \worldvolume{}, i.e.\ the pullback of the ($p+2$)-form $A_{(p+2)}$ on the transverse space is zero. In a less abstract formulation, letting $z_1\ldots z_{p+1}$ denote the coordinates on the \worldvolume{}, the only non-vanishing component of the gauge potential is $A_{z_1\ldots z_{p+1}}$.
\end{itemize}

We now discuss in detail the \ansatz{} and start with the metric.
The brane solutions we wish to consider can be stationary or time-dependent. The metric of a \emph{stationary $p$-brane} is given by
\begin{equation}\label{eq:stationary_metric}
\de s^2_D=f^2(r)\,\eta_{\mu\nu}\de x^{\mu}\de x^{\nu} + {g^2(r)}\,\de
r^2 + {h^2(r)}\,\de\Sigma_k^2\,,
\end{equation}
where $\eta$ is the usual Minkowski metric in $p+1$ dimensions,
$\eta=\text{diag}(-,+, \ldots, +)$, and $\de\Sigma_k^2$ is the metric
of a $d$-dimensional maximally symmetric space with unit curvature
$k=-1,0,1$, such that the Ricci scalar is given by $R_d = k d
(d-1)$.
 When $k=1$ the solutions possess a rotational
symmetry and can be asymptotically flat (in contrast to $k=-1$). {}For
$D=10$ and specific values of $a$ and $p$, the solutions can correspond
to D-branes in string theory. {}For example static, spherically symmetric \index{spherically symmetric} black holes in four dimensions fall under this class. They correspond to $D=4,\,p=0$ and $k=1$ (spherical symmetry).

The metric of the \emph{time-dependent $p$-branes} is similar
\begin{equation}\label{eq:timedependent_metric}
\de s^2_D=f^2(t)\,\delta_{\mu\nu}\de x^{\mu}\de x^{\nu} -g^2(t)\,\de
t^2 + {h^2(r)}\,\de\Sigma^2_k\,,
\end{equation}
where $\delta$ is the usual flat Euclidean metric in $p+1$
dimensions, $\delta=\text{diag}(+,+,\ldots,+)$. In the
$k=-1$ case the transverse space possesses a Lorentzian symmetry and
can be asymptotically flat (in contrast to $k=+1$ solutions). These
solutions are the spacelike branes (S-branes) introduced by Gutperle
and Strominger \cite{Gutperle:2002ai}, who conjectured that such
branes correspond to specific time-dependent processes in string
theory. As stressed before, these solutions should be understood in terms of their \worldvolume{}.

{}{}For the stationary solutions, the \worldvolume{} has Minkowski signature, which makes
the time direction stand out as a natural evolution parameter of a $p$-dimensional extended object. Time-dependent solutions,  on the other hand, have a Euclidean \worldvolume{}, which puts all the internal directions on the same footing. No unique coordinate can be picked out as an evolution parameter.  Instead, it is better to recognize the time-dependent solutions on the level of the \worldvolume{} and consider these objects to be ($p+1$)-dimensional. They can then be interpreted as `instantaneous' solutions stretching out in $p+1$ dimensions in space, but localized in time, much like stationary $p$-branes are localized in the radial direction.

Finally, note that the \ansatze{} for stationary branes \eqref{eq:stationary_metric} and time-dependent branes \eqref{eq:timedependent_metric} can be written in the same form, by including a parameter $\epsilon\pm 1$:
\begin{align}\label{eq:BJ-time+stat_metric}
\de s^2_D&={f^2(u)}\,\eta^\epsilon_{\mu\nu}\de x^{\mu}\de x^{\nu} +\epsilon{g^2(u)}\,\de u
^2 + {h^2(u)}\,\de\Sigma^2_k\,,\\
\eta^\epsilon&=\text{diag}(-\epsilon, +1,\ldots,+1) \nonumber\,.
\end{align}
Depending on the value of the parameter $\epsilon$, we describe the stationary or time-dependent case.
When $\epsilon = +1$, the coordinate $u = r$ is a radial coordinate and we revert to the stationary \ansatz{}
\eqref{eq:stationary_metric}. When $\epsilon = - 1$, the coordinate $u = t$ is a
timelike coordinate and we reproduce the time-dependent $p$-brane \ansatz{} \eqref{eq:timedependent_metric}.

{}From now on we call the stationary branes with spherical
slicing ($k=+1$) \emph{timelike} branes and the time-dependent
branes with hyperbolic slicing ($k=-1$) \emph{spacelike} branes. We focus on these cases, as these are the only ones with Minkowski asymptotics. We summarize in table \ref{tab:BJ-Slicings}.\footnote{This terminology refers to the \worldvolume{} of the object containing a time direction
or not and should not be confused with the type of coordinates these solutions depend on.
I.e.~timelike $p$-branes depend on a radial coordinate and have Minkowskian \worldvolumes{}s
in our terminology, while spacelike $p$-branes have Euclidean \worldvolume{}s and are time-dependent.}

\ctable[
cap =An overview of radial and time-dependent brane \ansatze{}.,
caption= {An overview of the different brane \ansatze{} covered by the metric \eqref{eq:BJ-time+stat_metric}. We call the case with $\epsilon=k=+1$ timelike branes and $\epsilon=k=-1$ spacelike branes.}
,
label = tab:BJ-Slicings,
width = \textwidth,
pos = ht!,
center
 ]
{l>{\small}X>{\small}X}{
}{ \FL
&$\epsilon=+1$& $\epsilon = -1$\NN
\cmidrule(r){1-1}\cmidrule(rl){2-2}\cmidrule(l){3-3}
Brane type& \textbf{STATIONARY}& \textbf{TIME-DEPENDENT} \NN
\cmidrule(r){1-1}\cmidrule(rl){2-2}\cmidrule(l){3-3}
Coordinate $u$&radial ($u=r$)&timelike ($u=t$)\NN[2mm]
\cmidrule(r){1-1}\cmidrule(rl){2-2}\cmidrule(l){3-3}
Transverse geom.\ ($\Sigma_k$)&&\NN 
$\qquad \qquad \qquad \quad k=-1$&Hyperboloid&$\boxed{\text{AdS}}$ (\textit{`spacelike branes'})\NN
%
\cmidrule(rl){2-2}\cmidrule(l){3-3}
$\qquad \qquad \qquad \quad k=0$&Euclidean plane&Minkowski\NN
\cmidrule(rl){2-2}\cmidrule(l){3-3}
$\qquad \qquad \qquad \quad  k=+1$&\framebox{Sphere} (\textit{`timelike branes'})
& dS\NN
\LL
}


All the other possible slicings are also covered here, but we choose to highlight only these two cases, as
they can have a natural interpretation in string theory (D-branes and S-branes, respectively).

\subsubsection{Towards a Bogomol'nyi approach to timelike and spacelike $p$-branes}
Now that we have set out the lines, we can go ahead and start looking for the possibility of applying a
Bogomol'nyi trick on the action \eqref{eq:BJ-action}, following the ideas of \cite{Miller:2006ay}.
This should then lead to first-order equations from which we can derive the explicit form of the $p$-brane solutions
(\ref{eq:BJ-time+stat_metric}). To build up to the main results of our paper
\cite{Janssen:2007rc}, we follow the logical line of the argument given in
the three points of the beginning of this section. First, we generalize the first-order equations for
the non-extremal \RN\index{\RN} solution (spacelike 0-branes in four dimensions) to timelike 0-branes,
in section \ref{ss:BJtimelike_0}. We immediately show that the same technique allows one to
rederive the S0-brane solution of Einstein--Maxwell theory \cite{Gutperle:2002ai}.
In the same section we obtain the straightforward generalization of these solutions to
Einstein-Maxwell theory coupled to a dilaton. In section \ref{ss:BJtimelike_p}
we consider a general number of $D$ \spacetime{} dimensions and consider a $p$-brane \ansatz{},
with $p\geq -1$. We explain how the Bogomol'nyi equations for the spacelike and timelike $p$-branes
can be obtained from  the BPS equations for $(-1)$-branes via an uplifting procedure along the brane's
\worldvolume{}.

\subsection{Generalizing to time-dependent dilatonic 0-branes\label{ss:BJtimelike_0}}
First we consider the inclusion of time-dependent 0-branes
 solutions to Einstein-Maxwell theory, and then we describe such solutions with an extra scalar (dilaton) in the theory.
 The novel feature here is that we are able to derive those solutions from first-order Bogomol'nyi equations, while the solutions themselves were previously known in the literature.

\subsubsection{Adding time dependence}
A first step towards a decription of first-order Bogomol'nyi equations for radial- and time-dependent $p$-brane solutions is to generalize the argument for \RN\index{\RN} black holes in Einstein-Maxwell theory to time-dependent 0-brane solutions.

Einstein-Maxwell theory in four dimensions is described by the action \eqref{eq:Maxwell-Einstein-action-4d} and has electric and magnetic $0$-branes solutions. Motivated by the discussion in section \ref{ss:BJ-PossGener}, we consider the following \ansatz{} for the $0$-brane metric which turns out to be useful (compare with \eqref{eq:BJ-time+stat_metric})
\begin{equation}
\de s^2 = -\epsilon\, \e^{2U(u)}\de z^2 +
\e^{-2U(u)}\Bigl(\epsilon \,\e^{2\C(u)}\e^{4B(u)}\de u^2 +\e^{2\B(u)}\de \Sigma_k^2
\Bigr)\,.\label{eq:BJ-Ansatz_Metric_0-brane_time+stat}
\end{equation}
The one coordinate $u$ on which the solution depends can be either a radial coordinate or a timelike coordinate. As in equation \ref{eq:BJ-time+stat_metric}, the parameter $\epsilon = \pm 1$ determines which of the two cases applies. If $\epsilon=+1$ then $z$ is time ($z=t$) and the metric is static\index{static} ($u$ is a spacelike coordinate). {}For spherical slicings ($k=+1$) this is the appropriate Ansatz for a black hole, where $u$ is then some function of the familiar radial coordinate $r$, cfr.~the black hole \ansatz{} \eqref{eq:MetricSinh}. When $\epsilon=-1$ the metric is time-dependent and for hyperbolic slicings $(k=-1)$ this is the appropriate Ansatz for an S$0$-brane \cite{Gutperle:2002ai} with a one-dimensional Euclidean worldvolume labelled by $z$, and $u$ is some function of the time-coordinate $\tau$ used in the Milne patch
of Minkowski \spacetime{}. 

The Ricci scalar\index{Ricci!scalar} for the metric \eqref{eq:BJ-Ansatz_Metric_0-brane_time+stat} is given by
\begin{equation}
R= 2\,\epsilon\, \e^{2(U-2\B-\C)}\Bigl(\ddot{U} - \dot{U}^2
 -2\ddot{\B} + 2\dot{\B}\dot{\C} +
\dot{B}^2 \Bigr) + 2 k \e^{2(U-\B)}\,,
\end{equation}
where a dot indicates a derivative with respect to $u$. The \ansatz{} for the gauge field is also the straightforward generalization of the \RN\index{\RN} case. {}For \emph{electrical} solutions, the Maxwell and Bianchi equations
are solved by
\begin{equation}
F = \QE \e^{2U+C} \de u \wedge \de z\,. \label{eq:BJ-mfs}
\end{equation}
As in the case of the \RN\index{\RN} black hole discussed in section \ref{ss:BJ-PossGener}, it is
possible to derive the equations of motion for $U,B,C$ from a one-dimensional effective action,
this time in terms of the coordinate $u$. The calculation goes as before and it is
straightforward to see that the equations of motion can be obtained
by varying the following action (up to boundary terms)
\begin{equation}\label{eq:BJ-blackhole-action}
S=\int \de u\,\,\left(\e^{-C}\Bigl(\dot{\B}^2-\dot{U}^2\Bigr)+\e^{\C}\Bigl(\epsilon\,
k\,\e^{2\B} - \epsilon \,\Conv^2 \QE^2\e^{2U}\Bigr) \right)\,.
\end{equation}
Again, the field $C$ acts as a Lagrange multiplier enforcing the following constraint
\begin{equation}\label{constraint}
\e^{-\C}\left(\dot{\B}^2-\dot{U}^2\right)-\e^\C\left(\epsilon\, k\e^{2B} -
\epsilon \,\Conv^2 \QE^2\e^{2U}\right)= 0\,.
\end{equation}
In the following we again choose the gauge $C=0$. Note that for the case $\epsilon = k = +1$,
we indeed find the effective action for the \RN\index{\RN} solutions of eq.~\eqref{eq:Uaction}.

It is straightforward to generalize the Bogomol'nyi rewriting
discussed in section \ref{ss:BJ-Review_MSW}, to include both stationary and
time-dependent configurations with arbitrary slicing of the
transverse space $k=0, \pm 1$. The action (\ref{eq:BJ-blackhole-action})
is, up to boundary terms, equivalent to
\begin{equation}
S=\int \de u\,\, \Bigl(\dot{\B}+\sqrt{\epsilon k \e^{2\B}
+\beta_1^2}\Bigr)^2 -\Bigl(\dot{U}+ \sqrt{\epsilon
\Conv^2\e^{2U}+\beta_2^2} \Bigr)^2\,, \label{}
\end{equation}
where $\beta_1$ and $\beta_2$ are constants. The Bogomol'nyi equations\index{Bogomol'nyi!equations} are
\begin{align}
\dot{\B}&=-\sqrt{\epsilon k \e^{2\B} +\beta_1^2}\,,\nonumber\\ 
\dot{U}&= -
\sqrt{\epsilon \,\Conv^2\QE^2\e^{2U}+\beta_2^2}\,.
\label{genbps}
\end{align}
The constraint (\ref{constraint}) implies that there is only one deformation parameter $\beta$:
\begin{equation}
\beta_1^2=\beta_2^2\equiv \beta^2\,.
\end{equation}
In section \ref{ss:BJ-WhyDoesItWork}, this is just the requirement of $(\beta_1,\beta_2)$ to be a null vector of the scalar metric. 

Note that for time-dependent solutions\index{time-dependent!solutions} with charge ($\epsilon=-1$, $\QE\neq 0$) the limit of
$\beta_{}\rightarrow 0$ does not exist, while for $\QE=0$ the limit
only exists for $k=-1$.

{}For later use, we give the general solutions to the first-order equations. The Bogomol'nyi equations\index{Bogomol'nyi!equations} are all of the form
\begin{equation}
 \dot D_\pm = - \sqrt{\beta^2 \pm K^2 \e^{2D_\pm}}\,,
\end{equation}
where $K$ is a constant, depending on the case under consideration.
The solutions to these equations are given by
\begin{equation}
\e^{-D_+} = K\frac{\sinh(\beta u + c_+)}{\beta}\,,\qquad\e^{-D_-} =
K\frac{\cosh(\beta u + c_-)}{\beta}\,,
\end{equation}
where $c_\pm$ are constants of integration. In the extremal limit
$\beta \rightarrow 0$ the solution for $D_+$ becomes
$\e^{-D_+}=Ku+c_+$.

\begin{itemize}
\item \underline{\emph{Rediscovering \RN\index{\RN} black holes (timelike 0-branes):}}
Putting $\epsilon = k =+1$ reproduces the results of Miller\index{Miller}, Schalm\index{Schalm} and Weinberg\index{Weinberg} for the \RN\index{\RN} black hole, see section \ref{ss:BJ-Review_MSW}.

\item \underline{\emph{Rediscovering spacelike 0-branes:}}
{}For spacelike branes $(\epsilon=k=-1)$ we find
\begin{equation}
\e^{-U} =  \QE \frac{\cosh(\beta u +
c_{U})}{\beta}\,,\qquad\e^{-\B} = \frac{\sinh(\beta u)}\beta\,.
\end{equation}
Once again, shifting the origin of the $u$-axis, the integration
constant in the equation for $B$ has been put to zero. Using the
coordinate transformation
\begin{equation}
t = \Conv \QE\frac{\cosh(\beta u + c_{U})}{\sinh\beta u
}\,, \hspace{2cm} x = \frac1{\Conv \QE}\frac{\beta}{\cosh c_U}
z\label{S0}\,,
\end{equation}
the solution then takes the following form
\begin{equation}
\de s^2 = G(t)\, \de x^2 - G(t)^{-1}\de t^2 + t^2\de
\mathbb{H}_2^2\,,\qquad F_{t x} = \frac{\QE}{t^2}\,,
\end{equation}
with
\begin{equation}
G(t) = 1 - 2\frac{\Conv \QE\sinh(c_U)}{t}
-\frac{\Conv^2 \QE^2}{t^2} \,,
\end{equation}
where we introduced the metric for the hyperboloid $\de
\mathbb{H}^2_2 = \de \Sigma^2_{-1}$. Again, this solution is
asymptotically flat. Moreover, we see that this reduces to the
metric for the S0-brane of \cite{Gutperle:2002ai} after a constant
rescaling of $x$ and $t$.
Taking the limit $\beta \rightarrow 0$, the metric is easily seen to
describe flat space in Milne coordinates.
\end{itemize}

\subsubsection{Addition of a dilaton}
We consider the coupling of the vector field to a dilaton, as
this is the generic situation in supergravity theories. The action
describing four-dimensional Einstein-Maxwell-dilaton theory is (remember we do not write the normalization $\frac{\gamma^2}{4\pi G_4}$)
\begin{equation}\label{Dilaton-Maxwell-Einstein}
S=\int \de x^4\sqrt{-g} \Bigr(\tfrac{1}{4\Conv^2}R
-\tfrac{1}{2}(\partial\phi)^2- \tfrac{1}{4}\e^{a\phi}F^2\Bigl)\,.
\end{equation}
The \ansatz{} for electrical solutions is now given by
\begin{equation}
  F = \QE \e^{2U+\C-a\phi}\de u \wedge \de z\,.\label{eq:BJ-Ansatz_GaugeField_DilatonBH}
\end{equation}
In the gauge $C=0$ the effective action becomes
\begin{equation}
S=-\int\de u\,\, \left(\dot{\B}^2-\dot{U}^2 - \Conv^2 \dot{\phi}^2 +
\epsilon\, k\e^{2\B} - \epsilon \,\Conv^2 \QE^2\e^{2U-a\phi} \right)\,.
\end{equation}
The potential for the metric field $U$ and the dilaton $\phi$ mixes both fields. In light of the conditions presented in section \ref{ss:BJ-WhyDoesItWork} under which the
Bogomol'nyi trick of \cite{Miller:2006ay} works, we see we need to rewrite the potential in terms of a sum of decoupled terms, while keeping the metric diagonal. Therefore we perform a rotation in $U,\phi$ space, which is orthogonal
w.r.t.~the scalar metric:
\begin{equation}
\begin{pmatrix}
  \tilde U\\\tilde \phi
\end{pmatrix}
=
\frac1b\begin{pmatrix}
  1 & -a/2\\ {a}/{2\Conv^2} & 1
\end{pmatrix}
\begin{pmatrix}
  U\\\phi
\end{pmatrix}
\,,\label{eq:BJ-DilBH_ChangeOfVars}
\end{equation}
where $b$ is  a normalization constant:
\begin{equation}
b^2 = \det \begin{pmatrix}
  1 & -a/2\\a/2\Conv^2& 1
\end{pmatrix} = 1 + \frac{a^2}{4\Conv^2}\,.
\end{equation}
This rotation is such that the potential is now a sum of a $B$-dependent term and a $\tilde U$-dependent term,
while the scalar metric remains diagonal and it even takes on the same form in the new basis
$\B,\tilde U,\tilde \phi$ as before, since the rotation matrix in equation \eqref{eq:BJ-DilBH_ChangeOfVars} is
 orthogonal w.r.t.~the original metric. By the criteria of section \ref{ss:BJ-WhyDoesItWork},
we conclude we can apply the Bogomol'nyi trick\index{Bogomol'nyi!trick}, similar
to the previous case without a dilaton.

Writing the action as a sum
of squares, we introduce three constants $\beta_1,\beta_2$ and
$\beta_3$:
\begin{equation}
S = -\int \de u\,\,\Bigl(\dot{\B}+\sqrt{\epsilon k \e^{2\B}
+\beta^2_1}\Bigr)^2 - \Bigl(\dot{\tilde U}+\sqrt{\epsilon
\Conv^2\QE^2\e^{2 b\tilde U}+\beta^2_2} \Bigr)^2
-\Conv^2(\dot{\tilde \phi} -
\beta_3)^2\,.\label{action+dilaton}
\end{equation}
In this case the equivalent of the constraint (\ref{constraint})
implies that only two of the three integration constants are
independent:
\begin{equation}
\beta_1^2 - \beta_2^2 -\Conv^2 \beta_3^2 =0\,.
\end{equation}
The Bogomol'nyi equations\index{Bogomol'nyi!equations} are those from before and an extra
equation for $\tilde \phi$:
\begin{align}
  \dot{\B}=-\sqrt{\epsilon k \e^{2\B} +\beta_1^2}\,,\qquad \dot{\tilde U}= -
\sqrt{\epsilon \,\Conv^2\QE^2\e^{2b\tilde U}+\beta_2^2}\,,\qquad
  \dot{\tilde \phi}=\beta_3
\label{eq:BJ-Dilaton-Einstein-Maxwell_FO_Eqs}
\end{align}

We can compare the solutions to the Bogomol'nyi equations to the literature, by plugging the solutions for $U, \B$
and $\phi$ back into the \ansatz. When $\epsilon=k=1$, we are describing the familiar dilatonic black hole solution
\cite{Gibbons:1982ih,Garfinkle:1990qj}. One then also notices that the two independent $\beta$-parameters appear in
a fixed combined way as to effectively form one deformation parameter, as follows. Naively, we have seven
integration constants: $\beta_1,\beta_2,\beta_3,M,\QE,\Sigma,\phi_\infty$, where $M$ is the ADM mass,
$\Sigma = \dot \phi|_{\infty}$ the dilaton charge and $\phi_\infty$ the dilaton at spatial infinity.
Only three of these constants are independent, since we have four equations to relate them (the
constraint equation and the three Bogomol'ny equations evaluated at spatial infinity). By fixing the
dilaton\index{dilaton} at spatial infinity, $\phi_\infty$, one obtains two free parameters $M,\QE$ and
effectively one deformation parameter relating those two in a BPS-bound. This is standard
convention in the literature, see for instance \cite{Garfinkle:1990qj}. When $\epsilon=k=-1$, we
are describing the dilatonic S0-brane, derived before by Kruzcensky, Myers and Peet \cite{Kruczenski:2002ap}.
{}For the other values of $\epsilon,k$, we have explicit solutions but are not aware of a good interpretation
in terms of known objects in the literature.

\outlook{Summary:}{All static, spherically symmetric black hole solutions and time-dependent\index{time-dependent}
0-brane solutions to Einstein-Maxwell theory\index{Einstein-Maxwell} in four dimensions coupled to a dilaton\index{dilaton}
allow for first-order equations \`a la Bogomol'nyi, by writing the action as a sum of squares. We reproduce the
known results for (dilatonic) black holes and S0-branes. }

\subsection{Generalizing to time-dependent dilatonic \texorpdfstring{$p$}{}-branes\label{ss:BJtimelike_p}}


In this section, we discuss the generalization of the first-order formalism for timelike and spacelike
0-branes in four dimensions discussed above to $p$-branes in $D$ dimensions.

Before we delve into the necessary calculations, let us take a step back and ask ourselves what was
essential in the successful application of the Bogomol'nyi trick in the four-dimensional setup.
First, we obtained an effective action as an integral over one
(radial or timelike) coordinate we called $u$. The dilaton and metric warp factors are scalars in this action,
while the gauge field is eliminated through its equation of motion. Second, we saw in section \ref{ss:BJ-WhyDoesItWork}
that the Lagrangian could be written as a sum over kinetic and potential terms for each scalar field separately.
This allowed us to rewrite each contribution (kinetic term $+$ potential) as a perfect square plus a boundary term.
We briefly discuss how these features can be applied to a more general $p$-brane \ansatz{}.
\begin{itemize}
 \item[\rab] \emph{One-dimensional effective action:}
 The following discussion applies to both the timelike and the spacelike solutions of section \ref{ss:BJ-PossGener}.
 We stick to the timelike case (black holes) for concreteness. {}For the four-dimensional black hole, the effective action is obtained by performing a reduction along the time direction (with coordinate $z$). The fields in the original theory are time-translation invariant: $\partial / \partial z$ is a Killing vector. A priori the dimensional reduction gives rise to a three-dimensional effective theory. The truncation to this three-dimensional theory is then in one-to-one correspondence to the black hole system in the original theory, because of time independence. It is only because we made a very restrictive \emph{static} \ansatz{}, that the three-dimensional theory also does not depend on the angular variables and is fully determined by its radial dependence (coordinate $u$). Therefore, the description is effectively one-dimensional.

To exploit this mechanism for a $p$-brane, notice that the time direction ($z$) of the black hole, along which the reduction was performed, is actually the one-dimensional \worldvolume{} of a $0$-brane. Similarly, when we demand that we have translation invariance along the directions of the ($p+1$)-dimensional \worldvolume{} of a $p$-brane, we can in principle perform a reduction along its \worldvolume{} to a ($D-p-1$)-dimensional system. Because of the translational invariance, solutions of the lower-dimensional system are then in one to one correspondence to those in $D$ dimensions. Finally, if we restrict the metric of transverse space such that it describes a static\index{static} solution, we should in principle be able to further reduce the problem to a one-dimensional one, in terms of one coordinate we  again denote as  $u$.

 \item[\rab] \emph{Sum of squares and decoupled first-order equations:}
Once we have obtained an effective one-dimensional description, the main question still stands if the action can
be written as a sum (difference) of squares. Again, we expect to extend the results of the calculation for
0-branes in four dimensions trivially. Indeed, by investigation of the action \eqref{eq:BJ-action}, one sees the gauge field
 only gives rise to one potential term in the effective action,
 which is a function of a certain combination of scalar fields. After a rotation in the scalar target space,
  such that the potential depends on one scalar only and we can readily apply the Bogomol'nyi trick, as in section \ref{ss:BJ-WhyDoesItWork}.
\end{itemize}

We see it should in principle be possible to find timelike and spacelike $p$-brane solutions of the form \eqref{eq:BJ-time+stat_metric} from a Bogomol'nyi trick\index{Bogomol'nyi!trick}, by performing a reduction over the brane \worldvolume{}. The dimensional reduction\index{dimensional reduction} of a $p$-brane along its \worldvolume{} is a ($-1$)-brane: a solution with a zero-dimensional \worldvolume{} (its transverse space covers the entire \spacetime{}). Essentially, this means we should perform the calculation of \cite{Miller:2006ay} for a ($-1$)-branes and can then derive the solution for a $p$-brane by uplifting the along the ($p+1$)-dimensional \worldvolume{}. Therefore, we first consider the calculation for $(-1)$-branes in $d= D-p-1$ dimensions and use this result to show that a first-order formalism \`a la Bogomol'nyi exists for $p$-brane solutions to the Einstein-Maxwell-dilaton theory \eqref{eq:BJ-action} in $D$ dimensions.

\subsubsection{Preliminary result: (-1)-branes in arbitrary dimensions}
A $(-1)$-brane\index{$p$-branes!(-1)-branes|imp} couples electrically to a 0-form gauge potential (a scalar)
often referred to as the axion. We call this scalar $\chi$. The \worldvolume{} is zero-dimensional and, in the case of a
timelike $(-1)$-brane, this implies that the whole space is Euclidean since it is entirely transverse, while for spacelike $(-1)$-branes, it has Lorentzian signature.
The action in $d$ dimensions is
\begin{equation}
S=\frac{\Conv_d^2}{4\pi G_d}\int\de^d x\sqrt{-g}\Big{(} \frac{1}{4\Conv_d^2}R_d-\tfrac{1}{2}(\partial\phi)^2
+\epsilon \tfrac{1}{2}\e^{b\phi}(\partial\chi)^2\Big{)}\,.
\end{equation}
Note the `wrong sign' kinetic term for the axion when $\epsilon=+1$,
which is normal for Euclidean theories.
{}From a higher-dimensional perspective, this wrong sign appears because the action above is obtained after a dimensional reduction over the  $p$-brane \worldvolume{}. The $p$-brane \worldvolume{} has a time direction and it is exactly the time component of the metric that contributes an extra minus sign to the axion kinetic term.

The entire solution is taken to depend on the coordinate $u$ only. In order to describe a $(-1)$-brane, we take the following \ansatz{} for the metric and the gauge field.
\begin{align}
\de s^2_d&=\epsilon \e^{2(d-1)B(u)}\e^{2\C(u)} \de u^2 + \e^{2B(u)}\de \Sigma_k^2\,,\\
\dot{\chi}&= \QE\, \e^{C(u)-b\phi(u)}\,,\hspace{5cm}
\,.
\label{eq:BJ-instanton_ansatz}
\end{align}
Again, a dot denotes differentiation w.r.t.~the coordinate $u$.
Notice that for $d=3$, this \ansatz{} reduces to the three-dimensional part  of the timelike and spacelike $0$-brane solutions we described in the last section, see eq.~\eqref{eq:BJ-Ansatz_Metric_0-brane_time+stat}.
The one-dimensional effective action that reproduces the equations
of motion for the metric fields $\B,\C$ and the dilaton $\phi$ is
\begin{equation}
S=\int \de u \,\left(\e^{-\C} \left(\tfrac12 {(d-1)(d-2)}\dot{\B}^2+\Conv^2\dot{\phi}^2
\right) + \e^{\C}\left(\epsilon k \e^{2(d-2)\B}
-\epsilon \e^{-b\phi}\Conv^2\QE^2\right)\right)\,.\label{eq:BJ-Instanton_effective_action}
\end{equation}
The field $\C$ is again not propagating and we can choose it at will;
we choose the gauge $C=0$. At $C=0$, the constraint arising from the equation of motion for $C$
\begin{equation}
\left(\tfrac12{(d-1)(d-2)}\dot{\B}^2+\Conv^2\dot{\phi}^2
\right) - \left(\epsilon k \e^{2(d-2)\B}
-\epsilon \e^{-b\phi}\Conv^2\QE^2\right)\,= 0\,.\label{eq:BJ-Constraint_(-1)brane}
\end{equation}

Up to boundary terms, the effective action is equivalent to the following BPS form (remember we choose to work in the gauge where $C=0$)
\begin{align}
S=\int\de u\,\Big{(}&\tfrac12{(d-1)(d-2)}\Bigl( \dot{\B}
+\sqrt{\epsilon k \e^{2(d-2)\B}+\beta_1^2}\Bigr)^2 \nonumber\\
&- \Conv^2\Bigl(\dot{\phi}+ \sqrt{\epsilon \QE^2\e^{-b\phi} +\beta_2^2} \Bigr)^2\Big{)}\,.\label{eq:BJ-instanton_action_squared}
\end{align}
The signs in the brackets are chosen such that we describe well-behaved, finite solutions. The first-order equations are:
\begin{align}
\dot{\B} &=
-\sqrt{\epsilon k \e^{2(d-2)\B}+\beta_1^2}\nonumber\\
\dot{\phi}
&= -\sqrt{\epsilon \QE^2\e^{-b\phi} +\beta_2^2}
\end{align}

When we evaluate the constraint equation \eqref{eq:BJ-Constraint_(-1)brane} on any solution to those first-order equations, we see there is only one effective deformation parameter:
\begin{equation}
\Conv^2 \beta_2^2=\tfrac12{(d-1)(d-2)}\beta_1^2\,.
\end{equation}

We give the explicit solution to the first-order equations found by putting the individual squared terms in the action
\eqref{eq:BJ-instanton_action_squared} to zero below. We choose to only highlight the timelike $(-1)$-brane ($\epsilon = k = 1$) and the spacelike $(-1)$-brane\index{$(-1)$-brane} ($\epsilon=k=-1$), since only for these cases we have a good interpretation and can compare with the literature.

\paragraph{Timelike case.}
We first solve the Bogomol'nyi equations\index{Bogomol'nyi!equation} for $k=\epsilon=1$. Now $u$ is a radial coordinate.

Consider the case with vanishing deformation parameters first: $\beta_1 = \beta_2 = 0$. If we define the coordinate $\rho$ via $\de \rho=-\e^{(d-1)\B(u)}\de u$, then the BPS equation, $\de B/\de u=-\e^{(d-1)\B}$, implies that $\rho=\e^{\B} + cst$. Shifting the origin of the $\rho$-axis we can put the constant in this equation to zero and we find that the metric describes the Euclidean plane\index{Euclidean plane} in spherical coordinates\index{spherical coordinates}:
\begin{equation}
\de s^2_d=\de \rho^2+\rho^2\de \Omega^2_{d-1}\,.
\end{equation}
The solutions for the scalar fields are
\begin{equation}
\e^{\tfrac{b}{2}\phi}=\frac{b |\QE|}{2(d-2)}\,\frac 1{\rho^{d-2}} +
\e^{\tfrac{b}{2}\phi_{\infty}}\,,\qquad \chi=-\frac{2|\QE|}{b \QE}
\,(\e^{-\tfrac{b}{2}\phi}  - \e^{-\tfrac{b}{2}\phi_{\infty}})+
\chi_{\infty} \,,
\end{equation}
where $\phi_\infty,\chi_\infty$ are integration constants denoting the values of the scalars at spatial infinity
($\rho =\infty$). This is indeed the extremal instanton solution, see \cite{Gibbons:1995vg, Gutperle:2002km, Bergshoeff:2004fq}.

{}For non-zero $\beta_{1},\beta_2$ the solution becomes 
\begin{align}
 \e^{-(D-2)\B(u)} &= \frac{1}{\beta_1}\sinh\left((d-2)\beta_1\,u+c_1\right)
\,,\label{eq:BJ-D-1metric}\\
 \e^{\frac{b}{2}\phi(u)}&= \frac{|\QE |}{\beta_2}\sinh \left(\tfrac{
b}{2}\beta_2\,u + c_2\right)\,,\\
\chi(u) &=
-\frac{2}{b\QE}\sqrt{\QE^2\e^{-b\phi(u)}+\beta_2^2} + c_3 = -\frac{2 \beta_2}{b |\QE|}\coth \left(\tfrac{
b}{2}\beta_2\,u + c_2\right)+c_3\,,
\end{align}
where $c_1, c_2$ and $c_3$ are arbitrary constants of integration.
These solutions correspond to the super-extremal instantons that
were constructed in \cite{Gutperle:2002km,Bergshoeff:2004fq}. In ref. \cite{Gutperle:2002km}, it was already noted that for extremal solutions, the second-order equations of motion can be written as first-order ones. This feature was correctly seen as a hint of supersymmetry. However, for the non-extremal case, as far as we know, a first-order description had not been given for the super-extremal instantons.

\paragraph{Spacelike case.}
Finally, the time-dependent S$(-1)$ brane solution (with
$k=\epsilon=-1$) that was first constructed in
\cite{Kruczenski:2002ap} can be rederived (again in the frame
$\C=0$)
\begin{align}
 \e^{(2-D)\B(u)} &=  \frac{1}{\beta_1}\sinh\left((d-2)\beta_1\,t + c_1\right)
\,,\\
 \e^{\frac{b}{2}\phi(u)}&= \frac{|\QE|}{\beta_2}\cosh\left(\tfrac{
b}{2}\beta_2\,u + c_2\right)\,,
\\\chi(u)&= -\frac{2}{b\QE}\sqrt{\QE^2\e^{-b\phi(u)}
- \beta_2^2} + c_3= -\frac{2 \beta_2}{b |\QE|}\tanh \left(\tfrac{
b}{2}\beta_2\,u + c_2\right)\label{eq:BJ-S-1axion}\,.
\end{align}
Again, altough the form of these solutions was known before, the derivation from first-order equations of Bogomol'nyi\index{Bogomol'nyi!equations} type is new.

\subsubsection{$p$-Brane \ansatz{}}
Now we come to the main point of this chapter. We can piece together previous elements and present the Bogomol'nyi trick\index{Bogomol'nyi!trick}
for timelike and spacelike $p$-brane\index{$p$-brane} solutions to Einstein-Maxwell-dilaton\index{Einstein-Maxwell-dilaton} theory in $D$ dimensions. The action for this setup in $D$ dimensions is given by \eqref{eq:BJ-action}, which we repeat for convenience (up to a proportionality factor $\gamma_D^2/4\pi G_D$):
\begin{equation}
S=\int \de^{D}x \sqrt{-g}\left(\tfrac{1}{4\Conv_D^2}R_D -
\tfrac{1}{2}(\partial\phi)^2 - \tfrac{1}{2(p+2)!}\e^{a\phi}F_{(p+2)}^2\right)\,. \label{action2}
\end{equation}
{}From now on we drop the factor of $1/16\pi G_D$ in front of the action. 

The field strength $F_{(p+2)}$ is related to a $(p+1)$-form potential $A_{(p+1)}$ as $F_{(p+2)} = \de A_{(p+1)}$
The corresponding $p$-brane solutions we wish to look at can all be reduced to
$(-1)$-brane solutions via reduction over their flat
\worldvolume{}s. Therefore we should be able to produce the Bogomol'nyi equations from the $(-1)$-brane calculation above, where the \spacetime{} dimension in which the $(-1)$-branes lives is now given as:
\begin{equation}
  d \equiv D - p - 1\,.
\end{equation}
A typical \emph{electrically charged} $p$-brane \ansatz{} takes the form
\begin{align}
& \de s^2=\e^{2\beta\,\varphi(u)}
\eta^\epsilon_{mn}\de z^m\de z^n+\e^{2\alpha\,\varphi(u)}\de s^2_{d}  \,, \hspace{1cm}\phi = \phi(u)\nonumber\\
& A_{p+1}(u)= \chi(u) \ \de z^{1}\wedge\de z^{2}\wedge\ldots\wedge \de
z^{p+1} \,. \label{eq:BJ-p-formAnsatz}
\end{align}
Let us give some more information about this \ansatz{} for the gauge field and the metric. {}For the $(p+1)$-form $A$, we have chosen to take an electric \ansatz{}.
The electric potential is given by the function $\chi(u)$, it will be a scalar in the effective action and play the role of
the axion. The metic $\de s^2_d$ of the $d$-dimensional transverse space to the brane \worldvolume{} is taken to be the $d$-dimensional $(-1)$-brane metric (\ref{eq:BJ-instanton_ansatz}). The coordinates $z^i \,(i=1\ldots p+1)$ parameterize the ($p+1$)-dimensional \worldvolume{} of the $p$-brane, which has a Minkowskian \emph{or} a Euclidean metric, determined by the value of the parameter $\epsilon = \pm1$:
\begin{equation}
  \eta^\epsilon=\text{diag}(-\epsilon, +1,\ldots,+1)\,.
\end{equation}
Again, choosing $\epsilon$ to be either $+1$ or $-1$, comes down to describing resp.~radial- or time-dependent solutions. Finally, the metric factor describing the warping of the \worldvolume{} w.r.t. transverse space,
is no longer called $U$, as in the four-dimensional description of section \ref{ss:BJ-Review_MSW},
but $\varphi$. We do this to have agreement with common notation used in Kaluza-Klein reductions. The constants $\alpha$ and $\beta$ are 
\begin{equation}
\label{eq:BJ-alpha and beta}
\alpha=-\frac1{2\Conv}\sqrt{\tfrac{p+1}{2(d+p-1)(d-2)}}\,,\qquad
\beta=\frac1{2\Conv}\sqrt{
\tfrac{d-2}{2(d+p-1)(p+1)}}\,.
\end{equation}
They are chosen such that $\varphi$ has a conventionally normalized kinetic term in the effective action obtained after
the dimensional reduction over the \worldvolume{}. The $p$-brane \ansatz{} \eqref{eq:BJ-p-formAnsatz} is a straightforward generalization of a 0-brane \ansatz{}
\eqref{eq:BJ-Ansatz_Metric_0-brane_time+stat} in the four-dimensional setup. Note that one should take $U = \frac12 \varphi$ to find agreement for $p=0,D=4$.

We now reduce the \ansatz{} (\ref{eq:BJ-p-formAnsatz}) over the \worldvolume{} coordinates $z$ and obtain a
lower-dimensional \ansatz{} of the form (\ref{eq:BJ-instanton_ansatz}). The equivalent reduction of the action
(\ref{action2}) leads to the $d$-dimensional action
\begin{equation}
S=\int \de^{d}x \sqrt{-g}\Bigl(\tfrac{1}{4\Conv_d^2}R_d -
\tfrac{1}{2}(\partial\varphi)^2 - \tfrac{1}{2}(\partial\phi)^2 +
\epsilon\tfrac{1}{2}\e^{a\phi+2(d-2)\alpha\varphi}(\partial\chi)^2\Bigr)\,.
\end{equation}
By a field rotation in the scalar target space,
we get that the potential term due to the electric charge depends on one scalar only. Therefore, define new scalars $\tilde \varphi,  \tilde \phi$ as linear combinations of $\varphi, \phi$ through:
\begin{equation}\label{eq:BJ-fieldredef}
\begin{pmatrix}
 \tilde \phi\\
\tilde \varphi
\end{pmatrix}
=\frac 1b
\begin{pmatrix}
 a&2(d-2)\alpha\\
-2(d-2)\alpha&a\end{pmatrix}
\begin{pmatrix}
\phi\\
\varphi
\end{pmatrix}
\,,\qquad
b^2 = a^2 + 4(d-2)^2\alpha^2\,.
\end{equation}
The rotation matrix is orthogonal, such that the kinetic terms of the scalars $\tilde \phi, \tilde \varphi$ are again canonically normalized. The scalar $\tilde \phi$ appears in the potential, while the scalar $\tilde \varphi$ decouples. The  action becomes:
\begin{equation}
S=\int \de^{d}x \sqrt{-g}\Bigl(\tfrac{1}{4\Conv_d^2}R_d -
\tfrac{1}{2}(\partial\tilde\varphi)^2 - \tfrac{1}{2}(\partial\tilde\phi)^2 +
\epsilon\tfrac{1}{2}\e^{b\tilde\phi}(\partial\chi)^2\Bigr)\,.
\end{equation}
It is clear that the discussion for the ($-1$)-brane is applicable. The only difference is the  extra decoupled dilaton $\tilde \varphi$ when compared to the $(-1)$-brane calculation of the previous section.
After we plug in the  ($-1$)-brane \ansatz{} \eqref{eq:BJ-instanton_ansatz} into the equations of motion, we get an effective one-dimensional action and a constraint, which are the extension of the eqs.~\eqref{eq:BJ-Instanton_effective_action},\eqref{eq:BJ-Constraint_(-1)brane} with the decoupled scalar $\tilde \varphi$. We refrain from giving these equations here, since the addition of the decoupled scalar follows trivially. Analogous to the $(-1)$-brane calculation, the effective action is then equivalent, up to boundary terms, to the Bogomol'nyi-form
\begin{align}\label{eq:BJ-effective-action2}
S=&\int\de z\,\Bigl(\tfrac12(d-1)(d-2)\Bigl(\dot{\B}+
\sqrt{\epsilon k \e^{2(d-2)\B}+\beta_1^2}\Bigr)^2 \nonumber\\&-
\Conv_d^2\left(\dot{\tilde{\phi}}+\sqrt{\epsilon
\QE\e^{-b\tilde{\phi}}+\beta_2^2} \right)^2
 -\Conv_d^2(\dot{\tilde{\varphi}}+\beta_3)^2\Bigr)\,,
\end{align}
where only two of the three deformation parameters $\beta_1$, $\beta_2$ and $\beta_3$
are independent due to the condition coming from the constraint equation:
\begin{equation}
\tfrac12{(d-1)(d-2)}\beta_1^2 - \Conv_d^2\beta_2^2 - \Conv_d^2\beta_3^2=0\,.
\end{equation}
The solutions to the BPS equations for $A$ and $\tilde{\phi}$ can be
found in the previous section in equations
(\ref{eq:BJ-D-1metric}-\ref{eq:BJ-S-1axion}), whereas the solution for the extra
field $\tilde{\varphi}$ is trivial:
\begin{equation}
 \tilde{\varphi}(z)=-\beta_3 z\,.
\end{equation}
{}From our \ansatz{} (\ref{eq:BJ-p-formAnsatz}) and the field redefinition
(\ref{eq:BJ-fieldredef}) we can immediately infer the timelike and
spacelike brane solutions in $D$ dimensions. We do not discuss these
solutions as they have been discussed in the literature in numerous
places. We just remark that both for timelike and spacelike solutions, we reproduce
the known results in the literature, see for instance the reviews of Youm \cite{Youm:1997hw} and Stelle \cite{Stelle:1998xg} on (radial dependent) $p$-branes and work of L\"u, Pope and collaborators on time-dependent brane solutions, for example in \cite{Lu:1996er} and references therein.

Finally, note that in our work \cite{Janssen:2007rc}, a slightly more general $p$-brane \ansatz{} was considered. In the cases discussed above, we considered non-extremal deformations by tweaking the warp factor of the $p$-brane \worldvolume{}  (encoded in the field $U$ or $\varphi$). However, the \worldvolume{} is always considered to be flat. In \cite{Janssen:2007rc}, also non-extremal deformations such that the brane \worldvolume{} is no longer flat are investigated. {}For these solutions, again the Bogomol'nyi trick follows immediately.

\section{Discussions and outlook\label{s:BJ-Discussion_Outlook}}
First, we give a short overview of the key results. Then we discuss the most important findings in depth. Finally, we broaden our scope and consider pointers for future work, explaining also how the research questions raised by the work reported in this chapter naturally leads to the our work presented in chapter \ref{c:JP-Gradient_Flow}.
\subsection{Summary}
We started this chapter with a historical overview of first-order formalisms that are found by writing a functional of the fields as the energy, or the action, as a sum of squares. Mostly, such a feature is specific to supersymmetric or at least extremal solutions. The result of Miller\index{Miller}, Schalm\index{Schalm} and Weinberg\index{Weinberg} \cite{Miller:2006ay} was a surprise in this respect: it showed that Bogomol'nyi type\index{Bogomol'nyi!type equations} first-order equations exist for the non-extremal \RN\index{\RN} black hole\index{\RN} in four dimensions.

In this chapter, we have discussed the extension of that setup in three ways: (1) to  Einstein-Maxwell theory coupled to a dilaton, (2) for $p$-branes in arbitrary dimension $D$, (3) these $p$-branes can be stationary (and non-extremal), but also time-dependent.
We have shown that all known brane-type solutions of an Einstein-dilaton-$p$-form theory can be found from decoupled first-order equations, thereby extending the results of \cite{Miller:2006ay} to arbitrary dimensions and time-dependent cases. By brane-type solutions we mean solutions with a \spacetime{} \ansatz{} given by \eqref{eq:BJ-time+stat_metric}. The key point is that these solutions depend on one coordinate and therefore can be constructed from a one-dimensional effective action, as was first discussed for black holes in \cite{Ferrara:1997tw}. By rewriting the action as sums and differences of squares we arrive at first-order equations \`a la Bogomol'nyi.

\subsection{Discussion}
\underline{\emph{Application to time-dependent solutions.}} It is important to stress that the application of these ideas to time-dependent brane solutions (S-branes) is a \emph{less trivial extension} of \cite{Miller:2006ay}. One possible way to understand why it was to be expected that a similar first-order formalism exists for time-dependent branes, stems from the known fact that non-extremal stationary branes can be analytically continued to time-dependent solutions, which is impossible for extremal branes \cite{Lu:1996er}. As explained in the introduction, this first-order formalism\index{first-order!formalism} for time-dependent $p$-branes is the natural generalization of the so-called pseudo-BPS equations\index{pseudo-BPS} for FLRW-cosmologies\index{FLRW-cosmology} \cite{Skenderis:2006fb, Skenderis:2006jq}.

\underline{\emph{Is there a hidden supersymmetry at work?}} That first-order equations are found for some extremal timelike brane solutions was maybe no big surprise as at least some of them can be seen as supersymmetric solutions when embedded into an appropriate supergravity theory. In \cite{Miller:2006ay} the question was raised as to whether also for the non-extremal case, the deformed Bogomol'nyi equations\index{Bogomol'nyi!equation} could be understood from the point of view of supersymmetry. One may imagine that the bosonic Lagrangian (\ref{eq:Maxwell-Einstein-action-4d}) could be embedded into a different (non-standard) supergravity theory for which the non-extremal solutions preserve some fraction of supersymmetry. However, there is  an obstruction to even defining Killing spinors\index{Killing!spinor} which implies that the non-extremal solutions\textit{ cannot preserve supersymmetry}. 
Of course one should repeat the same calculations of \cite{Miller:2006ay} for the case of $p$-branes with $p>0$, but we believe that the same {negative answer} will be found.

\underline{\emph{What type of solutions did we miss?}} We did not completely exhaust all possible brane solutions in our analysis, as we \emph{did not consider branes with codimension less than three }. When the codimension is one, the stationary branes are domain walls\index{domain wall} and the time-dependent branes are FLRW-cosmologies, for which the fake supergravity\index{fake supergravity} and pseudo-supersymmetry\index{pseudo-supersymmetry} formalism is by now well developed. However, the case of branes with codimension-two\index{codimension-two} cannot be included as these solutions depend on one \emph{complex} coordinate rather than on one real coordinate.

\subsection{Outlook}
Let us take a step back and reconsider the underlying principle that made it possible to apply the method of \cite{Miller:2006ay} to the Einstein-dilaton-Maxwell theories described in this chapter. (See also section \ref{ss:BJ-WhyDoesItWork}.) All solutions covered by an \ansatz{} of the form \eqref{eq:BJ-time+stat_metric}, are described by an effective action with a potential $V(q)$,
where $q=q^i$ is a vector whose entries are the (scalar) fields of the effective description. Whenever the potential is written as $V(q) = \cg^{ij} \partial_i W( q)\partial_j \cw( q)$, where $\cg_{ij}$ is the metric appearing in the kinetic term of the scalars, the function $\cw$ determines the form of the scalar equations of motion as a gradient flow, such that $\dot q^i = \cg^{ij}\partial_j \cw(q)$. The stationary and time-dependent solutions to Einstein-Maxwell theory with one scalar (a dilaton), which were discussed above, are of this form. However, they are very restrictive: the first-order equations decouple and the function $\cw$ is a sum of terms that depend on one scalar $q^i$ each. These systems are fully integrable, in the sense that we can perform $n$ independent integrations, $n$ being the total number of scalars $q^i$ in the effective description. Theories with more scalars are a natural complication to investigate, since supergravity theories can describe many scalar fields. We then do not expect the equations of motion to be equivalent to decoupled first-order equations any more, but equations of the form $\dot q^i = G^{ij}\partial_j \cw(q)$, for a more general function $\cw$, may still exist. In chapter \ref{c:JP-Gradient_Flow}, we investigate the most general form of the function $\cw$ for  black hole solutions to $D$-dimensional supergravity.

Of related interest is an alternative way to understand the existence of  first-order equations  for stationary and time-dependent brane solutions, by the approach of mapping $p$-branes to $(-1)$-branes. If one does not solve for the gauge field (axion), but keeps this as a scalar in the effective theory, then the $(-1)$-brane solutions are solely carried by the metric and scalar fields. It is then easy to observe that the scalar fields only depend on one coordinate and describe a geodesic motion on moduli space. In fact, for many cases this moduli space is a symmetric space, for which it is known that the geodesic equation of motion can easily be integrated to first-order equations (see for instance \cite{Gaiotto:2007ag}). {}From this we expect that there exist BPS equations for all extremal and non-extremal black holes in theories that have a symmetric moduli space after reduction over one dimension. In chapter \ref{c:JP-Gradient_Flow}, we relate the two forms of the effective action (one with a potential, one describing geodesics), see section \ref{ss:Two_Effective_Action}. We also investigate supergravity theories with symmetric moduli spaces, thereby providing several examples of the geodesic approach.

\cleardoublepage
\chapter{
Gradient flows for non-super\-symmetric \mbox{black holes}\label{c:JP-Gradient_Flow}}
\punchline{{}For both \textit{extremal} and \textit{non-extremal} spherically symmetric black holes\index{spherically symmetric} in theories with massless scalars and vectors coupled to gravity, we derive a general form of first-order flow equations\index{first-order!equation}, equivalent to the equations of motion. We look for the most general flow equations that are fully determined by the  gradient of one function defined on the scalar manifold  and no longer restrict to the subset of decoupled equations as described in chapter \ref{c:BJ}.  The existence of the most general gradient flow\index{gradient flow!equations} turns out to be equivalent to  having an effective description which is integrable in the sense of Liouville (i.e.~there are as many constants of motion as scalars, which furthermore commute under the standard Poisson brackets). {}For theories that have a symmetric moduli space\index{symmetric space} after a dimensional reduction over the timelike direction, we provide examples of such a gradient flow\index{gradient flow}. This chapter reviews and reassesses the results of the author's article \cite{Perz:2008kh} with Jan Perz\index{Perz}, Paul Smyth\index{Smyth} and Thomas Van Riet\index{Riet} and takes into account recent advances in this research field with appropriate comments. This chapter can be read independently of the previous chapter.}

\section{Introduction and overview}

In this chapter, we return to the study of black holes. Black hole solutions to extensions of general relativity, such as the various kinds of supergravity naturally occurring in the low-energy effective description of superstrings, often exhibit features unknown from pure Einstein's theory. One such feature, distinctive for extremal black holes in gravity coupled to scalar and vector fields, is the attractor phenomenon, see section \ref{s:BJ-History_FoForm_BHs} for more information, or the original references \cite{Ferrara:1995ih,Ferrara:1996dd,Ferrara:1996um,Ferrara:1997tw}. It causes the end-points of the radial evolution of the scalars -- their values on the event horizon -- to be determined by the charges associated with the vectors and to be insensitive to the values of the scalars at spatial infinity. Another feature is that, for solutions that are supersymmetric, the evolution is governed by first-order (so-called BPS or Bogomol'nyi) equations\index{BPS!equations}\index{Bogomol'nyi!equations}: the scalar fields follow a gradient flow\index{gradient flow} in target space. If one parameterizes the scalar manifold (including now both the original scalar target space and additional metric functions) in terms of coordinates $q^i$, such a gradient flow\index{gradient flow} can be written as
\begin{equation}
 \dot q^i = \cg^{ij}\pf \cw {q^j}\,,\label{eq:JP-GradientFlow_Intro}
\end{equation}
where $\cg_{ij}(q)$ is the metric on the scalar manifold, $\cw(q)$ is a function of the coordinates (scalars) $q^i$ and a dot denotes differentiation w.r.t.~the appropriate radial parameter.\footnote{ {}For supersymmetric spherically symmetric black hole solutions to $\Cn=2$ supergravity\index{supergravity!$\cN=2$}, we had flow equations in chapter \ref{c:BJ} of the form $\dot U \sim - \e^U |Z|\,,~\dot z^\alpha \sim-  G^{\alpha\bar\beta}\partial_{\bar \beta} |Z|$  in terms of the central charge $Z(z,\bar z)$. We show below in detail how we can write this as in eq.~\eqref{eq:JP-GradientFlow_Intro} with $\cw(U,z) \sim \e^U |Z|$.}

However, also for non-supersymmetric\index{non-supersymmetric!black holes} setups, first-order equations of the form \eqref{eq:JP-GradientFlow_Intro} have been written down. This has for instance been the subject of the fake supergravity\index{fake supergravity} formalism for domain walls\index{domain wall} \cite{Freedman:2003ax,Celi:2004st,Skenderis:2006jq, Papadimitriou:2006dr,Afonso:2006gi,Elvang:2007ba, Sonner:2007cp}. In case of black hole solutions, as was pointed out in chapter \ref{c:BJ}, it was recently noticed that non-supersymmetric, extremal black hole solutions may also obey first-order gradient flows of the form \eqref{eq:JP-GradientFlow_Intro}, see \cite{Ceresole:2007wx,Andrianopoli:2007gt,Cardoso:2007ky}. Examples of first-order equations have even been found for some non-extremal (and hence neither supersymmetric nor attractive) black holes. The first of these examples was already constructed in 2003, by Lu et al.~\cite{Lu:2003iv}, but it was not until 2006 before a specific construction method was given by Miller, Schalm and Weinberg\index{Miller}\index{Schalm}\index{Weinberg} \cite{Miller:2006ay} to obtain first-order equations of the form \eqref{eq:JP-GradientFlow_Intro} for non-extremal \RN black holes. The latter article formed the starting point of the author's work \cite{Janssen:2007rc} described in the previous chapter, where that result was extended to stationary and time-dependent $p$-brane solutions\index{$p$-brane!time-dependent}\index{$p$-brane!stationary} to gravity in arbitrary dimensions coupled to  a gauge field\index{gauge field} and a dilaton\index{dilaton}. This situation is sketched in table \ref{tab:BHstuff}.

\ctable[
caption = {Overview of the status on the existence of $\cw$ in 2007 (prior to \cite{Perz:2008kh}).},
label = tab:BHstuff,
width = 120mm,
pos = ht!,
center
]{ c>{\small}X>{\small}X}{
\tnote{At least for uneven $\cN$, see \cite{Andrianopoli:2007gt}.}
}{ 
\FL
\multicolumn{3}{c}{\textit{Existence (and form) of $\cw$ in 2007 for black holes (BH)}} \ML
\multicolumn{2}{c}{\bf EXTREMAL}& \multicolumn{1}{c}{\textbf{\mbox{NON-EXTREMAL}} }\NN
\cmidrule(rl){1-2}\cmidrule(l){3-3}
\multicolumn{1}{c}{Supersymmetric} &\multicolumn{1}{c}{Non-supersymmetric}&\multicolumn{1}{c}{\mbox{non-supersymmetric}}\NN
\multicolumn{1}{c}{$\cw \sim \e^U |Z|$}&\multicolumn{1}{c}{$\cw \sim \e^U W$}&\multicolumn{1}{c}{$\cw = ?$}\NN
\multicolumn{1}{c}{Always exists} &\multicolumn{1}{c}{Exists in principle} &\multicolumn{1}{c}{Only one example} \NN
&\multicolumn{1}{c}{for symmetric spaces}\tmark&\multicolumn{1}{c}{(charged dilatonic BH)}\LL
}
{}Note that the work cited above on \textit{extremal} black hole solutions, was done for a subset of four-dimensional (super)gravity theories, namely those for which the scalar manifold is a symmetric space. In these cases the black hole equations of motion are explicitly integrable\index{inegrable} \cite{Breitenlohner:1987dg, Gaiotto:2007ag, Gunaydin:2005mx,Bergshoeff:2008be} and more accessible for study. Such symmetric spaces appear in all supergravity theories which have a certain amount of extended supersymmetry: theories with more than 8 real supercharges\index{supercharge} ($\Cn>2$ in four-dimensional terminology) and some with 8 real supercharges or less ($\cN\leq 2$). Also the research of the previous chapter on \textit{non-extremal} solutions (refs.~\cite{Miller:2006ay} and \cite{Janssen:2007rc}) actually hinged on symmetric moduli spaces\index{symmetric space}, which furthermore had some very particular simplifying properties, see section \ref{ss:BJ-WhyDoesItWork}.

It is the purpose of this chapter to investigate all possibilities of first-order equations determined by the gradient of a scalar function $\cw$ on target space, \textit{without a priori demanding any peculiarities} the effective description should obey. In particular, two main question arise, which one would like to answer for non-supersymmetric solutions, both extremal and non-extremal ones:
\begin{itemize}
\item What is the general form of first-order flow equations\index{first-order equation} for black holes?
\item What are the necessary and sufficient conditions for a gradient flow\index{gradient flow} \eqref{eq:JP-GradientFlow_Intro} to exist?
\end{itemize}
We name the function $\cw$ that determines the gradient flow\index{gradient flow} in eq.~\eqref{eq:JP-GradientFlow_Intro} the `generalized superpotential'\index{generalized superpotential}. Note that the existence of a superpotential\index{superpotential!generalized} and gradient flow\index{gradient flow!equations} of the form \eqref{eq:JP-GradientFlow_Intro} goes much further than just exhibiting a symmetry property of the equations of motion. The equations can be interpreted as renormalization group\index{renormalization group} flow equations in a dual field theory, where the radial parameter corresponds to an energy scale, see for instance \cite{Hotta:2009bm} for extremal solutions. Moreover, having a first-order description at hand for non-extremal solutions might shed light on open problems concerning the relation between the scalar charges and the entropy\index{entropy} of non-extremal black holes\index{non-extremal!black hole} \cite{Ferrara:2008hw}.


In \cite{Perz:2008kh}, we addressed the two questions raised above for general, (non-)extremal solutions. However, a concrete condition for a generalized superpotential\index{superpotential!generalized} to exist was again only provided for theories whose moduli space\index{moduli space} is a symmetric space\index{symmetric space}. What about more general theories, not necessarily with symmetric moduli spaces? 
Motivated by results of the Torino group published in \cite{Andrianopoli:2009je}, we can explicitly give the necessary and sufficient condition for a superpotential and hence first-order flow equations of the form \eqref{eq:JP-GradientFlow_Intro} to exist, for general (non-)extremal solutions. This condition is the main point of this chapter, but did not appear in our published work. We illustrate it with examples taken from \cite{Perz:2008kh}. 

\textit{Note added:} after this manuscript was finished, \cite{Chemissany:2010zp} appeared, where it was shown that for theories with symmetric moduli spaces, the superpotential can be constructed for all (regular) extremal and non-extremal spherically symmetric black holes. This work is again mentioned in the conclusions to this chapter.\\

\outline{This chapter is structured as follows. We begin our discussion by recalling the necessary background material (section \ref{s:History_GradientFlows}). In particular, we recall what was known about the construction of a black hole effective action\index{effective action} and first-order flow equations from the existing literature, predating the work \cite{Perz:2008kh} with an overview of earlier work on gradient flow\index{gradient flow!equations} for extremal black holes. This should be seen as an addition to the discussion of the first-order equations in chapter \ref{c:BJ}, where the accent is shifted towards the specific form of the equations in terms of a superpotential. 

In section \ref{s:JP-gradient_from_sum}, we start from such a one-dimensional effective action with a black hole potential\index{black hole potential}, in an arbitrary number of space-time dimensions, and explain how to find the most general first-order flow equations from a `generalized superpotential', assuming that it exists.  The existence criterion for such a superpotential to exist is briefly touched upon, but not (yet) dug out in detail.

We then come to the main point of this chapter. In section \ref{s:JP-Integrability_GradientFlow}, we show how the effective black hole system is described as a Hamiltonian system\index{Hamiltonian system}, such that the radial variable plays the role of time. {}From classical mechanics, it is straightforward to show that a gradient flow\index{gradient flow} of the form \eqref{eq:JP-GradientFlow_Intro} exists if and only if the system is Liouville integrable\index{Liouville integravility}: there must be $n$ constants of motion in involution\index{involution} under the Poisson brackets\index{Poisson brackets}, where $n$ is the total number of scalar fields in the effective description. These considerations have not appeared in \cite{Perz:2008kh}. We comment on the results of \cite{Andrianopoli:2009je} linking the superpotential to Hamilton-Jacobi\index{Hamilton-Jacobi} theory.

Two introductory examples, the \RN\index{\RN black hole} and dilatonic black holes\index{dilatonic black hole}, illustrate the formalism in section \ref{s:JP-EasyExamples}. These examples are taken from chapter \ref{c:BJ} and are rewritten in a form appropriate to the discussion in the present chapter.

In section \ref{s:JP-Examples_Symm_modSpaces}, we give the main results of the author's work \cite{Perz:2008kh} and explain how to obtain a free-geodesic form of the effective action by timelike dimensional reduction\index{dimensional reduction!timelike --}, for systems whose scalar manifold\index{scalar manifold} is a symmetric space\index{symmetric space} after the reduction. {}From this action we derive the first-order equations and discuss a method of principle to check the existence of a superpotential\index{superpotential!existence of --}. This method is investigated for a single-scalar example (the dilatonic black hole, once more) and a two-scalar one.

This chapter ends with a discussion of our results and comments on new results that appeared after \cite{Perz:2008kh} in section \ref{s:JP-Conclusions}.
}

\section{History of first-order gradient flows\label{s:History_GradientFlows}}
Before setting out for our contribution to the research field, it is useful to explain in detail a few underlying principles, and to give a detailed overview of the literature on first-order flows for scalars in black hole backgrounds that are  determined by a gradient on the scalar target space.

\outline{We start with specifying the class of $D$-dimensional gravity theories, coupled to abelian vectors and neutral scalars we wish to study, as well as the details of the \ansatz{} in section \ref{ss:BH_D_dimensions}. In section \ref{ss:Two_Effective_Action}, we explain in detail how one can obtain an effective action that describes static, spherically symmetric black holes in $D$ dimensions\index{$D$ dimensions}. Two techniques are widely used: either one writes down an effective action  with a potential for the black hole warp factor and the scalars of the theory.  We already encountered such an effective action\index{effective action} for the four-dimensional case in section \ref{s:BJ-History_FoForm_BHs}. Another, equivalent, effective action can be obtained by performing a dimensional reduction over time. This (one-dimensional) action describes geodesics on an enlarged target space\index{target space!enlarged --}, containing the scalars of the $D$-dimensional theory, the metric warp factor \emph{and} electric (and possibly magnetic) potentials\index{potential!electric}\index{potential!magnetic}. We use both effective actions interchangeably in the remainder of the chapter. In section \ref{ss:FlowEqsLiterature}, we give a detailed overview of the literature on first-order gradient flows, both for black holes and domain walls and cosmologies. The discussion for black holes is complementary to the remarks made about first-order equations for supersymmetric black holes in four dimensions in chapter \ref{c:BJ}.
}
\subsection{Black holes in \texorpdfstring{$D$}{} dimensions\label{ss:BH_D_dimensions}}

\subsubsection{Bosonic action}
\index{Gradient flow|defn}
Our main motivation before was to look at black hole solutions to (super)gravity in four dimensions. We now extend this to $D$ dimensions, as for instance ($D=5)$-dimensional black holes can also be of considerable interest.  Therefore we consider static, spherically symmetric black hole solutions in gravity coupled to a number of neutral scalars $\phi^a$ and vector fields in $D$ dimensions that are solutions to an action of the form
\begin{equation}\label{eq:JP-Sugra-action}
S = \frac1{16\pi G_D}\int \de^{D}x \sqrt{-g_D}\Bigl(R_{D} -
\tfrac{1}{2}G_{ab}\partial_\mu\phi^a\partial^\mu\phi^b
\Bigr) + S_V\,.
\end{equation}
The influence of the vector fields is captured by the action $S_V$ and is illustrated below. {}For comparison, note that we fix the convention-dependent constant $\Conv_D$ appearing in chapter \ref{c:BJ} to $\Conv_D^2 = 1/4$ and we choose a notation $\phi^a$ for the scalars.\footnote{The reader can reinstate all factors of $\Conv_D$ by rescaling the gauge fields and charges $F\to 4\Conv_D F\,, (q,p)\to 4\Conv_D(q,p)$}.  The scalar metric $G_{ab}$ is a function that depends on the scalar fields $\phi ^ a$. Greek indices are raised and lowered with the \spacetime{} metric $g_{\mu\nu}$ and $g=\det g_{\mu\nu}$.  We  always write the vector fields as a collection $A ^I, I = 1 \ldots n$, where $n$ is the total number of vector fields. The  exact form of the action for the vector fields $S_V$ depends on the dimensionality of \spacetime{} and the class of theories we wish to consider. For example, in four dimensions, we write:
\begin{equation}
 S_V= \frac1{16\pi G_4}\int \de^{4}x \Bigl(\tfrac{1}{4}\mu_{IJ}(\phi) \, F^I_{\mu\nu} F^{J\,\mu\nu}-\tfrac14
\nu_{IJ}(\phi)\, F^I_{\mu\nu}(\star F^J)^{\mu\nu}\Bigr)\,,
\end{equation}
in terms of the Abelian field strengths $F^I_{\mu\nu} = \partial_\mu A^I_\nu - \partial_\nu A^I_\mu$ and scalar dependent gauge couplings $\mu(\phi)$ and $\nu(\phi)$.
In five dimensions, on the other hand, we can have:
\begin{equation}
 S_V= \frac1{16\pi G_5}\int \de^{5}x \Bigl(\tfrac{1}{4}\mu_{IJ}(\phi)\, F^I_{\mu\nu} F^{J\,\mu\nu}-\tfrac16 D_{IJK}\epsilon^ {\mu\nu\rho\sigma\tau}\,
A^{I}_{\mu}F^{J}_{\nu\rho}(\star F^K)_{\sigma\tau}\Bigr)\,,
\end{equation}
where the gauge coupling $\mu(\phi)$ is again scalar dependent, but the 3-tensor $D_{IJK}$ is constant, such that the second term in the vector action is a gauge invariant, topological term. To keep our discussion as general as possible we make no further assumptions about the combined gravity-scalar-vector\index{vector}\index{scalar}\index{gravity} theory: we make no specific assumptions about the form of the scalar dependent functions $G(\phi),\mu(\phi),\nu(\phi)$. The reader should notice that it is of the appropriate form to describe the bosonic sector of ungauged supergravity in $D$ dimensions, for example by comparing to the action of $\cN =2$ supergravity in four dimensions \eqref{eq:N-=2SUGRA_2}\index{supergravity!$\cN=2$, four dimensions} and in five dimensions. 

\subsubsection{Spherically symmetric \ansatz{}}
We now specifiy the \ansatz{} for the metric of a spherically symmetric\index{spherically symmetric} black hole solution. In the next section we then treat the effective action for the fields in the \ansatz{}. Spherically symmetric black hole solutions of the theory described by the action (\ref{eq:JP-Sugra-action}) have a metric of the form 
\begin{equation}
\de s^2_{D} =  - \e^{2\beta\varphi(\tau)}\de t^2
+ \e^{2\alpha\varphi(\tau)} \de s_d^2\,,
\label{eq:JP-BH_metric_Ddim}
\end{equation}
For notational simplicity, we  henceforth use the notation $d$ for $D-1$:
\begin{equation}
 d=D-1\,.
\end{equation}
The $d$-dimensional metric of the transverse space is given as
\begin{equation}
 \de s_d^2=\Bigl(\e^{2(d-1)\B(\tau)}\de\tau^2
+ \e^{2\B(\tau)}\de\Omega_{d-1}^2\Bigr),\quad
\e^{-(d-2)\B}=\frac{\sinh\left((d-2)c\tau\right)}c\label{eq:JP-BH_metric_ddim}\,.
\end{equation}
The constants $\alpha,\beta$ were introduced in equation \eqref{eq:BJ-alpha and beta} and are chosen such that the warp factor\index{warp factor} $\varphi$ has a conventionally normalized kinetic term in the effective action: 
\begin{equation}
\alpha=-1/\sqrt{2(d-1)(d-2)}, \qquad\beta=-(d-2)\alpha = \sqrt{{(d-2)}/{2(d-1)}}\label{eq:JP-beta_and_alpha}\,.
\end{equation}
(Remember we put $\Conv_D=1/2$ to compare to the epxressions for $\alpha$ and $\beta$ in the previous chapter, eq.~\eqref{eq:BJ-alpha and beta}). The constant $c$, as before, signals if the black hole is extremal ($c = 0$) or non-extremal ($c\neq0$). Under the requirement that the solution is static\index{static} and spherically symmetric, the vector fields can (but need not) be eliminated in terms of their electric/magnetic conserved charges. Finally, the scalars depend solely on the radial coordinate: $\varphi = \varphi(\tau)$, $\phi^a = \phi^a(\tau)$. 

The above \ansatz{}\index{\ansatz{}} generalizes the four-dimensional one in chapter \ref{c:BJ} for the discussion of the attractor mechanism\index{attractor mechanism}. Note that from now on, we denote the `black hole warp factor' as $\varphi$. In most black hole literature and in earlier discussions in this text, the warp factor in four and five dimensions is denoted in terms of a function $U$, related to $\varphi$ as $U = -\alpha \varphi$, i.e.~in the four-dimensional case ($d+1=4$) $U=\varphi/2$, while in five dimensions ($d+1=5$) $U= \varphi/\sqrt{3}$.  The notation $\varphi$ is also in accordance with standard Kaluza-Klein literature\index{Kaluza-Klein!literature}.

\subsection{Two effective actions\label{ss:Two_Effective_Action}}
The approach we take to study scalar flows in black hole backgrounds,  always starts from a one-dimensional effective action that encompasses the scalar dynamics. In the literature, one typically writes down two different effective actions for the radial dependent fields coupled to the black hole. We have encountered the first one before, it is obtained after eliminating the gauge fields in terms of their conserved charges. This gives rise to an action describing a particle in an enlarged target space $\mathbb{R}\times \cm$ subject to a potential, where $\mathbb{R}$ describes the metric warp factor and $\cm$ is the target space of the scalars $\phi^a$. To write down the second type of effective action, we do not integrate out the vector fields. Instead, one considers a dimensional reduction over time, a Killing direction. The dynamics of the gauge fields are given by their electric (and possibly magnetic) potentials, which appear as additional scalar fields in the dimensionally reduced theory. Moreover, the effective action does not have a potential, but describes geodesics on an enlarged target space $\mathbb{R}\times \tilde\cm$, where $\tilde\cm$ the target space containing the original scalars $\phi^a$ and electric/magnetic potentials. Note that this space does not factorize in general ($\tilde \cm \neq \cm\times \ck$, with $\ck$ the space of electric/magnetic potentials), but couples the scalars $\phi^a$ and the electric/magnetic potentials in a non-trivial manner. We give examples of this in sections \ref{s:JP-EasyExamples} and \ref{s:JP-Examples_Symm_modSpaces}. Both effective actions are supplied by a constraint, fixing one integration constant in terms of the non-extremality parameter $c^2$.

\subsubsection{Action with black hole potential}
\index{black hole potential}
The first technique to construct the effective action
\cite{Gibbons:1996af, Ferrara:1997tw} expresses the Maxwell field
strengths in terms of the electric charges, and possibly also magnetic charges in $D=4$ dimensions, via the respective
equations of motion (and Bianchi identities).\footnote{In more than four dimensions, a 0-brane can only source vectors through an electric coupling. 
}  The one-dimensional effective action obtained as explained above turns out to be that of a particle subject to an
external force field given by the effective black hole potential $V$ (we drop a proportionality factor $1/16\pi G_D$):
\begin{equation}
S=-\int \de \tau \,\Bigl(\tfrac 12 \dot \varphi^2
+\tfrac{1}{2}G_{ab}(\phi)\dot{\phi}^a\dot{\phi}^b +
\e^{2\beta\varphi}V(\phi)\Bigr),\label{eq:JP-action1}
\end{equation}
where a dot means differentiation with respect to the radial parameter $\tau$. The configuration space of this `fiducial' particle is a direct product $\mathcal{M}\times \Real$ where $\mathcal{M}$ is the scalar target space with metric $G_{ab}$, and $\Real$ represents the warp factor. The `mass parameters' in the black hole potential $V$ are given by the electric and magnetic charges. Solutions to this action have to obey a constraint, stemming from part of the information in the $D$-dimensional Einstein equations which cannot be derived from the effective action:\footnote{The constraint can be found from the effective action by introducing an `einbein' encoding the repa\-ra\-metrizations of the radial coordinate. The
einbein acts as a Lagrange multiplier enforcing the
constraint. See section \ref{s:BJ-MainResults}, more specifically the metric \ansatz{} \ref{eq:MetricSinh}, where $\e^\C$ denoted the einbein.}
 \begin{equation}
{(d-1)(d-2)}c^2=\tfrac 12 \dot \varphi^2
+\tfrac{1}{2}G_{ab}(\phi)\dot{\phi}^a\dot{\phi}^b -
\e^{2U}V(\phi)\label{eq:JP-HamiltonianI}
\end{equation}
In the case $D=4$, we have the identifications $V = 2V_{BH}$ introduced in chapter \ref{c:BJ}, $\varphi = U/2$ and this constraint reverts to the equation \eqref{eq:HamConstraintN=2} in the discussion of the four-dimensional black hole. We elaborate on the use of this effective action for $D$-dimensional black holes in section \ref{s:JP-gradient_from_sum}.

\subsubsection{Action describing geodesics on an enlarged target space}
The second technique for constructing a one-dimensional effective action, first described in the $D = 4$ case in
\cite{Breitenlohner:1987dg}, is based on the observation that a static solution in $D$ dimensions can be dimensionally reduced over time to a Euclidean $d=(D-1)$-dimensional instanton solution. Because of the assumed spherical symmetry, the resulting instanton solutions are carried only by the metric and the scalars in $D$ dimensions. 
Since the reduction is performed over a Killing direction, the $d$-dimensional solutions fully specify the solutions in $D$ dimensions. As explained in \cite{Breitenlohner:1987dg, Bergshoeff:2008be} the equations for the $D$-dimensional metric decouple and are easily solved. 

The scalar field equations of motion are found from the following effective one-dimensional action (again, dots are $\tau$-derivatives)
\begin{equation}
S=\int \de \tau \, \tilde \cg_{ij}\dot{\tilde q}^i
\dot{\tilde q}^j\,,\label{eq:JP-action2}
\end{equation}
which describes the free geodesic motion of a particle in an enlarged target space of scalar fields $\tilde q^i$ that contain the scalar fields $\phi^a$ of the $D$-dimensional theory plus axion-type scalar fields arising from the reduced vector potentials. The choice of notation $\tilde q^i,\tilde \cg_{ij}$ becomes clear later. This action has to be complemented by the Hamiltonian
constraint
\begin{equation}
(d-1)(d-2)c^2 =  \tfrac12\tilde \cg_{ij}\dot{\tilde q}^i
\dot{\tilde q}^j\,.\label{eq:JP-HamiltonianII}
\end{equation}
In the remainder of this chapter, we always use the notation $\tilde \cg$ for the metric of the moduli space  in the reduced (Euclidean) gravity theory. Note that in this procedure the vectors (or equivalently, the axions) are not eliminated by their equations of motion. These axionic scalars have the opposite sign for their kinetic term, which causes the metric $\tilde{\cg}$ to have indefinite signature. If we were to integrate out those axions, we would find the other black hole effective action (\ref{eq:JP-action1}).


\subsubsection{Choosing which is best}
The question of which technique (or effective action) is best suited for the given task depends on the theory one considers and on which aspects of black hole solutions one wishes to investigate. For instance, if the scalar target space in the effective action of the second type    is a symmetric space then the geodesic equations are manifestly integrable and can be used to construct explicit solutions, see for instance \cite{Bergshoeff:2008be} for more details. When one is interested in studying supersymmetry and the black hole attractor mechanism, as in chapter \ref{c:BJ}, the first approach is more commonly used. 

\outlook{Outlook:}{Having explained two possible ways of writing down a one-dimensional effective action for black hole solutions in $D$ dimensions, we continue with a short review of cases known in the literature where the scalars follow a gradient flow\index{gradient flow} on the target space. After this review, we ask the question: under what circumstances does such a gradient flow\index{gradient flow} exist (for both extremal and non-extremal solutions) in section \ref{s:JP-gradient_from_sum}.}

%

\subsection{Flow equations in the literature \label{ss:FlowEqsLiterature}}
We recall the appearance of gradient flow\index{gradient flow!equations} in the literature. First, we consider the main point of interest and focus on black hole solutions. 
Afterwards, we also discuss the superpotential\index{superpotential} and corresponding gradient flow equations\index{gradient flow!equations} for the scalars in domain wall\index{domain wall} and cosmology\index{cosmology} solutions. 


\subsubsection{Black holes}
We now recast the discussion of section \ref{s:BJ-History_FoForm_BHs}, which concentrated on supersymmetric black holes in $D=4$ dimensions, to arbitrary \spacetime{} dimension $D=d+1$. The discussion of such flow equations follows naturally from a certain rewriting of the black hole potential. Therefore we start from the first type of effective action, see \eqref{eq:JP-action1}. Suppose the black hole potential $V(\phi)$ can be written in terms of a function $W(\phi)$ as:\footnote{Note that in our conventions, the constant $\Conv_D^2$ in the Einstein--Hilbert term $\tfrac{1}{4\Conv_D^2}\sqrt{-g_D}R_D$ introduced in earlier chapters is set to $\tfrac{1}{4}$. This influences the coefficients in \eqref{eq:JP-centralcharge}.}
\begin{equation}\label{eq:JP-centralcharge}
V=\frac14\left(W^2 + \frac 1 {\beta^2}G^{ab}\partial_a W\partial_b W\right)\,.
\end{equation}
Remember that the constant $\beta$ appears in the metric \eqref{eq:JP-BH_metric_Ddim} and is given as $\beta^2 = \frac 12 \frac{d-2}{d-1}$. Whenever the potential allows such a rewriting in terms of $W(\phi)$, the equations of motion for the warp factor and the scalars  reduce to\footnote{To compare to the form \eqref{eq:JP-GradientFlow_Intro} of the flow equations, we can write $q^i=(\varphi,\phi^a)$ and $\cw(q) = \e^{\beta\varphi}W/2\beta$. This rewriting is worked out further from section \ref{s:JP-gradient_from_sum} onwards.}
\begin{align}
\dot{\varphi}&=-\frac{1}2\e^{\beta\varphi}W\nonumber\,,\\
\dot{\phi}^a &=-\frac1{2\beta}\e^{\beta\varphi}G^{ab}\partial_{b}W\label{eq:JP-FLOW'III}\,.
\end{align}
These equations are called BPS or \emph{gradient flow equations}\index{gradient flow}, and they describe an \emph{attractor flow}\index{attractor flow} if there is an attractive fixed point (that is, when the black hole potential has a minimum).
The real function $W(\phi)$ is often called the superpotential\index{superpotential}. We have seen some of its properties before. Evaluated at the black hole horizon, it determines the horizon area and hence the Bekenstein-Hawking entropy\index{entropy!Bekenstein-Hawking --} of the black hole as $S_{BH} = a W^{\frac{d-1}{d-2}}\big{|}_H$, where $a$ is some $d$-dependent constant. {}For instance for $d=3$, the case of the four-dimensional black hole, we have $S_{BH}=\left. \pi W^2\right|_H$. When evaluated at spatial infinity\index{spatial infinity}, the superpotential $W$ gives the ADM mass\index{ADM mass}, $M\sim \left.W\right|_\infty$, while its derivatives give the scalar charges\index{scalar charge} $\dot \phi^a|_\infty \sim \partial_a W|_\infty$.  We distinguish three cases:
\begin{itemize}
  \item \underline{\textit{Supersymmetric black holes.\index{black hole!supersymmetric}} \index{superpotential!for supersymmetric black holes}} {}For a supersymmetric (BPS) black hole \ansatz, the first-order, integrated form of the second-order equations of motion in terms of the $W(\phi)$ correspond to the first-order Killing spinor equations in $D=d+1$ dimensions \cite{Ferrara:1995ih,Strominger:1996kf,Ferrara:1996dd}. In this case we also have that the superpotential $W$ is equal to the modulus of the complex central charge function:
\begin{equation}
 W(\phi) = |Z(\phi)|\,.
\end{equation}
Note that the attractor values\index{attractor values!of scalars} of the scalars $\phi$ minimize the black hole potential\index{black hole!potential}. {}For supersymmetric solutions, these are the values of the scalars such that also $W = |Z|$ is minimal, i.e.~$\partial |Z|/ \partial \phi = 0$ at the horizon.

\item \underline{\textit{Extremal, non-supersymmetric black holes.}}\index{black hole!extremal, non-supersymmetric} \index{superpotential!for extremal  black holes} {}For non-supersymmetric black holes the first-order equations are no longer guaranteed to exist. Nonetheless, some years ago, Ceresole\index{Ceresole} and Dall'Agata\index{Dall'Agata} \cite{Ceresole:2007wx} showed that it is possible for some extremal non-supersymmetric black holes to mimic the BPS equations\index{BPS equation} of their supersymmetric counterparts, exactly in the sense of the first-order equations (\ref{eq:JP-FLOW'III}). Shortly after, Andrianopoli\index{Andrianopoli} et al.~\cite{Andrianopoli:2007gt} gave the explicit form for the superpotential $W$ for well-nigh all extremal solutions (supersymmetric and non-supersymmetric alike) to extended four-dimensional supergravity, when $\Cn>2$.\footnote{{}For even $\cN$ there is a minor issue with finding a manifestly duality-invariant\index{duality-invariant} form of the superpotential.} {}For $\Cn=2$ some explicit cases were treated. The rewriting in terms of a superpotential $W$ is not necessarily a remnant of supersymmetry in one dimension higher, see \cite{Cardoso:2007ky}.

Necessarily, extremal black holes that are not supersymmetric have a superpotential that is not equal to the central charge function:
\begin{equation}
 W(\phi) \neq |Z(\phi)|\,.
\end{equation}
Also for these solutions the  attractor values of the scalar fields minimize the potential. These solutions share the property that again $\partial W /\partial \phi= 0$ at the horizon. However, note that $W\neq|Z|$ for non-supersymmetric solutions and \mbox{$\partial |Z|/ \partial \phi \neq 0$} at the black hole horizon. 

Subsequent work has provided further examples of the hidden structure in non-supersymmetric extremal solutions, see e.g.~\cite{Andrianopoli:2007gt,Ferrara:2007qx,Ferrara:2008hw, Ferrara:2008ap,Bellucci:2008sv} and references therein. Of most direct relevance for the author's results described in this chapter, is the work of Andrianopoli et al. \cite{Andrianopoli:2007gt}.
  
  \item \underline{\textit{Non-extremal black holes.}}
While non-extremal black holes\index{black hole!non-extremal} are of considerable interest, little is known about their possible interpretation as solutions of first-order equations\index{first-order equation}. We can say for sure that a rewriting of the black hole potential in terms of a superpotential $W$ as in equation \eqref{eq:JP-centralcharge} is impossible, as follows. If we were to follow that route, the first-order equations, which give $\dot \varphi,\dot \phi$ in terms of the function $W$, would follow. Plugging those first-order relations into the Hamiltonian constraint\index{Hamiltonian constraint} \eqref{eq:JP-HamiltonianII}, would then imply that $c^2 = 0$ -- the prescription with the superpotential $W$ above is only consistent for extremal black holes.

Strangely enough, we saw in the last chapter that also for non-extremal black hole solutions, a first-order description can exist (albeit necessarily of a different form than for the extremal case). It is rather surprising that some non-extremal solutions can be found from first-order equations derived from a superpotential, see \cite{Lu:2003iv} and more recently Miller et al.~\cite{Miller:2006ay}, as explained in section \ref{ss:BJ-Review_MSW}. The latter authors provided the simplest possible example--the non-extremal Reissner-Nordstr\"om black hole in $D=4$ dimensions--by making use of Bogomol'nyi's trick of completing the squares from non-gravitational field theories. It was pointed out in \cite{Miller:2006ay} that the coupling to gravity introduces at least one term with a relative minus sign. Exactly the relative minus sign makes the rewriting of the action as a sum (difference) of squares non-unique and allows one to introduce a one-parameter deformation. This leads to the non-extremal version of first-order equations\index{first-order equation}, with the deformation parameter\index{deformation parameter} measuring the deviation from extremality.

The Bogomol'nyi\index{Bogomol'nyi} approach has been generalized to include the non-extremal dilatonic black hole \index{dilatonic black hole} and $p$-brane solutions, as well as time-dependent (cosmological) solutions in arbitrary dimensions \cite{Janssen:2007rc} and non-extremal black holes in gauged supergravity\index{supergravity!gauged --} \cite{Lu:2003iv, Cardoso:2008gm}.  The explicit structure of non-extremal flow equations in theories with more complicated scalar matter couplings\index{matter coupling} is not known, although some suggestions were made in \cite{Andrianopoli:2007gt}.
\end{itemize}

\subsection{Outlook}
We have given an overview of the literature on known instances where the scalars of an effective supergravity theory follow a gradient flow, determined by a scalar function $W$ on the scalar manifold. In the following sections, the main objective is to obtain the most general form of such gradient flow\index{gradient flow!equations} for \emph{black hole systems in $D$ dimensions which are not supersymmetric}. Doing so, we hope to extend the known results discussed above for the non-supersymmetric extremal solutions and see what can be said about non-extremal solutions.  The main focus of the gradient flow\index{gradient flow} formalism will be on finding an answer to the questions: when does a gradient flow\index{gradient flow} exist? And is a possible rewriting unique? 
\section{Gradient flow from sum of squares\label{s:JP-gradient_from_sum}}
In this section, we return to spherically symmetric\index{spherically symmetric} black hole solutions to the $D$-dimensional action given in eq.~\eqref{eq:JP-Sugra-action} with a metric \ansatz{} as before \eqref{eq:JP-BH_metric_Ddim}. We have seen that the metric and scalars\index{metric}\index{scalar} in such an \ansatz{} can be described by two types of effective action\index{effective action}. Here, we choose to work with the effective action with a black hole potential\index{black hole potential}.

As we have seen in the discussion of the previous section, a successful proposal of rewriting the black hole potential $V(\phi)$ in terms of a real superpotential $W(\phi)$ for some \emph{extremal} solutions has been made before. We now make a proposal that generalizes the form of the relation \eqref{eq:JP-centralcharge} to \textit{non-extremal} black holes. 

Before we continue, we rewrite the black hole effective action with a potential, in order to make the discussion more transparent. We arrange the black hole warp factor $\varphi$ and the scalars $\phi^a$ of the $D$-dimensional theory into one coordinate vector $q^i(\tau)$ as
\begin{equation}
 q^i(\tau) = (\varphi(\tau),\phi^a(\tau))\,.\label{eq:JP-qi_defn}
\end{equation}
Then the effective action \eqref{eq:JP-action1} is written as:
\begin{align}
\cl 
&= \tfrac{1}{2}\cg_{ij}(q)\dot{q}^i\dot{q}^j +
\cv(q)\,.\label{eq:JP-action1_bis}
\end{align}
The potential $\cv(q)$ and the metric $\cg$ on the target space with coordinates $q^i$ are the natural extension of the original potential and metric and can be read off from  \eqref{eq:JP-action1}: 
\begin{align}
 {\cal G}_{ij}(q) = \begin{pmatrix}1 &0\\0&G_{ab (\phi)} \end{pmatrix}\,,\qquad \cv(q) = \e^{2\beta \varphi} V(\phi)\,.\label{eq:JP-CalG_NormG}
\end{align}
The Hamiltonian constraint\index{Hamiltonian constraint} \eqref{eq:JP-HamiltonianI} is written as 
\begin{equation}
\tfrac{1}{2}\cg_{ij}(q)\dot{q}^i\dot{q}^j -
\cv(q)={(d-1)(d-2)}c^2\,.\label{eq:JP-Hamiltonian_q_first}
\end{equation}
The above notation in terms of $q^i$ is used throughout this chapter. 

\subsection{Generalized superpotential\label{ss:gener_superpot}}
%
\index{superpotential!generalized --|see{generalized superpotential}}
\index{generalized superpotential|defn}
\index{generalized superpotential|(imp}
Let us assume that there exists a function ${\cal W}(q^i) = \cw(\varphi,\phi^a)$, which we call the `generalized superpotential', such that\footnote{Remark on notation: In the article \cite{Perz:2008kh}, we denoted the generalized superpotential as a function $Y$. However, personally I prefer to follow reference \cite{Andrianopoli:2009je} and denote the superpotential as $\cw$.}
\begin{equation}
\cv(q)=\frac12\cg^{ij}\partial_i \cw \partial_j \cw - \En\,,
\label{eq:JP-V_intermsof_W}
\end{equation}
where we write $\partial_i = \partial/\partial q^i$ and $\En$ is a constant to be determined later (eq.~\eqref{eq:JP-gammadeltaE} below). The effective action (\ref{eq:JP-action1_bis}) can then be written in the following form, up to boundary terms:\footnote{In fact a minus sign is also possible within the squares, but this choice amounts to a redefinition of $\tau$ and $\cw$, so without loss of generality we may choose plus.}
\begin{align}
S=&-\frac12\int\de\tau\,\cg_{ij}(\dot q^i+\cg^{ik}\partial_k \cw)(\dot q^j+\cg^{jk}\partial_k \cw),\label{eq:SumSquares1}
\end{align}
The first-order form of the equations of motion is  obtained by putting the terms within brackets in (\ref{eq:SumSquares1}) to zero, giving a stationary point of the action:
\begin{align}
\dot q^i = -\cg^{ij}\partial_j {\cal W} \qquad\Leftrightarrow\qquad 
\dot{\varphi}&=-\tfrac12\partial_{\varphi}\cw\nonumber\,,\\
\dot{\phi}^a &=- \tfrac1{2\beta}G^{ab}\partial_{b}\cw\label{eq:JP-gen_flow}\,.
\end{align}
We call these first-order equations \emph{generalized flow equations}. Note that the Hamiltonian constraint (\ref{eq:JP-Hamiltonian_q_first}) fixes the constant
$\En$ appearing in (\ref{eq:JP-V_intermsof_W}) to be
\begin{equation}\label{eq:JP-gammadeltaE}
\En ={(d-1)(d-2)}c^2\,,
\end{equation}
or in other words, the constant $\En$ is only non-zero for non-extremal solutions. A few comments regarding our proposal are in order.
\paragraph{Remarks}
\begin{itemize}
\item[\ra] {\bf The extremal case.}
We compare our proposal for a generalized gradient flow\index{gradient flow} in terms of the superpotential $\cw$ to the proposal of \cite{Ceresole:2007wx,Andrianopoli:2007gt} for extremal black holes\index{extremal!black hole} (see section \ref{ss:FlowEqsLiterature}). Only in the extremal case ($\En=c^2=0$), can we search for a factorized form of the generalized superpotential $\cw(\varphi, \phi^a)$ as:
\begin{equation}
\cw(\varphi,\phi^a)=\frac1{2\beta}\e^{\beta\varphi}W(\phi^a)\,,\label{factorization}
\end{equation}
such that the black hole potential assumes the familiar form \eqref{eq:JP-centralcharge}
 \begin{equation}
V(\phi) =\frac14( W^2+\frac1{\beta^2}\partial_a W\partial^a W)\,,
\end{equation}
and the flow equations\index{flow equation} become the known expressions for extremal black hole\index{extremal!black hole} solutions (see equations (\ref{eq:JP-FLOW'III})). For non-extremal solutions, such a factorization \textit{cannot} hold.

Note that it is a priori possible to have also extremal solutions for which the generalized superpotential does not factorize as $\cw(\varphi,\phi)=\frac1{2\beta}\e^{\beta \varphi} W(\phi)$. This corrects a mistake in \cite{Perz:2008kh}, where it was claimed that extremality implies factorization of the superpotential. The correct statement would be that factorization only follows for extremal black holes which are non-singular (non-zero horizon area, ill-behaved scalars\ldots).
\item[\ra] {\bf Comparison to the first-order equations of chapter \ref{c:BJ}.} The first-order equations derived in the author's work \cite{Janssen:2007rc} and discussed in chapter \ref{c:BJ}, can also be written in terms of a gradient flow\index{gradient flow}, see section \ref{ss:BJ-WhyDoesItWork}: they form a certain subset of the most general gradient flow\index{gradient flow!equations} considered here. The only difference is that the effective description in chapter \ref{c:BJ} contained a second metric function, denoted $\B$. Eliminating that field by its equations of motion, exactly gives rise to the constant $\En=(d-1)(d-2)c^2$ in the expression relating the potential $\cv$ to the generalized superpotential\index{generalized superpotential} $\cw$ (eq.~\eqref{eq:JP-V_intermsof_W}).  In chapter \ref{c:BJ}, only the first-order equations were given, but $\cw$ was not constructed. We  perform this latter calculation for two examples of chapter \ref{c:BJ} in section \ref{s:JP-EasyExamples}.
\item[\ra] {\bf Comparison to an earlier proposal in the literature.} The form of the flow equations for non-extremal solutions presented here differs somewhat from the conjecture made in \cite{Andrianopoli:2007gt}, which proposes to preserve the form of the flow equations from the extremal case by putting $\cw(\varphi,\phi) = \frac1{2\beta}\e^{\beta \varphi} W$, but allows $W$ to explicitly depend on $\tau$: $W(\phi,\tau)$. Noting that explicit $\tau$-dependence can locally be rephrased as $\varphi$-dependence, with $\tau$ then considered as a function of $\varphi$, one sees that this is in the same vein as our proposal. The two are not equivalent, however, as in \cite{Andrianopoli:2007gt} the dependence of $W$ on $\tau$ is of a very specific kind: $\partial_{\tau}W\sim-\En \e^{-\varphi/2}$. Using the example of the dilatonic black hole\index{dilatonic black hole} in section \ref{s:JP-EasyExamples}, we shall show that the flow equations we present here (based exclusively on the premise that the effective action should be written as a sum and difference of squares\index{sum of squares}) are more general.
\end{itemize}

\subsection{Existence of a generalized superpotential, first remarks}\label{ss:Existence}

\index{superpotential!existence of generalized --}
Note that the proposal we made for the form of a scalar gradient flow\index{gradient flow} that can also hold for non-extremal black holes, is valid \textit{under the condition the superpotential $\cw$ exists}. We initiate the discussion of the existence\index{superpotential!existence of --} of the generalized superpotential\index{generalized superpotential} $\cal W$ now and make a thorough analysis in the next section.

We reformulate the discussion of the previous section. The question is to see if the scalar equations of motion\index{equations of motion!for scalars} are equivalent to a scalar flow of the form 
\begin{equation}
\cg_{ij} (q) \dot q^j = f_i(q)\label{eq:JP-flow_eqs_fi}\,, 
\end{equation}
such that the functions $f_i$ are the gradient of the superpotential $\cw(q)$
\begin{equation}
f_i(q) = \pf {\cw(q^j,C_j)}{q^i}\,. \label{eq:JP-f_i-W}
\end{equation}
(compare eqs.~\eqref{eq:JP-gen_flow}.) Since such a rewriting as first-order equations requires an integration of the original second-order equations of motion, we include an explicit dependence on integration constants\index{integration constant} $C_j$.
The necessary and sufficient condition for this to hold locally is, by Poincar\'e's lemma\index{Poincar\'e's lemma}, that $\partial_{i} f_{j}(q) = \partial_j f_i(q)$, or in other words, that the one-form $f = f_i(q) dq$ is closed:
\begin{equation}\label{eq:BJ-curl}
\partial_{i} f_{j}=\partial_j f_i \qquad \Leftrightarrow\qquad d f = 0\,.
\end{equation}
This should be read as a condition on the velocity field\index{velocity field} $\dot q ^i = \cg^{ij}f_j(q)$. 
Whether or not the field $\cw(q)$ is defined over the whole target space\index{target space} depends on the cohomology of target space. We allow the possibility that $\cw$ can only be constructed piecewise on the scalar manifold\index{scalar manifold} (i.e.~defined on different patches, but not globally).

\begin{figure}
\centering
\subfigure[The flow generated by $\vec f = (q_2,-q_1)$, for which $\partial_1 f_2 - \partial_2 f_1 =2 \neq 0$. ($\cw$ does not exist)]{
\begin{picture}(160,160)
\put(0,160){\epsfig{figure=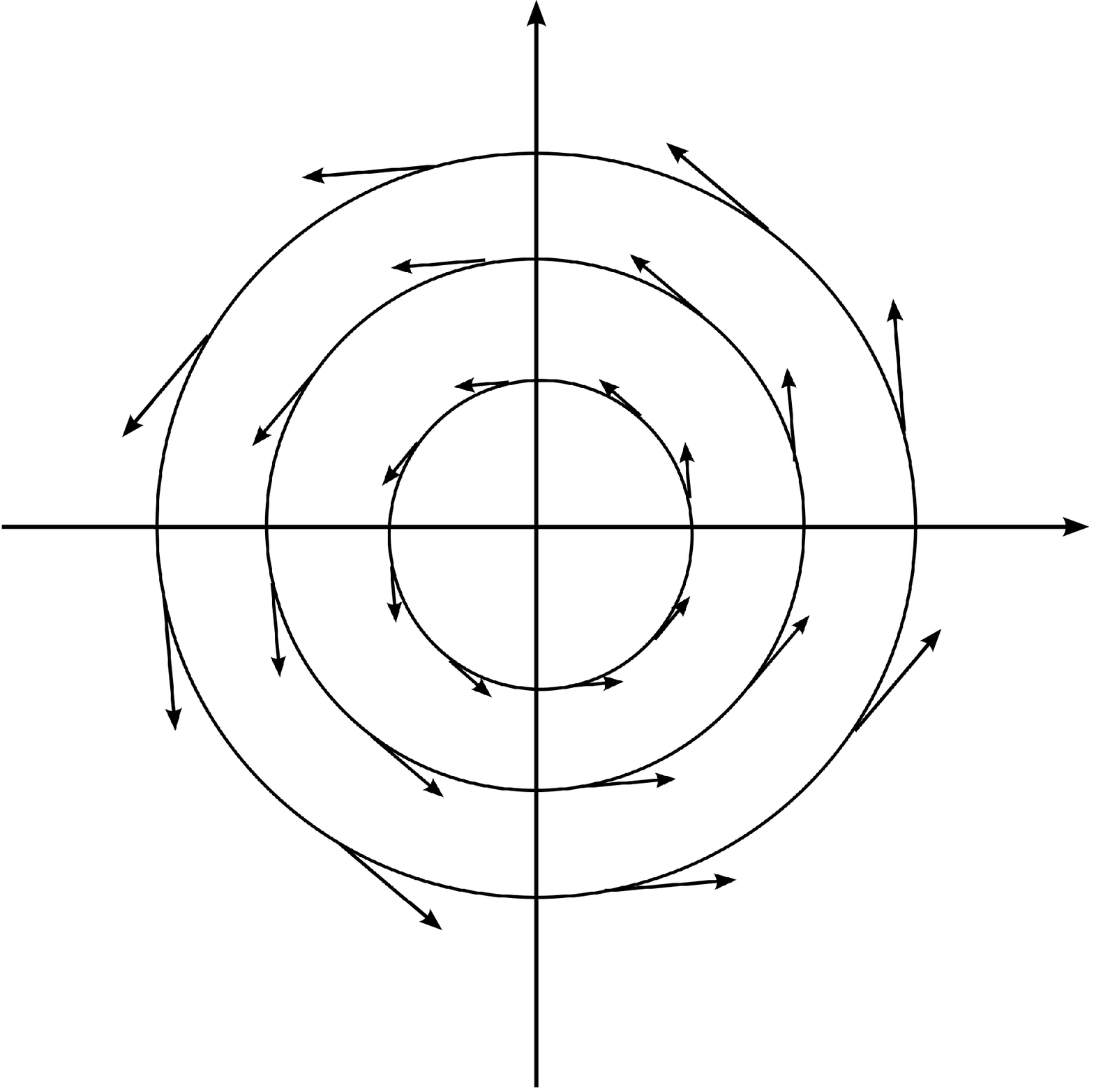,scale=.28,angle=-90}}
\put(63,153){$q^2$}
\put(147,92){$q^1$}
\end{picture}
}
\hspace{.5cm}
\subfigure[Flow generated by $\vec f = (q^2,q^1))$, for which $\partial_1 f_2 - \partial_2 f_1 = 0$. ($\cw = q^1 q^2$)]{
\begin{picture}(160,130)
\put(0,7.5){\epsfig{figure=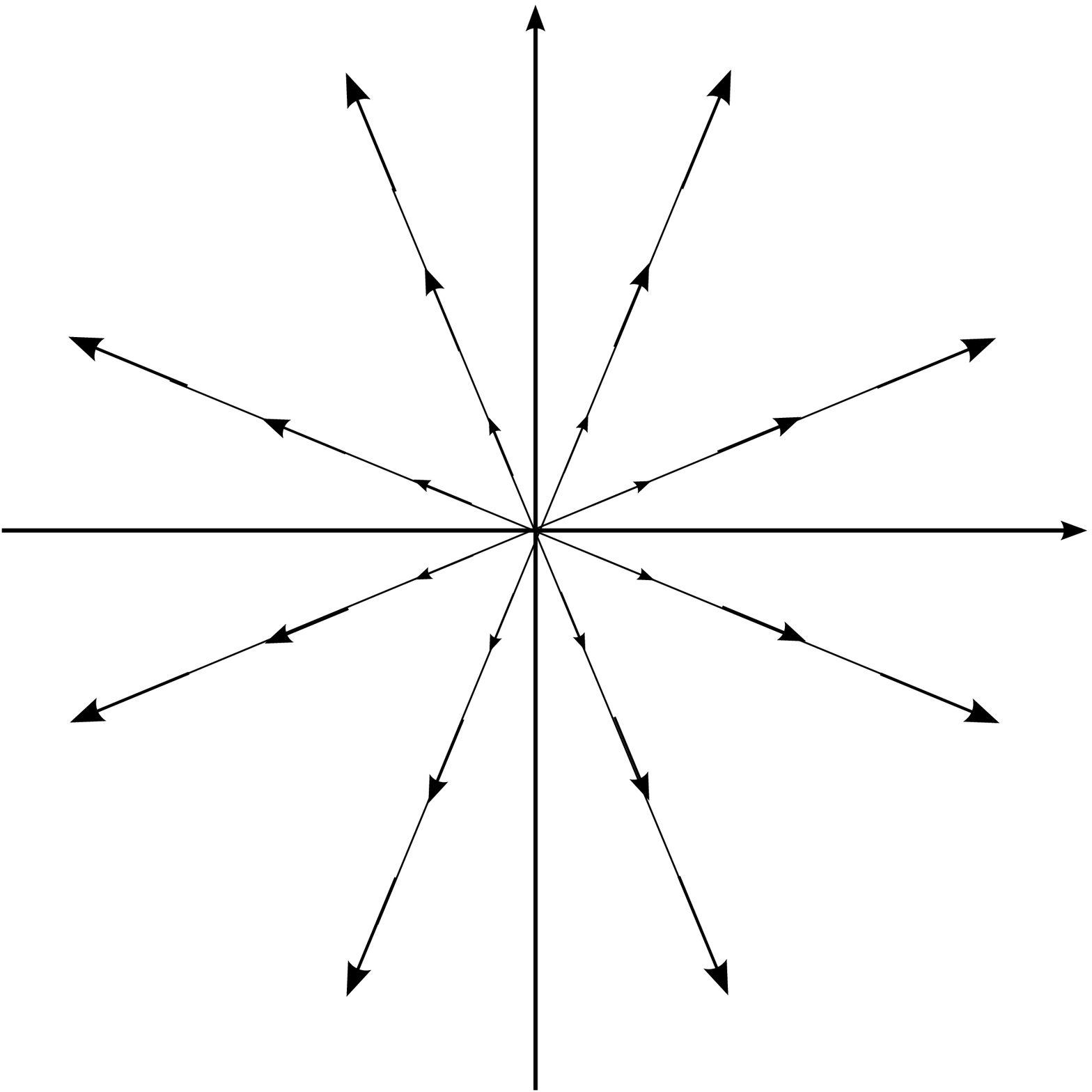,scale=.28,angle=0}}
\put(63,153){$q^2$}
\put(147,92){$q^1$}
\end{picture}
}
\caption[Depiction of a (gradient) flow in a two-dimensional example.]{Depiction of a flow in a two-dimensional example. On the axes, we have $q^1,q^2$, the arrows represent the vector $\vec f = (f_1(q),f_2(q))$. The filled lines represent the flow, which can be integrated from $f_i(q) = \cg_{ij}\dot q^j$.\label{fig:JP-GradientFlow} }
\end{figure}

In general the existence of $\cw$ is a hard question to answer. In fact, we are asking about the properties of a velocity field\index{velocity field} on the manifold the scalars $q$ parametrize.  Demanding that the velocity field is fully determined by one function $\cw$ puts severe restrictions on its form. We represent the gradient flow\index{gradient flow} pictorially in figure \ref{fig:JP-GradientFlow}. 
This can be contrasted to electrostatics\index{electrostatics}, for example. The electric field $E_i = \partial_i V(\vec x)$ is given as the gradient of an electrostatic potential\index{electrostatic potential}. However, not any vector field is of this form, of course, one needs that the vector $\vec E$ is rotation-free $\vec\nabla \times \vec E=0$.  In our setup, the analogous condition is \eqref{eq:BJ-curl}.

The \textit{original motivation} of our work \cite{Perz:2008kh}, was to provide a proof of the existence of a gradient flow, or equivalently, a proof that \eqref{eq:BJ-curl} holds. However, a general proof was not found, instead a method of principle was given to check the existence of gradient flow equations\index{gradient flow!equations} for supergravity theories\index{supergravity!theories} for which the scalar manifold\index{scalar manifold} is a symmetric space\index{symmetric space}. In later literature, the existence of a gradient flow\index{gradient flow!equations} was answered more generally. Even though it could be more logical to present the existence criterion\index{existence criterion!for generalized superpotential} first, we keep to the historical flow of events. Therefore, we first have a short overview of the arguments made in \cite{Perz:2008kh}. In the next section we go into a detailed analysis of eq.~\eqref{eq:BJ-curl} and more recent literature.

\subsubsection{Black holes with a single scalar field}
The argument for the existence of a superpotential for regular and \emph{extremal} black hole solutions involving one scalar field is taken from the fake supergravity formalism for single scalar domain walls \cite{Skenderis:2006jq} and proceeds as follows. Assume that the extremal solution exists, such that we know $\varphi(\tau)$.  As explained above, for a non-singular extremal solution, we have $\cw(\varphi,\phi) = \frac1{2\beta} \e^{\beta\varphi}W$ and $\dot \varphi = -\frac12\e^{\beta \varphi} W$ (see (\ref{eq:JP-FLOW'III}) above). This can be used to give $W$ in terms of the radial parameter $\tau$, i.e.~this defines the function $W(\tau)$. Since the black hole is supported by a single scalar $\phi$, we have that $W$ depends only on $\phi$. Locally we can always invert the function $\phi(\tau)$ to $\tau(\phi)$ and this defines $W(\phi)$.

Having constructed the fake superpotential $W(\phi)$ for the extremal solution, we could then attempt the deformation technique of \cite{Miller:2006ay} to obtain the function $\cw(\varphi,\phi)$ in the \textit{non-extremal} case. However, this approach requires the Lagrangian to satisfy specific simplifying conditions (see section \ref{ss:BJ-WhyDoesItWork} for details), which are not satisfied in general. More thought has to be given to this case, we come back to this in the next section.

\subsubsection{Black holes with multiple scalar fields\label{sss:MultiScalar}}
When a black hole solution is carried by multiple scalars, the above argument for the existence of \emph{extremal} flow equations does not apply\footnote{Unless some complicated conditions are satisfied, as explained for domain walls in \cite{Celi:2004st, Sonner:2007cp}.} and also for non-extremal solutions we do not have an immediate starting point. The idea of our work \cite{Perz:2008kh}, was to first perform a dimensional reduction over time. In the dimensionally reduced theory, all fields of interest (original scalars $q^i$ and gauge field potentials) are scalars, and we studied the construction of a gradient flow from that `enlarged' scalar manifold. This method is explained in section \ref{s:JP-Examples_Symm_modSpaces}, but first we give the existence criterion for $\cw$ to exist starting from the original scalar manifold with coordinates $q^i$ in the following section (which did not appear in \cite{Perz:2008kh}).


\index{generalized superpotential|)}

\section{Existence criterion for a gradient flow from Hamiltonian dynamics}\label{s:JP-Integrability_GradientFlow}
This section is a combination of new results in the literature that appeared after \cite{Perz:2008kh} and some unpublished ideas, inspired mostly by Andrianopoli et.~al.~\cite{Andrianopoli:2009je}. The aim is to investigate what condition(s) the effective black hole description has to obey, such that the integrability condition $\partial_i f_j = \partial_j f_i$ is satisfied and the scalars thus follow a gradient flow\index{gradient flow} in target space. 

\outline{
We first show in section \ref{ss:JP-Eff_BH_Hamiltonian} that the effective black hole description with a potential, see eq.~\eqref{eq:JP-action1}, can be seen as a Hamiltonian system $(q^i,p_i)$, where $q^i$ consist of the metric warp factor and the scalars in the original supergravity theory and $p_i\equiv \cg_{ij}\dot q^j$ are the associated momenta ($\cg$ is the appropriate metric on the scalar target space, see eq.~\eqref{eq:JP-CalG_NormG}). 
Using standard classical mechanics, we show in section \ref{ss:JP-Liouville_Existence} that a superpotential and the matching gradient flow\index{gradient flow!equations} exist \emph{if and only if} there are $n$ constants of motion in involution\index{involution} (i.e.~they are all commuting under the Poisson brackets\index{Poisson brackets}), $n$ is the number of scalars $q^i$. One says the Hamiltonian system is Liouville integrable.\index{Liouville!integrability} 

It has been realized before that the \textit{Hamilton-Jacobi formalism}\index{Hamilton-Jacobi} can be a helpful tool for constructing a superpotential $\cw$. The equivalence of Hamilton's characteristic function appearing in the standard treatment of the Hamilton-Jacobi equation and the superpotential $\cw$, has been shown in the case of domain walls in \cite{deBoer:1999xf,Verlinde:1999xm,Fukuma:2002sb}, where $\cw$ was also linked to the study of the renormalization group flow in the dual CFT, and in the context of the domain wall/cosmology correspondence in \cite{Skenderis:2006rr,Townsend:2007aw, Townsend:2007nm} and references therein. Only recently was the relation to black hole solutions exploited, first for extremal solutions in \cite{Hotta:2009bm} and later for non-extremal solutions as well in \cite{Andrianopoli:2009je}. We comment on this in section \ref{ss:JP-Hamilton-Jacobi}. Note that such a discussion remains on the formal level: one can always find a superpotential locally by using the Hamilton-Jacobi formalism, but this does not mean that a superpotential $\cw$, and thus a gradient flow\index{gradient flow} of the form $p_i = \partial \cw / \partial q^i$ exists globally. In order to be sure that a gradient flow and a superpotential $\cw$ exist on the entire  moduli space (or a certain patch thereof), one still needs to assume the system is Liouville integrable. In classical mechanics literature, there has been some misconception about this issue, where people have mistakenly assumed all Hamiltonian systems to be solvable or even integrable in the sense of Liouville. Also on this we comment in section \ref{ss:JP-Hamilton-Jacobi}. 

Finally, following \cite{Andrianopoli:2009je}, the superpotential is shown to be invariant under the action of the electric-magnetic duality group in section \ref{ss:W_dual_invariant}.
}

\index{integrability!Liouville|defn}


\subsection{Effective black hole description as a Hamiltonian system\label{ss:JP-Eff_BH_Hamiltonian}}
We show how the equations of motion corresponding to the effective action with a black hole potential\index{black hole!potential} \eqref{eq:JP-action1_bis}, describing static, spherically symmetric black hole solutions in $D$ dimensions, fit in a Hamiltonian description. The radial coordinate\index{radial coordinate} $\tau$ plays the role of time. 

We repeat the effective black hole Lagrangian\index{effective Lagrangian} \eqref{eq:JP-action1_bis}  for convenience:
\begin{align}
\cl=\tfrac{1}{2}\cg _{ij}(q)\,\dot{q}^i\dot{q}^j +
\cv (q)\,.\label{eq:JP-Lagrangian_q}
\end{align}
The momenta $p_i$ conjugate to $q^i$ are defined as
\begin{equation}
 p_i \equiv \pf{\cl}{\dot q^i} = \cg_{ij} \dot q^j\,.
\end{equation}
The Hamiltonian, with $\tau$ playing the role of time, is constant:
\begin{align}
 \ch &\equiv p_i\dot q^i - \cl\\
 &=\tfrac{1}{2}\cg^{ij}\,p_i p_j -
\cv (q) = \En\,.\label{eq:JP-Hamiltonian_q}
\end{align}
The value of the `energy' $\ch = \En$ is fixed by noting that this is just the Hamiltonian constraint of before, see e.g.~\eqref{eq:JP-Hamiltonian_q_first}. (Recall $\En$ is proportional to the non-extremality parameter $c^2$ by \eqref{eq:JP-gammadeltaE}. The fact that $E$ is minimal for extremal solutions fits nicely with the intuition that a supersymmetric, and hence extremal, solution is stable.)  The Euler-Lagrange equations of motion are equivalent to the Hamiltonian equations\index{Hamiltonian!equations}:
\begin{equation}
 \dot q^i = \pf{\ch}{p_i}\,,\qquad \dot p_i = - \pf{\ch}{q^i}\,.\label{eq:JP-Hamilton_eqs}
\end{equation}
We see that the effective black hole description corresponds to a Hamiltonian system\index{Hamiltonia! system} $(\ch,p^i,q_j)$, with $\tau$-independent Hamiltonian $\ch(q,p)$.

\subsection{Existence  of the superpotential from Liouville integrability\label{ss:JP-Liouville_Existence}}
Up to now, we have done nothing new. We have just rewritten the equations of motion in terms of variables $(q,p)$. We now turn to the question when a superpotential $\cw(q)$ exists such that the equations of motion are rewritten as in  eqs.~\eqref{eq:JP-flow_eqs_fi}, \eqref{eq:JP-f_i-W}. In terms of $(q_i,p^i)$, eqs.~\eqref{eq:JP-flow_eqs_fi}, \eqref{eq:JP-f_i-W} become:
\begin{equation}
 p_i =  f_i(q) = \pf {\cw(q,C)} {q^i} \,.\label{eq:JP-grad_flow_cond_p}
\end{equation}
As discussed before, this only holds when the $p$'s can be written as functions of the $q$'s ($p = f(q)$) \'and those functions satisfy the integrability condition $\partial_i f_j = \partial_j f_i$. For a general Hamiltonian system\index{hamiltonian system}, the coordinates $p$ and $q$ are independent and the integrability condition does not hold. We now discuss which Hamiltonian systems do allow for equations of the form \eqref{eq:JP-grad_flow_cond_p}. We take the argument from the classical mechanics literature, see textbooks \cite{Goldstein:1972,Arnold:1978,MacCauley}.\footnote{A pedagogical reference is \cite{MacCauley} by MacCauley, as that book is mainly about (integrability) of first-order flows. Also, MacCauley\index{MacCauley|fn}  points out several misconceptions in previous literature and is especially careful in questions regarding integrability and the connection to Hamiltonian systems and Hamilton-Jacobi theory.  We follow \cite{MacCauley} in the remainder of this section.}

\subsubsection{The criterion}

Denote the number of coordinates (scalars) $q^i$ by $n$. Assume we have $n$ constants of motion $G_i(q,p,\tau)\,(i=1\ldots n)$ that commute w.r.t.~the Poisson brackets\index{Poisson brackets}:
\begin{equation}
 \{G_i,G_j\} \equiv\sum_k\left( \pf {G_i} {q^k} \pf {G_j} {p_k} -\pf{ G_i} {p_k}\pf{G_j} {q^k} \right)=0\,.\label{eq:JP-involution_G}
\end{equation}
One says the $G_i$ are \textit{in involution}. By constants of motion, we mean that the total $\tau$-derivatives of $G_i$ are zero, in principle they can have an explicit time-dependence:
\begin{equation}
 \ta{G_i}{t} = \pf{G_i}{t} + \{\ch,G_i\} = 0\,.
\end{equation}
This means the expressions $G_i(q,p,\tau)$ can be equated to constants $C_i$:
\begin{equation}
 G_i(q,p,\tau) = C_i =cst\,.\label{eq:JP-COM}
\end{equation}
In order to rewrite the momenta $p$ as functions of the coordinates $q$, we furthermore make the assumption that the $n$ constants of motion are isolating. By this we mean that we can solve the equations \eqref{eq:JP-COM} for the momenta $p$ as:
\begin{equation}
 p_i = f_i(q,C,\tau)\,.\label{eq:JP-p_f}
\end{equation}
It then readily follows that the expressions $f_i$ are the gradient of a scalar function of the coordinates $q$. Indeed, since the $n$ constants of motion $G_i$ are in involution, one can check that consequently the expressions $p_i-f_i$ are in involution as well:
\begin{equation}
  \{p_i-f_i,p_j-f_j\}=0\,.
\end{equation}
(In this equation, $f_i$ should be seen as functions of the $q$'s and the $p$'s, and the $q$'s should be read as independent variables, i.e.~we do not use \eqref{eq:JP-p_f}.) It immediately follows that $ \pf {f_i}{q^j} =\pf {f_j}{q^i}$ and the functions $f_i(q,C,\tau)$ can be written in terms of one function $\cw (q,C,\tau)$ as in \eqref{eq:JP-grad_flow_cond_p}.

The above argument can be traced back in the other direction. We conclude that \textit{a gradient flow\index{gradient flow} of the form $p_i = \partial_i \cw(q)$ exists if and only if their are $n$ Poisson commuting constants of motion $G_i(q,p,\tau) = C_i$ which can be solved in terms of the momenta $p$.} This is the content of Liouville's involution theorem. \index{Liouville!involution theorem|defn}

\subsubsection{Comments}
We comment on the above result. First, when there are $n$ constants of motion in involution and $\cw$ can be constructed, we see that its exterior derivate can be written as
\begin{align}
 \de \cw &= p_i \de q^i\,.\label{eq:JP-dW_pdq}
\end{align}
Moreover, this implies that in principle, we can find the superpotential $\cw$ as the integral \cite{Andrianopoli:2009je}:
\begin{align}
 \cw &= \cw_0 +\int_{q_0,\tau_0}^{q,\tau}p_i \dot q^i\de \tau= \cw_0+2\int_{q_0,\tau_0}^{q,\tau}(\En + \cv(q))\de \tau\label{eq:JP-Superpot_Integral}\,,
\end{align}
where $q_0,\tau_0$ are initial conditions, $\cw_0$ is a constant and we used the form \eqref{eq:JP-Hamiltonian_q} of the constant Hamiltonian $\ch = \En$. 

Second, in the examples of sections \ref{s:JP-EasyExamples} and \ref{s:JP-Examples_Symm_modSpaces} we will see that the constants of motion $C_i$ correspond to `non-extremality parameters', which are zero for an extremal solution. In the context of the restrictive set of non-extremal examples of the previous chapter, these are exactly the integration constants (then called $\beta_i$) introduced to derive the first-order form of the equations of motion.


\outlook{Outlook}{In the following section \ref{ss:JP-Hamilton-Jacobi}, we link the superpotential $\cw$ to Hamilton-Jacobi theory. That section can be skipped without loss of the logical flow of the text. We continue with examples for black hole systems in section \ref{s:JP-EasyExamples}}

\subsection{Hamilton-Jacobi theory\label{ss:JP-Hamilton-Jacobi}}
Now we show that the superpotential $\cw$, \textit{if it exists}, is the generating function for a specific type of canonical transformation one encounters in Hamilton-Jacobi theory. By the Hamilton-Jacobi method of constructing the generating function, we can in principle give a method of constructing $\cw$. Andrianopoli et al. advocate the Hamilton-Jacobi approach in the context of black holes in \cite{Andrianopoli:2009je}. Note that the condition for existence of $\cw$ remains the same as above (Liouville integrability): the Hamilton-Jacobi approach is only useful in that case.  {}For a complete review of Hamilton-Jacobi theory,  the reader is referred to the classic references \cite{Goldstein:1972,Arnold:1978} and to \cite{MacCauley} for the relation to gradient flows. We briefly mention the main points of interest.

\subsubsection{Canonical transformations}
Considerations of Hamilton-Jacobi theory start in the context of canonical transformations, from one Hamiltonian system to another. Consider new phase space variables $(Q_i(q,p,\tau),P^i(q,p,\tau))$ and a function $\ck(Q,P,\tau)$. We ask when $(\ck,Q,P)$ defines a  Hamiltonian system, such that the equations \eqref{eq:JP-Hamilton_eqs} hold for the variables $Q,P$ replacing $q,p$ and with Hamiltonian $\ck$ instead of $\ch$. When the data $(\ck,q,p)$ define such a Hamiltonian system, the transformation to these variables is called canonical. It can be shown that therefore the one-forms $p \de q - H \de \tau$ and $P \de Q  -\ck \de \tau$ defined on the cotangent space of the phase space must be equivalent up to a closed form. This is equivalent to demanding that
\begin{equation}
  p \de q - H \de t + Q \de P + \ck \de t=\de F_2\,.\label{eq:JP-S_defn}
\end{equation}
for some function $F_2(q,P,\tau)$, which is often called a generating function `of the second kind', hence the subscript 2.\footnote{With this definition of $F_2$, the one-forms $p \de q - H \de t$ and $Q \de P  -\ck \de t$ are equivalent up to the closed form $\de(F_2 + PQ)$.} Note that $F_2$ is \emph{not} a function of $p$ nor of $Q$, because of the structure of the left-hand side of \eqref{eq:JP-S_defn}. \index{Hamilton-Jacobi equation|(imp} Instead, we should view $(q,P)$ as the independent variables. {}From eq.~\eqref{eq:JP-S_defn}, we see that $Q$ and $p$ are found as the partial derivatives of the function $F_2$:
\begin{equation}
p_i = \pf {F_2} {q^i}\,,\qquad Q^i = \pf {F_2} {P_i}\label{eq:JP-S_conditions}
\end{equation}
while the function $F_2$ itself is the solution to the differential equation
\begin{equation}
  \ck(\pf {F_2} P, P,\tau)- \ch(q,\pf {F_2} q,\tau)=  \pf {F_2} \tau\,.
\end{equation}
A word of warning: the function $F_2$ can only be defined (or: a solution to the Hamilton-Jacobi equation only exists) when a canonical transformation exists from the system $(\ch,q,p)$ to $(\ck,Q,P)$. This is not possible for all choices of coordinates $(Q,P)$!
\index{Hamilton-Jacobi equation|)}

\subsubsection{Comparison to superpotential approach}
Inspection of the first of the defining conditions \eqref{eq:JP-S_conditions} for the generating function\index{generating function} $F_2$ seems to suggest we have to identify $F_2$ with the generalized superpotential $\cw$, as defined in eq.~\eqref{eq:JP-grad_flow_cond_p}.  However, we need to specify the coordinates $(Q,P)$ a little more: $F_2$ is a function of the $q$'s and the $P$'s and possibly has an explicit dependence on the radial coordinate $\tau$, while $\cw$ should only depend on the coordinates $q$. In order for $F_2$ to be a function of $q$ only, with no explicit $\tau$-dependence, we demand that a canonical transformation $(q,p) \to (Q,P)$ exists such that the momenta $P$ are constant and the new Hamiltonian is equal to the old one:
\begin{align}
&P_i = \alpha_i= cst \nonumber\\
&\ck(P,Q) = \ch(p(P,Q),q(P,Q))\label{eq:JP-conds_ActionAngle}
\end{align}
It immediately follows from these conditions that the coordinates $Q$ are linear in $\tau$, since $\dot Q = cst$ by the Hamiltonian equations of motion (the eqs.~\eqref{eq:JP-Hamilton_eqs} with $(q,p,\ch)$ replaced by $(Q,P,\ck)$. If the conditions \eqref{eq:JP-conds_ActionAngle} are met, the defining equation for the function $F_2$ \eqref{eq:JP-S_defn} becomes:
\begin{equation}
 \frac{\de F_2}{\de \tau} = p_i \frac{\de q^i}{\de \tau}\,,
\end{equation}
and by comparison to \eqref{eq:JP-dW_pdq}, we conclude that indeed the generalized superpotential as a function of $\tau$ is  given by $\cw(q(\tau),C) = F_2(q(\tau),\alpha)$, up to a cosntant. The constants of motion $C_i$ are functions of the constant momenta $P_i=\alpha_i$.

\subsubsection{Hamilton-Jacobi theory}

In classical mechanics, the generating function for the transformation that meets the conditions \eqref{eq:JP-conds_ActionAngle} is known as \textit{Hamilton's characteristic function} in Hamilton-Jacobi theory. We discuss Hamilton-Jacobi theory in two steps.

First, consider a canonical transformation as in eq.~\eqref{eq:JP-S_conditions}, such that the new Hamiltonian $\ck$ is equal to zero and  the new coordinates $(Q,P)$ are  \emph{constant}:\footnote{Note that the generating function $F_2$ for this transformation is not the same the function $F_2$ for the generating function discussed above, which had $\ck = \ch$.} 
\begin{align}
Q^i &= \beta^i=cst\nonumber\\
P_i&=\alpha_i=cst\,.\label{eq:JP-alpha_beta}
\end{align}
In this case, it is customary to denote the solution to equation \eqref{eq:JP-S_conditions} for the generating function  as $S$ and to call it \textit{Hamilton's principal function}\index{Hamilton's principal function}. The generating function is now defined through the differential
\begin{equation}
 \de S = p_i \de q^ i - \ch \de \tau = \cl \de \tau\,\label{eq:JP-dS=Ldt}
\end{equation}
in terms of the integrand of the Lagrangian. When the action $\int \cl \de\tau$ is path independent, it can be integrated and gives the unambiguous form of the  function $S$. Hamilton's principal function $S$ is the solution to the so-called Hamilton-Jacobi equation (compare to  \eqref{eq:JP-S_conditions})
\begin{equation}
 \ch(q,\pf S q,\tau) + \pf S \tau = 0\,.\label{eq:JP-HJ_Equation}
\end{equation}

Second, note that the Hamilton-Jacobi equation for Hamilton's principal function $S$ can be simplified when the original Hamiltonian is independent of the coordinate $\tau$, as is the case for the effective black hole description. In this case, the $\tau$-dependence of $S$ can be separated and the function $S$ takes on the form:
\begin{equation}
 S(q,\alpha,\tau) = \cw(q,\alpha) - \En(\alpha) \tau\,,\label{eq:JP-S_and_W}
\end{equation}
for some constant $\En$ that is a function of the constant momenta $P_i=\alpha_i$.
Substituting this \ansatz{} into the Hamilton-Jacobi equation \eqref{eq:JP-HJ_Equation} gives
\begin{equation}
\ch(q,\pf \cw q) = \En\,.
\end{equation}
This is just the Hamiltonian constraint encountered before, with the same interpretation for constant $\En$  (proportional to the extremality parameter $c^2$, see eqs.~\eqref{eq:JP-Hamiltonian_q}, \eqref{eq:JP-gammadeltaE}). The function $\cw$ is referred to as \textit{Hamilton's characteristic function}. Not that $\cw$ is the generating function for a canonical transformation to $Q^i=\beta^i\tau +cst, P_i=\alpha_i$, as in eqs. \eqref{eq:JP-conds_ActionAngle}.

Caution has to be taken when applying Hamilton-Jacobi theory, since this method of solution  has been the source of some confusion in the literature on the nature of the constants $(Q^i,P_i)=(\beta^i,\alpha_i)$ of \eqref{eq:JP-alpha_beta}. In classical mechanics textbooks, it is sometimes advocated that the constants $(\beta^i,\alpha_i)$ are equal to (or functions of) the initial conditions $(q_0,p_0)$ of the motion in terms of the original coordinates $(q,p)$. This would suggest $S$ can always be defined \textit{globally}. However, the conditions for $S$, and thus $\cw$ to exist globally, are the same as in section \ref{ss:JP-Liouville_Existence}: we need $n$ Poisson commuting constant of motion.\footnote{When the system is not Liouville integrable, the function $S=\int \de_{q_0,p_0} \tau \cl$ is path-dependent. We can then in principle take $(\beta^i,\alpha_i)$ to be the initial conditions $(q_0,p_0)$ to construct $S$, but if it exists at all, it will be limited to very small distances $\tau$, i.e.~we can only define $S$ \textit{locally}.} For a general Hamiltonian system, one will not find $n$ Poisson commuting constants of motion among the $2n$ constants $(q_0,p_0)$ and therefore $S$ and $\cw$ is not always globally defined. In conclusion, this means that at most $n$ of the $2n$ constant $(P,Q)$ can be given by the initial conditions, but at least $n$ have to be Poisson commuting constants of motion, in order for the function $S$ (or $\cw$) to exist.

\outlook{Upshot:} {In principle, \textit{if} the superpotential $\cw$ exists, one can use Hamilton-Jacobi theory to find $\cw$ as the generating function for the canonical transformation to constant phase space variables as in eq.~\eqref{eq:JP-S_and_W}. The condition for existence remains the same as before, namely Liouville integrability.
}

\subsection{Duality invariance of the superpotential\label{ss:W_dual_invariant}}
We now go back to $D=4$ dimensions. In  \cite{Andrianopoli:2009je}, it was proven that the superpotential $\cw$ is invariant under the global symmetry group of the four-dimensional equations of motion. These global symmetries consist of the isometries of the scalar manifold, combined with a simultaneous symplectic action on the field strengths $F^I$ and their duals $G_I$. (See equation \eqref{eq:DualFieldStrength} for the definition of the dual field strengths.)  

The question of whether or not the superpotential is duality invariant had been an outstanding issue. {}For supersymmetric black holes, the superpotential is given in terms of the complex central charge $Z$ defined in eq.~\eqref{eq:BJ-Z} as $\cw = \e^{\beta \varphi }|Z|/2$. Since $|Z|$ is a symplectic invariant, $\cw$ is naturally invariant in this case. {}For extremal black holes, the question had only been partially answered and for non-extremal solutions nothing was known in this respect. Some evidence had been given in \cite{Ceresole:2007wx}, and in \cite{Andrianopoli:2007gt}, a search was initiated to find $\cw$ from duality invariant expressions. Explicit forms for the superpotential were found for all cases in extended supersymmetry with $\Cn>2$, while the $\Cn=2$ case was incomplete.

We sketch the proof of \cite{Andrianopoli:2009je}, which solved these outstanding issues. The effective Lagrangian $\cl$ \eqref{eq:JP-Lagrangian_q} is invariant under the global symmetries of the four-dimensional equations of motion. This is because the black hole potential $\cv$ is a symplectic invariant expression in terms of the black hole charges, while the kinetic terms are invariant under the isometries of the target space by construction. Since the solution $\cw(q,C)$ from eq.~\eqref{eq:JP-Superpot_Integral} (or equivalently $S(q,P)$ from \eqref{eq:JP-dS=Ldt}) is defined in terms of the integral of the effective Lagrangian and the black hole potential, it follows that the superpotential $\cw$ is invariant under the duality group.

The proof of \cite{Andrianopoli:2009je} that the superpotential is duality invariant, makes the problem of finding an explicit solution for $\cw$ more accessible: only duality invariant functions should be investigated. This was an idea that led to a successful construction of $\cw$ for symmetric moduli spaces\index{symmetric moduli space|see{symmetric space}} in work of Bossard et.~al. \cite{Bossard:2009at,Bossard:2009my} and Ceresole et.~al.~\cite{Ceresole:2009iy,Ceresole:2009vp} and subsequent results. We mention that work in more detail in the conclusion to this chapter.

\subsection{Preliminary conclusion/Outlook\label{ss:DimRedux+HJ}}
We have derived an existence criterion for first-order flows to exist for the effective Lagrangian \eqref{eq:JP-Lagrangian_q} with a potential.  The equations of motion for the $n$ scalars $q^i$ follow $\cg_{ij}\dot q^j = \partial \cw / \partial q^i$, whenever there are $n$ Poisson commuting constants of motion. The function $\cw(q)$, called (generalized) superpotential, is seen to be equal to Hamilton's characteristic function.

In the following, we give explicit examples to illustrate the discussion above. First, we use methods of chapter \ref{c:BJ} to write down the superpotential for the \RN and dilatonic black holes in four dimensions in section \ref{s:JP-EasyExamples}. Second, we consider more intricate setups in section \ref{s:JP-Examples_Symm_modSpaces} by performing a dimensional reduction over time.

\section{Examples inspired by chapter \protect{\ref{c:BJ}}\label{s:JP-EasyExamples}}
In this section and section \ref{s:JP-Examples_Symm_modSpaces}, we give examples illustrating the Hamiltonian description and the subsequent existence criterion for a superpotential to exist. The current section is dedicated to constructing the superpotential for two examples that were treated in the previous chapter, section \ref{s:BJ-MainResults}. 

In principle, all examples of that chapter (black holes and both stationary and time-dependent $p$-brane solutions coupled to a dilaton) allow for a rewriting in terms of a superpotential $\cw$, since the first-order equations were decoupled and hence satisfy the integrability condition $\partial_i f_j = \partial_j f_i$. We only perform the construction of the superpotential for the two four-dimensional black hole examples of section \ref{s:BJ-MainResults}: the \RN and dilatonic black holes. We do this following the Hamiltonian approach advocated in section \ref{s:JP-Integrability_GradientFlow}. The \RN black hole has one scalar (the warp factor) and one constant of motion (the Hamiltonian), so is trivially amenable to the methods outlined above. For the dilatonic black hole (two scalars: warp factor and dilaton), we show that there are two Poisson commuting constants of motion and give the superpotential. Note that the Bogomol'nyi trick of section \ref{ss:BJtimelike_0} was only given for electrically charged dilatonic black holes. We also give the superpotential for the case with magnetic charge and show that only for a specific value of the dilaton coupling ($a=1$), the method of section \ref{ss:BJtimelike_0} can be used for solutions with both electric and magnetic charges. 

Note that in four dimensions, the metric \ansatz{} becomes 
\begin{equation}
 \de s_4^2 = \e^{\varphi} \de \tau^2 + \e^{-\varphi}\left(\frac{c^4}{\sinh^4 c\tau}\rmd\tau^2 +\frac{c^2}{\sinh^2 c\tau}(\rmd\theta ^2 + \sin^2\theta \rmd\varphi ^2)\right)\,, 
\end{equation}
(i.e.~the constants $\alpha$ and $\beta$ are $\beta = -\alpha = 1/2$) and the effective Lagrangian reads
\begin{equation}
 {\cl} = \tfrac 12\dot \varphi^2 + \tfrac12  G_{ab} \dot\phi^a\dot \phi^b + \cv(\varphi,\phi)\,,
\end{equation}
where $\phi^a$ are the scalars in four dimensions and $\cv = \e^\varphi V(\phi)$, with $V(\phi)$ the black hole potential.

\subsection{The \RN black hole\label{eq:JP-BJ}}
The \RN black hole is a solution to the Einstein-Maxwell action:
\begin{equation}
S=\frac{1}{16\pi G_4}\int d^4 x \sqrt{-g} \left(R - \frac14 F_{\mu\nu}F^{\mu\nu}\right)\label{eq:JP-EinsDil_Action}
\end{equation}
In case of the \RN black hole, we eliminate the gauge field in terms of its electric charge $\QE$ and magnetic charge $\QM$. 
We have one scalar, $\varphi$,	 and the effective Lagrangian for the \RN black hole is\footnote{This should be constrasted to the effective action \eqref{eq:Uaction} obtained in the previous chapter. Remember we have chosen $\e^B = \sinh(c\tau)/c, U=\varphi/2$ and put the constant coupling $\gamma^2=1/4$. }
\begin{equation}
 \cl = \frac12\dot \varphi^2 + \frac12 (\QE^2+\QM^2) \e^{\varphi}\,.
\end{equation}
(See table \ref{tab:ReissnerNordstrom} to compare to the $\Cn=2$ supergravity action.) The Hamiltonian is
\begin{equation}
 \ch = \frac12 p_\varphi^2 - \frac12(\QE^2+\QM^2) \e^{\varphi} = 2c^2\,.\label{eq:JP-RN-HamConstr}
\end{equation}
where $p_\varphi =\dot \varphi$. Remember $\En=(d-1)(d-2)c^2 =2c^2$  for $D=d+1=4$ to compare to \eqref{eq:JP-Hamiltonian_q}. There is one scalar ($\varphi$) and one constant of motion (Hamiltonian), so we can solve $p_\phi$ in terms of $\varphi$ and the constant of motion $c$ as:
\begin{align}
p_\varphi = f_\varphi(\varphi,c) = \pf {\cw(\varphi,c)} \varphi
\end{align}
with
\begin{align}
 f_\varphi(\varphi,c) =  2\sqrt{\tfrac14(\QE^2+\QM^2) \e^\varphi +  c^2}\,.\label{eq:JP-RN_Hamilton_fo_eq}
\end{align}
The superpotential $\cw$ is
\begin{align}
\cw& = \int \de \tau \,2\sqrt{\tfrac14(\QE^2+\QM^2) \e^\varphi +  c^2}  \nonumber\\
&=4\sqrt{\tfrac14(\QE^2+\QM^2) \e^\varphi +  c^2} -8c\log\left(2c + \sqrt{\tfrac14(\QE^2+\QM^2) \e^\varphi +  c^2} \right) + 4 c \varphi\nonumber
\end{align}
We can interpret the Hamiltonian constraint as the Hamilton-Jacobi equation \cite{Andrianopoli:2009je}:
\begin{align}
 \frac12 \left(\pf \cw \varphi\right)^2 - \frac12(\QE^2+\QM^2) \e^{\varphi} = 2c^2\,,
\end{align}
which gives, of course, the same solution for $\cw$.

\subsection{The dilatonic black hole}\index{dilatonic black hole}
The simplest theory involving scalar fields admitting charged black hole solutions is given by the Einstein-dilaton-Maxwell action
\begin{equation}
S=\int \de^4 x\sqrt{|g|}\Bigl(
R-\tfrac{1}{2}(\partial \phi)^2-\tfrac{1}{4}\e^{a\phi}F^2\Bigr)\,,\label{eq:EMDaction}
\end{equation}
where $a$ is a non-zero constant, called the dilaton coupling\index{dilaton!coupling}. In the previous chapter, the author's work in \cite{Janssen:2007rc} was reviewed, where the Bogomol'nyi equations for the purely electric, extremal and non-extremal solutions of this theory were given by writing the action as a sum and difference of squares, generalizing the results on the pure \RN black hole in \cite{Miller:2006ay}. In the following we reconsider these results in the language of sections \ref{s:JP-gradient_from_sum} and \ref{s:JP-Integrability_GradientFlow} and extend to the full dyonic solution. {}For dyonic solutions, however, we will notice that only in the $a=1$ case can we easily find the superpotential. In section \ref{ss:dil_BH_revisited} we return to this subject and discuss the $a=\sqrt{3}$ example. We refer the reader to \cite{Gibbons:1982ih} for the original treatment of dilatonic black hole solutions.
Following the language of sections \ref{s:JP-gradient_from_sum} and \ref{s:JP-Integrability_GradientFlow}, we consider the first-order equations and the construction of a generalized superpotential for the dilaton $\phi$ and the warp factor $\varphi$ appearing in the metric (\ref{eq:JP-BH_metric_Ddim}). As explained above, the equations of motion for $\varphi$ and $\phi$ can be derived from a one-dimensional action. The black hole effective potential is now given by
\begin{equation}
V(\phi) = \tfrac12 \QE^2\e^{-a\phi} + \tfrac12 \QM^2\e^{+a\phi}\,,\label{eq:JP-Dolaton_BH_Potential}
\end{equation}
where $\QE$ is the electric charge and $\QM$ is the magnetic charge (in what follows, we assume both to be non-negative), such that the effective Lagrangian is
\begin{align}
\cl =\frac12 \dot \varphi^2 + \frac12 \dot \phi^2 + \e^{\varphi} V(\phi) \,.\label{eq:JP-Lagrangian_DIlBH}
\end{align}
In terms of the Hamiltonian description of section \ref{ss:JP-Eff_BH_Hamiltonian}, we have $q^i=(\varphi,\phi),~p_i = (\dot \varphi,\dot\phi)$ and the Hamiltonian \eqref{eq:JP-Hamiltonian_q} becomes
\begin{align}
\ch &= \tfrac12\cg^{ij}p_i p_j - \cv(q)\nonumber\\
&=\tfrac12 p_\varphi^2 + \tfrac12 p_\phi^2 - \e^{\varphi} V(\phi)  = 2c^2\,,\label{eq:JP-HamConstr_DIlBH}
\end{align}
We distinguish two cases. Either one of the charges $\QM,\QE$ is zero and we can use the results of the previous chapter (chapter \ref{c:BJ}). There it was shown that the equations of motion for the scalars $q^i=(\varphi,\phi)$ can be written as decoupled first-order equations \eqref{eq:BJ-Dilaton-Einstein-Maxwell_FO_Eqs}. It trivially follows the velocity field $\dot q^i$ follows a gradient flow.
We construct the generalized superpotential for this flow below and show that there are 2 constants of motion in involution as required by the existence criterion of section \ref{ss:Existence} for the superpotential $\cw$ to exist. Afterwards we investigate the dyonic case ($\QE$ and $\QM$ both non-zero) and see that we can only apply the Bogomol'nyi trick of the previous chapter when the dilaton coupling  $a=1$. 

\subsubsection{Purely electric or magnetic solutions}
In section \ref{ss:Existence}, we discussed how a minimal set of Liouville commuting constants of motion can lead to the construction of a flow $p_i = f_i (q,C) = \partial _i \cw(q,C)$.\footnote{{}From now on we drop the explicit $\tau$-dependence in the superpotential $\cw$, since the Hamiltonian and flow equations are $\tau$-independent.} In the case at hand, we could search for two such constants of motion to find the superpotential $\cw$. However, we can follow an easier route, since first-order equations of the form $p_i = f_i(q,C)$ were already constructed in section \ref{ss:BJtimelike_0}. The Poisson commuting constants of motion $C_i = G_i(p,q)$ were in that case denoted $\beta_i$. The only task left to us is to find the superpotential $\cw$ from the flow equations \eqref{eq:BJ-Dilaton-Einstein-Maxwell_FO_Eqs}.

The first-order equations \eqref{eq:BJ-Dilaton-Einstein-Maxwell_FO_Eqs} found in chapter \ref{c:BJ} for \emph{purely electric solutions} are written in terms of the coordinates $q^i = (\varphi,\phi), p_i=(\dot\varphi,\dot\phi)$ as (remember $U=\varphi/2$ for comparison)
\begin{align}
p_\varphi&\equiv
f_{\varphi}(\varphi,\phi;\beta_i)=-\tfrac{2}{\sqrt{1+a^2}}\sqrt{\tfrac14\QE^2\e^{
\varphi-a\phi}+\beta_2^2}- \tfrac{2a}{\sqrt{1+a^2}}\beta_3\,,\\
p_{\phi}&\equiv f_{\phi}(\varphi,\phi;\beta_i)
=+\tfrac{2a}{\sqrt{1+a^2}}\sqrt{\tfrac14\QE^2\e^{\varphi-a\phi}+\beta_2^2}-\tfrac{2}{\sqrt{1+a^2}}\beta_3\,.
\end{align}
The constancy of the Hamiltonian \eqref{eq:JP-HamConstr_DIlBH} requires the constants $(\beta_2,\beta_3)$ to obey
\begin{equation}
c^2=\beta_2^2+\beta_3^2\,.
\end{equation}
We rescaled $\beta_3\to 2\beta_3$ when compared to section \ref{ss:BJtimelike_0}.

One immediately verifies that the condition \eqref{eq:BJ-curl} holds: $\partial_{\phi}f_{\varphi} = \partial_{\varphi}f_{\phi}$.
Therefore a generalized superpotential $\cw$ exists and the above two-dimensional flow must be a gradient flow. According to the existence criterion of section \ref{ss:JP-Liouville_Existence}, this means that there are at least 2 Poisson commuting constants of motion. It is straightforward to check that the deformation parameter $\beta_3 =-\frac{1}{2\sqrt{1+a^2}}(ap_\varphi+ p_\phi )$ and the Hamiltonian $\ch = \En$ are two such constants, since:
\begin{equation}
\{ap_\varphi+p_\phi,\ch\} = \tfrac12\QE^2\e^{\varphi-a\phi}\{ap_\varphi+p_\phi,\varphi- a\phi\}=0\,.
\end{equation}
As in the general expression \eqref{eq:JP-p_f}, the $f_i$ have an explicit dependence on the Poisson commuting constants of motion $\beta_i$ (denoted before as $C_i$).

It is not difficult to construct the generalized superpotential explicitly,
\begin{equation}
\cw(\varphi,\phi;\beta_i)=-\frac{2}{1+a^2}\Bigl(2\surd{s_\mathrm{e}} -2\beta_2\log(\beta_2+\surd{s_\mathrm{e}}) + \beta_2(\varphi-a\phi) + \beta_3(a\varphi + \phi)\Bigr)\,,\label{eq:ElDilBHSupPot}
\end{equation}
where $\surd{s_\mathrm{e}}$ is shorthand for $\sqrt{\tfrac{1}{4}\QE^2 \e^{\varphi-a\phi} + \beta_2^2}$. Extremality ($\En=\beta_2=\beta_3=0$) implies that the superpotential $\cw$ factorizes according to (\ref{factorization})
\begin{equation}
\cw(\varphi,\phi;0)=\e^{\varphi/2}\bigl(-\tfrac{2}{\sqrt{1+a^2}}\QE\e^{-a\phi/2}\bigr)
\equiv\e^{\varphi/2}W(\phi)\,.
\end{equation}

We now come back to the third and last remark made at the end of section \ref{ss:gener_superpot}. If we compare with the flow equations of \cite{Andrianopoli:2007gt}, by locally inverting $\varphi(\tau)$, we do not find the form of $\partial_\tau W(\phi,\varphi(\tau))$ suggested in \cite{Andrianopoli:2007gt}, unless $\gamma = \beta_2 = \beta_3 = 0$. However, from the expression (\ref{eq:ElDilBHSupPot}) for the generalized superpotential, we see that we can explicitly construct $Y$ for all possible values of the non-extremality parameters $\beta_2, \beta_3$.

In the case of \emph{purely magnetic charge}, the above equations hold when the following electromagnetic duality rule is imposed:
\begin{equation}
\QE\leftrightarrow \QM\,,\qquad\phi\leftrightarrow -\phi\,,\qquad
\beta_2\leftrightarrow\beta_3\,.\label{eq:JP-DIlBH_EM_duality}
\end{equation}
This shows the invariance under duality rotations as explained in section \ref{ss:W_dual_invariant}.

\subsubsection{Dyonic solutions\index{dyonic}}

The dyonic\index{dyonic!dilatonic black hole} case with arbitrary dilaton coupling\index{dilaton coupling} $a$ is more involved. In general, it is not easy (or possible) to construct two constants of motion in involution. In this text, we focus on the two cases for which this can be readily done. In this section we treat $a=1$, the case $a=\sqrt{3}$ is treated in section \ref{ss:dil_BH_revisited}, involving dimensional reduction\index{dimensional reduction} and group theory\index{group theory} methods.

The theory with $a=1$ is the simplest.\index{dilatonic black hole!for $a=1$} We construct the superpotential $\cw$ by applying the deformation trick of \cite{Miller:2006ay} described in chapter \ref{c:BJ} (section \ref{ss:BJ-WhyDoesItWork}). There it was explained how we can find non-extremal first-order equations\index{first-order!equations} from the extremal flow equations through a constant-parameter deformation when two conditions are met: it must be possible to find a basis on the scalar manifold such that we can rewrite the potential as a sum of decoupled terms and diagonalize the (constant) metric simultaneously. {}For the dilatonic black hole\index{dilatonic black hole}, these conditions are only met when $a=1$.\footnote{Those two conditions were that the potential is a sum of decoupled terms (each depending on one scalar only) and that the constant scalar metric is diagonalized in the same basis. By examining the effective action for the dilatonic black hole, we see that the potential is a sum of two terms depending on $\varphi \pm a\phi$ respectively. The metric $\cg_{ij}$ can only be diagonalized in terms of $\varphi \pm a\phi$ when $a=1$.}  We construct the flow equations $p_i = f_i(q,C)$ for the extremal case and then apply the deformation trick of chapter \ref{c:BJ} to find the non-extremal flow equations. The non-extremality parameters in these equations correspond to two Poisson commuting\index{Poisson commuting} constants of motion. The resulting superpotential\index{superpotential} $\cw$ is given in eq.~\eqref{eq:JP-DyonicBH_Superpot} below.

Put the dilaton coupling\index{dilaton coupling} $a=1$. In the \textit{extremal case}, it is not difficult to see that the momenta $p_i= f_i(q)$ can be found by summing the electric and magnetic ones
\begin{align}
f_{\varphi}(\varphi,\phi) &=
-\tfrac{1}{\sqrt{2}}\QE\e^{(\varphi-\phi)/2}-
\tfrac{1}{\sqrt{2}}\QM\e^{(\varphi+\phi)/2}
\,,\\
f_{\phi}(\varphi,\phi) &=
+\tfrac{1}{\sqrt{2}}\QE\e^{(\varphi-\phi)/2}
-\tfrac{1}{\sqrt{2}}\QM\e^{(\varphi+\phi)/2}\,.
\end{align}
The corresponding superpotential $\cw$ is
\begin{equation}
\cw(\varphi,\phi)= -\e^{\varphi/2}\sqrt{2}\left(\QE\e^{-\phi/2} + \QM \e^{+\phi/2}\right)\equiv \e^{\varphi/2} W(\phi),
\end{equation}
and is the sum of the pure electric and magnetic superpotentials. An extremum of the superpotential $W(\phi)$, and consequently of the black hole potential\index{black hole potential} $V(\phi)$,  only exists in the dyonic case, corresponding to the fact that an attractive $AdS_2$ horizon exists only in the extremal dyonic case.

Let us now extend to \textit{non-extremal solutions} using the technique of \cite{Miller:2006ay}, as explained for the purely electric case ($\QM =0$) above. This gives
\begin{align}
f_{\varphi}(\varphi,\phi;\beta_i)/\sqrt2
&=-\sqrt{\tfrac{1}{4}\QE^2\e^{\varphi-\phi}+\beta_2^2}
-\sqrt{\tfrac{1}{4}\QM^2\e^{\varphi+\phi}+\beta_3^2}\,,\\
f_{\phi}(\varphi,\phi;\beta_i)/\sqrt2
&=+\sqrt{\tfrac{1}{4}\QE^2\e^{\varphi-\phi}+\beta_2^2}
-\sqrt{\tfrac{1}{4}\QM^2\e^{\varphi+\phi}+\beta_3^2} \,,
\end{align}
where $\beta_2,\beta_3$ are again integration constants. The corresponding generalized superpotential $\cw$ reads
\begin{equation}
\begin{split}
\cw(\varphi,\phi;\beta_i)={}&-2\surd{s_\mathrm{e}}
+2\beta_2\log(\beta_2+\surd{s_\mathrm{e}})-\beta_2(\varphi-\phi)\\
&-2\surd{s_\mathrm{m}}
+2\beta_3\log(\beta_3+\surd{s_\mathrm{m}})-\beta_3(\varphi+\phi)\,,
\label{eq:JP-DyonicBH_Superpot}
\end{split}
\end{equation}
where $\surd{s_\mathrm{e}}=\sqrt{\tfrac{1}{4}\QE^2 \e^{\varphi-\phi} + \beta_2^2}$ is defined as in the electric case and $\surd{s_\mathrm{m}}$ is shorthand for $\sqrt{\tfrac{1}{4}\QM^2\e^{\varphi+\phi}+\beta_3^2}$. Note that the superpotential is invariant under the action of the (discrete) electric/magnetic duality transformation \eqref{eq:JP-DIlBH_EM_duality}, as it should from the argument of section \ref{ss:W_dual_invariant}.

We have not been able to integrate the second-order equations for
$\varphi$ and $\phi$ when $a\neq 1$. However, we demonstrate
in the next section that the case $a=\sqrt{3}$ can also be
solved explicitly with the aid of group-theoretical methods.

\section{Examples with symmetric moduli spaces\label{s:JP-Examples_Symm_modSpaces}}
In this section, we treat some more complicated examples to clarify the formalism outlined in the previous section. The examples are taken from the author's work \cite{Perz:2008kh}. In order to study the gradient flow of the scalars in the $D$-dimensional theory, we have seen above that a study of the effective action with a black hole potential is useful. Liouville integrability of the system ensures that the scalars, denoted $q^i$ as in \eqref{eq:JP-qi_defn}, follow a gradient flow of the form $\cg_{ij}\dot q^j = \partial_i \cw$. In further examples, we prefer to discuss the generalized superpotential from the point of view of the second type of effective action of section \ref{ss:Two_Effective_Action}, namely the geodesic action that appears after a dimensional reduction over time of the $D$-dimensional system. 

\outline{
First, in section \ref{ss:BlackHoles_geodesics}, we give some more details on the construction of an effective action for black holes by dimensional reduction over time. This effective action was already mentioned in section \ref{ss:Two_Effective_Action} and describes geodesics on an enlarged target space with coordinates $\tilde q = (q,\chi^\alpha)$, where $q  = (\varphi, \phi^a)$ are the generalized coordinates introduced in \eqref{eq:JP-qi_defn} describing the warp factor and $D$-dimensional scalar fields $\phi^a$, and $\chi^\alpha$ are the electric (and possibly magnetic) potentials. We concentrate on the case where the target space with coordinates $\tilde q$ is a symmetric space, since in this case the equations of motion are explicitly solvable, see \cite{Fre:2009et,Chemissany:2009hq,Fre:2009dg,Chemissany:2009af}.\footnote{After our work \cite{Perz:2008kh}, it was even shown that the equations of motion are Liouville integrable in this case \cite{Chemissany:2010zp}.} An illustration of the geodesic approach for symmetric spaces is given in a second consideration of the \RN black hole in section \ref{ss:JP-ReissnerNordstrom_revisited}. In section \ref{ss:dil_BH_revisited} we apply these ideas in a second study of the four-dimensional dilatonic black hole. In section \ref{ss:Example_KK_Bh} we treat an example with two scalars, the Kaluza-Klein black hole in five dimensions.
}

\subsection{Black holes and geodesics\label{ss:BlackHoles_geodesics}}
\subsubsection{Timelike dimensional reduction}\index{dimensional reduction!over time}
We give some more details on how to obtain the geodesic action\index{geodesic action} \eqref{eq:JP-action2} by dimensionally reducing the system described by the action and spherically symmetric metric of section \ref{ss:BH_D_dimensions}, over the time direction $t$. Such a dimensional reduction\index{dimensional reduction!over time} was already briefly discussed in section \ref{ss:Two_Effective_Action}.
Notice first that the \ansatz{} \eqref{eq:JP-BH_metric_Ddim} for stationary black holes can always be interpreted as the \ansatz{} for the dimensional reduction over time. We again write $D=d+1$:
\begin{align}
\de s^2_{d+1} &= -\e^{2\beta\varphi(\tau)}(\de t - \omega)^2 + \e^{2\alpha\varphi(\tau)}\de s^2_d\,,\\
A^I &= \chi^I(\tau)\,(\de t -\omega) + A^I_{(d)}\,,
\end{align}
where $\omega$ and $A^I_{(d)}$ are one-forms in the $d$-dimensional Euclidean space and $\chi^ I$ are the electric potentials, which are scalars in the dimensionally reduced theory. The vector associated to the one-form $\omega$ is also called the Kaluza--Klein (KK) or Taub-NUT vector\index{Kaluza-Klein!vector}\index{Taub-NUT!vector}, while the warp factor\index{warp factor} $\varphi$ is in this context often referred to as the (KK) dilaton\index{Kaluza-Klein!dilaton}.  By comparison to the metric \eqref{eq:JP-BH_metric_Ddim} the KK vector $\omega$ needs to be truncated, in order to describe solutions of the form  specified before. We comment on this below. Normalizations are chosen such that the $d$-dimensional theory is in the Einstein frame\index{Einstein frame} and $\varphi$ is canonically normalized.

If we now perform the dimensional reduction over time, we get a $d$-dimensional theory involving scalars ($\varphi,\phi^a,\chi^I$) and vectors ($\omega,A^I_{(d)}$) coupled to Euclidean gravity. We restrict to the scalar part of the theory only. Note that in the special case of $d=3$ (the $D=4$-dimensional black holes) we have a theory involving only scalars, since we can dualize vector fields to scalars: we dualize the one-forms $\omega$ and $A^I_{3}$ to axionic scalars $\tilde \chi^0$ and $\tilde \chi_I$. One can verify that the kinetic terms of the axions $\chi^I$ and $\tilde \chi_I$ appear with the opposite sign \cite{Breitenlohner:1987dg}.

When we restrict to  spherically symmetric solutions, we truncate the Kaluza--Klein vector $(\omega=0)$.
In $D= d+1=4$ we make an exception. When $d=3$, the Taub-NUT vector $\omega$ can be dualized to a scalar and it is part of the scalar manifold. We  make use of the group structure associated to this manifold and truncate scalar dual to $\omega$ at the end of the calculation. 

The nature of the enlarged scalar manifold depends on the number of dimensions $d$. Denote the collection of scalar fields in the dimensionally reduced theory as $\tilde q^i(\tau)$. We can then make the distinction:
\begin{itemize}
\item $D=4$. In this case, $d=3$ and the electric and magnetic potentials $\chi^I,\tilde \chi_I$ as well as the scalar $\tilde \chi^0$, dual to the Kaluza-Klein vector are present. We combine all scalars as
\begin{equation}
 \tilde q^i = (\varphi,\phi,\chi,\tilde \chi)\,.
\end{equation}
\item $D>4$. The magnetic vector field and the Kaluza-Klein vector cannot be dualized to scalars. Only the electric potentials $\chi^I$ appears in the enlarged scalar manifold (we restrict to electrically charged black holes). We write:
\begin{equation}
  \tilde q^i = (\varphi,\phi,\chi)\,.
\end{equation}
\end{itemize}

Restricting to the scalar part of the theory in $d$ dimensions, the effective action takes the form of a non-linear sigma model coupled to (Euclidean) gravity.
\begin{equation}
 \int \sqrt{g_d}\left( R_d -\frac12 (g_d)^{\tau\tau} \cg_{ij}\partial_\tau \tilde q^i \partial_\tau\tilde q^j\right)\,,
\end{equation}
We can decouple the sigma model from gravity, by choosing the affine paramatrization $\tau$ such that $\sqrt{g_d}(g_d)^{\tau\tau}=1$. This is precisely achieved by the $d$-dimensional metric \ansatz{} \eqref{eq:JP-BH_metric_ddim}. The action is then equivalent to (denoting $\tau$-differentiation again with dots)
\begin{equation}
S=\int \de \tau \, \tilde \cg_{ij}\dot{\tilde q}^i
\dot{\tilde q}^j\,,\label{eq:JP-action2-bis}
\end{equation}
and describes geodesics on a manifold with coordinates $\tilde q^i$ and metric $\tilde \cg _{ij}$ (a dot denotes differentiation w.r.t.~$\tau$). The velocity along the geodesic is constant:
\begin{equation}
 \tilde \cg_{ij}\dot{\tilde q}^i
\dot{\tilde q}^j = v^2 = cst\,.\label{eq:JP-Hamiltonian2-bis}
\end{equation}
From the Einstein equations for the $d$-dimensional metric, one relates the velocity $v$ to the non-extremality parameter $c$ of the metric \eqref{eq:JP-BH_metric_ddim} as $v^2 =2(d-1)(d-2)c^2$.  Because the metric $\cg_{ij}$ is of indefinite signature, the affine velocity of the geodesic can be positive, negative or zero. In the latter case (null geodesics), we are describing an extremal solution; $v^2>0$ corresponds to non-extremal black hole solutions, while $v^2<0$ is not a good black hole solution \cite{Breitenlohner:1987dg}.

\subsubsection{Extracting first-order equations}

We want to show how we can extract the superpotential $\cw$, defined in terms of the scalars $q$ of the $D$-dimensional theory, from the larger manifold with coordinates $\tilde q$ of the dimensionally reduced theory in $d=D-1$ dimensions.

Suppose we can explicitly write down the velocity vector field $\tilde f$ on the enlarged scalar manifold in $d=D-1$
dimensions
\begin{align}
\tilde f^i(q,\chi)&=\tilde \cg_{ij}\dot{q}^j\,,\\
\tilde f^{\alpha}(q,\chi)&=\tilde \cg_{i\alpha}\dot{\chi^{\alpha}} \,,
\end{align}
where the $\chi^{\alpha}$ are the scalars descending from the vector potentials upon dimensional reduction. When there are enough `integrals of motion' to fully eliminate $\chi^{\alpha}$ in terms of the $q^i$, one can write down a velocity field on the target space in $D$ dimensions:
\begin{equation}\label{velocity2}
f_i(q)\equiv \cg_{ij}\dot q^j = \tilde f_i(q,\chi(q))\,.
\end{equation}
Having obtained the velocity field on the moduli space in $D$ dimensions, it suffices to show that the velocity one-form $f_i$ is locally exact
\begin{equation}\label{curl}
f_i(q) = \tilde f_i(q,\chi(q))=\partial_i
\cw(q)\,,
\end{equation}
leading to an explicit algorithm to check the condition for a gradient flow to exist.

We  use this method to check the existence of $\cw$ for several examples below. In order to assure that the scalar manifold with coordinates $\tilde q$ has enough integrals of motion to eliminate the electric and magnetic potentials, we make the assumption that is describes a \textit{symmetric space}. For these systems, we can using the integrability (i.e.~solvability) of the effective action to obtain the velocity field $f_i(q)$.

\subsection{Geodesics on symmetric spaces}\index{symmetric space}
The assumption we make is that the target space in $d$ dimensions is a symmetric coset space $G/H$, where $G$ is a Lie group and $H$ some subgroup subject to certain conditions that we state below. This assumption is always valid when the scalar manifold of the original $(d+1)$-dimensional gravity theory is a symmetric space, which happens for supergravity theories with more than eight supercharges and for some theories with less supersymmetry. Nevertheless, our analysis here is independent of any supersymmetry considerations. 

We briefly summarize the defining properties of a (locally) symmetric space. The reader unfamiliar with the group-theorical notion of a symmetric space, is referred to the standard in-depth treatment of Helgason \cite{Helgason:1978}. A shorter and less technical treatment can for instance be found in \cite{Magnea:2002xx}.  The Lie algebras associated to $G$ and $H$ are denoted by $\mathfrak{g}$ and  $\mathfrak{h}$ respectively. The defining property of a locally symmetric space is that there exists a Cartan decomposition
\begin{equation}
\mathfrak{g} = \mathfrak{h} + \mathfrak{f}\,,
\end{equation}
with respect to the Cartan automorphic involution\index{Cartan involution} $\theta$, such that $\theta (\mathfrak{f})= - \mathfrak{f}$ and $\theta(\mathfrak{h}) = + \mathfrak{h}$. Take a coset representative\index{coset representative} $L(\tilde q) \in G$. We first define the group multiplication from the left, $L\rightarrow gL$, $\forall g\in G$, and we let the local symmetry act from the right $L\rightarrow Lh \sim L $, $\forall h \in H$.
From the Cartan involution we can construct the symmetric coset matrix $\Mg=LL^{\sharp}$, where $\sharp$ is the` generalized transpose'\footnote{This terminology stems from the use of the operation $\sharp$ in (special) orthogonal groups, where it corresponds to taking the transpose: $L^\sharp = L^T$.}
\begin{equation}
L^{\sharp}=\exp[-\theta(\log L)]\,.\label{eq:JP-GenTranspose}
\end{equation}
The matrix $\Mg$ is invariant under $H$-transformations that act from
the right on $L$. Under $G$-transformations from the left, $\Mg$
transforms as $\Mg \rightarrow g \Mg g^{\sharp}$.

With the aid of the matrix $\Mg$ the line element on the space $G/H$ with coordinates $\tilde q^i$ can be written as
\begin{equation}\label{eq:JP-SS_Moduli_metric}
\de s^2 = \tilde \cg_{ij}\de\tilde q^i\de\tilde q^j =
-\tfrac12\tr\bigl(\de \Mg \de \Mg^{-1}\bigr)\,.
\end{equation}
Both the \emph{local} action of $H$ on $L$ from the right and the \emph{global} action of $G$ on $L$ from the left leave the metric invariant. The latter implies that $G$ is indeed the isometry group of $G/H$. The action \eqref{eq:JP-action2-bis} of the dimensionally reduced theory then describes the geodesic curves on $G/H$ and the resulting equations of motion are
\begin{equation}\label{eq:scalarEOM1}
\tfrac{\de}{\de \tau}(\Mg^{-1}\tfrac{\de}{\de \tau}\Mg)=0\qquad
\Rightarrow\qquad \Mg^{-1}\tfrac{\de}{\de \tau}\Mg=\Qc\,,
\end{equation}
with the matrix of Noether charges $\Qc$ being a constant matrix in some
representation of $\mathfrak{g}$. We now see that the geodesic
equations are indeed integrable and their general solution is
\begin{equation}
\Mg(\tau)=\Mg(0) \e^{\Qc\tau}\,.\label{SOLUTION}
\end{equation}
The affine velocity squared of the geodesic curve is
\begin{equation}\label{eq:affinev}
\tilde \cg_{ij}\dot{\tilde q}^i\dot{\tilde q}^j =
\tfrac{1}{2}\tr(\Qc^2) = 2\En\,,
\end{equation}
and coincides with the Hamiltonian constraint.

An integrable geodesic motion on an $n$-dimensional space is
specified by $2n$ constants: the initial position and velocity
of the geodesic curve. So the geodesic motion on $G/H$ is
specified by \mbox{$2(\dim G-\dim H)$} integration constants. In
eq.~(\ref{SOLUTION}) $\Mg(0)$ contains $(\dim G-\dim H)$
constants corresponding the initial position. The number of
arbitrary constants in $\Qc$ (the initial velocity) is reduced from $\dim G$ to $(\dim G-\dim H)$
through the constraint $\Mg^{\sharp}(\tau)=\Mg(\tau)$, which gives
\begin{equation}
\theta(\Qc)=-\Mg(0)^{-1}\Qc\Mg(0)\label{involution1} 
\end{equation}


\subsubsection{First-order equations for the scalars $\tilde q^i$}
The first-order equation $\Mg \frac {\de \Mg^{-1}}{\de \tau} = \Qc$
can be written in terms of the scalars  as 
\begin{equation}
(\Mg^{-1}\partial_j \Mg)
\dot{\tilde q}^j=\Qc\,. 
\end{equation}
Multiplying this equation with $\frac12 (\Mg^{-1}\partial_i \Mg)$ and taking the trace, we obtain
\begin{equation}
\tilde f_i(\tilde q) \equiv \tilde \cg_{ij}\dot{\tilde q}^j  = \tfrac12\tr (\Mg^{-1}\partial_i \Mg
\cdot \Qc)\,,\label{eq:fo_eqn-general}
\end{equation}
where we  used the expression \eqref{eq:JP-SS_Moduli_metric} for the scalar metric $\tilde \cg _{ij}$. This expression indeed depends on the scalars $\tilde q$ only. However, remember that we have $\tilde q = (q,\chi)$, where $q$ are the scalar fields in $D$ dimensions and $\chi$ are the electric/magnetic potentials.  In terms of the discussion in the previous section, we still need to eliminate the electric/magnetic potentials $\chi$.

Remark therefore that \eqref{eq:fo_eqn-general} are only $(\dim G-\dim H)$ equations, while $\Mg^{-1} \tfrac{\de}{\de \tau} \Mg = \Qc$ has $\dim G$ independent components. By substituting
\eqref{eq:fo_eqn-general} into $\Mg^{-1} \tfrac{\de}{\de \tau}\Mg = \Qc$, we obtain
\begin{equation}
 \tilde \cg^{ij}(\Mg^{-1}\partial_i \Mg) \tilde f_j(\tilde q) = \Qc\,.
\end{equation}
This matrix equation constitutes of $\dim H$ independent components, which are \textit{non-differential} equations. This shows the
power of \eqref{eq:scalarEOM1}: we split the $\dim G$ differential equations in $\Mg^{-1}\tfrac{\de}{\de \tau}\Mg=\Qc$ into $(\dim G-\dim H)$ first-order equations and $\dim H$ equations without any derivatives. In the context of the previous section, these non-differential
equations are precisely what is needed to eliminate the additional scalars resulting from dimensional reduction, so that we obtain first-order equations in terms of the scalars in $D$ dimensions, as in eq.~\eqref{velocity2}. From the flow $f_i(q)$ for the scalars $q$ in $D$ dimensions, we can then examine if $\cw$ can be constructed. We put this method into use in three examples in the following sections.

\subsection{The \RN black hole revisited \label{ss:JP-ReissnerNordstrom_revisited}}
We use the \RN example to illustrate the use of symmetric spaces. It is possible to relate the Einstein-Maxwell action (\ref{eq:JP-EinsDil_Action}) to a symmetric
moduli space, after dimensional reduction over time. It can be shown that the most general timelike reduction of the four-dimensional Einstein-Maxwell system gives
rise to an  $\bSU(2,1)/{\rm S}(\bU(1,1)\times \bU(1))$-sigma model, coupled to Euclidean gravity in three dimensions. {}For simplicity we restrict to the subset of \textit{static} solutions  only carrying \textit{electric} charge. This restriction  corresponds to an $\bSLn 2 R / \SO(2)$ sigma model in three dimensions and we show below
the solution corresponds to the \RN black hole. We follow the procedure outlined in the previous sections. First, we make a (static) ansatz for the metric and gauge field. Then we reduce over the black hole `world-volume', i.e.~the time direction.  We then solve
the geodesics on the $\bSLn 2 R / \SO(2)$ symmetric space.

We first require \spacetime{} to be static and spherically symmetric.  Then we know that we can write the
metric in the form:
\begin{equation}
 \de s^2_4 =-\e^{\varphi(\tau)}\de t^2 + \e^{-\varphi(\tau)}\de s_3^2\,,\label{SSeq:MetricAns}
\end{equation}
We write the four-dimensional gauge field in form notation as:
\begin{equation}
A = \chi(\tau)\de t\,,\label{SSeq:GaugeFieldAns}
\end{equation}
where we introduced a scalar $\chi(\tau)$, dubbed the axion. The axion is just the electric potential (the $A^0$ component of the gauge field).

If we now perform a dimensional reduction over a time, we obtain a three-dimensional theory, whose solutions are in one to one correspondence to the four-dimensional ones \cite{Breitenlohner:1987dg}. Also, we are left with a theory in three dimensions which consists of gravity coupled to scalar fields only: the warp factor $\phi$ appearing in the four-dimensional metric and the electric axion $\chi$.

Using the \ansatz{} for the metric (\ref{SSeq:MetricAns}) and gauge field (\ref{SSeq:GaugeFieldAns}), we see that the field equations can be derived from the three-dimensional action:
\begin{equation}
 S = \int \de\tau \left(- \tfrac12(\partial\varphi)^2+\tfrac12\e^{-\varphi}(\partial \chi)^2\right)\label{SSeq:3DEMaction1}
\end{equation}
Note the `wrong sign' kinetic term of the axion $\chi$. This is a reflection of the timelike reduction we performed and it indicates we are dealing with a {pseudo-Riemannian} symmetric space. 

The scalars in the above action parameterize an $\bSLn 2 R/ \SO(1,1)$ symmetric space. To see this, define the matrix $L(\varphi,\chi) \in \SL (2,\Real)$ as:
\begin{equation}
L = \begin{pmatrix}\e^{\varphi/4} & \e^{-\varphi/4}\chi\\ 0&\e^{-\varphi/4}   \end{pmatrix}
\end{equation}
The Cartan involution that singles out an $\SO(1,1)$ subgroup is given in terms of the generalized transpose as $L^\sharp = -\eta L^T \eta$, where $\eta = \diag (1,-1)$. 
We define the matrix $\Mg$ in a slightly different way as before, to make later calculations a little easier:
\begin{equation}
 \Mg = L L^\sharp \eta = L \eta L^T =\begin{pmatrix}   \e^{\varphi/2} -\e^{-\varphi/2}\chi^2 &\e^{-\varphi/2}\chi\\\-e^{-\varphi/2}\chi &-\e^{-\varphi/2}  \end{pmatrix}\,,
\end{equation}
such that $\Mg$ is symmetric. The effective action for $\varphi$ and $\chi$ is then:
\begin{equation}
 S = \int \de \tau \tfrac12 \Tr\left(\ta{\Mg }\tau \ta{\Mg^{-1}}\tau \right)\label{SSeq:3DEMaction2},
\end{equation}
Note that this action is invariant under $\Mg\to \Mg\eta$, justifying the alternative definition for $\Mg$.

\paragraph{Affine parametrization and EOM.}
The solution to the equations of motion, geodesics given in terms of the matrix $\Mg$, are determined by the constant matrices $\Mg(0)$ and $\Qc$:
\begin{equation}
\Mg = \Mg(0) \e^{\Qc\tau}\,.\label{SSeq:M-1dMQ}
\end{equation}
Note that $\Mg(0)$ is an element of the group $\SL(2,\Real)$, while $\Qc$ sits in the Lie algebra $\mathfrak{sl}(2,\Real)$. We discuss the integration constants in those matrices. $\Mg(0)$ corresponds to the initial position of the geodesic,
while $\Qc$ contains information on the initial velocity. We restrict $\Mg$ to describe geodesics through the origin. This corresponds
to choosing $\Mg(0) =\eta$, or equivalently, considering asymptotically flat \spacetime{} ($\phi,\chi \to 0$ as $\tau \to 0$). We can also
cut down the number of components in $\Qc$. First, it is an element of the Lie algebra of $\SL(2,\Real)$ and therefore traceless. Since $\Qc$ is defined through (\ref{SSeq:M-1dMQ}), it moreover follows that $\Qc = \Qc^T$ (see eq.~\ref{involution1}). We can write:
\begin{equation}
  \Qc = \begin{pmatrix}-M & \QE\\ \QE & M\end{pmatrix}\,,
\end{equation}
where $M$ and $\QE$ are constants. By examining the form of $\Mg = \Mg(0)\e^{\Qc\tau}$ at spatial ininity ($\tau \to 0$):
\begin{align}
  \Mg = \unit_2 + \QE\tau + O(\tau^2)\quad\Rightarrow\quad &\varphi = -2 M \tau + \co(\tau^2)\\
  &\chi = \QE\tau +\co(\tau^2)\,.
\end{align}
we see that $M$ is the ADM mass of the solution and $\QE$ is the electric charge (by pluggin this asymptotic behaviour in the form of the \ansatz{} (\ref{SSeq:MetricAns},\ref{SSeq:GaugeFieldAns})). We can of course go further and use the full solution $\Mg = \Mg(0)\e^{\Qc\tau}$ to infer the form of $\varphi$ and $\chi$, and we would find the electrically charged \RN solution. However, we want to elucidate the appearance of first-order equations and therefore we follow the route outlined in the previous sections.  

Using \eqref{eq:fo_eqn-general} we see the equations of motion are written as first-order equations:
\begin{align}
\tilde \cg_{ij}\dot{\tilde q}^j  = \tfrac12\tr (\Mg^{-1}\partial_i \Mg
\cdot \Qc) \quad \Leftrightarrow \quad \dot \varphi &= -2 M + \frac12 \QE \chi\,,\\
\dot \chi    &= \frac12(\QE + 2 M \chi +\QE(\e^{2\phi} + \chi^2))\,.\nonumber
\end{align}
Plugging these solutions into the Hamiltonian constraint \eqref{eq:affinev} (the expression of constant affine velocity), we get
\begin{equation}
 \frac14\e^{-\varphi}(-\QE + 2M\chi - \QE (\e^{2\phi} -\chi^2))=0\,.
\end{equation}
We can use this constraint to simplify the first-order equations, by solving for $\chi$:
\begin{align}
\dot \phi &= -\sqrt{\frac14\QE^2 \e^{2\phi} +c^2}\,,\nonumber\\
\dot \chi    &= -\QE \e^{2\varphi}\,.
\end{align}
The first of these equations is none other than the usual first-order equation for the warp factor of the \RN black hole, derived in two different contexts before (see section \ref{ss:BJ-Review_MSW} and section \ref{eq:JP-BJ}). Remembering that the electric potential $\chi$ is related to the field strength as $F =  \dot \chi \de t $, the second of these equations is recognized as the normal ansatz for the gauge field of the spherically symmetric \RN black hole, see for instance eq.~\eqref{eq:BJ-RN_GaugeField_Ansatz} in another notation. 


\subsection{The dilatonic black hole re-revisited\label{ss:dil_BH_revisited}\index{dilatonic black hole}}
When the Einstein-dilaton-Maxwell \index{Einstein-dilaton-Maxwell}action (\ref{eq:EMDaction}) has the specific dilaton coupling\index{dilaton coupling} $a=\pm\sqrt{3}$, a symmetry enhancement takes place upon dimensional reduction over a (timelike or spacelike) circle. This can be explained by the fact that for $a=\pm\sqrt{3}$ (\ref{eq:EMDaction}) is the action obtained from reducing five-dimensional gravity over a spacelike circle and subsequent reduction should display at least the $\GL(2,\Real)$ symmetry of the internal torus. Furthermore, if the 3d vectors are dualized to scalars the $\GL(2,\Real)$-symmetry turns out to be part of a larger $\SL(3,\Real)$-symmetry. In this section, we choose $a=-\sqrt{3}$ for concreteness. The three-dimensional Euclidean action then describes gravity minimally coupled to the $\SL(3,\Real)/\SO(2,1)$ sigma model with coordinates $\tilde q^i =(\varphi,\phi,\chi^0,\chi^1,\chi^2)$:
\begin{equation}
\begin{split}
\tilde \cg_{ij}\de\tilde q^i\de\tilde q^j={}&(\de\varphi^1)^2 + (\de\phi)^2 -
\e^{-\sqrt{3}\phi+\varphi}(\de\chi^0)^2 + \e^{2\varphi}(\de\chi^2)^2\\
&-\bigl[\e^{+\sqrt{3}\phi+\varphi} - \e^{2\varphi}(\chi^0)^2\bigr](\de
\chi^1)^2 + 2\chi^0 \e^{2\varphi} \de\chi^2\de\chi^1 \,.\label{sigma}
\end{split}
\end{equation}
The details of the Kaluza--Klein reduction\index{Kaluza--Klein!reduction} can be found in the appendix of our work \cite{Perz:2008kh}; for details on the $\SL(3,\Real)/\SO(2,1)$ sigma model, see appendix \ref{app:SS}. The scalar $\phi$ is the four-dimensional dilaton, $\varphi$ the black hole warp factor of the metric. The scalars $\chi^0$ and $\chi^1$ are the electric and magnetic potentials, $\chi^2$ comes from the dualisation of the KK vector in the reduction from four to three dimensions and is hence related to the NUT charge $\QT$ via
\begin{equation}
\label{eq:NUT} \QT \sim \dot{\chi}^2 + \chi^0\dot{\chi}^1 \qquad \text{as } \tau\to0\,.
\end{equation}
As explained before, to achieve spherical symmetry we put $\QT=0$, leading to a truncated target space, where $\chi^0$ and $\chi^1$ have a shift symmetry
\begin{equation}
\de s^2=(\de\phi)^2 + (\de\varphi)^2 - \e^{-\sqrt{3}\phi+\varphi}(\de\chi^0)^2 -\e^{+\sqrt{3}\phi+\varphi}(\de \chi^1)^2 \,.
\end{equation}
If one were to eliminate the electric and magnetic potentials $\de \chi^1$ and $\de\chi^2$ by their equations of motion, one would obtain the black hole potential \eqref{eq:JP-Dolaton_BH_Potential} for $\phi$ and $\varphi$, illustrating the relation between the effective action with a potential and the geodesic action.

Let us discuss the geodesic equations of motion for the full sigma model (\ref{sigma}). The charge matrix $\Qc\in \mathfrak{sl}(3)$ that specifies a geodesic solution contains a priori eight arbitrary parameters $\Qc^\Lambda$ defined through $\Qc = \Qc^\Lambda T_\Lambda,\Lambda = 1\ldots 8$, where the eight generators of $\mathfrak{sl}(3)$, denoted $T_{\Lambda}$, are given in the appendix, eq.~\eqref{eq:T}. We  reduce the number of integration constants to four. First, we restrict the solution to be asymptotically flat. This means that the geodesic curve must pass through the origin which, using \eqref{involution1}, gives an involution condition on $\Qc$
\begin{equation}\label{eq:InvCond}
\Qc = -\theta(\Qc)\,,
\end{equation}
and requires identifying
$\Qc^3=-\Qc^6\,, \Qc^4=-\Qc^7\,, \Qc^5=\Qc^8$\,.
This condition is gauge-equivalent to the most general expression for $\Qc$. It amounts to fixing the $U(1)$-gauge transformation and the boundary conditions for the black hole warp factor and dilaton at spatial infinity. This specification is without loss of generality.\footnote{We put $\phi(r\to\infty)=\varphi(r\to\infty)=0$. This condition on the warp factor $\varphi$ can always be achieved by a coordinate transformation. The condition for the dilaton cannot be changed, but any other boundary value is equivalent upon a shift of the dilaton and accordingly a compensating rescaling of magnetic and electric charge.}. Secondly, we restrict ourselves to solutions with a vanishing NUT charge, which amounts to $\Qc^5=\Qc^8=0$, so that we are left with four independent integration constants to describe the dilatonic black hole solutions: the ADM mass, the electric and magnetic charges and the scalar charge. Upon demanding a regular horizon one can write the scalar charge in terms of the three others \cite{Gibbons:1982ih}, but we do not make that restriction here for the sake of generality. In terms of those four constants, the constant matrix $\Qc$ is then written as:
\begin{equation}
\Qc =\Qc^\Lambda T_\Lambda=\begin{pmatrix}-\tfrac{1}{\sqrt{3}}\Sigma+2M & -\QM & 0\\
\QM & \tfrac{2}{\sqrt{3}}\Sigma & -\QE\\
0& \QE& -\tfrac{1}{\sqrt{3}}\Sigma -2M\end{pmatrix}\,.\label{eq:charge1}
\end{equation}
below we show that the four integration constants $M,\QE,\QM,\Sigma$ that make up the matrix $\Qc$ are the ADM mass, electric-magnetic charges and the dilaton charge.

We now have sufficient information to construct the velocity vector field $f^i(\tilde q^j) = \dot{\tilde q}^i$ for the charge configuration (\ref{eq:charge1})
\begin{align}
f^{\varphi}&=-2M -\tfrac12( \QE \chi^0 + \QM \chi^1)\,,\label{eq:JP-fo_varphi}\\
f^{{\phi}}&=\Sigma + \tfrac{\sqrt{3}}{2}(\QE \chi^0 - \QM
\chi^1)\,,\\
f^{{\chi}^0}&=
\QE \e^{\sqrt{3}\phi - \varphi}\,,\label{eq:fo_chi0}\\
f^{{\chi}^1}&=
\QM \e^{-\sqrt{3}\phi- \varphi}
\,,\label{eq:fo_chi1}\\
f^{{\chi}^2}&=-\chi^0 f^{\chi^1}\,.\label{eq:JP-fo_chi2}
\end{align}
Note that we have already used the component of the velocity field for the Taub-NUT scalar ($\dot \chi^2 = f^{\chi^2}$) to eliminate $\dot \chi^2$ from the other components of the velocity field via eq.~\eqref{eq:NUT} with $Q_\mathrm{T}=0$. {}From the asymptotic behaviour of the velocity field we can then identify the integration constants: $M = -\tfrac12\dot \varphi|_{\tau=0}$ is the ADM mass, $\Sigma = \dot \phi|_{\tau=0}$ is the dilaton charge, while $\QM$ and $\QE$ are equal to the magnetic and electric charge respectively (remember that $\tau=0$ corresponds to spatial infinity). 

In order to find the  superpotential in terms of the original scalars $q=(\varphi,\phi)$, we need to eliminate the axions $\chi$ in terms of $(\varphi,\phi)$, following the logic at the end of  section \ref{ss:BlackHoles_geodesics}. Therefore, note that aside from the explicit expression for the velocity field, there is more information in the \emph{eight} first-order equations $\Mg^{-1}\dot{\Mg}=\Qc$. The velocity field uses five out of these eight. The remaining three equations are non-differential and we call them \emph{constraint equations}:
\begin{align}
0&=\QM + e^{\sqrt{3} \phi+\varphi} \bigl[\bigl(\sqrt{3}
\Sigma+2M\bigr) \chi^1-\QM \bigl(1+(\chi^1)^2\bigr)-\QE
\chi^2\bigr]\,,\\
0&=\QE+e^{-\sqrt{3} \phi+\varphi} \bigl[\bigl(-\sqrt{3}
\Sigma+2M\bigr)\chi^0 -\QE \bigl(1+(\chi^0)^2\bigr) +\QM( \chi^2
+\chi^0\chi^1)\bigr]\,,\\
\begin{split}
0&= 2 Q^2(2\chi^2 + \chi^0\chi^1)
-\QM\bigl[\bigl(1+e^{-\sqrt{3} \phi-\varphi}\bigr)\chi^0 + \chi^1
\chi^2\bigr]\\
&\quad+\QE \bigl[\bigl(1+e^{\sqrt{3} \phi-\varphi}\bigr)
\chi^1-\chi^0 (\chi^2+\chi^0 \chi^1)\bigr]\,.
\end{split}
\end{align}
Note that we already used the constraint equations to simplify the first-order derivatives of $\chi^0$ and $\chi^1$, eqs.~(\ref{eq:fo_chi0}) and (\ref{eq:fo_chi1}). The constraint equations (at least theoretically) enable one to extract the functional dependence of the $\chi^{\alpha}$ on the $\phi$ and $\varphi$, such that we can write $f^{\phi}(q,\chi)$ and $f^{q}(\phi,\chi)$ purely in terms of $\phi$ and $\varphi$: 
\begin{equation}
f^{\phi}(\phi)\equiv f^{\phi}[q,\chi(q)]\,,\qquad
f^{\varphi}(\phi)\equiv f^{\varphi}[q,\chi(q)]\,.
\end{equation}
The condition for the existence of a first-order gradient flow then
becomes
\begin{equation}\label{curl2}
\partial_{[\phi}f_{\varphi]} = -\tfrac{1}{4}\QE(\sqrt{3}\partial_{\varphi}+\partial_{\phi})\chi^0(\phi,\varphi) + \tfrac14\QM(\sqrt{3}\partial_{\varphi}- \partial_{\phi})\chi^1(\phi,\varphi)=0\,,
\end{equation}
and we can evaluate under which conditions on the charges $\Qc^\Lambda$ the
expression (\ref{curl2}) holds.

In principle, we have three constraint equations at our disposal to
eliminate the three axions
$\chi_\alpha(\phi,\varphi),\alpha=0,1,2$, but in practice this
would require solving relatively complicated non-linear simultaneous
equations, which is not straightforward. {}Fortunately, the curl
(\ref{curl2}) requires only a knowledge of the derivatives of the
axions with respect to the dilatons, i.e.~the Jacobian matrix
$[J^{\alpha}_i](\phi)\equiv \partial_{\phi^i} \chi^{\alpha}$. It
turns out that the inverse Jacobian matrix $[J^i_{\alpha}](\chi)
\equiv \partial_{\chi^{\alpha}}\phi^{i}$ is easily computable using
the constraint equations. If we then use
\begin{equation}
[J^{\alpha}_i](\phi(\chi))=[J_{\alpha}^i]^{-1}(\chi)\,,
\end{equation}
where the inverse is with respect to the whole matrix, we can
evaluate condition \eqref{curl2} in terms of the fields $\chi_i$. An explicit
calculation shows that condition is satisfied. We thus conclude that
\emph{when $a=\sqrt{3}$, all the dilatonic black holes with
arbitrary mass, electric, magnetic and scalar charge possess a
generalized superpotential.}

We have not attempted the construction of the generalized
superpotential for arbitrary solutions with $a = \sqrt{3}$, but
rather only for extremal cases. 
Even in this simplified
setting the result is very long compared to the $a=1$ case and not
illuminating, we therefore refrain from quoting it here.

\subsection{The Kaluza-Klein black hole\label{ss:Example_KK_Bh}}
Let us now consider  black holes carried by multiple scalars and
vectors. In $D=d+1=5$ dimensions  there is an example for which we can use the
same hidden symmetry as for the dilatonic black hole, namely
$\SL(3,\Real)$. This theory is obtained by reducing gravity in seven dimensions on a
two-torus. This gives a five-dimensional  theory with two vectors and three
scalars: an axion-dilaton system and an extra dilaton $\tilde\varphi$
\begin{equation}\label{eq:KKBHaction}
S=\int\de^5
x\sqrt{|g|}\Bigl(\mathcal{R}-\tfrac{1}{2}
(\partial\tilde\varphi)^2+\tfrac { 1 } { 4 }\tr(\partial
K\partial K^{-1})
-\tfrac{1}{4}\e^{\sqrt{\frac{5}{3}}\tilde\varphi}K_{mn}F^{(m)}
F^{(n)}\Bigr) ,
\end{equation}
where the matrix $K$ defines the $\SL(2,\mathbb{R})$ axion-dilaton
system and indices $m,n=1,2$. Details on this action and the reduction to $d=4$
dimensions are found in the appendix of \cite{Perz:2008kh}.

This Lagrangian is a consistent truncation of maximal and
half-maximal supergravity in $D=5$. Upon reduction over time one
obtains four-dimensional Euclidean gravity coupled to a set of
scalars that span the coset $\SO(1,1)\times \SL(3,\Real)/\SO(2,1)$.
The dynamics of the decoupled scalar (the $\SO(1,1)$ part) is
trivial and the $\SL(3,\Real)/\SO(2,1)$ part differs from the previous
example only in that this coset has a different $\SO(2,1)$ isotropy
group embedded in $\SL(3,\Real)$. We again use coordinates $\tilde q^i =(\phi^1,\phi^2,\chi^0,\chi^1,\chi^2)$. The effect of this is purely a
matter of signs, as can be seen in the metric on the moduli space
\begin{equation}
\begin{split}
\tilde \cg_{ij}\de\tilde q^i\de \tilde q ^j={}&(\de\phi^1)^2 + (\de\phi^2)^2 +
\e^{-\sqrt{3}\phi^1+\phi^2}(\de\chi^0)^2 - \e^{2\phi^2}(\de \chi^2)^2\\
&-\bigl[\e^{+\sqrt{3}\phi^1+\phi^2} + \e^{2\phi^2}(\chi^0)^2\bigr](\de
\chi^1)^2 - 2\chi^0 \e^{2\phi^2} \de\chi^1\de\chi^2 \,.\label{sigma2}
\end{split}
\end{equation}
The scalars of the five-dimensional theoy are linear combinations of $\chi^0,\phi^1,\phi^2$, while $\chi^1,\chi^2$ are the electric potentials of the gauge fields $F^{(1)},F^{(2)}$.
This sigma model can be obtained from (\ref{sigma}) through the
analytic continuation
\begin{equation}
\chi^0\rightarrow \im\chi^0\,,\qquad \chi^2 \rightarrow \im\chi^2\,.
\end{equation}
The representative $\tilde L$ of the full coset $\SO(1,1)\times
\SL(3,\Real)/\SO(2,1)$ is given by
\begin{equation}
\tilde L = \e^{\phi^0/\sqrt{6}} L\,,
\end{equation}
where $L$ is the $\SL(3,\Real)/\SO(2,1)$ coset representative
\eqref{L}.

We again assume that the
charge matrix describes only the geodesics that go through the
origin. As before we can justify this restriction by proper field
redefinitions and coordinate transformations of the general
solution. The matrix $\Qc$ is written in terms of 9 constants $\Qc^\Lambda$ as $\Qc=\Qc^\Lambda T_\Lambda$, where now $\Lambda = 0,\dotsc,8$, $T_0$ is the three-dimensional
identity matrix generating the decoupled $\SO(1,1)$ part and the
remaining generators $T_\Lambda$ are as before. The Cartan involution $\theta(\Qc)= -\Qc$ condition implies
\begin{equation}
\Qc^3 = -\Qc^6, \qquad \Qc^4 = \Qc^7,  \qquad \Qc^5 = -\Qc^8\,,
\end{equation}
so that
\begin{equation}
\Qc = \Qc^\Lambda T_\Lambda =
\begin{pmatrix}
\Qc^0-\frac{\Qc^1}{\sqrt{3}}-\Qc^2 & -\Qc^6 & -\Qc^8 \\
\Qc^6 & \Qc^0+\frac{2 \Qc^1}{\sqrt{3}} & \Qc^7 \\
\Qc^8 & \Qc^7 & \Qc^0-\frac{\Qc^1}{\sqrt{3}}+\Qc^2
\end{pmatrix},
\end{equation}
The constants $\Qc^0,\Qc^1,\Qc^2$ are certain linear combinations of the ADM mass and scalar charges for the two five-dimensional scalars, while the parameters $\Qc^6$ and $\Qc^8$ can be identified with the electric charges in $D=5$. This can be seen from the first-order equations below and the relations of the basis of scalars given here as opposed to the original fields in five dimensions.

To obtain the first-order velocity field for the effective action with
the black hole potential one needs to eliminate
$\chi^1$ and $\chi^2$ in terms of the remaining scalars using the
constraint equations $ \tilde \cg^{ij}(\Mg^{-1}\partial_i \Mg) \tilde f_j(\tilde q) = \Qc$.
Unlike in the dilatonic black hole example, there are more
constraints than variables to eliminate, unless specific choices for
the charges make fewer of them independent. Using different
combinations of constraint equations to eliminate $\chi^1$ and
$\chi^2$ leads to different velocity fields in five dimensions.
Although they become equivalent upon using the Hamiltonian
constraint (which is exactly the remaining constraint equation), the
expression for the integrability condition $\partial_i f_j = \partial_j f_i$ is not unique. One preferred form should
however distinguish itself, namely that not containing second-order
integration constants. Finding such a combination of constraint
equations is a technically complex task, as it involves relaxing
the boundary conditions ($\Mg(0)=1$) in order to distinguish first- and
second-order integration constants.\footnote{The first-order
integration constants in $\Qc$ and the second-order integration
constants in $\Mg(0)$ are intertwined through the involution condition
$\Mg=\Mg^\sharp$,  making it difficult to distinguish them in the
coset matrix formalism.} {}For this reason we have not pursued it
further. In light of the Hamiltonian formalism presented above, it is then not immediately clear what the Poisson commuting constants of motion are from the $(D=5)$-dimensional perspective. The ambiguity in eliminating the extra scalars of the $(d=4)$-dimensional theory, is just a translation of that fact. We comment on this in the conclusions of this chapter.

\section{Conclusion, literature survey and outlook\label{s:JP-Conclusions}}
\subsubsection{What was the research question?}
In this chapter, we considered spherically symmetric black hole solutions with metric \ansatz{} \eqref{eq:JP-BH_metric_Ddim} to theories of (super)gravity in $D$ dimensions coupled to neutral scalar fields and Abelian vector fields, described by the action \eqref{eq:JP-Sugra-action}. After integrating out the vector fields by their equations of motion, an effective action can be constructed in terms of the radial variable $\tau$ for the black hole warp factor $\varphi$ and scalars $\phi^a$. This action contains a potential that depends on the scalars and the electric-magnetic charges. We asked ourselves the question if first-order equations exist for generic extremal and non-extremal black holes, that mimic the form of the attractor flow equations of the previous chapter which are known to exist for supersymmetric and certain non-supersymmetric extremal black hole solutions. The flow equations would then take the form
\begin{equation}
  \cg_{ij} \dot q^j = \pf {\cw (q)}{q^i}\,,\label{eq:JP-Flow_Eqs_Concl}
\end{equation}
where the coordinates $q$ contain the warp factor and the scalars of the $D$-di\-men\-sio\-nal theory, $q^i = (\varphi,\phi^a)$. We call the function $\cw$ a \emph{generalized superpotential}\index{superpotential!generalized}.

\subsubsection{What is our answer?}
We have shown that the most general form of the flow equations as in \eqref{eq:JP-Flow_Eqs_Concl} follow from rewriting the effective action as a sum of squares.  The generalized superpotential $\cw$ is related to the black hole potential by eq.~\eqref{eq:JP-V_intermsof_W} as $V(\phi) = G^{ij}\partial_i \cw \partial_j\cw -\En$, $\En$ being a constant that is (non-)zero for (non-)extremal black hole solutions. The above gradient flow\index{gradient flow!equations} equations are equally applicable to extremal (whether supersymmetric or not) as well as non-extremal black holes (necessarily non-supersymmetric). They naturally encompass previously known partial results.\footnote{\ldots although they differ from the form conjectured in \cite{Andrianopoli:2007gt}.}

{\underline{\emph{Existence and uniqueness.}} }We were able to present an existence criterion for flow equations of the form \eqref{eq:JP-Flow_Eqs_Concl}, based on Liouville integrability of the effective black hole description. The effective action with a black hole potential can be written as a Hamiltonian system, with coordinates $q^i=(\varphi,\phi^a),p_i=\cg_{ij}\dot q^j$. {}From standard classical mechanics, it follows that a superpotential $\cw$ and thus flow equations of the form \eqref{eq:JP-Flow_Eqs_Concl} exist \emph{if and only if the system is Liouville integrable }(i.e.~contains at least $n$ constants of motion that commute under the Poisson brackets, where $n$ is the total number of scalars $q^i$). We have linked this to the observations of ref. \cite{Andrianopoli:2009je}, where it was noticed the superpotential is actually Hamilton's characteristic function of standard Hamilton-Jacobi theory.

{\underline{\emph{Examples.}} }The existence criterion presented above is a keen theoretical result, but does not mean we have a good method to construct the superpotential in practice. All examples given in the previous chapter allow for a straightforward construction of the superpotential $\cw$. We have exploited this to construct the superpotential for the \RN black hole and the dilatonic black hole, with either only electric or magnetic charges. However, those examples are of a restrictive kind. We considered then a more general set of examples, for theories with scalar manifolds being symmetric spaces after a timelike dimensional reduction. Those systems are explicitly integrable (solvable), see \cite{Fre:2009et,Chemissany:2009hq,Fre:2009dg,Chemissany:2009af}. We gave a method of principle to verify whether a generalized superpotential exists. It relies on the fact that black hole solutions trace out geodesics on the moduli space of the theory when reduced over time.   We provided examples of extremal and non-extremal solutions with a generalized superpotential. We applied our formalism to the dilatonic black hole\index{dilatonic black hole} in four dimensions (one scalar field) and a Kaluza--Klein black hole in five dimensions (multiple scalars). {}For the dilatonic black hole with dilaton coupling\index{dilaton coupling} $a=\pm 1$, we were able to show by direct integration that the generalized flow equations exist in all situations (both electric and magnetic charges turned on), using the procedure of deforming the extremal solution employed in chapter \ref{c:BJ}. When $a=\pm\sqrt{3}$ we were able to show the same, using group-theoretical tools to integrate the second-order equations of motion to first-order equations. {}For all other values of $a$ we derived the existence of a fake superpotential\index{superpotential!fake} in the \emph{extremal} case, using the argument applied for single scalar domain walls \cite{Skenderis:2006jq}.\footnote{Although the $a=1$ case was easy to integrate by hand, this is also an example for which we could have constructed the flow using group theory. The reason is that this case is embeddable in an $\cN=4$ action, which has a symmetric moduli space after timelike reduction: $\SO(8,8+n)/[\SO(6,2)\times \SO(6+n,2)]$ (see for instance \cite{Bergshoeff:2008be}).} The existence of generalized flow equations for non-extremal black holes with arbitrary dilaton coupling is not known to us. The investigation of the Kaluza--Klein black hole in five dimensions, in turn, showed the shortcomings of the dimensional reduction approach. There were several ways of eliminating the extra scalars that appear after dimensional reduction, reflecting different possibilities for the expression $f_i(q^j)=\cg_{ij} \dot q^j $. This is a reflection of the fact that the $f_i$ have an explicit dependence on integration constants $C_j$ as $f_i(q^j,C_j)$. Different ways of eliminating the extra fields correspond to finding different expression in terms of different sets of constants of motion $C_i$. The preferred form would be the one where all the constants of motion commute under the Poisson brackets. However, in the dimensionally reduced theory, it is not a priori clear which are the preferred expressions $C_j$ to use, as we explained above. This question merits further research.

%
%
%
%
%
%
%
%
%
%

\subsubsection{What do our colleagues say in the literature?}
Since the appearance of our work \cite{Perz:2008kh}, fruitful research has been done in this subject by several authors. 
First, as we discussed in the main text, the Torino group used Hamilton-Jacobi methods to write down the general form of this superpotential for static solutions and proved that it is duality invariant  \cite{Andrianopoli:2009je}. However, they did not comment on the existence of the superpotential and the connection to Liouville integrability\index{Liouville!integrability}.\footnote{At the moment of writing, part of this group was working out those details in a large collaboration \cite{Chemissany:2010zp}, which appeared after finishing this work, see `outlook'.} Also, a way of constructing the potential in practice was not supplied.  Since then, most attention has gone to theories for which the scalars parametrize a symmetric space. We noted before that such systems are explicitly solvable \cite{Fre:2009et,Chemissany:2009hq,Fre:2009dg,Chemissany:2009af} and presumably also Liouville integrable, leading to the assumption that at least for these theories a superpotential can be written down. And indeed, the superpotential has been succesfully written down, at least for \textit{extremal} black holes. This generalizes the method we gave above to a real proof for those cases. {}For symmetric moduli spaces, the superpotential has been given in great detail for (BPS and non-BPS) extremal solutions, with one or multiple centers, by Bossard et al. \cite{Bossard:2009at,Bossard:2009my,Bossard:2009bw,Bossard:2009we}, using group theory machinery.
Soon after, Ceresole and Dall'Agata used the knowledge that the superpotential was duality invariant to write down a solution from a simpler method for the non-BPS branch of extremal static  solutions in terms of duality invariants in \cite{Ceresole:2009vp,Ceresole:2009iy}. This was done in four-dimensional $\Cn=2$ supergravity for the $t^3,st^2$ and $stu$ models.
More recently, the Torino group investigated aspects of the structure of the superpotential \cite{Andrianopoli:2010bj}, e.g.\ they showed it has the properties of a Liapunov's function and that $\cw$ can always be constructed for small black holes (with vanishing horizon area).
\subsubsection{Outlook}

In light of the existence criterion presented above, the obvious option would be to have a classification of systems which are Liouville integrable and those which are not. We advocate a clean proof to discern which supergravity theories give rise to Liouville integrable effective descriptions and which do not. Since all supergravity theories with more than 8 real supercharges (e.g. $\Cn>2$ supergravity in four dimensions) and some theories with 8 or less supercharges have symmetric moduli spaces, it is of prime interest to find a proof of Liouville integrability for such systems. Since these systems are explicitly integrable by means of constructing a Lax pair representation, they seem like a sure bet. Current research   suggests such systems are Liouville integrable and thus also for non-extremal solutions a superpotential can, in principle, be constructed for those cases. However, for supergravity theories with less extended supersymmetry (as the $\Cn=2$ supergravity theory in four dimensions which is of considerable interest), the scalar manifold is not a symmetric space and no results are known to date. Again, research by the authors of \cite{Chemissany:2010zp} shows that when the moduli space is a symmetric space,  the superpotential can always be constructed for both extremal and non-extremal spherically symmetric black hole solutions. Furthermore, \emph{multi-center solutions} are certainly of interest in this context. Some ideas toward that question were given in \cite{Galli:2009bj}, by extending Denef's construction of multi-center solutions, but a clear overview is not available.


%
%
\index{magnetic potential|see{potential!magnetic}}
\index{electric potential|see{potential!electric}}
\index{existence!of gen.~ superpotential|see{superpotential!existence of --}}

\cleardoublepage
\part{Entropy in supergravity: a search for microstates}
\label{pt:Entropy}
\cleardoublepage
\chapter{
Where are the microstates of a black hole?}\label{c:FB}

\punchline{
This chapter tries to shed a light on the question what the microstates of a black hole are. The existence of such microstates is suggested by the fact that in the classical gravity description, a black hole has a Bekenstein-Hawking entropy. We give a short overview of the  literature on the construction of black hole microstates in a gravity description. We start with a discussion of the \textit{fuzzball proposal}, stating that for every black hole with Bekenstein-Hakwing entropy $S_{BH}$, there exists a number $\e^{S_{BH}}$ of horizonless, non-singular solutions that look like the black hole asymptotically, but differ from the black hole on horizon-scale. In a second part, we discuss a related way of explaining black hole entropy of a four-dimensional black hole in terms of \textit{multi-center configurations} with a scaling property. Section \ref{s:FB_proposal} (fuzzball proposal) forms useful background for chapter \ref{c:JR}, section \ref{s:FB_scaling} gives motivation for chapter \ref{c:TL} and discusses multicenter black hole configurations and their relation to black hole microstates.}

\section{The fuzzball proposal for black holes\index{fuzzball!proposal|(imp}}\label{s:FB_proposal}

As we saw in chapter \ref{c:BH_Playground}, the laws of black hole thermodynamics suggest that the horizon area behaves as an entropy. Semiclassical arguments show that this is more than just an analogy. A full quantum theory of gravity must explain the origin of this Bekenstein-Hawking entropy $S_{BH}$, preferably by identifying a large number $N = \e^{S_{BH}}$  of microstates, much like the large number of configurations of a gas in a bounded volume gives rise to a large entropy for the macroscopic description.

The entropy issue has been partially addressed in string theory, mainly for supersymmetric black holes. One can identify `microstates' of the black hole by studying the theory in a dual regime where (super)gravity is no longer valid, (e.g.~through details of brane physics at low string coupling, or through the AdS/CFT correspondence). For supersymmetric solutions, the number of microstates is invariant under variations of the string coupling constant, because it is protected by supersymmetry.
Cranking up the value of the string coupling, such that supergravity becomes a good approximation,  shows that the number of microstates $N$ counted in the other regime of the theory reproduces the black hole entropy $S_{BH}$ as $N =\e^{S_{BH}}$ (to first-order in the charges the black hole carries). Even though this approach has been extremely successful in reproducing the entropy from a microscopic picture,
other open issues remain. First, we do not know what an individual microstate is in the supergravity picture. Related to this, we do not have a handle on how to understand or resolve the information paradox -- the fact that information falling into the black hole does not seem to come out, at least not in the semiclassical approximation to black holes in general relativity. And what about the singularity that is hidden behind a horizon? We would expect that in a good quantum theory of gravity, quantum effects would resolve the singularity somehow. These questions must be dealt with in a regime where (super)gravity is valid: from the dual (field theory) point of view, this is an impossible task.

\begin{figure}[ht]
\centering
\subfigure[center][Black hole]{
\includegraphics[scale=.12]{BH_Singularity.ps}
\label{fig:FB-intro_FuzzBall_a}
}
\hspace{3cm}
\subfigure[center][Fuzzball]{
\includegraphics[scale=.12]{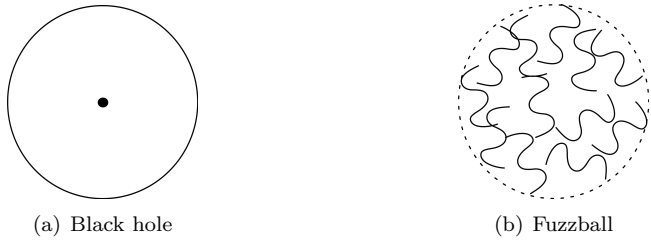}
\label{fig:FB-intro_FuzzBall_b}
}
\caption[Black hole vs.~fuzzball: singularity/quantum `fuzz'.]{Classically, a black hole is a singularity cloaked by a horizon (left). A typical fuzzball state has no horizon, only a `quantum fuzz' stretching out to horizon size (right). (Figures inspired by \cite{Mathur:2004sv}.)\label{fig:FB-throat}}
\label{fig:FB-intro_FuzzBall} 
\end{figure}

In order to answer these questions, it is thus natural to ask the following question:
\begin{center}
\emph{What do microstates look like in the regime where the black hole exists?}
\end{center}
In other words, can we find an interpretation of black hole microstates, computed in a dual theory, on the gravity side? Or are such microstates, if they can be constructed, necessarily stringy in origin? This question and a tentative answer forms the basis of the fuzzball proposal, advocated by Mathur and collaborators and taken up by many different groups afterwards. According to this proposal, \textit{for a black hole with an associated entropy $S_{BH}$, there exist $\e^{S_{BH}}$ smooth horizonless solutions (specified by a metric and other fields), called fuzzballs, which differ from the black hole on horizon-size scales}.\footnote{These fuzzballs should be solutions to the full quantum gravity (in casu string theory) and should be seen as the duals to the individual microstates discussed above. The hope is that most of these fuzzball geometries that can account for the Bekenstein-Hawking entropy, can be found as solutions to the (super)gravity approximation of string theory.} This is illustrated in figure \ref{fig:FB-intro_FuzzBall}. 

\outline{
We give a more precise definition of the fuzzball proposal in section \ref{ss:FB_FuzzballsIntro} with references to the original literature. We also motivate the proposal from AdS/CFT and hint at how it could solve the issues raised in the previous paragraph. In sections \ref{ss:FB-Two_charge_system} and \ref{ss:FB-Three_charge_system}, we briefly discuss the construction of fuzzball solutions in five dimensions, carrying two and three charges respectively, since it is in the five-dimensional picture that the fuzzball proposal has received most attention. We discuss the relation to four-dimensional black holes in the latter section. We end with a conclusion in section \ref{ss:FB-4D/5D}.}

\subsection{The fuzzball programme\label{ss:FB_FuzzballsIntro}}

\subsubsection{The fuzzball proposal for black holes}
Consider a black hole with Bekenstein-Hawking entropy $S_{BH}$. The fuzzball proposal states that \textit{there exist $\e^{S_{BH}}$ solutions that look like the black hole asymptotically}, but differ from the black hole \spacetime{} on a scale typically of the order of the horizon scale. In particular, the fuzzball solutions carry the same asymptotic charges (electric/magnetic charges, angular momentum) as the original black hole. The black hole should then correspond to an average description, in analogy with the thermodynamic description of a gas as opposed to statistical mechanics in terms of microscopic configurations.  This idea is depicted in figure \ref{fig:FB-throat}. Originally, the fuzzball proposal was put forth by Mathur and Lunin in \cite{Lunin:2001fv,Lunin:2001jy,Lunin:2002qf} and was taken up afterwards by many researchers. In particular, good reviews from different groups have appeared over the last years, see for instance \cite{Mathur:2005zp,Bena:2007kg,Skenderis:2008qn,Mathur:2008nj,Balasubramanian:2008da} and references therein.

These $\e^{S_{BH}}$ states that are conjectured to exist\ldots
\begin{itemize}
\item \ldots have no horizon. If such a solution would have a horizon, we could associate a Bekenstein-Hawking entropy to it and would write it in terms of a number of other, fundamental, microstates.
\item \ldots are supposed to be non-singular. 
\end{itemize}
Below, we give typical fuzzball solutions and we will see they differ from the classical black hole solution on the horizon scale, the black hole should be replaced by some `fuzzy' solution.
\begin{figure}[ht!]
\centering
\includegraphics[scale=.4,angle=-90]{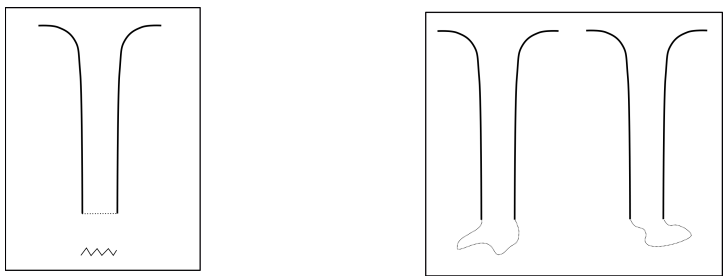}
\caption[Black hole and two fuzzballs as throats.]{Sketches of black hole and fuzzball near horizon regions, depicted as throats. For a black hole (left) 
this throat is infinite and ends on the horizon (dotted line), behind which sits a singularity. For typical fuzzball solutions (as the two on the right),
 the throat is very long but not infinite and ends smoothly, there is no horizon.\label{fig:FB_throat}}
\end{figure}	

Black holes are found as solutions to general relativity or supergravity, so we could ask if all fuzzball solutions are solutions to supergravity. This is a hope that has turned out to be false: for a typical black hole, one cannot construct $\e^{S_{BH}}$ fuzzball solutions that are \textit{all} well-described by the supergravity equations of motion.  Instead, we should consider supergravity as a low energy effective description of string theory and consider fuzzball solutions to the latter, more general, theory. This means concretely that we can have an interpretation of fuzzball solutions  in terms of a metric and other fields, but the \spacetime{} metric need not always be a solution to supergravity (for instance, because the curvature becomes string scale and the supergravity equations of motion are no longer valid) and we can have excitations of many other fields (which need not all be supergravity fields, but can a priori also be other string excitations).

\subsubsection{Fuzzball solutions from the dual CFT}
The fuzzball proposal says that a black hole should be seen as some ensemble of horizonless solutions. In principle, such fuzzball solutions are conjectured to exist for all black hole solutions, whether they are supersymmetric, extremal or non-extremal. In practice, however, it is a formidable task to construct fuzzball solutions for a generic black hole solution. Instead, most successes of the fuzzball proposal have been achieved exactly for extremal solutions. The reason is that these solutions develop an $\AdS{}$ region as one approaches the horizon and the fuzzball states can be compared to  states in a conformal field theory, dual to the Anti-de Sitter space, by the AdS/CFT correspondence.

\subsubsection{Implications on black hole physics}
The original motivation for the fuzzball proposal lies in trying to solve the information paradox (see section \ref{s:BH-infoparadox}). So far, the fuzzball idea seems indeed on its way to offer a solution. The origin of the \textit{information paradox} lies in the study of black hole \spacetime{} near the black hole horizon. For a (large) black hole, the region near the horizon is a region of low curvature. It is therefore justified to study particle production in the black hole background, which is kept fixed. This leads to the celebrated Hawking radiation. The fact that the radiation reveals no details of what formed the black hole, leads to the information puzzle.  Intuitively, one can understand that quantum effects that only change the black hole near the singularity, would not be able to resolve the information paradox, as the region near the horizon remains quasi unaltered and the same arguments apply. This can be made more precise, see for instance \cite{Mathur:2000ba,Lunin:2001jy,Mathur:2002ie}. However, the requirement that a black hole should be replaced by some ensemble of fuzzball states, that have a non-trivial structure all the way up to the horizon, would evade these arguments. 
Computation of particle scattering off black holes interpreted as thermal ensembles of fuzzball states motivate that the thermal radiation emitted by a black hole reveals details (information) of all matter that fell into the hole, by correlations with the fuzzball states (see \cite{Mathur:2008nj,Mathur:2008wi}). Moreover, for black holes with AdS regions, this can be further motivated by the  AdS/CFT correspondence: since the evolution in the dual CFT is unitary, so should the physics on the AdS side be and we expect the information paradox to be evaded.

\subsubsection{Results and outlook}
The main focus in the literature so far has been on constructing fuzzball states in supergravity in five dimensions. Using one charge only (e.g.~a number of strings or D-branes of one type wound around a compact direction), we cannot model a five-dimensional black hole, so searches for fuzzball solutions start with two-charge solutions. For black holes with two asymptotic charges (for instance, D1-D5 black holes in five dimensions), one has been able to construct all corresponding fuzzball solutions, for large charges. Many of them are well-described in supergravity, but a large subset are definitely stringy in origin. However, the two-charge system in five dimensions is only a toy model, as the two-charge black hole has vanishing horizon area and Bekenstein-Hawking entropy in the supergravity approximation. To describe a well-behaved black hole in supergravity with non-vanishing horizon area, we need to add at least a third charge. For three-charge systems in five dimensions, the results are incomplete so far and searches for fuzzball solutions have always started by constructing large classes of horizonless solutions in supergravity, carrying the same asymptotic charges as the black hole. These supergravity solutions typically account for only a finite part of the Bekenstein-Hawking entropy, but not all of it.\footnote{In four dimensions, a black hole with non-vanishing horizon area has at least four different charges. It can be related to five-dimensional solutions through the 4D-5D connection, see below.}

\subsection{The two-charge system\label{ss:FB-Two_charge_system}}

Ten-dimensional solutions carrying two charges, namely D1- and D5-brane charge, have been a useful test case for the fuzzball proposal. Originally, a large set of horizon-free supergravity solutions carrying the same charges as the \textit{supersymmetric} D1-D5 black hole, have been found in \cite{Lunin:2001jy} and were proposed to correspond to microstates of the CFT. However, only a finite fraction of the black hole entropy can be reproduced from these solutions. Later, the entire class of solutions that reproduces the entropy were found in \cite{Lunin:2002iz,Taylor:2005db,Kanitscheider:2007wq}. All these solutions preserve 1/4 of the supersymmetry, since each type of brane breaks half the amount of supersymmetry. The discussion of this section follows references \cite{Lunin:2001fv,Skenderis:2006ah,Skenderis:2008qn}.

\subsubsection{The naive D1-D5 black hole geometry}
We consider black holes in five dimensions, in compactified type IIB string theory, as this is the setup in which most tests of the fuzzball programme have been performed.  It is useful to briefly repeat how we can make a black hole out of fundamental objects as strings and D-branes. Take ten-dimensional \spacetime{} to be of the form $M_{1,9} = M_{1,4} \times X_5$, where $M_{1,4}$ is five-dimensional \spacetime{} (with Lorentzian signature) and $X_5$ is a compact space. By wrapping branes on the internal space $X_5$, we can build solutions that are pointlike in $M_{1,4}$. Different type of branes give different types of electric/magnetic charges in five dimensions. 

We take $X_5$ to describe a five-torus, and write it as a four-torus times a circle, such that ten-dimensional space-time is compactified as  $M_{1,9} = M_{1,4} \times T^4\times S^1$.
We consider $N_5$ D5-branes wrapped on $T^4\times S^1$, $N_1$ D1 branes wrapped on $S^1$ (i.e.~the D1 branes are embedded in the D5-branes). 
We consider the corresponding supergravity solution. Denote again with $y$ the coordinate on the circle and $(t,x^i)$ the coordinates on $M_{1,4}$. Following the harmonic function rule (see for instance \cite{Cvetic:1995bj,Tseytlin:1996bh}), the ten-dimensional metric (in string frame) is 
\begin{align}
\de s_{10d}^2 &= \frac 1{\sqrt{H_1 H_5}} (-\de t^2 + \de y^2) + \sqrt{H_1 H_5} (\de x^i)^2 + \sqrt{\frac{H_1}{H_5}} \de s^2(T^4)\,,\label{eq:FB-Metric_D1D5_BH}
\end{align}
where $\de s^2(T^4)$ is a metric on $T^4$ and $H_1$, $H_5$ are harmonic functions living in $\mathbb{R}^4$:
\begin{equation}
 H_1(r) = 1 + \frac{Q_1}{r^2}\,,\qquad H_5(r) = 1 +\frac{Q_5}{r^2}\,,\qquad r^2 = \sum_{i=1}^4(x^i)^2\label{eq:FB-D1D5_harm_functions}
\end{equation}
The charges $Q_{1},Q_5$ (with dimensions of length squared) are related to integers $N_1,N_5$ counting the number of D1 and D5-branes (see \cite{Becker:2007zj} for a pedagogical derivation)
\begin{equation}
 Q_1 = \frac{g_s (\alpha')^3}{V} N_1\,,\qquad Q_5 = g_s \alpha' N_5\,,\label{eq:FB-integer_charges}
\end{equation}
where $(2\pi)^4 V$ is the volume of $T^4$ as measured with the metric $\de s^2(T^4)$. After compactification, the five-dimensional metric in Einstein frame is given by:
\begin{equation}
 \de s^2 = -f(r)^2 \de t^2 + f(r)^{-1} \,\de x^i \de x^i\,,\qquad f(r) = \sqrt{H_1 H_5} = \sqrt{(1 + \frac{Q_1}{r^2})(1 + \frac{Q_1}{r^2})}\label{eq:FB-5d_EinsteinMetric}
\end{equation}
This metric has no horizon, but only a (naked) singularity as $r\to 0$. 
We conclude the  D1-D5 black hole does not have a macroscopic horizon in supergravity. However, 
one can identify microstates in the dual CFT description carrying the same total \mbox{D1-D5} charge, with a microscopic entropy 
\begin{equation}
S =  2\pi\sqrt2\sqrt{N_1 N_5} \label{eq:FB-D1D5_micro_entropy}\,,
\end{equation}
to leading order in the charges. The appearance of a microscopic entropy is unclear from the supergravity solution, since the curvature stays well behaved near the singularity $r=0$ and higher-order corrections in the curvature do not generate a horizon. We show that we should replace the naive black hole geometry by smooth `fuzzball' solutions.

\subsubsection{D1-D5 fuzzballs}
\begin{figure}[ht]
\centering\subfigure[]{
\epsfig{figure=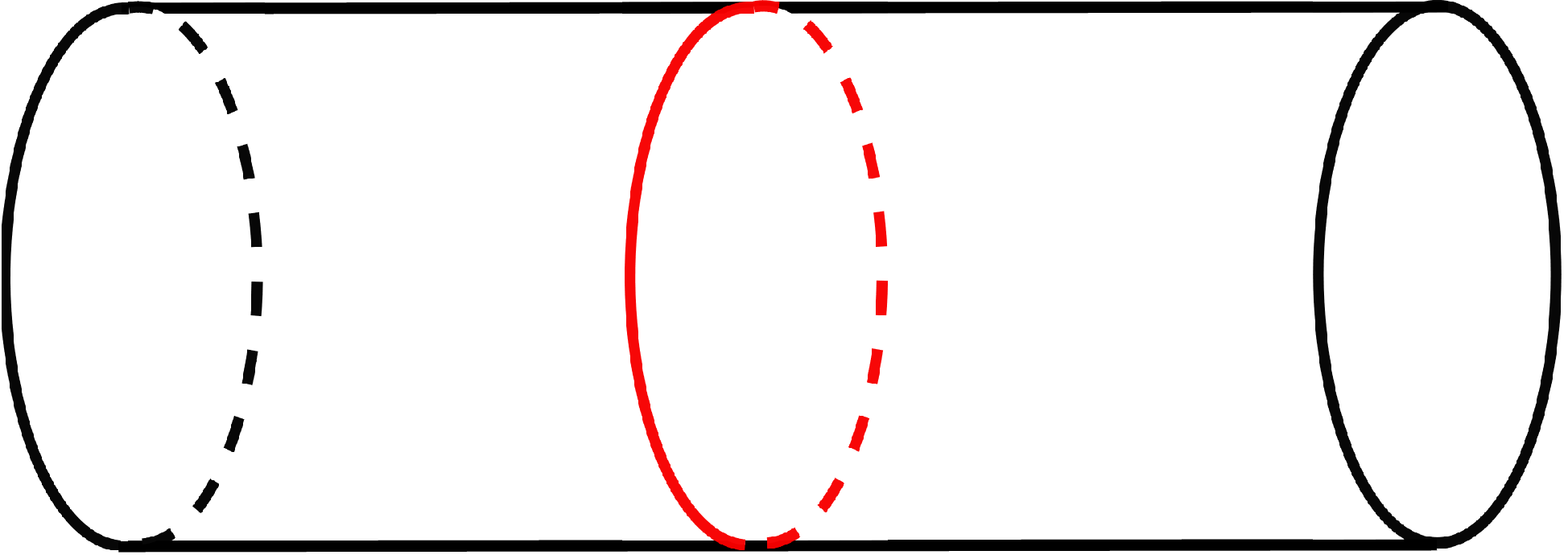,angle=-90,scale=.25}
\label{fig:FB-string_wound_a}
}
\hspace{1cm}
\subfigure[]{
\epsfig{figure=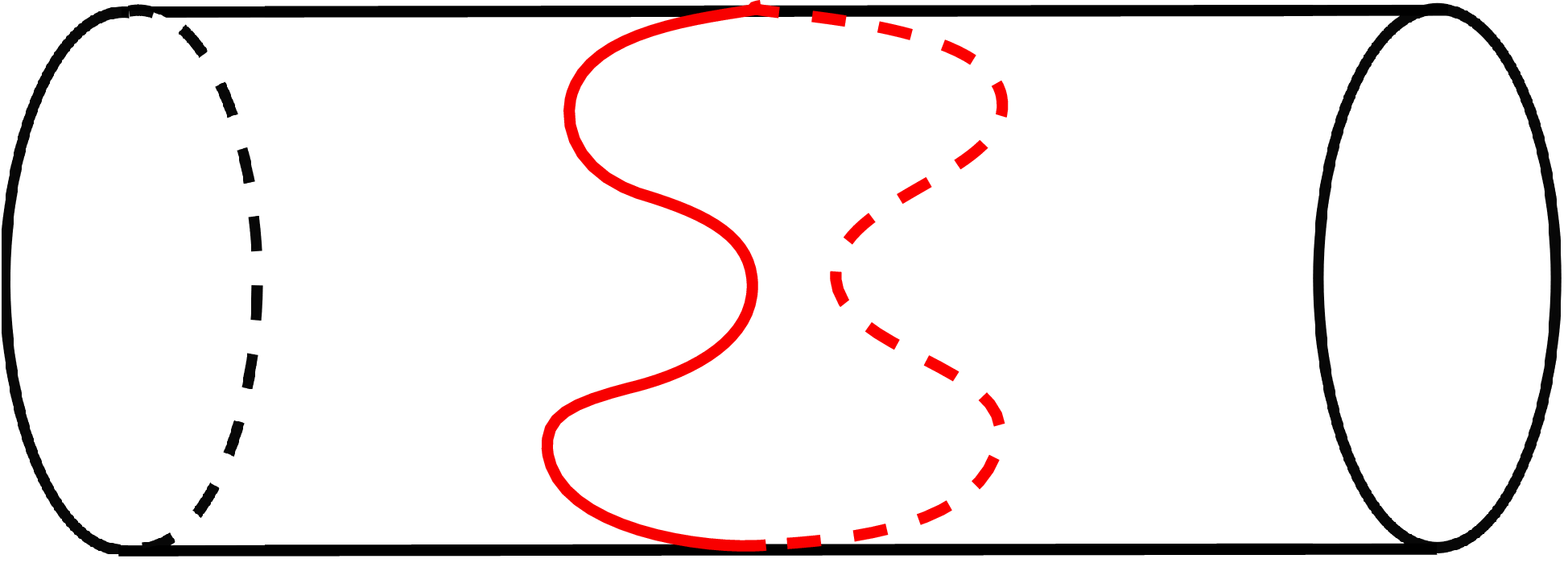,angle=-90,scale=.25}
\label{fig:FB-string_wound_b}
}
\caption[A cartoon of the F1-P system.]{
A cartoon of the F1-P system. In figure \subref{fig:FB-string_wound_a}, we see the naive way of wrapping a string, figure  \subref{fig:FB-string_wound_b} shows how a string with added momentum has a non-trivial profile in the transverse directions.\label{fig:FB-string_wound}}
\end{figure}

The construction of fuzzball states, preserving 1/4 of the supersymmetries, for the D1-D5 system is given by considering a set of S- and T-dualities to the F1-P system. This is a system consisting of a fundamental string (F1) carrying momentum (P) along the internal directions. In particular, under the relevant dualities, the D1-D5 system is mapped to a string wound $N_5$ times around the $y$-circle, carrying $N_1$ units of momentum along that circle.
As Lunin and Mathur pointed out in \cite{Lunin:2001fv,Lunin:2002qf}, the important thing to note is that a string has no longitudinal vibrations: the momentum the string carries is realized through travelling waves giving  only \textit{transverse} excitations, figure  \ref{fig:FB-string_wound}. A string carrying momentum is thus not a point in the transverse directions, but has a finite size.
For a string in static gauge, with worldsheet coordinates $(\tau = t,~\sigma = y)$, the excitations are in the eight dimensions transverse to the $t,y$ directions. Lunin and Mathur only consider excitations in the four-dimensional non-compact transverse space with coordinates $x^i$ (and not along the internal $T^4$). Such solutions are determined by a vibration profile $\vec F(v) = F^i(v)$, with $v = t+y$ the lightcone coordinate, appropriate for left-moving excitations.\footnote{In general, we would expect the profile $F^i$ to be a generic function of both $t+y$ and $t-y$. However, a solution having both left-and right-moving modes does not have minimal energy for a given momentum charge $P$ and is therefore not extremal and a fortiori not supersymmetric.}

The metric for these solutions is smooth and has no horizon: the black hole is replaced by some `fuzzball'.  Due to the string carrying only transverse excitations, this should be seen as a `blow-up' of the naive black hole solution \eqref{eq:FB-Metric_D1D5_BH}. After performing the appropriate dualities, the metric in the \mbox{D1-D5} frame  is 
\begin{align}
\de s_{10d}^2 = \frac 1{\sqrt{H_1 H_5}} (-(\de t + A)^2 + (\de y+B)^2) + \sqrt{H_1 H_5} (\de x^i)^2 + \sqrt{\frac{H_1}{H_5}} \de s^2(T^4\,,\label{eq:FB-D1D5_FB_metric}
\end{align}
where now $H_1$ and $H_5$, as well as the one-forms $A = A_i \de x^i\,,~B=B_i\de x^i$  are obtained by smearing the harmonic functions discussed for the naive black hole solution over a region in the four-dimensional transverse $\mathbb{R}^4$ through the profile $\vec F$:
\begin{align}
H_5(\vec x)&=1+\frac{Q_5}L\int_0^L\frac{\de v}{|\vec x - \vec F(v)|^2}\,,\nonumber\\
H_1(\vec x)&=1 + \frac{Q_5}L\int_0^L\frac{|\dot{\vec {F}}(v)|^2}{|\vec x - \vec F(v)|^2}\de v\,,\nonumber\\
A_i(\vec x)&=-\frac{Q_5}L\int_0^L\frac{\dot F_i(v)}{|\vec x - \vec F(v)|^2}\de v\,,\qquad
B = -\star_4 \de A\,,
\end{align}
where a dot means a derivative w.r.t. $v$ and $\star_4$ denotes the Hodge dual in the four-dimensional non-compact transverse space $\mathbb{R}^4$ with coordinates $x^i$.  Note that $H_5$ is obtained by a uniform smearing, while $H_1$ corresponds to a distribution weighted with the first derivative of $F^i$. Finally, the D1 charge is related to the D5 charge as
\be Q_1 = \frac{Q_5}{L} \int_0^L |\dot {\vec{F}}(v)|^2
dv\,. \ee

In later work, fuzzball states with angular momentum \cite{Lunin:2002iz} and  with internal excitations along both the $x^i$ and $T^4$ directions have been constructed, as well as states with fermionic excitations, see \cite{Taylor:2005db,Kanitscheider:2007wq}. It is believed that these constitute all microstates of the D1-D5 system, as we briefly motivate below.

\subsubsection{Motivation for the D1-D5 fuzzball picture}
We summarize the evidence supporting the fuzzball idea for the D1-D5 system.

The naive black hole has zero-size horizon and hence no Bekenstein-Hawking entropy. This is in seeming contradiction with the non-zero degeneracy one finds from the D1-D5 brane physics or dual F1-P system. We can intuitively solve this puzzle by looking at the \textit{typical size} $r \simeq r_{fuzz}$ where two fuzzball states start differing from each other. In the F1-P picture, this radius is of order string scale ($l_s =\sqrt{\alpha'}$). After dualizing to the D1-D5 system, one finds that the area $A_{fuzz}$ of the locus $r=r_{fuzz}$ calculated with the metric \eqref{eq:FB-D1D5_FB_metric} behaves as expected for a Bekenstein-Hawking entropy:
\begin{equation}
 S_{BH} = \frac {A_{fuzz}}{4 G_{5}} \sim \sqrt{N_1 N_5}
\end{equation}
This is very satisfactory: the area of the `fuzzball region' satisfies a Bekenstein-Hawking type relation, agreeing with the microscopic entropy. It also follows that a typical fuzzball profile is bounded  as $|\vec F(v)| < r_{fuzz}$. When $r\gg r_{fuzz}$, we see that the fuzzball geometry \eqref{eq:FB-D1D5_FB_metric} takes the same form as the naive geometry \eqref{eq:FB-Metric_D1D5_BH}: asymptotically, a fuzzball state looks like the black hole. 

A priori, in the classical supergravity theory, there are infinitely many such fuzzball states, one for each choice of the profile $\vec F$. In order to associate a finite set of states, one needs to follow an appropriate quantization scheme. In \cite{Rychkov:2005ji}, this was done by Rychkov through \textit{geometric quantization} of the classical supergravity solution. 
In the end, the degeneracy of states can be found by a partitioning argument as the degeneracy of the $N_1N_5$ energy level in a system of four chiral bosons:
\begin{equation}
S_{Rychkov}=2\pi\sqrt{\frac c{6} N_1N_5}\,,\qquad c = 4\,.
\end{equation} 
This correctly reproduces the entropy for bosonic excitations in the transverse $\Real^4$. It is interesting -- but unclear how -- to see what geometric quantization for the most general fuzzballs \cite{Lunin:2002iz,Taylor:2005db} would give, especially since the internal excitations (along $T^4$) are not visible in supergravity. One expects to find the result for the full microscopic entropy, accounting for 8 bosonic and 8 fermionic modes (along $\Real^4\times T^4$), giving a total of $c=12$ and agreeing with \eqref{eq:FB-D1D5_micro_entropy}.

The best testing ground for the fuzzball proposal in the D1-D5 system, has been by \textit{comparison with the dual CFT}. As a first step, one can perform a Fourier expansion as above for the most general profile $F^I(v)$, where $I$ now also runs over the internal $T^4$ directions and all fermionic modes. The oscillators (Fourier coefficients) can be mapped directly to the creation operators for string oscillations in the F1-P system, and by duality to (superpositions) of ground states in the Ramond sector of the D1-D5 CFT. The latter is the conformal field theory living on the $1+1$-dimensional \worldvolume{} of string formed by the common D1-D5 direction. Alternatively, this CFT can be seen as living on the boundary of the $\AdS 3$ throat of the black hole geometry \eqref{eq:FB-5d_EinsteinMetric}. By the AdS/CFT dictionary, one has been able to compare the physics on the gravity side and the CFT side. For instance, one-point functions on both sides agree \cite{Skenderis:2006ah,Kanitscheider:2006zf,Kanitscheider:2007wq}, as well as absorption cross sections and travel times of particles in the fuzzball background/dual CFT state. 

\outlook{Conclusion}{For the two-charge D1-D5 system, there is plentiful evidence for the fuzzball conjecture. However, there is no black hole with a regular horizon for these charges in supergravity. We can try to add more charges for a realistic test of the conjecture.}

\subsection{The three-charge system\label{ss:FB-Three_charge_system}}
We give a very brief overview of research in identifying fuzzball states for three-charge black holes in five dimensions.

Above we discussed the D1-D5 system because it is best under control: there are maps from all microstates to the dual conformal field theory. 
Only the averaging procedure that should reproduce the black hole properties from the microstates is less understood. Note that there is one important caveat: the black hole one would naively construct out of D1 and D5 branes has no horizon itself, but corresponds to a naked singularity.  

In order to describe black holes with regular, macroscopic, horizons in supergravity in five dimensions, we need to add another charge. Most original literature looked at the D1-D5 system with added momentum (P) on the common direction of the D1 and D5 branes. Some success was achieved in this way. Two different paths have been followed since then in trying to construct microstates in the supergravity regime. The first one starts from a five-dimensional black hole or black ring with macroscopic entropy and tries to find smooth, horizonless solutions by tweaking the original solution, for instance by taking a certain limit. Of particular interest is the work of \cite{Bena:2005ay}, which is discussed in the next chapter in the light of our work \cite{Raeymaekers:2008gk}. The second path has by far been most succesful. It is inspired by the classification of supersymmetric solutions of minimal ungauged supergravity in five dimensions of \cite{Gauntlett:2002nw}. Through explicit solutions of the general equations of motion of five-dimensional maximal supergravity, one has been able to construct large classes of smooth so-called `bubbling' solutions with three asymptotic charges. The epithet `bubbling' refers to the non-trivial structure of these solutions. The four-dimensional spacelike part of the geometry of these solutions has non-trivial two cycles (`bubbles') which are of order horizon-size. The collaboration around Bena and Warner has provided many examples, see \cite{Bena:2007kg} for a review and reference therein for related work. A lot of examples have been constructed for `scaling solutions', where the several centers between which the bubbles form limit to a single point. We discuss similar solutions in four dimensions in section \ref{s:FB_scaling}.

\subsection{Relation to four-dimensional physics: 4D-5D connection}\label{ss:FB-4d/5d-connection}

\begin{figure}
\centering
\begin{picture}(100,70)
\put(-35,60){\epsfig{figure=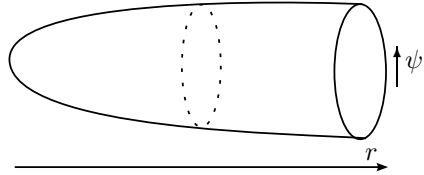,width=.18\textwidth,angle=-90}}
\put(100,0){$r$}
\put(112,28){\vector(0,1){15}}
\put(115,35){$\psi$}
\end{picture}
\caption[Four dimensional Taub-NUT space as a `cigar'.]{Four dimensional Taub-NUT space as a `cigar'. Coordinates $(\theta,\phi)$ are suppressed.}
\label{fig:FB-TN} 
\end{figure}

Of special interest is the translation of the five-dimensional three-charge systems to four-dimensional setups in order to find `microstates' of four-dimensional black holes. The guide is the so-called`4D-5D connection' to map three-charge solutions in five dimensions to four-dimensional solutions with four charges. The idea is to replace the four-dimensional $\Real^4$ spatial part of the five-dimensional geometry by a Taub-NUT space. Four-dimensional (Euclidean) Taub-NUT space is given as a circle fibration (with coordinate $\psi\sim \psi +2\pi$) over three-dimensional flat space (with standard spherical coordinates $r,\theta,\phi$):
\begin{equation}
\de s^2_{TN} = V(r)(\de r^2 + r^2(\de \theta^2 + \sin^2 \theta \de \phi^2)) + \frac1{V(r)} R^2( \de \psi + N(1+\cos \theta)\de \phi)^2\,,\label{eq:FB-TaubNUT}
\end{equation}
where $R$ is a constant and $N$ an integer, while $V(r)$ is a harmonic function in $\mathbb{R}^3$:
\begin{equation}
 V(r) = 1 + \frac{NR}{2 r}\,.
\end{equation}
To picture the Taub-NUT geometry, note that the actual radius of the $\psi$-circle in the metric \eqref{eq:FB-TaubNUT} is $\tilde R(r) = V(r)^{-1/2}R$, which approaches $R$ as $r\to \infty$ and $0$ as $r\to 0$. 
This geometry can be pictured as a cigar, see figure \ref{fig:FB-TN}. 
Far from the tip of the cigar, the spatial part of the geometry looks like $\Real^3\times S^1$ and near the tip of the cigar as $\Real^4$. By varying the value of $R$, we can interpolate between a four-dimensional description ($R \to 0$) and a five-dimensional one ($R\to \infty$). Note that when $R$ is small, we can perform a Kaluza-Klein reduction over the $\psi$-circle. The lower-dimensional theory includes a magnetic monopole field $A = -N(1+\cos\theta)\de \phi$. For this reason, this geometry is also referred to as the (five-dimensional) Kaluza-Klein monopole. We see $N$ measures units of monopole charge.

By placing a three-charge black hole in a Kaluza-Klein monopole background, one finds a four-charge solution in four dimensions where the additional charge is given by the monopole charge. This approach has been used in references \cite{Berglund:2005vb,Saxena:2005uk,Balasubramanian:2006gi}. In four dimensions, one obtains horizonless solutions, which are not necessarily smooth: we conclude that fuzzball solutions in different duality frames may become singular. As long as they are entropyless (no horizons), we still accept them as good microstates in light of the fuzzball proposal. 

In the remainder of this thesis, we always work with the four-dimensional setup in mind. In the next section, we discuss a way of finding supergravity microstates in the four-dimensional picture, which forms the basis of the research presented in chapter \ref{c:TL}. In the next chapter, we work out  a particular relation between supergravity microstates of the five-dimensional D1-D5-KK system and multi-center configurations in four dimensions. (The D1-D5-KK system consists of the D1-D5 system discussed above, placed in a Kaluza-Klein (KK) monopole background, which can be related to a four-dimensional system through the standard 4D-5D connection.)
\subsection{Summary\label{ss:FB-4D/5D}}
We have seen in previous chapters how black hole solutions in general relativity and extensions to supergravity give rise to several unexplained phenomena in the semiclassical approximation: How can we explain the Bekenstein-Hawking entropy? Does the information paradox lead to problems in defining a well-behaved quantum theory? In this chapter, we have hinted how these issues can be understood trough the fuzzball proposal. Evidence for the proposal was provided for the two-charge D1-D5 system of type IIB string theory. However, a lot of work still needs to be done for black holes with at least three different charges, in order to describe non-singular black holes with non-vanishing horizon area. 

One expects that a full set of solutions accounting for the Bekenstein-Hawking entropy of the black hole includes solutions which cannot be described by supergravity alone, but need the full string theory. However, so far fuzzball solutions for three-charge systems are only found in supergravity. Luckily, there are arguments that we can extract useful information about  microstates even if only small part of the total microstates of the full theory are known (namely, the supergravity states), if they are suitably dense in the full Hilbert space \cite{Bena:2008nh}. Nevertheless, for the supergravity solutions constructed it remains to precisely relate them to the dual CFT descriptions, as no general detailed map has been worked out.

The questions raised above motivate further study for black hole microstates in supergravity. In the following, the focus is shifted towards  four-dimensional black hole solutions. The reader can continue with the next chapter, where a map between four-dimensional multi-center solutions and five-dimensional fuzzballs is explained. The next section can be skipped for now. It gives an alternative way  of describing `microstates' of four-dimensional black holes, motivating the research of chapter \ref{c:TL}.

\index{fuzzball!proposal|)}
\section{Black hole microstates from multi-center configurations}\label{s:FB_scaling}
\begin{figure}[ht!]
\centering
\begin{picture}(300,88) 
\put(0,-4){\includegraphics[width=300pt]{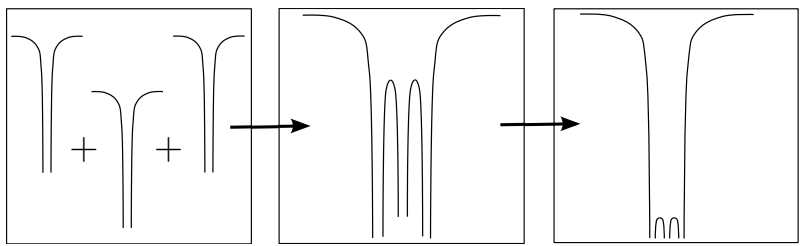}}
\put(24,28){\Large $+$}
\put(56,28){\Large $+$}
\end{picture}

\caption[The scaling limit for a solution with three centers.]{The scaling limit for a solution with three centers. 
By taking the centers to approach each other, the solution resembled more and more a single center black hole.\label{fig:Scal-MultiThroat}}
\end{figure}

In this section, we return to four-dimensional black holes. The appearance of the black hole solutions in four-dimensional supergravity with multiple centers opens up a new interesting possibility of discussing the microscopic structure of black holes with one center. For certain sets of electric and magnetic charges $\Gamma = (p^I,q_I)$, the supergravity equations of motion allow  for both a single center solution and one (or several) solutions with multiple centers, each with charges $\Gamma_i$, such that their total charge adds up to $\Gamma$.
For certain charges $\Gamma_i$, one can find solutions such that the different centers can approach each other arbitrarily closely: for an observer outside the horizon(s), the multi-center solution becomes indistinguishable from the single center black hole with total charge $\Gamma = \sum_i \Gamma_i$. The fine structure dissappears behind a single horizon, see figure \ref{fig:Scal-MultiThroat}. 
This behaviour is given the name \textit{scaling behaviour}.
%
%
%
Even though invisible for an outside observer, there is still a structure of the different centers deep down the black hole `throat', associated with choosing relative positions of the different centers and the ways one can distribute charges $\Gamma_i$ over the centers, for a given total charge $\Gamma$.

The structure of the several centers down the throat leads to the possibility of associating an entropy to these configurations, by counting in some way the number of possible distributions of charges among the several centers. This idea has been used to explain how to account for the entropy of the four-dimensional black hole with total D0-D4 charges. However, two different ways of calculating the entropy for this system give different results: the first way, discussed in \cite{Denef:2007yt,Gimon:2007mha}, can account for the total Bekenstein-Hawking entropy, but treats most of the charged centers only in a probe approximation, not taking the backreaction on the metric and other fields into account. In \cite{Gimon:2007mha}, one does include the backreaction, but can only account for a small part of the entropy. We  discuss these two results, since they formed the basis of our work \cite{Levi:2009az}, where we provided first steps in understanding this discrepancy. Similar scaling solutions have been constructed for related five-dimensional systems, see e.g.~\cite{Bena:2007qc,Bena:2006is,Cheng:2006yq} and \cite{Bena:2007kg} for a review.

\subsection{Scaling solutions with total D0-D4 charge}

\subsubsection{D0-D4 black hole revisited}
Consider type IIA string theory compactified on a Calabi-Yau. A charge vector of the four-dimensional gauge fields is written as  $\Gamma = (p^I,q_I)$. For a state with D0 charge $q_0$ and D4 charge $p^A$, we then have $ \Gamma = (0,p^A,0,q_0)$. For large charges $(q_0,p^A\gg 1)$, gravity becomes a good description and a black hole is formed. The Bekenstein-Hawking entropy of this black hole is given as
\begin{equation}
 S_{BH} = 2\pi\sqrt{\frac {q_0 p^3}6}\,,\label{Bekenstein-Hawking}
\end{equation}
where $p^3 = D_{ABC}p^A p^B p^C$ and $D_{ABC}$ are the intersection numbers of the Calabi-Yau.

\subsubsection{Deconstructing charges}
We show how charges can be distributed over different centers giving the same total charge as the D0-D4 black hole. We take one center with D6 brane charge $+1$ and \worldvolume{} flux inducing lower charges, one center with D6 charge $-1$ and \worldvolume{} flux and a number of centers carrying D0 brane charge to match up to the total charge vector $\Gamma = (0,p,^A0,q)$ of the D0-D4 system:
\begin{equation}
 \Gamma = \Gamma_{D6} + \Gamma_{\Db 6} + n \Gamma_{D0,a}\,,
\end{equation}
where $\Gamma_{D6}, \Gamma_{\Db 6}$ describe the charge vectors of a D6 brane and a D6 anti-brane with \worldvolume{} flux inducing lower brane charges
\begin{align}
\Gamma_{D6} &= (1,\frac12p^A,\frac18D_{ABC}p^B p^C,\frac1{48}D_{ABC}p^A p^B p^C)\,,\nonumber\\
\Gamma_{\Db 6}&= (-1,\frac12p^A,-\frac18D_{ABC}p^B p^C,\frac1{48}D_{ABC}p^A p^B p^C)\,.
\end{align}
and in addition, we have a number $n$ of D0 branes of charge $\Gamma_{D0,a} = (0,0,0,q_{0,a})$, with total charge $N = \sum_a  q_{0,a}$, such that the  total D0 charge of the D0-D4 system is
\begin{equation}
 q_0 = N - D_{ABC}p^A p^B p^C/24 = N - p^3/24\,,
\end{equation}
%
where we defined $p^3 = D_{ABC}p^A p^B p^C$. The given charges of the $D6-\Db 6-D0$ system have two regimes, depending on the relative value of the parameters $N$ and $p^3/12$ . In particular, when $N \geq p^3/12$, the centers can approach each other arbitrarily closely and this charge regime is referred to as the scaling regime. For $N<p^3/12$, scaling is not possible.

\paragraph{Counting the entropy using probe quantum mechanics}
In the scaling  regime with ($N \gg p^3/12$) (the D0 charge is parametrically larger then the total D4 charge), the authors of \cite{Denef:2007yt} showed that the quantization of these multi-center configurations gives a number of states $N=\e^ {S_{micro}}$ with
\begin{equation}
S_{micro} = 2\pi \sqrt{\frac{q_0 p^3}6}\,,
\end{equation}
reproducing the Bekenstein-Hawking entropy \eqref{Bekenstein-Hawking}.
However, this calculation makes two crucial assumptions: first, the D0 branes are assumed to blow up to D2 branes through a version of the Myers effect -- this corresponds to D0 branes that have non-Abelian degrees of freedom. Second, one performs a `probe' counting of quantum mechanical ground states that ignores the backreaction of the D0 (D2) branes on the geometry. 

\paragraph{Counting the entropy using geometric quantization}
In \cite{deBoer:2009un}, a geometric quantization of the solution space of the multi-center configurations above was considered. The solution space consists of all possible ways one can arrange the various D6 and D0 centers.\footnote{Not every combination of charges and positions of the centers gives a bound state and describes a well-behaved multi-black hole solution. The constraints the charges and centers have to obey for the $D6-\Db 6-D0$ system to describe a bound state, can be found in \cite{Denef:2007yt}.} Classically, this can be done in an infinite amount of ways, but by performing a proper quantization procedure, one can find a measure for the number of states and hence the entropy of the one-center black hole with the same total charge as the multi-center configuration, the D0-D4 black hole. 
The results gives an entropy of the form
\begin{equation}
 S_{micro} \sim \sqrt[3]{q_0p^3}\,.
\end{equation}
For large charges, this is very small compared to the Bekenstein-Hawking entropy. Hence, this suggests that counting supergravity microstates \textit{alone}, cannot reproduce the entropy of the D0-D4 black hole. This way of reproducing the microscopic entropy does take the backreaction of the D0 branes into account, but does not consider non-abelian degrees of freedom.

\subsection{Outlook}

We have seen that scaling solutions and more generally multi-center configurations can be used to obtain supergravity interpretations of black hole microstates. However, it is unclear that the entropy can be reproduced from these states. It seems that non-abelian degrees of freedom play an important role. In chapter \ref{c:TL}, we re-investigate the multi-center configurations above. The purpose is to have both the effect of non-abelian degrees of freedom and the full backreaction of all centers on the geometry, to see if one can find more information on the black hole entropy.

\cleardoublepage
\chapter{
5D fuzzball geometries and 4D polar states\label{c:JR}}

\punchline{The goal of this chapter is to give an explicit mapping between  supertube solutions arising in the fuzzball
picture in five dimensions and multi-centered solutions in four dimensions under the 4D-5D connection, and to
interpret the resulting configurations using the tools developed by Denef and Moore \cite{Denef:2007vg}.  In five dimensions, we consider Kaluza-Klein monopole supertubes with circular profile which represent microstates of a small black ring. The resulting four-dimensional configurations are, in a suitable duality frame, polar states consisting of stacks of D6 and anti-D6 branes with \worldvolume{} flux. We argue that these four-dimensional configurations represent zero-entropy constituents of a 2-centered configuration where one of the centers is a small black hole. We also discuss how spectral flow transformations in five dimensions, leading to configurations with momentum, give rise to four-dimensional D6 anti-D6 polar configurations with different flux distributions at the centers.}

\section{Introduction}

Recent years have seen a significant progress in the understanding of the
supergravity description of BPS states of string theory, both in four and five noncompact
dimensions. In four dimensions, it has been established that BPS states of a given
charge are often realized as multi-centered solutions in supergravity \cite{Denef:2000nb,Denef:2001xn,Bates:2003vx,Denef:2007vg,deBoer:2008fk}.
An important class of multi-centered configurations are the `polar' states for which no single-centered solution exists and  which
contribute to the polar part of the OSV partition function \cite{Ooguri:2004zv} regarded as a
generalized modular form. From the knowledge of their microscopic degeneracies, the full partition function was  reconstructed in \cite{Denef:2007vg},
leading to a derivation of an OSV-type relation.
Another important type of configurations  are the so-called `scaling' solutions, which carry the same charges as a (large)
black hole and can be seen as a deconstruction of the black hole into zero-entropy constituents \cite{Denef:2007yt} (see also chapters \ref{c:FB} and \ref{c:TL}).

On the five-dimensional side as well, the BPS objects are not restricted to single-centered
black holes. There also exist supersymmetric black rings and  black hole-black ring composites  \cite{Elvang:2004rt,Bena:2004de,Elvang:2004ds}, see  \cite{Emparan:2006mm} for review
and a more complete list of references.
There are also Kaluza-Klein monopole supertube
solutions which carry the charges of a black hole or black ring and are smooth and horizonless \cite{Maldacena:2000dr,Lunin:2001jy,Lunin:2002qf,Lunin:2002iz,Lunin:2004uu,Giusto:2004ip,Giusto:2004id,Giusto:2004kj,
Bena:2005ay,Taylor:2005db,Saxena:2005uk,Giusto:2005ag,Giusto:2006zi,Kanitscheider:2007wq}.
These can be seen as gravity duals to individual microstates in the CFT description of the black hole, leading to
the `fuzzball' picture proposed by Mathur and collaborators (see chapter \ref{c:FB} and refs.~\cite{Mathur:2005zp,Bena:2007kg,Skenderis:2008qn,Mathur:2008nj,Balasubramanian:2008da} for reviews and further references). As was discussed in the previous chapter, in this proposal, the black hole horizon is an artefact of
 an averaging procedure over an ensemble of such smooth solutions.

These zoos of four and five-dimensional BPS configurations are not unrelated, and it is often possible to
continuously interpolate between 4D and 5D configurations using the `4D-5D connection' \cite{Gaiotto:2005gf,Gaiotto:2005xt,Elvang:2005sa,Bena:2005ni,Behrndt:2005he,Ford:2007th}, see section \ref{ss:FB-4d/5d-connection}.
Five-dimensional configurations can often be embedded in Taub-NUT space in a supersymmetric manner.
The spatial geometry of Taub-NUT space interpolates between ${\mathbb{R}}^4$ near the origin and ${\mathbb{R}}^3 \times S^1$ at infinity.
By varying the size of the $S^1$, one can then interpolate between effectively five and four-dimensional configurations.
Under this map, a point-like configuration at the center of Taub-NUT space becomes a 4D pointlike solution with
added Kaluza-Klein monopole charge. A ring-like configuration at some distance from the center goes over into a two-centered
solution where one center comes from the wrapped ring and the other contains Kaluza-Klein monopole charge. Angular momentum
 in 5D goes over into linear momentum along $S^1$ in four dimensions.

The goal of the current work is to analyze the map between a class of `fuzzball' solutions in five dimensions and four-dimensional multicentered solutions under the 4D-5D connection, and to interpret the resulting configurations in the framework of \cite{Denef:2007vg}.
We work in toroidally compactified  type II string theory, and consider a symmetric class of 2-charge  supertubes which are described
by a circular profile \cite{Maldacena:2000dr,Lunin:2001jy,Lunin:2002qf,Lunin:2002iz}, as well as 3-charge solutions obtained from those under spectral flow
\cite{Lunin:2004uu,Giusto:2004ip,Giusto:2004id,Giusto:2004kj}.
Placing such  supertubes  in Taub-NUT space gives the solutions that were constructed
in \cite{Bena:2005ay,Saxena:2005uk}. Applying the 4D-5D connection, we show that,
in  the standard type IIB duality frame, one obtains  4D solutions which  are two-centered
 Kaluza-Klein monopole-antimonopole pairs  carrying flux-induced D1 and D5-brane charge and momentum.
 These solutions can be described within an STU-model truncation of $\cN=8$ supergravity
and  can be seen as simple examples of `bubbled' solutions \cite{Bena:2005va,Berglund:2005vb,Bena:2006is,Balasubramanian:2006gi,Bena:2006kb,Cheng:2006yq,Bena:2007ju,Gimon:2007mha,Bena:2007qc}
(for a review, see (\cite{Bena:2007kg}).
To make contact with the techniques developed for analyzing multi-centered  configurations in Calabi-Yau
compactifications, we
transform these configurations to a type IIA duality frame where all charges and dipole moments carried
arise from a D6-D4-D2-D0 brane system.
In this duality frame, the relevant configurations are two stacks of D6-branes and anti-D6 branes
with worldvolume fluxes turned on.
Those configurations fall into the class of `polar' states in 4D for which no single centered solution exists.

\outline{
In section \ref{s:JR-TypeII}, we give the four-dimensional supergravity theory that results from the torus reduction of string theory that we consider, namely the STU model. We discuss how it fits to describe both a reduction of type IIA supergravity and type IIB supergravity. In section \ref{s:JR-IIA}, we briefly discuss the construction of the multi-centered solutions we wish to consider, carrying total D2 and D4 charges, in the STU-model in the type IIA frame and construct multi-center solutions with total D2-D4 charge. We mention these are polar states. In section \ref{s:fuzzballs_in_FrameB}, we transform to a U-dual type IIB duality frame and give the lift of our solutions to 10 dimensions. We show that the solutions represent supertubes embedded in Taub-NUT space, and discuss the limit to five non-compact dimensions, through the 4D-5D connection, to interpret our solutions as fuzzball states. For simplicity, the calculations of the previous sections are for solutions without D0 charge in the type IIA picture (or equivalently, with no momentum (P) along the internal directions in the type IIB picure). We show how to add D0 charge/momentum in section \ref{s:JR-D0charge}. In section \ref{s:JR-micro}, we discuss the microscopic interpretation of our configurations from the 4D and 5D points of view. We end with some prospects for future research in section \ref{s:JR-concl}.
}

\section{Type II string theory on \texorpdfstring{$T^6$}{}}\label{s:JR-TypeII}
We consider type II string theory compactified on a six-torus $T^6$. For our purposes, it is sufficient to consider a consistent truncation to a sector where only gravity and 3 vector multiplets are excited. This sector is described by the well-known STU model \cite{Duff:1995sm,Behrndt:1996hu} of  $\cN=2$ supergravity coupled to 3 vector multiplets.
The bosonic part of the action is given by
\bea   S &=& \frac{1}{ 16 \p
G_4 } \int d^4 x \sqrt{-g}\Big[ R -\half \sum_{A=1}^3 \frac{\pa_\m z^A
\pa^\m \bar z^A}{({\rm Im} z^A)^2}\nonu
&& + \frac{\norm^2 }{ 4 } {\rm Im}
\cn_{IJ} F^I_{\m\n}F^{J\;\m\n}  + \frac{\norm^2 }{ 8 } {\rm Re}
\cn_{IJ}\e^{\m\n\r\s}
F^I_{\m\n}F^J_{\;\r\s}\Big]\label{STUaction} \,.\eea
with $z^A \equiv a_A + i b_A,\ A =1,2,3,\ I = 0,1,2,3$ and $\e^{0123
}\equiv 1$. We have left an arbitrary normalization constant $\norm$ in front of the
kinetic terms of the $U(1)$ fields for easy comparison with different conventions used in the
literature. 

{}Below, we consider charged BPS states from D-branes in 10 dimensions wrapping internal cycles. These give rise to four-dimensional states carrying  electric charges $Q_I$ and magnetic charges $Q^I$, with  $Q^I = \frac1{4\pi}\int_{S^2} F^I$ and $Q_I=\frac1{4\pi}\int_{S^2} G_I$\,, (where $G_I = {\rm Im}  \cn_{IJ} \star F^J + {\rm Re}  \cn_{IJ} F^J$ and $\star$ denotes the Hodge dual). For later convenience, as we will be taking the size of one of the internal directions to infinity,  it will be useful  to work in conventions  where we do not fix the coordinate volume of the internal cycles. The components of $Q$ are then given in terms of integers $q_I,p^I$ counting the numbers of wrapped branes as\footnote{To find agreement with
chapter \ref{c:BJ}, one should take  the coordinate volume of all cycles equal to one in units of $2\p \sqrt{\a'}$.
In that case, the relation between $Q$ and $\G = (p,q)$ is $Q = \frac{ \sqrt{G_4} }{ \norm }\G$.}
  \be Q^I = \frac{ 2  }{\norm}
T^I V^I G_4 p^I\,, \qquad Q_I = {2 \over \norm } T_I V_I G_4 q_I  \label{norm}\,.\ee
where $T^I, T_I$ are the tensions  of the  branes in table
\ref{chargesframeA}
and the $V^I, V_I$ are the coordinate volumes of the cycles they are wrapping.

\subsection{Frame A}
We show how the STU model above follows from toroidally compactified IIA string theory. We choose the six-torus to be metrically a direct product of three 2-tori and choose one  2-torus  to be rectangular, denoting its two circles by $S^4 , S^5$:
\begin{equation}
 T^6= T_1 \times T_2 \times T_3\,,\qquad T_1 = S^4\times S^5\,,\label{eq:JR-T6}
\end{equation}
The 10-dimensional origin of the fields in (\ref{STUaction}) then is the following. 
%
%
The vector multiplet scalars $z^A$ describe complexified K\"ahler deformations of the tori  $T_A$:
\be B + i J = z^A \o_A\,, \ee
where $B$ is the Kalb-Ramond two form, $J$ the K\"ahler form of $T^6$ and $\o_A$  are normalized volume forms on $T_A$ satisfying
$\int_{T_A} \o_B = \d^A_B$. The gauge fields arise from reduction of the RR-sector of type IIA supergravity. In particular, the charges $q_I,p^I$ correspond to D-branes wrapped on the various non-trivial cycles in the internal $T^6$ geometry. In particular, we
write a general charge vector $\G$ either by a row vector or an element of the even  cohomology of $T^6$: \be \G = (p^0,p^A,q_A,q_0) = p^0 + p^A \o_A + q_A \o^A + q_0 \o_{\rm vol}\,,\label{eq:JR-CHargeVector}\ee with $\o^A = 3 D_{ABC} \o_B \wedge \o_C$ and $ \o_{\rm vol} = \o_1 \wedge  \o_2 \wedge  \o_3$ and $A= 1, \ldots, 3$ (The constants $D_{ABC}$  are proportional to the intersection numbers 
 $D_{ABC} = {1 \over 6 } \int \o_A \wedge \o_B \wedge \o_C$). Taking into account charge quantization, the components $p^I, q_I$ should be integers or $\G \in H^{\rm even}(T^6, {\mathbb{Z}})$. For the orthogonal choice \eqref{eq:JR-T6}, the D-brane interpretation of the charges is given in table \ref{chargesframeA}. 
\begin{table}[ht!]
\centering
\begin{tabular}{cl|cl}
$q_0:$& $D0      $      & $p^0:$& $D6(T_1 \times T_2 \times T_3)$\\
$q_1:$& $D2(T_1) $      & $p^1:$& $D4(T_2 \times T_3)$\\
$q_2:$& $D2(T_2) $      & $p^2:$& $D4(T_1 \times T_3)$\\
$q_3:$& $D2(T_3) $      & $p^3:$& $D4(T_1 \times T_2)$\\
\end{tabular}
\caption[D-brane charges in type IIA.]{Type IIA D-brane charges carried by  our configurations. We have denoted  the submanifold wrapped by the
brane in brackets.}\label{chargesframeA}\end{table}

Below, we consider the construction of general \textit{multi-centered BPS solutions} in the STU model above, along the lines of Bates and Denef \cite{Bates:2003vx}. 

%

\subsection{Frame B}
We make a duality transformation to a type IIB frame such that multi-center configurations of IIA theory with D0, D2 , D4 and D6 charges $q_I,p^I$ can correspond to typical fuzzball geometries in five dimensions that carry D1 and D5 charges, and possibly momentum (P) along one or more compact directions. Moreover, to interpolate between four and five-dimensional geometries, we want to place these solutions in a Kaluza-Klein monopole background.\footnote{Recall that a Kaluza-Klein monopole in 10D is a 5+1-dimensional object whose transverse 4-dimensional space has Taub-NUT geometry or,
in the case of several centers, a Gibbons-Hawking space.} We can then use the standard 4D-5D connection (see section \ref{ss:FB-4d/5d-connection}).

For these reasons, we go to a duality frame where  $p^0$ becomes a Kaluza-Klein monopole charge with Taub-NUT circle $S^4$, $p^1$ becomes a Kaluza-Klein monopole charge with Taub-NUT circle $S^5$, $p^2$ becomes the charge of a $D1$-brane wrapped on $S^4$ and $p^3$ becomes the charge of a $D5$-brane wrapped on $S^4\times T_2 \times T_3$. This is accomplished by making a  U duality transformation consisting of a T duality along $S^4$, followed by  S-duality and 4  T-dualities  along $T_1\times T_3$, as illustrated in table \ref{Udual}. 
\begin{table}[ht!]
\centering
\small
\begin{tabular}{lclclcl}
IIA (frame A)& & IIB &  & IIB & &IIB (frame B)\\
D6 ($T^6$) & & D5 & &   NS5 & &  KK5 ($S^5 \times T_2 \times T_3$)\\
D4 ($T_2 \times T_3$) & T ($S^4$) & D5& S &  NS5&T ($S^4,S^5,T_3$)&  KK5 ($S^4 \times T_2\times T_3$)\\
D4 ($T_1 \times T_3 $) & $\longrightarrow$ &  D3& $\longrightarrow$ & D3&$\longrightarrow$&  D1 ($S^4$)\\
D4 ($T_1 \times T_2$) & &  D3 & &  D3& & D5 ($S^4\times T_2 \times T_3$)
\end{tabular}
\caption{U-duality transformation from frame A to frame B.}\label{Udual}\end{table}
This new duality frame is denoted `frame B'. In this frame, the vector multiplet scalars $z^1, z^2, z^3$ represent the complex structure modulus of $T_1$, the 4D axion-dilaton and the (complexified) K\"{a}hler modulus of $T_1$ respectively. The $U(1)$ fields $\ca^0$ and $\ca^1$ are Kaluza-Klein gauge fields from the metric components $g_{\m 4}$ and $g_{\m 5}$ respectively, while $\ca^2$ and $\ca^3$ arise from the RR two form components $C_{\m 4}$ and $C_{\m 5}$. The 10-dimensional origin  of the full set of charges  in this frame is given in table \ref{frameBcharges}. Frame B is  naturally suited for discussing the relation to five-dimensional fuzzball solutions, this will become clearer in section \ref{s:fuzzballs_in_FrameB}.
\begin{table}[ht!]
\begin{center}
\begin{tabular}{cl|cl}
$q_0$& P ($S^4$)        & $p^0$& KK5 ($S^5\times T_2\times T_3$)\\
$q_1$& P ($S^5$) & $p^1$& KK5 ($S^4\times T_2\times T_3$)\\
$q_2$& D5 ($S^5\times T_2 \times T_3$)       & $p^2$& D1 ($S^4$)\\
$q_3$& D1 ($S^5$)& $p^3$& D5 ($S^4\times T_2 \times T_3$)\\
\end{tabular}\end{center}\caption{10D origin  of the charges in frame B}\label{frameBcharges}\end{table}

\section{Class of polar states in frame A}\label{s:JR-IIA}
We describe a particular set of 2-centered solutions where the centers are stacks of D6 and anti-D6 branes with worldvolume fluxes turned on.  It can be shown that for these configurations no single-centered solutions with the same total charge exist.  In the language of \cite{Denef:2007vg}, these  are `polar states'. 

We consider \textit{two classes of  polar states}: the first class carries no D0-brane charge and has four net D4-D2 charges $p^1, p^2,p^3,q_1$. Through a  spectral flow transformation (see section \ref{s:JR-D0charge}), we can obtain a  second class of solutions which carry the above four charges as well as D0-brane charge $q_0$. In section \ref{s:fuzzballs_in_FrameB} we  show that these  two classes of configurations, after a U-duality transformation to frame B, give rise to smooth `fuzzball' solutions placed in a Taub-NUT background. The  solutions without D0-charge map to fuzzball solutions with D1-charge and D5-charge  in Taub-NUT space  while  the solutions carrying D0-charge  map to fuzzball solutions with D1-D5 charge and momentum P in Taub-NUT.

For simplicity, we only consider the class without D0 charge in detail. The case with D0 charge and its translated to frame B is treated briefly in section \ref{s:JR-D0charge}.


\subsubsection{Configurations without D0-charge}\label{solswoD0}
The first class of solutions we want to consider consists of a stack of $n$ D6 branes and a stack of $n$ anti-D6 branes. Each stack of branes has $U(n) = U(1) \times SU(n)$ gauge fields living on the worldvolume. We turn on $U(1)$ worldvolume fluxes so that each stack carries  lower dimensional D-brane charges as well. The fluxes we turn on are characterized by three numbers which, for later convenience, we  label $N_K, N_1, N_5$. In the vector notation of eq.~\eqref{eq:JR-CHargeVector}, i.e.~$\Gamma= (p^0,p^1,p^2,p^3;q_1,q_2,q_3,q_0)$,
the charges at the centers are 
\bea\label{twocenters}
 \G_1 = 
\left(-n,N_K,0,0;0,0,0,0\right)\,,\qquad
 \G_2 = 
\left(n,0,N_1,N_5;{N_1 N_5 \over n}, 0,0,0\right)\,.
\eea
Charge quantization restricts $n, N_K, N_1, N_5$ to be integers
and   $n$ to be a divisor of $N_1 N_5$. These configurations carry
4 nonzero net charges $p^1, p^2,p^3,q_1$:
\be
\G_{\rm tot} = \left(0, p^1, p^2, p^3;q_1,0,0,0\right)=\left(0, N_K, N_1, N_5;{N_1 N_5\over n},0,0,0\right) \,.\label{totchargeclass1}
\ee
We showed in our work \cite{Raeymaekers:2008gk} that no single centered solution with this total charge $\G_{\rm tot}$ exists, using split attractor flow techniques. The states $\G_1$ and $\G_2$ are thus `polar states'.

One can then find the explicit expressions for the metric, scalar fields and $U(1)$ fields in terms of a set of harmonic functions that are sourced at the two centers with charges $\G_1,\G_2$ (see \cite{Bates:2003vx}). We choose coordinates on ${\mathbb{R}}^3$ such that the first
center $\G_1$ is located at the origin and $\G_2$ lies on the positive $z$-axis at $z=a$. The harmonic functions are
\be\begin{array}{lll}
 H^0 = -{Q_1 Q_5 \over Q_n Q_K} - {Q_n \over r} + {Q_n \over r_+ }\,,&\qquad  &  H_0 = -1\,,\\
H^1 = 1 + {Q_K  \over r }\,,& \qquad & H_1 = {Q_1 Q_5 \over Q_n Q_K} + {Q_1 Q_5  \over Q_n r_+ }\,, \\
H^2 = 1 + {Q_1  \over r_+ }\,,& \qquad & H_2 = 0\,, \\
H^3 = 1 + {Q_5  \over r_+ }\,,& \qquad & H_3 = 0\,. \\
\end{array}\label{harmfctnsclass1}
\ee We have defined $r_+$ to be the radial distance (in flat $\mathbb{R}^3$) to the second
center : \be r_+ \equiv \sqrt{r^2 + a^2 - 2 a r \cos \theta}\,,\ee
and using (\ref{norm}), the normalizations in the harmonic functions are given by\footnote{From now on, we choose the normalization constant $\norm$ in (\ref{STUaction}) to be $\norm=1$.}
\be
\begin{array}{lcl}
Q_n = \half \sqrt{\a'} g n & \qquad &Q_K = { (2 \p)^2 (\a')^{3 \over 2} g \over 2 V_{T_1}}N_K\\
Q_1 = { (2 \p)^2 (\a')^{3 \over 2} g \over 2 V_{T_2}}N_1 & \qquad &Q_5 = { (2 \p)^2 (\a')^{3 \over 2} g \over 2 V_{T_3}}N_5
\end{array}
\label{normframeA}
\ee
where $g$ is the 10D  string coupling constant. The metric is then given as
\bea 
ds_4^2&=&-\frac1{ \S(H) }(dt+  \omega )^2+{\S(H) } d \vec x^2 \,,\\
\S &=& \sqrt{ - 4 H_0 H^1 H^2 H^3 - ( H_0 H^0 - H_1 H^1)^2}\,,\nonumber\\
\o &=& {Q_K Q_1 Q_5  \over 2 a Q_n}
\left( {r + a \over r_+} -1 \right) ( \cos
\theta - 1 ) d\f\,.
\label{omegasol} \eea
The solution for the $U(1)$ fields and scalars can be found in \cite{Raeymaekers:2008gk}. There is a  constraint on the distance between the centers (see \cite{Bates:2003vx})
\be a = {Q_K Q_1 Q_5  \over  Q_n^2-  Q_1 Q_5
  }\,.\label{posconstr}
\ee
Finally, note that the solution carries angular momentum, due to crossed electric and magnetic fields:
\be J_z = {N_K N_1 N_5 \over 2 n} . \label{angmomclass1}\ee


\section{Class of fuzzballs in frame B}\label{s:fuzzballs_in_FrameB}
In this section, we  make contact between the polar solutions
constructed above and  various horizonless supertube solutions in five noncompact dimensions that are central to the fuzzball proposal advocated by Mathur and collaborators.

We briefly review these configurations. Fuzzball solutions in five noncompact dimensions can be seen as Kaluza-Klein (KK) monopole supertubes where the KK monopole charge is sourced along a contractible curve in 4  noncompact  directions.
One of the compact directions, which becomes $S^4$ in our conventions  (recall that we denoted $T_1 = S^4 \times S^5$), is
a Taub-NUT circle that pinches off at every point of the curve. By adding flux to the KK-monopole, one can source the charge of D1 and D5-branes wrapped around the $S^4$ circle. For a circular curve, one can place this configuration in a Taub-NUT space with a different Taub-NUT circle, $S^5$ in our conventions, and interpolate between five and four  dimensions by varying the size of $S^5$. We show that the four-dimensional configurations obtained in this manner are U-dual to the D6-anti D6 polar solutions we discussed above.

In frame B, our class of  polar solutions with charges (\ref{twocenters}) (no D0 charge) corresponds to  two stacks of $n$ KK monopoles and anti-KK monopoles with Taub-NUT circle $S^4$ carrying flux-induced charges of D1, D5, momentum and KK monopoles wrapped on the $S^4$ circle. 
The more general solutions discussed below (with D0 charge), see (\ref{chargesclass2}), carry momentum along $S^4$ as well.  

Such solutions are smooth, and, as we show now, have the interpretation of KK monopole supertubes embedded in Taub-NUT space.

\subsection{Lift of polar states without D0 charge}
In frame B  these states correspond,  according to table \ref{frameBcharges},  to  two stacks of $n$ KK monopoles and anti-KK monopoles with Taub-NUT circle $S^4$ which carry flux induced  D1, D5 and KK monopole charges    wrapped on the $S^4$ circle. We show that from a 10D point of view, these charges precisely correspond to the Kaluza-Klein monopole supertubes in Taub-NUT space that were constructed by Bena and Kraus in \cite{Bena:2005ay}. 

\begin{figure}[ht!]
\setlength{\unitlength}{.8cm}
\centering
\begin{picture}(15,5.8)
\put(0,0) {\includegraphics[width=0.45\textwidth,angle=0]{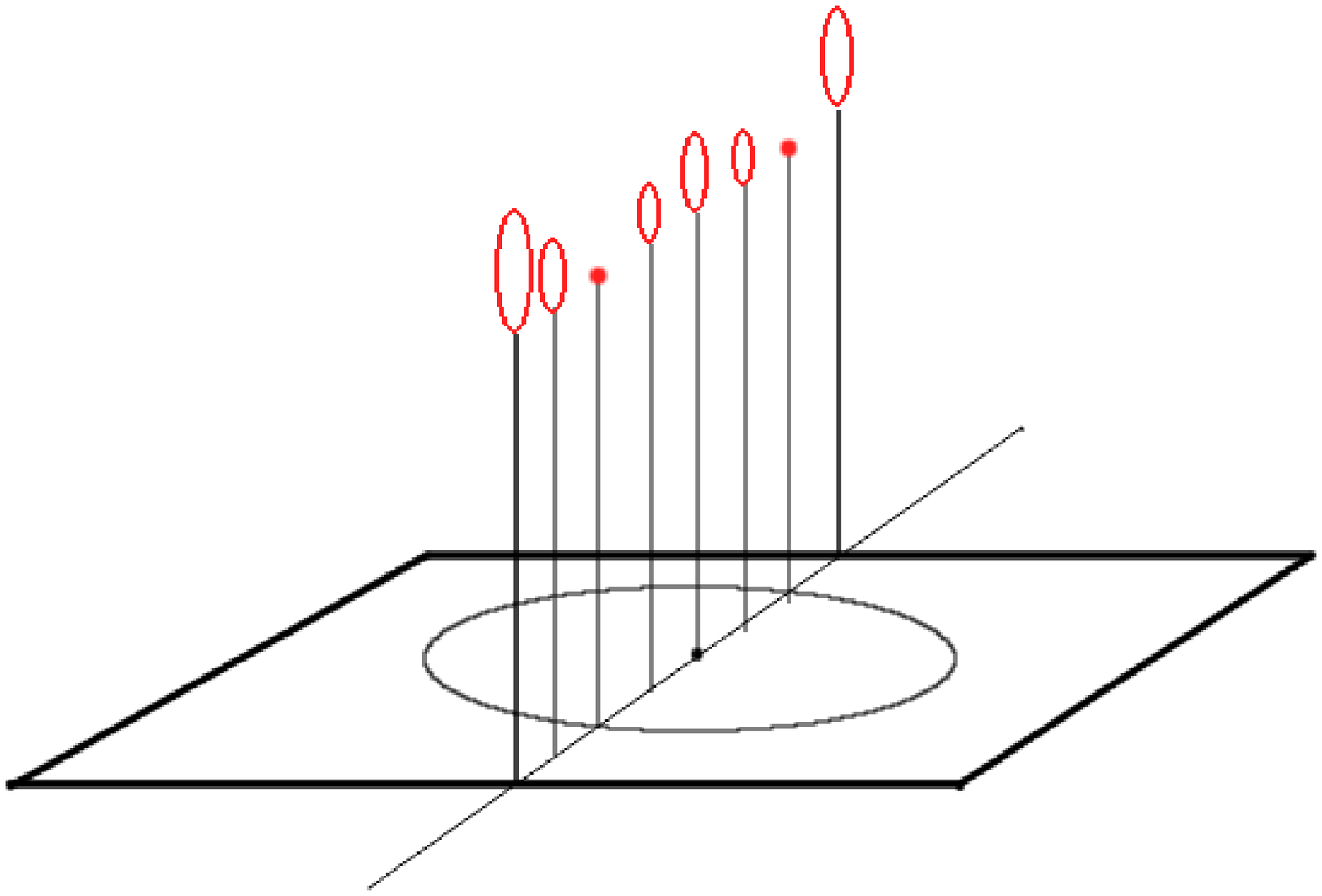}}
\put(8,0) {\includegraphics[width=0.45\textwidth,angle=0]{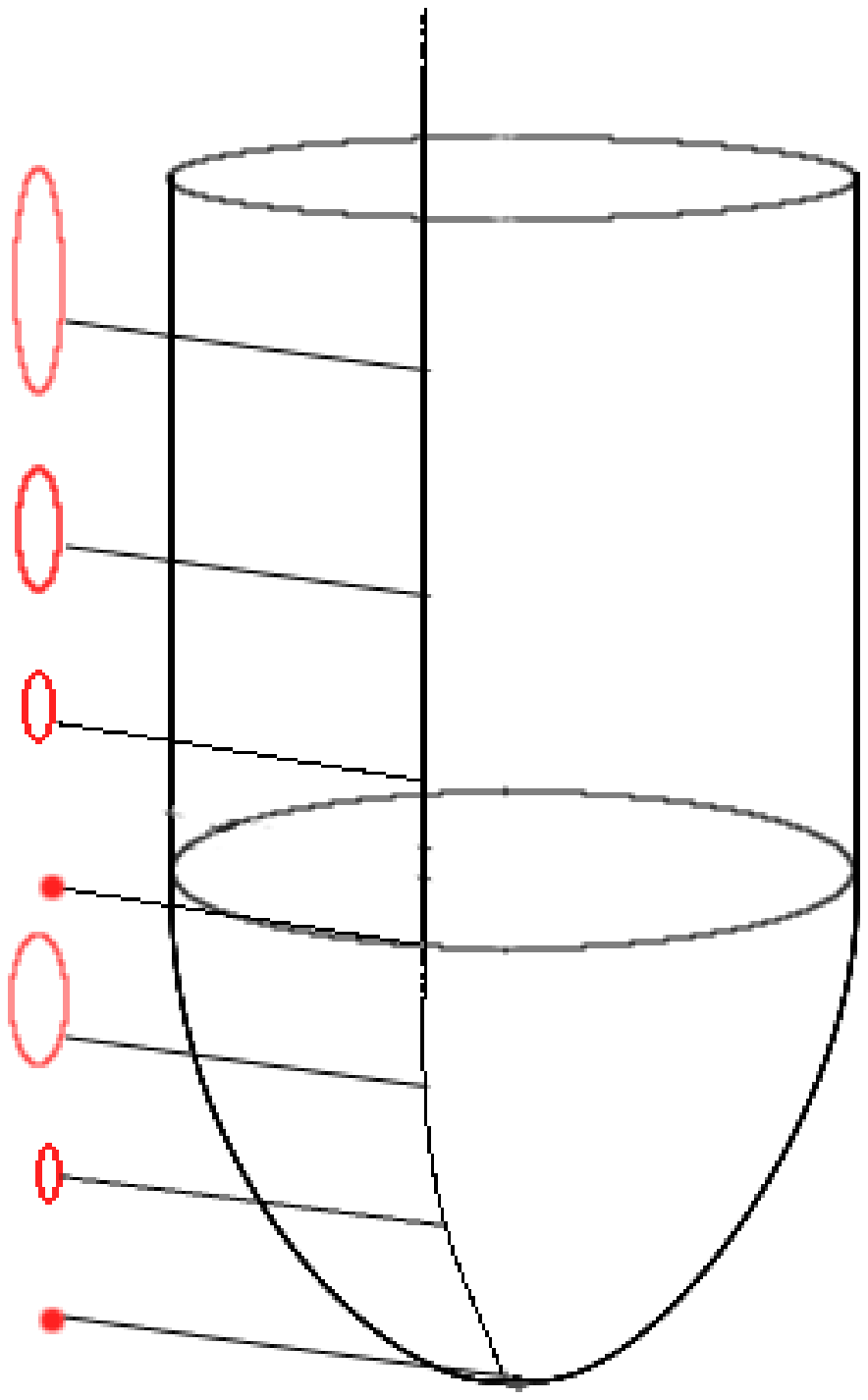}}
\put(4.5,4.5) {$S^4$}
\put(9.3,4.5) {$S^4$}
\put(12,4.68) {$S^5$}
\end{picture}
\caption[KK monopole with circular profile, in $\mathbb{R}^4$ and $\mathbb{R}^3\times S$ asymptotics.]{\label{pinchpict} Left: The black circle represents a KK monopole supertube   with  a circular profile of radius $a$ in 5 dimensions. At every point of the
curve, the internal circle $S^4$ (drawn in red) pinches off to zero size. Right: After placing another KK monopole
wrapped on $S^4$
in the origin, the asymptotic geometry becomes ${\mathbb{R}}^4 \times S^5$. As argued in the text, the $S^4$ circle pinches off along the curve as well as in the origin. }
\end{figure}

The harmonic functions of the solution are again given by (\ref{harmfctnsclass1}). However,the normalizations in the current duality frame are changed, according to (\ref{norm}),
\be
\begin{array}{lcl}
Q_n ={n R_4\over 2}\,,& \qquad&Q_K = {N_K R_5\over 2}\,,\\
Q_1 =  {(2 \p)^4 g \a'^3 \over 2 R_5 V_{T_2\times T_3} }N_,\,.& \qquad &Q_5 = {g \a' \over 2 R_5} N_5\,.
\end{array}
\label{normframeB}
\ee
Defining $\tilde H^1 = 1 + {Q_K \over a}$, the constraint on the distance between the centers (\ref{distconstr1}) is
\be Q_n = \sqrt{Q_1 Q_5 \tilde H^1 }\,, \label{distconstr1}\ee
We give the ten-dimensional metric in type IIB. Choose coordinates $x^4$ along  $S^4$ and $x^5 = R_5 \psi$ along $S^5$. Making a coordinate transformation $x^4 \to x^4  + t$, the ten-dimensional metric in frame B becomes 
\bea
ds^2 &=& { 1 \over \sqrt { H^2 H^3} } \left[ - (dt + k)^2  + ( dx^4 - s -  k)^2\right]
+ \sqrt { H^2 H^3}  ds^2_{\rm TN} + \sqrt{ H^2 \over H^3} ds^2_{T_2 \times T_3}\,,\nonu
ds^2_{\rm TN} &=& {1 \over H^1} (R_5 d \psi  + Q_K  \cos \theta  d \f)^2 +  H^1 d {\vec x}^2\,,
\label{6Dformclass1}
\eea
where $d{\vec x}^2 = (d x^1)^2+(d x^2)^2 + (d x^3)^2$. This metric describes four non-compact directions with coordinates $(t,x^1,x^2,x^3)$.
The one-forms $k$ and $s$ have components along
$\f$ and $\psi$ and, using the distance constraint (\ref{distconstr1}), can be written as
{\small
\be
\begin{array}{ll}
k_\psi = {R_5 Q_n  Q_K \over 2 a r r_+ \tilde H^1 H^1} \left[ r_+ - r - a - {2 a r \over Q_K}
\right]\,, ~~ &
k_\f = {Q_n  Q_K \over 2 a  r_+ \tilde H^1 } \left[ r_+ - r - a + { r - a - r_+ \over H^1}
\cos \theta\right]\,, \\
s_\psi = { R_5 Q_n \over r  r_+  H^1 } \left[ r - r_+ - { r r _+ \over Q_K \tilde H^1}
\right]\,, &
s_\f = {Q_n   \over   r_+  }\left[ a + { r_+ - r - {r_+ \over \tilde H^1} \over H^1}\cos \theta
\right]\, . \\[-28pt]
\end{array}
\label{metricclassI}
\ee
}
{}\\
As we have argued, the above solutions represent the lift of a two-centered
KK-monopole anti-monopole system in frame B (or a D6 anti-D6 system in frame A),
where the Taub-NUT circle for these KK monopoles is the $S^4$. The KK monopoles sit
at radius $r = r_+$, while the anti-monopoles sit at the origin $r=0$. At the position of these centers,
the $S^4$ circle pinches off (has zero size). 
This is illustrated in figure \ref{pinchpict}.

These are precisely the solutions constructed by Bena and Kraus
\cite{Bena:2005ay}\footnote{To make contact with the conventions
in \cite{Bena:2005ay}, one has to make a further coordinate transformation
$\f \to -\f, \theta \to \p - \theta$.}. They represent Kaluza-Klein
monopole supertubes which have been embedded into a Taub-NUT space
which has the asymptotic spatial geometry  ${\mathbb{R}}^3 \times S^5$.
By varying the radius $R_5$ of the circle $S^5$ we can interpolate
between solutions in 4 and in 5 noncompact dimensions; this
procedure goes under the name of the `4D-5D connection'
\cite{Gaiotto:2005gf,Gaiotto:2005xt}.

\subsection{4D-5D connection and 5D  fuzzball geometries}
We use the 4D-5D connection (see section \ref{ss:FB-4d/5d-connection}) to show that the ten-dimensional solutions in frame B of the previous section above with four non-compact dimensions to obtain  solutions with five non-compact dimensionsand show these are highly symmetric fuzzball solutions where the curve that defines the supertube is circular. 

We take the  $R_5 \to \infty$ limit keeping the following quantities fixed:
\be
 2 r R_5 \equiv \tilde r^2 \,, \qquad 2 a R_5 \equiv \tilde a^2/n^2\,.\label{4D5Dlimit}
\ee
After taking this limit, the $p^1$ charge $N_K$ of our configuration becomes a deficit angle
and one obtains a configuration embedded in an   orbifold space ${\mathbb{R}}^4/{\mathbb{Z}}_{N_K}$.
\textit{We therefore specialize to the case $N_K=1$  from now on}, so that we obtain solutions
in asymptotically flat space.
We  define charges $\tilde Q_1 , \tilde Q_5 $ which remain finite in the limit (\ref{4D5Dlimit}) and
are the correctly normalized D1 and D5-brane charges in 5 noncompact dimensions:
\bea
\tilde Q_1 = 2 R_5 Q_1 = { g (2 \p)^4 \a'^3 N_1 \over V_{T_2 \times T_3}}\,,\qquad
\tilde Q_5 = 2 R_5 Q_1 = g \a' N_5\,.
\eea
The solution (\ref{metricclassI}) can, in this limit, be written as a  fuzzball solution with a circular profile function \cite{Maldacena:2000dr,Lunin:2001jy,Lunin:2002qf,Lunin:2002iz}:
\bea
ds^2 &=& { 1 \over \sqrt { H^2 H^3} } \left[ - (dt + k)^2  + ( dx^4 - s -  k)^2\right]
+ \sqrt { H^2 H^3} d{\vec x}^2 + \sqrt{ H^2 \over H^3} ds^2_{T_2 \times T_3}\,,\nonumber
\eea
where $d{\vec x}^2 = (d x^1)^2+(d x^2)^2 + (d x^3)^2$ and the harmonic functions and one-foms $k,s$ take the form
\bea
 H^2 &=& 1 + {\tilde Q_5 \over L} \int_0^L {d v \over |{\vec x} - {\vec F}|^2}\,, \qquad
 H^3 = 1 + {\tilde Q_5 \over L} \int_0^L {|\dot {\vec F}|^2 d v \over |{\vec x} - {\vec F}|^2}\,,\\
s &=& {\tilde Q_5 \over L} \int_0^L {d v F^a \over |{\vec x} - {\vec F}|^2 } d x^a\,,\qquad
d(s + k) = - \star_4 d s\,. \label{5Dsol}
\eea
The profile function ${\vec F}(v)$ describes a circular profile in the $x^1-x^2$ plane:
\be
\begin{array}{lcl}
F^1 = {\tilde a \over n}\, \cos {2 \p n \over L} v ,& \qquad &F^3  = 0\,,\\
F^2 = {\tilde a \over n}\, \sin {2 \p n \over L} v, & \qquad &F^4
=0 \,.
\end{array}\label{circprofile}
\ee
 where
$ L \equiv {2 \p \tilde Q_5 \over R_4} $. The averaged length of
the tangent vector to the profile should be proportional to the
D1-brane charge:\footnote{As a consistency check, one sees this is
the case using the constraint (\ref{distconstr1})  on the distance between the centers, which reads $R_4 = {\sqrt{ \tilde Q_1 \tilde Q_5}\over \tilde a}$ in the limit \ref{4D5Dlimit}.} \be Q_1 = {Q_5 \over L} \int_0^L |\dot {\vec F}|^2
dv\,. \ee
%

\section{Solutions with D0 charge/momentum}\label{s:JR-D0charge}
We are also interested in solutions with D0 charge in frame A, which correspond to fuzzballs with momentum (P) along the internal directions in frame B. We obtain these solutions by a spectral flow transformation on the charge vector.  These transformations consist of adding a harmonic form $S$ to the Born-Infeld flux on the worldvolumes of the two stacks of D6 branes in frame A, while in frame B they correspond to a spectral flow transformation in the dual CFT. We choose this specific way of discussing the addition of D0 charge, since it leaves the discussion about polarity unchanged: whether a given charge is polar or not, is left unaffected by spectral flow, see e.g.~\cite{Denef:2007vg}.

\subsection{Spectral flow and solutions with D0 charge}

The second class of solutions we are interested in is obtained from the ones
considered above by a spectral flow transformation $ \G \to e^S \G .$\footnote{We consider the addition of this charge by the effect of large gauge transformations of the $B$-field,  under which  the $B$-field shifts with a harmonic form. Gauge invariance requires this is accompanied by a
shift in the \worldvolume{} flux, resulting in a transformation of the charge vector $B \to B + S,\, \G \to e^S \G$.
Large gauge  transformations change the boundary conditions at infinity and, in the dual CFT have the effect of inducing a spectral flow \cite{deBoer:2006vg,deBoer:2008fk,Bena:2008wt}.}
We can choose $S$ such that the new configuration carries nonzero $p^1, p^2,p^3,q_1$ charges as well as D0-charge $q_0$, while keeping $q_2$ and $q_3$ zero. There is a one-parameter family of spectral flows $S$ which does the job and which we label by a parameter $m$:
\be S = -m N_K  \o_1 +m N_1  \o_2 + m N_5  \o_3\,.\ee 
When taking charge quantization into account, the parameter $m$ could be fractional but such that  $m$ is a common multiple of $1/N_1, 1/N_5 $ and $1/N_K$. The charges carried by the two centers are then: \bea \G_1 &=&- n e^{-\left( m + {1\over n} \right)N_K \o_1 + m N_1 \o_2 + m N_5  \o_3} \,,\nonu \G_2 &=&  n e^{-m N_K \o_1 + \left(  m + {1 \over n}\right)N_1 \o_2 +  (m+ {1\over n}) N_5 \o_3}\,,\label{chargesclass2}
\eea
and the total charge of the solution is
\be
\G_{\rm tot} = \left( 0,N_K,N_1,N_5,\left(2m + {1\over n}\right)N_1 N_5,0,0,-
m\left(m+{1\over n}\right)N_K N_1 N_5\right)\,. \label{totchargeclass2}\ee
The angular momentum  of these configurations is independent of the parameter $m$ and still given by (\ref{angmomclass1}).
For $m=0$ we recover the configurations of section \ref{s:JR-IIA}.

\subsection{Spectral flow and fuzzball solutions with momentum}
In the dual frame B, the solutions of frame A with added D0 charge through a spectral flow transformation labeled by a parameter $m$, carry nonzero momentum charge P on the $S^4$ circle, see the identification of charges of table \ref{frameBcharges}.   When we take the special case $Q_1 = Q_5$, we find  precisely the solutions constructed in \cite{Saxena:2005uk} representing fuzzball geometries with momentum placed in a Taub-NUT space.

We can again take the  5D limit $R_5 \to \infty$ as discussed above. Taking again $N_K=1$ to get solutions in flat space,  one obtains the five-dimensional fuzzball solutions with momentum that were constructed in \cite{Lunin:2004uu,Giusto:2004ip,Giusto:2004id,Giusto:2004kj}. These  solutions were originally obtained by applying a spectral flow transformation in CFT to the five-dimensional solutions without momentum (\ref{5Dsol}). They carry the 5D charges
\be \begin{array}{lcl}
J^3 = - {N_1 N_5 \over 2} \left( 2 m + {1\over n} \right)\,, & \qquad & \bar J^3 = -{N_1 N_5 \over 2 n}\,,\\
P = N_1 N_5 m \left(m+ {1 \over n}\right),& &
\end{array}
\ee
where $P$ denotes the momentum on the $S^4$ circle. Flux quantization imposes that the parameter $m$ should be an integer.

\section{Microscopic interpretation}\label{s:JR-micro}
We  now discuss the microscopic interpretation of the solutions we considered both from the 4D and 5D point of view.

\paragraph{Solutions \textit{without} D0 charge (frame A)/ no momentum (frame B).}  Let us start with the configurations (\ref{twocenters}) without D0-charge in frame A. We showed that these arise, through the 4D-5D connection, from 5D fuzzball solutions with  circular profile which  carry macroscopic angular momentum   $J_{12} =  N_1 N_5 / n $ and are placed in a Taub-NUT geometry.

The first thing to note is that we should see our 4D solutions as coming from \textit{small black ring microstates in five dimensions}.\footnote{A priori, the 4D solutions could be zero-entropy constituents of a spinning  black hole \textit{or} of a  black ring  in five dimensions. The 4D configuration corresponding to a spinning black hole in 5D, would be a small black
hole with charges $(0,N_K, N_1, N_5,  N_1 N_5/n,0,0,0)$. However, this is a polar charge for which there cannot
exist a single center black hole solution. This
interpretation also corresponds
to the one argued   in \cite{Lunin:2002qf,Iizuka:2005uv,Dabholkar:2005qs,Balasubramanian:2005qu}.}
We can then ask which \textit{states in the dual CFT} correspond to the configurations (\ref{twocenters})  from the 5D point of view.
The D1-D5 CFT is a deformation of a symmetric product  CFT with target space $({T_2 \times T_3})^{N_1 N_5} / S_{N_1 N_5}$ (see \cite{David:2002wn} for a review). The CFT  states corresponding to (\ref{twocenters}) are ground states in the R sector that carry the quantum numbers 
\be \begin{array}{lcl}
L_0 = {N_1 N_5\over 4}\,, & \qquad & \bar L_0 = {N_1 N_5\over 4}\,, \\
J^3 = - {N_1 N_5 \over 2 n} \,, & \qquad & \bar J^3 = -{N_1 N_5 \over 2 n}\,,\\
P = L_0 - \bar L_0 = 0\,. & &
\end{array}
\ee
The above states belong to a `microcanonical' ensemble of R ground states at fixed
D1-charge $N_1$, D5-charge $N_5$, and angular momenta\footnote{A different ensemble, where the angular momenta are not fixed, was advocated in the light
of the OSV conjecture in \cite{Dabholkar:2005qs}} $J_{12} = N_1 N_5/n,\ J_{34} = 0$.
When $n\gg 1$, $J_{12}$ is sufficiently far from the maximal value $N_1 N_5$, and there is  an exponential degeneracy of states carrying these quantum numbers, leading to a microscopic entropy \cite{Balasubramanian:2005qu} 
\be S_{\rm micro} = 2\sqrt{2}\p \sqrt{N_1 N_5 - J_{12} } = 2\sqrt{2}\p \sqrt{N_1 N_5(1 - {1 \over n})}\,.\label{Smicro} \ee 
It is expected on the basis of general arguments \cite{Sen:1995in} that, after including higher derivative corrections to the effective action,  there exists a black ring solution with a matching macroscopic entropy. It is an open problem to explicitly compute such corrections in toroidal compactifications, unlike the case where  the four-torus $T_2 \times T_3$ is  replaced with  $K_3$ \cite{Iizuka:2005uv,Dabholkar:2005qs,Dabholkar:2006za}.

Finally, we comment on \textit{what system these are microstate of}. Consider therefore the case $N_K=1$, corresponding to one unit of Taub-NUT charge. When a small black ring is placed in Taub-NUT space with $N_K=1$ and the radius of the Taub-NUT circle is decreased, one obtains a 4D configuration consisting of two centers. One center, coming from the wrapped ring itself,  becomes a small black hole in 4D, while the other center,  coming from the Taub-NUT charge, is a KK monopole carrying zero entropy \cite{Iizuka:2005uv,Dabholkar:2005qs}. In our duality frame A, the first center is a small $D4-D2$ black hole with charge $(0,0, N_1, N_5, N_1 N_5/n,0,0,0)$ and entropy given by (\ref{Smicro}) and the second center is a pure D4-brane with charge $(0,1,0,0,0,0,0,0)$. The total charge of this system is exactly the total charge of the D6-anti D6 configurations (\ref{twocenters}) (for $N_K=1$) and therefore we can see our 4D polar D6-anti D6 configurations as zero-entropy constituents of this two-centered configuration.

\paragraph{Solutions \textit{with} D0 charge (frame A)/ momentum (frame B).} 
A similar discussion shows their CFT counterparts have quantum numbers\footnote{They are related to the CFT states above by an additional left-moving spectral flow with parameter $2m$.}
\be \begin{array}{lcl}
L_0 = N_1 N_5 \left(m^2 + {m\over n} + 1/4\right)\,, & \qquad & \bar L_0 = {N_1 N_5\over 4}\,, \\
J^3 = - {N_1 N_5 \over 2} \left( 2 m + {1\over n} \right)\,, & \qquad & \bar J^3 = -{N_1 N_5 \over 2 n}\,,\\
P = L_0 - \bar L_0 = N_1 N_5 m \left(m+ {1 \over n}\right)\,. & &
\end{array}
\ee
In the CFT, the parameters $n$ and $m$ should be quantized such that $n$ is a divisor of $N_1 N_5$ and $m$
is an integer. This matches with the  conditions we found from charge quantization in the
corresponding D-brane configurations.
These states are part of an ensemble of CFT states with fixed D1-D5 charges, angular momenta $J^3, \bar J^3$ and momentum $P$. The degeneracy is  again given by (\ref{Smicro}).

\section{Conclusions and outlook}\label{s:JR-concl}

In this chapter we identified four-dimensional multi-center D-brane configurations that correspond
to a class of fuzzball solutions in five noncompact dimensions under the 4D-5D connection. In a
type IIA duality frame where all the charges come from D6-D4-D2-D0 branes, the relevant 4D configurations are
two-centered D6-anti D6 solutions with fluxes  corresponding to polar states .

The fuzzball solutions considered here are highly symmetric, where the
 profile function that defines the solution is taken to be a circular
curve. Let us
first comment on the fate of more general fuzzball solutions under the
4D-5D connection. A fuzzball solution arising from a generic curve
 typically does not have enough symmetry to be written as a torus fibration
over a four-dimensional base and can hence not be
given a four-dimensional interpretation. However, according to the proposed
dictionary between microstates and fuzzball solutions  in \cite{Skenderis:2006ah,Kanitscheider:2006zf},
the subclass of fuzzball solutions that semiclassically represent eigenstates of the R-symmetry group
should possess $U(1)\times U(1)$ symmetry and are represented by (possibly disconnected)
circular curves in the $x^1-x^2$ and $x^3-x^4$ planes in the coordinates . Such solutions
have isometries $\pa / \pa \f$ and $\pa / \pa  \psi $ as well as
along the Taub-Nut direction $\pa / \pa x^4$, and should therefore be the lift of axially symmetric solutions
in four dimensions. When the quantum numbers are chosen appropriately, these would describe
other constituents of the 4-dimensional 2-centered system with entropy (\ref{Smicro}). It would be interesting to explore this ensemble of four-dimensional
configurations.

We would also like to comment on the relation between  the present work and black hole deconstruction \cite{Denef:2007yt}, see section \ref{s:FB_scaling}. In four dimensions, say in our frame A, there exist multi-centered `scaling' solutions with  centers  so close  that their throats have `melted' together and which are asymptotically  indistinguishable from single centered solutions. Such solutions can carry the same charges as a large single-centered D4-D0 black hole, and can be seen as a  deconstruction of such a black hole into zero-entropy constitutents. The scaling solutions
consist of a `core' D6 anti-D6 system with flux, and a `halo' of D0-brane centers added to it (again, see \cite{Denef:2007vg} for more details on the formalism of `cores' and `halos').
The scaling limit consists of taking the total D0-charge to be parametrically larger than the magnetic charge
$p^1 p^2 p^3$. The entropy of the black hole  in this limit can be understood by treating the D0-branes as probes
and counting the supersymmetric ground states of the probe quantum mechanics \cite{Gaiotto:2004ij}.
The `core' D6 anti-D6 system in these configurations is precisely of the kind  that we studied
in this chapter and mapped to 5D fuzzball solutions. Indeed, for the special values $n=1,\  m = -1/2 $ of our parameters we obtain the following charges at the centers
\bea \G_1&=& \left(-1,\frac{N_K}{2},\frac{N_1}{2},\frac{N_5}{2},-\frac{N_1N_5}{4},-\frac{N_KN_5}{4},-\frac{N_KN_1}{4},
\frac{N_KN_1N_5}{8}\right),\nonumber\\
\G_2&=&\left(1,\frac{N_K}{2},\frac{N_1}{2},\frac{N_5}{2},\frac{N_1N_5}{4},\frac{N_KN_5}{4},\frac{N_KN_1}{4},
\frac{N_KN_1N_5}{8}\right)\,.
\eea
These are precisely  the charges that appear in the core of the scaling solutions in \cite{Denef:2007yt}. It seems
natural to expect that, for the other values of our parameters $m$ and $n$, our configurations can serve as
the core system for the deconstruction of a black hole with added D2-charge.

The relation to deconstruction could have interesting implications in five dimensions as well. If we take
a scaling solution in four dimensions, dualize it to  frame B and take the 4D-5D limit, we should end
up with a  configuration carrying the charges of a large D1-D5-P Strominger-Vafa \cite{Strominger:1996sh} black hole.
The scaling limit implies that we  have $P \gg N_1 N_5$, which is equivalent to the Cardy
limit $\L_0 \ll c$  where the CFT microstate counting is performed. Therefore such configurations
would be candidates for describing typical microstates of the D1-D5-P black hole, and
it would be interesting
to study such solutions in more detail.
\cleardoublepage
\chapter{
G\"odel space from wrapped M2 branes}\label{c:TL}\label{C:TL}

\punchline{This chapter can be read by different types of readers, depending on their interests: (1) interest towards black hole entropy in supergravity, (2) to have some intuition about the G\"odel \spacetime{}, a \spacetime{} with closed timelike curves, (3) to see a supersymmetric embedding of three-dimensional G\"odel \spacetime{} in string theory. The outline at the end of the introduction is a helpful guideline. In particular, we show that M-theory admits a supersymmetric compactification to the three-dimensional G\"odel universe of the form G\"odel$_3\times\sS 2\times\CY 3$. We interpret this geometry as coming from the backreaction of M2-branes wrapping the $\sS 2$ in an $\AdS 3 \times \sS2 \times\CY 3$ flux compactification. In the black hole deconstruction proposal given in chapter \ref{c:FB},
similar states give rise to the entropy of a D4-D0 black hole. By dualities, the system can be obtained as a compactification of either type IIA/IIB string theory and F-theory. This chapter summarizes the work with Thomas Levi, Joris Raeymaekers, Dieter Van den Bleeken and Walter Van Herck in \cite{Levi:2009az}.}

\section{Introduction}
BPS states have played a major role in the successes of string
theory, from the understanding of black hole microstates to
nonperturbative checks of dualities. An interesting set of BPS states is that of supersymmetric
D-branes in an $\AdS q \times \sS p$ background  (see e.g
\cite{Witten:1998xy,Bachas:2000ik,Simons:2004nm,Raeymaekers:2006np,Raeymaekers:2007vc}
and references therein). Such states are of interest for the
AdS/CFT correspondence in general. Furthermore, in the special
case where the background geometry corresponds to the near horizon of an
extremal black hole, string or ring, there are strong indications
that such BPS states, formed by wrapping branes around the S$^p$
part of the geometry correspond to black hole (string or ring)
microstates \cite{Gaiotto:2004pc,Gaiotto:2004ij,Denef:2007yt,Gimon:2007mha}.
The study of such
sphere-wrapping branes has so far been performed purely in the
probe approximation
\cite{Simons:2004nm,Gaiotto:2004pc,Gaiotto:2004ij,Das:2005za}, see chapter \ref{c:FB} for a specific example.
There are however some indications that these branes strongly
backreact on the background geometry, and that some of their
properties can only be fully understood once these effects are
properly taken into account.

In this chapter we take a first step at studying the fully
backreacted geometries corresponding to such wrapped branes. We
 specialize to the M-theory flux compactification
$\AdS 3\times\sS 2\times \CY 3$ and construct supergravity
solutions corresponding to M2-branes wrapped around the $\sS 2$.
Note however, that by taking the $\CY 3$ to be T$^6$ or
K3$\times$T$^2$ and applying U-dualities these solutions can be
mapped to similar configurations in type IIA/IIB string theory or
F-theory.

We start our search for these solutions by noting that all the
dynamics can be captured by a reduction to three
dimensions and performing a consistent truncation to the fields of
interest. As we will discuss in detail, the problem can be brought
back to studying three-dimensional gravity with a negative
cosmological constant, coupled to an axion-dilaton system: \be
{S_{3d}}=\frac{1}{16\pi G_3}\int
dx^3\,\sqrt{-g}\left(R+\frac{2}{l^2}-\frac{1}{2}\frac{\partial_\m
\tau\partial^\m \bar \tau}{\tau_2^2}\right)\,. \ee To our knowledge this three-dimensional theory has never before been studied
in the literature. This is somewhat surprising as three-dimensional 
gravity with a negative cosmological constant is a
surprisingly rich gravitational theory that is well explored and
remains the subject of present investigations (see
\cite{Witten:2007kt,Li:2008dq} and references therein).
Furthermore, the above theory without a cosmological constant was
the subject of the classic paper \cite{Greene:1989ya}, and is very
closely related to F-theory.

Due to its embedding in eleven-dimensional supergravity the above bosonic action
is naturally completed into a supersymmetric theory. We  show
that this theory has 1/2-BPS solutions that are all locally
G\"odel space\footnote{G\"odel's original \spacetime{} was
four-dimensional, but it is nothing but the direct product of a
non-trivial three-dimensional \spacetime{} with a space-like line. It is
this three-dimensional \spacetime{} that we refer to as G\"odel
space.}:
\be
ds^2_3=\frac{l^2}{4}\left(-(dt+\frac{3}{2}\frac{dx}{y})^2+\frac{3}{2}\frac{dx^2+dy^2}{y^2}\right) \,,\qquad
\tau=x+iy . \ee The full eleven-dimensional solution can be read
off by substituting this metric in formula (\ref{11Dreduction})
below. G\"odel space \cite{Godel:1949ga} has a long history and
this work provides a new supersymmetric embedding into
string/M-theory. For an earlier example see \cite{Israel:2003cx}.
More precisely our work shows that M-theory has a compactification
of the form G\"odel$_3\times\sS 2\times \CY 3$ that preserves 4
out of the 32 supersymmetries. For other embeddings of spaces with
closed timelike curves in string theory, see e.g.
\cite{Gauntlett:2002nw,Compere:2008cw}.

G\"odel space suffers from closed timelike curves (CTCs). We
study some simple domain wall configurations, made out of smeared
M2-branes, that allow us to glue G\"odel space to $\AdS 3$. The initial hope was that this would eliminate the CTCs as it does for a similar system dubbed the `hypertube' \cite{Gimon:2004if}. Unfortunately, at present we have not been able to find a global solution that fully eliminates CTCs. We remain optimistic that a future treatment, either with another patching or a smooth resolution of the patching such as  was found for the hypertube  \cite{Berglund:2005vb,Bena:2005va} will resolve this issue. Nevertheless, the glued geometries we have found seem very interesting from the point of view of holography.\footnote{\textit{Note}: we give an alternative way of gluing G\"odel space to an asymptotically AdS space in the next chapter. The remarkable observation is that in that setup, the requirement of unitarity in the CFT dual to the AdS space, is tantamount to having no more closed timelike curves in the geometry.}

\outline{
In the next section, we briefly review
and discuss the supersymmetric properties of the wrapped M2-brane
states that motivate our study and detail
our flux compactification of M-theory to three dimensions. This section is of interest for readers with an interest in questions regarding black hole entropy.
In section \ref{s:GB-EOM}, we give the \ansatz{} and equations of motion, appropriate for supersymmetric solutions.
The solution we find, three-dimensional G\"odel space, is detailed in section \ref{s:GB-Soln}.
Section \ref{s:GB-Godel_AdS} covers how the G\"odel space can be
supersymmetrically glued to AdS space through the introduction of
an appropriate domain wall. Finally, section \ref{sec-con} presents some
discussion and suggestions for future directions. 
}

\section{Motivation}\label{s:Motive}

\subsection{Probe M2 branes and black hole entropy}

Before embarking on the construction of backreacted solutions of
$\sS 2$-wrapping M2-branes, let us review these BPS objects in the probe approximation
\cite{Simons:2004nm,Gaiotto:2004pc,Das:2005za} as a motivation. 

Our starting background is M-theory compactified on $\AdS 3 \times \sS 2 \times
\CY 3$ and we assume that the anti-de-Sitter factor is global
$\AdS 3$ and not a local solution such as a BTZ black hole. Such a
background arises e.g. in a certain limit of a D6-anti-D6
configuration when lifted to M-theory \cite{Denef:2007yt,deBoer:2008fk} (see below). Introducing coordinates $\rho$ (radius), $\sigma$ (time) and an angle $\psi \sim \psi + 2\pi$, the $\AdS 3$ part of the metric is 
\be
ds_3^2=l^2\left[-\cosh^2\r\,d\sigma^2+d\r^2+
\sinh^2\r\,d\psi^2\right]\, \label{globalmetr} \ee and the isometry group in three-dimensions is
$SL(2,\R )_L \times SL(2,\R )_R$.

\begin{wrapfigure}{R}{0.45\textwidth}
  \centering
\vspace{-.02\textheight}
    \includegraphics[width=0.43\textwidth]{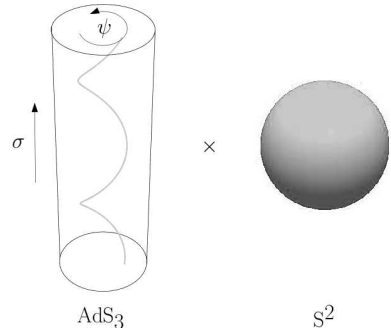}
  \caption[Helical wordline of M2 brane probe.]{The BPS M2-brane wraps the $\sS 2$ and has a helical worldline, $\psi=\sigma+$constant,  in the external $\AdS 3$ space.}
\label{probebr}
\end{wrapfigure}
We add to this background a probe M2-brane (or anti-M2-brane) wrapped on $\sS 2$, which behaves as
a massive point particle in $\AdS 3$. As can be seen from (\ref{globalmetr}), we cannot place
a static particle  (with respect to the global time $\s$) at finite $\r$, as it experiences a gravitational potential and will fall towards $\r = 0$.
A spinning particle however, obeying $\psi = \s + cst $, can stay at any fixed constant radius $\r = \r_0$. The $\r_0$-dependent momentum conjugate to
$\psi$ determines the D0-brane charge after compactification on a circle  to type IIA in ten dimensions.
Such spinning M2-branes are BPS states
and are the objects we  study, see \figref{probebr}.

As explained in chapter \ref{c:FB}, these wrapped M2 brane probes in global $\AdS 3 \times \sS 2$ play an important role in counting the entropy of the D0-D4 black hole in the deconstruction proposal. In this proposal, the D0-D4 charges are redistributed over multiple centers, one carrying D6 brane charge, one carrying D6 anti-brane charge and a large number of D0 centers (D4 charge is induced through \worldvolume{} flux on the D6-branes). Counting the number of ways one can distribute the D0 centers to form a total D0-D4 black hole in this way, gives a measure of entropy. Two different approaches give different results, see section \ref{s:FB_scaling}:
\begin{enumerate}
\item In \cite{Denef:2007yt}, the authors conjecture the D0 branes carry non-abelian degrees of freedom and should be described in the M-theory, by exactly the M2-brane probes discussed above. Counting the number of ground states for such M2 branes, gives the correct number for the D0-D4 black hole entropy. However, one needs to consider the M2-branes as probes (backreaction on the metric is not considered).
\item In \cite{deBoer:2009un}, an alternative way of counting the number of states for the deconstructed setup is given, that does not reproduce the entropy, only a parametrically small part of it. This second approach did take the \textit{backreaction} of the branes into account, but did not imply using non-ableian degrees of freedom.
\end{enumerate}
It seems that the non-abelian nature of the D0 branes is crucial in deriving the entropy. We thus find the motivation for studying the M2 brane probes above: we want to study the backreaction of M2 probe branes on the $\AdS 3 \times \sS 2 \times \CY 3$ background, in hope to find a backreacted version of the entropy arguments above, that includes the effect of non-abelian degrees of freedom. The work in this chapter should be seen as a first step in that direction: we were not able to attack the entropy issue, but did find an interesting solution representing the backreaction of the sphere-wrapped M2 brane probes.

\subsection{Backreaction of M2 branes: fields and effective description}
The arguments of the last section lead us to search for supersymmetric solutions where the fields that are sourced by the sphere-wrapped M2-branes under consideration are turned on and to consider how they backreact on the $\AdS 3 \times \sS 2 \times \CY 3$ geometry.

\begin{figure}[ht!]
\subfigure[]{
\begin{tabular}{ccccc}
 \begin{tabular}{c}\epsfig{figure=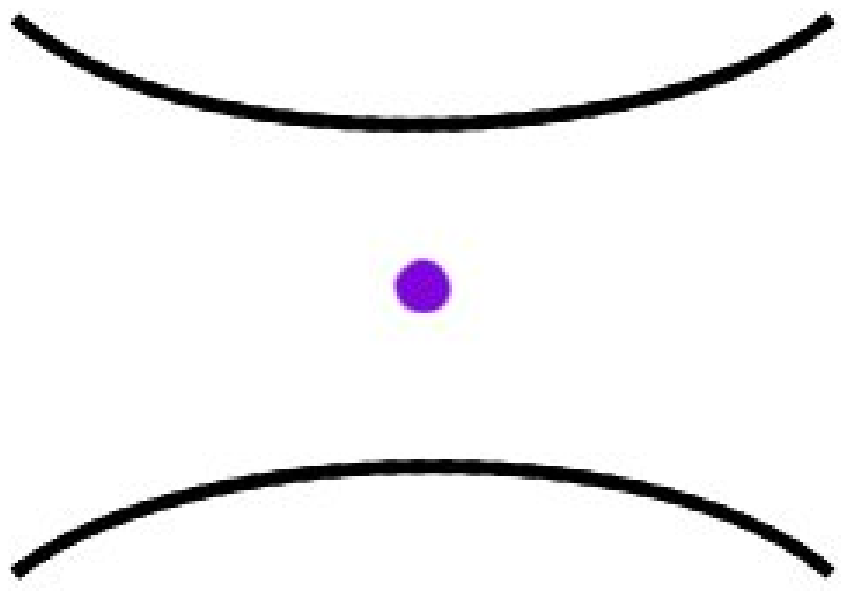,angle=-90,scale=.15}\\[2cm]$\AdS 3$\end{tabular}&\hspace{-.7cm}\begin{tabular}{c}{\Large $\times$}\\\end{tabular}\hspace{-.7cm}\hspace{-2cm}&\begin{tabular}{c}\epsfig{figure=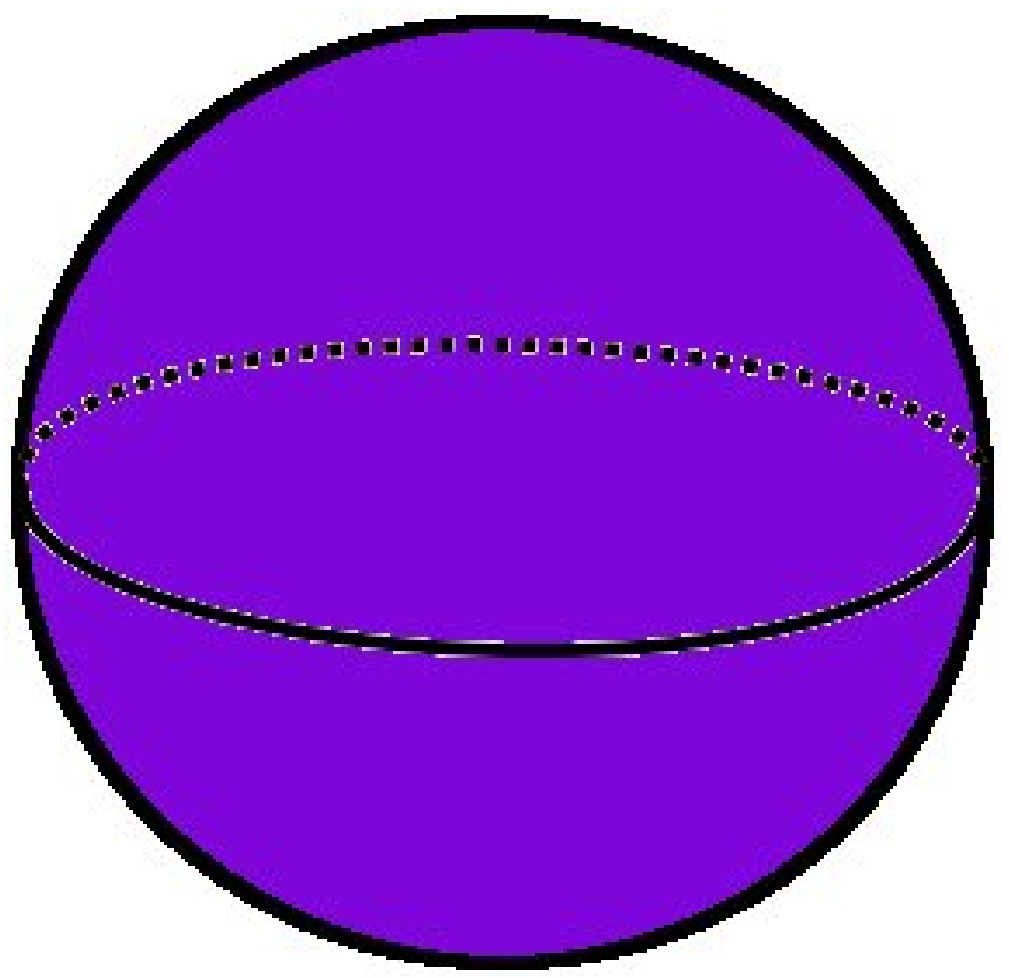,angle=-90,scale=.15}\\[2cm]$\sS 2$\end{tabular}&\hspace{-.7cm}\begin{tabular}{c}{\Large $\times$}\\\end{tabular}\hspace{-.7cm}\hspace{-2cm}&\begin{tabular}{c}\includegraphics[height=2cm]{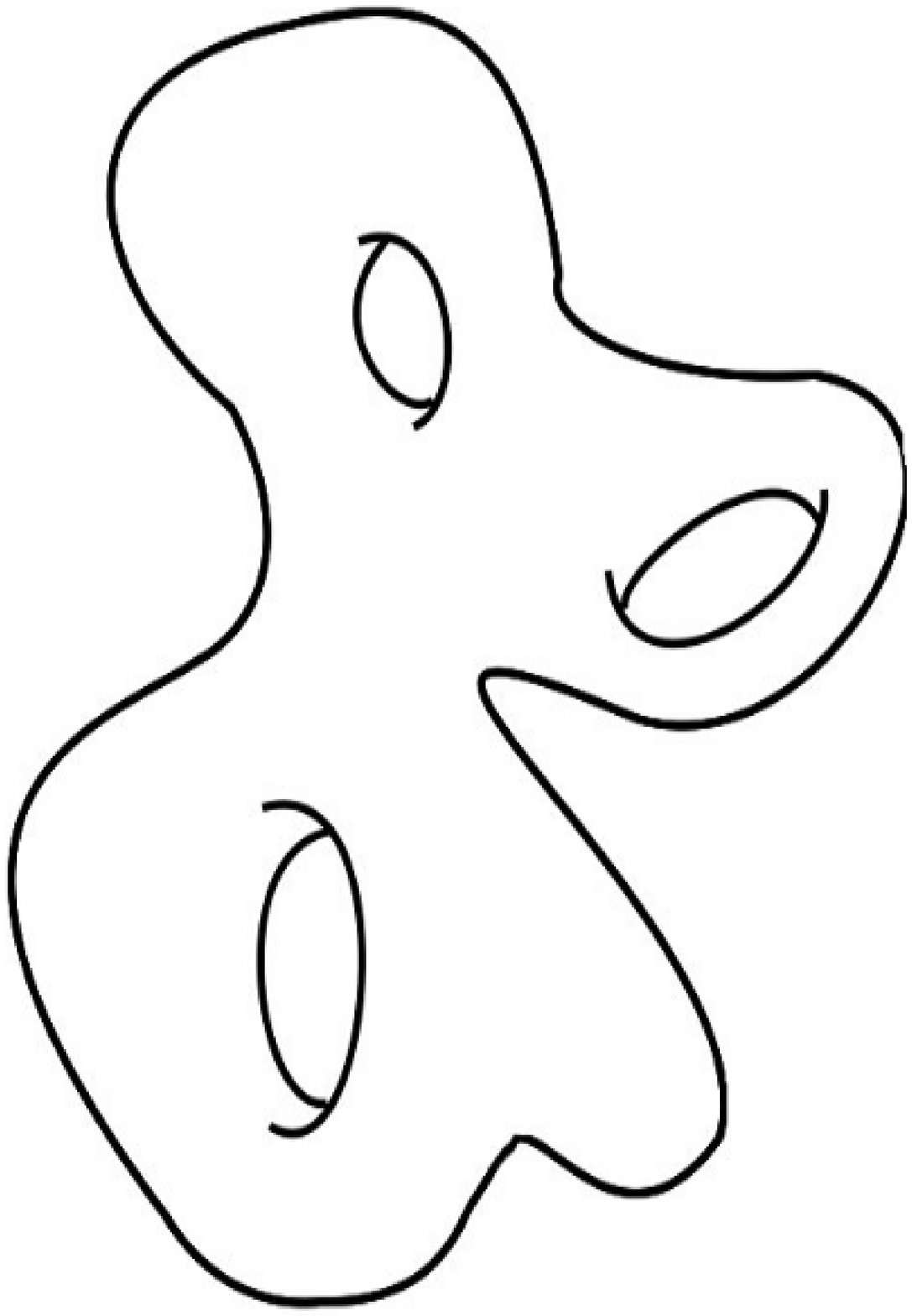}\\$\CY 3$\end{tabular}
\end{tabular}
\label{fig:GB-AdS3_S2_CY-a}
}
\subfigure[]{
\begin{tabular}{ccccc}
\begin{tabular}{c}\epsfig{figure=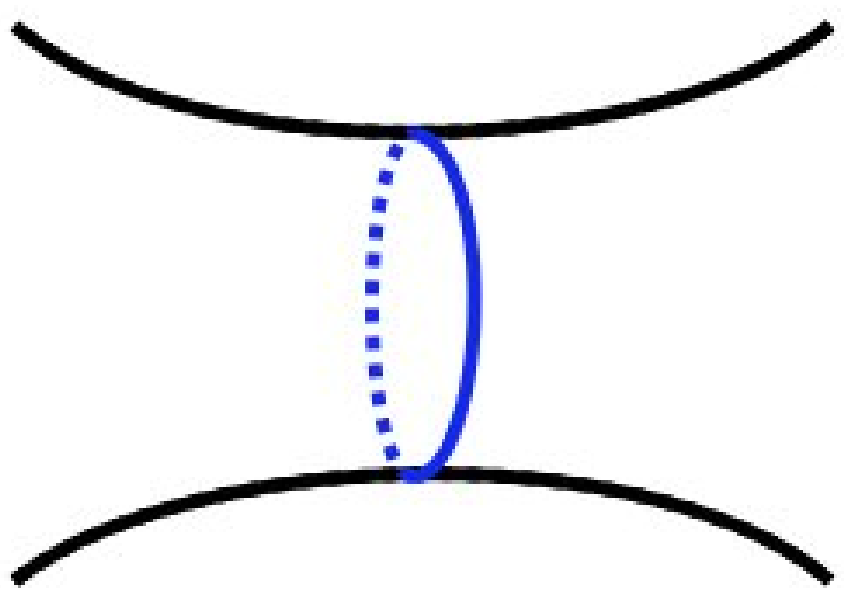,angle=-90,scale=.15}\\[2cm]$M_3$\end{tabular}&\hspace{-.7cm}\begin{tabular}{c}{\Large $\times$}\\\end{tabular}\hspace{-.7cm}\hspace{-2cm}&\begin{tabular}{c}\epsfig{figure=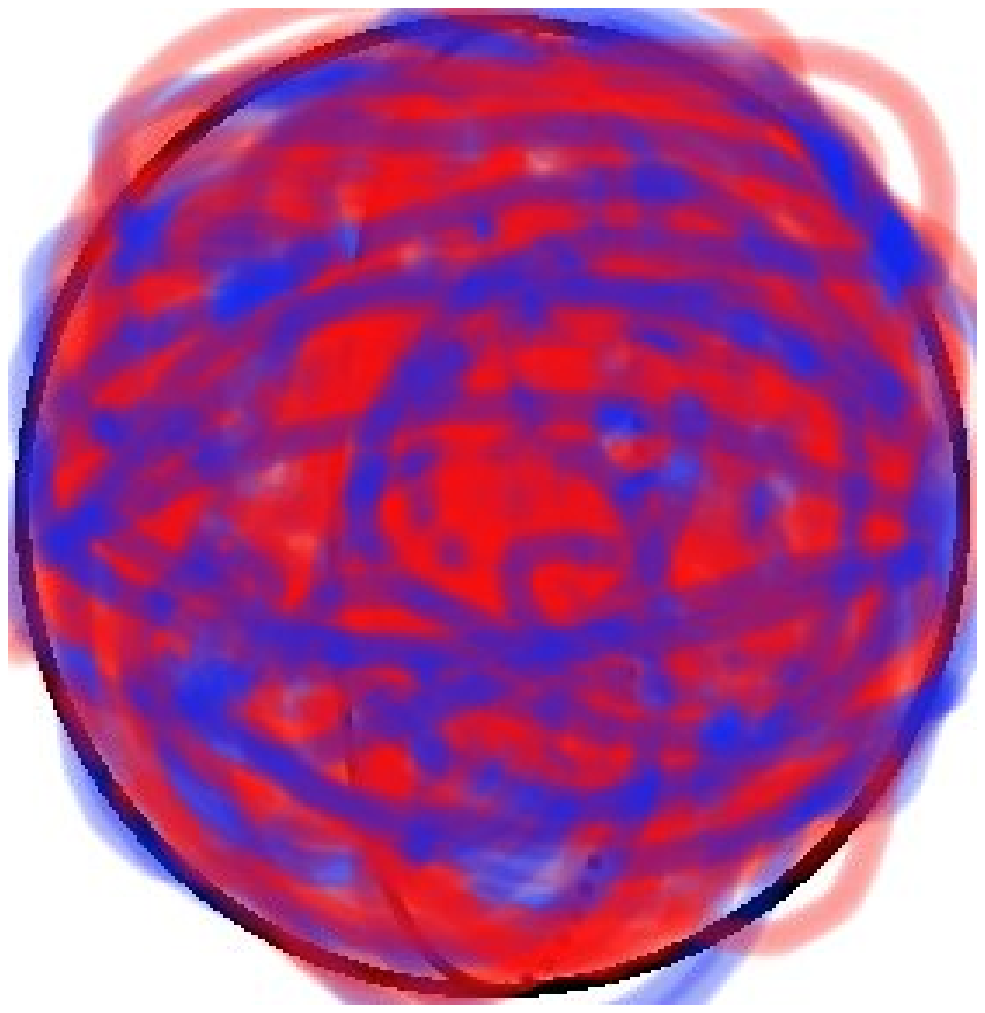,angle=-90,scale=.15}\\[2cm]$\sS 2$\end{tabular}&\hspace{-.7cm}\begin{tabular}{c}{\Large $\times$}\\\end{tabular}\hspace{-.7cm}\hspace{-2cm}&\begin{tabular}{c}\includegraphics[height=2cm]{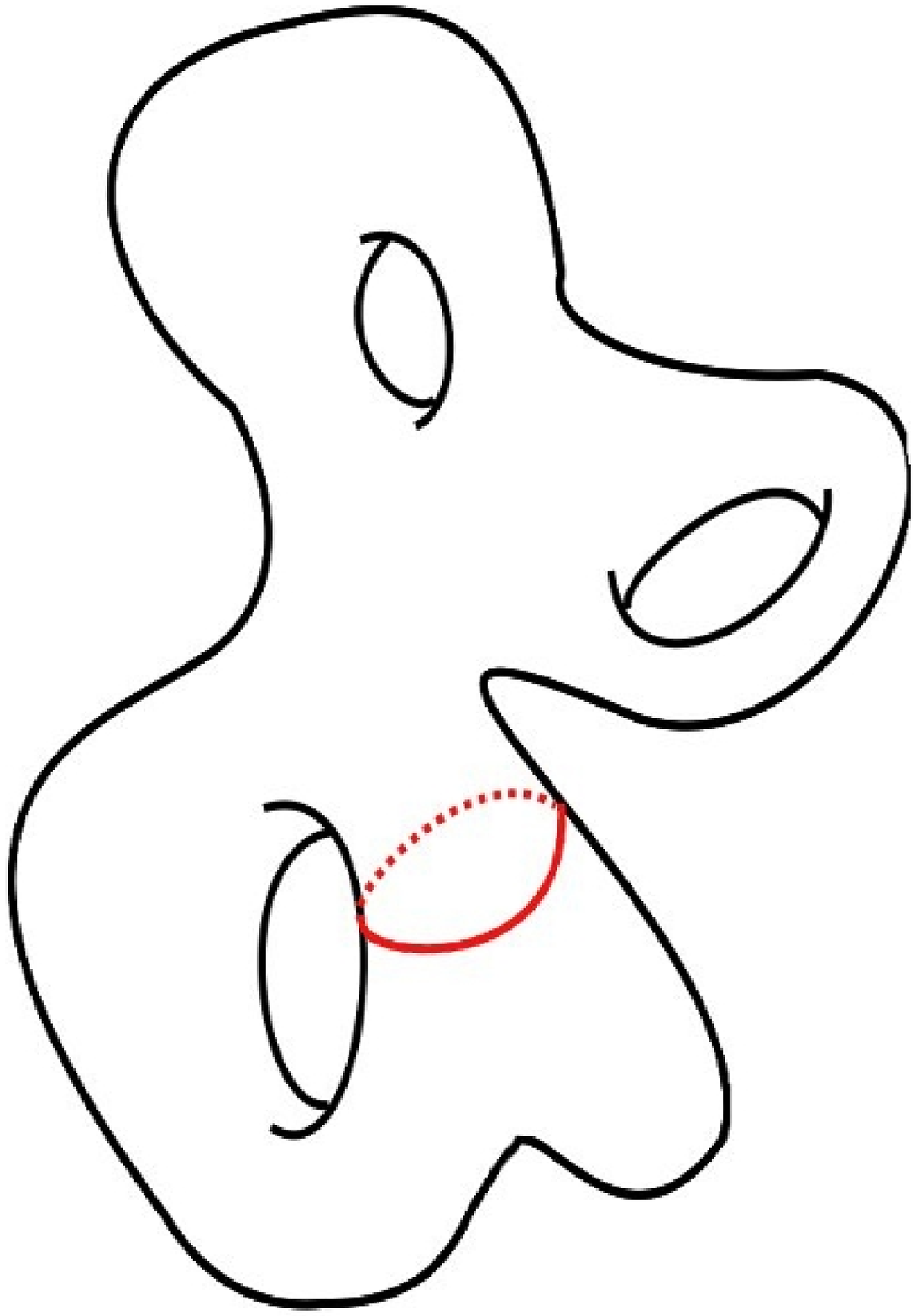}\\$CY 3$\end{tabular}
\end{tabular}
\label{fig:GB-AdS3_S2_CY-b}
}
\caption[M2 brane probes and sourced fields on  $\AdS 3 \times \sS 2 \times \CY 3$.]{In figure \subref{fig:GB-AdS3_S2_CY-a}, we sketch the form of the   $\AdS 3 \times \sS 2 \times \CY 3$ background, with M2 brane probes wrapped on the $\sS 2$ (in purple). Figure \subref{fig:GB-AdS3_S2_CY-b} shows pictorially how we expect the M2 branes to affect the background: the three-dimensional geometry changes to some manifold $M_3$ (to be calculated) and $F_4$ couples to the brane \worldvolume{} (blue). (In red, we have the part of the flux $F_4$ needed to support the background.)}\label{fig:GB-AdS3_S2_CY}
\end{figure}

We begin with the bosonic part of the eleven-dimensional supergravity (M-theory) action. In our conventions it is given by
\be
\frac{S_{M}}{2\pi}=\frac{1}{l_{M}^9}\int d^{11}x\sqrt{-g}R-\frac{1}{2l_{M}^3}\int F_4\wedge\star F_4+\frac{1}{6}\int A_3\wedge F_4\wedge F_4, \label{Maction}
\ee
we put the eleven-dimensional Planck length to one and write $F_4=dA_3$. We seek solutions sourced by M2-branes that wrap the two-sphere in the  flux compactification of M-theory on $\sS 2\times \CY 3$. To consider how these M2 branes backreact on the geometry, we need to specify how the metric and four-form $F_4$ are affected. We assume that only the three-dimensional part of the geometry is changed, while the four-form $F_4$ couples to the M2 brane \worldvolume{}. See figure \ref{fig:GB-AdS3_S2_CY}.

We make a consistent reduction for such solutions using the ansatz 
\bea
ds^2_{11}&=&\tau_2^{-2/3}\left(ds^2_3+d s^2_{S^2}\right)+\tau_2^{1/3}d s^2_{\CY 3}\,, \label{11Dreduction}\\
F_4&=&d A\wedge\o_{2}+\o_{2}\wedge J_{\CY 3}\,.\label{F4}
\eea%
The dynamic Calabi-Yau volume (measured in units of 11-dimensional Planck length) is parametrized with a scalar $\tau_2$. The volume form on the $\sS 2$ with constant radius is $\o_2$, and $J_{\CY 3}$ is the K\"ahler form on the unit volume Calabi-Yau. In this ansatz we allow two contributions to the M-theory gauge field $F_4$: the part along $\omega_2\wedge J_{\CY 3}$ describes the flux needed to support the background (red cycles in figure \ref{fig:GB-AdS3_S2_CY-b}), while the first term in \eqref{F4} is the one sourced by sphere-wrapped M2-branes (blue cycles in figure \ref{fig:GB-AdS3_S2_CY-b}). 

The \textit{degrees of freedom} in this \ansatz{} are the components of the three-dimensional metric $g_3$, the Calabi-Yau volume $\tau_2$ and the electric field $A$ in three dimensions. We describe the one-form $A$ in terms of a real scalar field $\tau_1$, as this is more natural from the three-dimensional point of view, using the Hodge duality:
\begin{equation}
dA = -\frac{\star_3 d\tau_1}{\,\tau_2^2}\,,\label{Fvstau1}
\end{equation}
where $\star_3$ is taken w.r.t.~ the three-dimensional metric $\de s^2_3$. Finally, it is natural in three dimensions to combine the dualized electric field $\tau_1$ and the Calabi-Yau volume $\tau_2$ into one complex scalar $\tau$ as:
\beq
\tau = \tau_1 + i\tau_2\,.
\eeq
One can then check that for an \ansatz{} of the form specified above, the eleven-dimensional equations of motion become equivalent to those of the three-dimensional action ($G_3$ is Newton's constant in three dimensions):
\begin{equation}
 S_{3d}=\frac{1}{16\pi G_3}\int dx^3\,\sqrt{-g}\left(R+\frac{2}{l^2}-\frac12\frac{\partial_\m \tau\partial^\m \bar \tau}{\tau_2^2}\right)\,.\label{3daction}
\end{equation}

\noindent We point out a few interesting facts about this reduction:%
\begin{itemize}
\item Although we are mainly interested in the M-theory origin
of the three-dimensional system  (\ref{3daction}) discussed above, solutions to
(\ref{3daction}) can of course  be embedded in any higher
dimensional theory that allows (\ref{3daction}) as a consistent
truncation. Of special physical interest are embeddings in a type
IIB string theory on $\sS 3 \times \calm_4$ (with $\calm_4$ either
$K_3$ or $T_4$), which is the near horizon limit of the D1-D5
system. In the probe approximation, branes wrapping the $\sS 3$ have
been conjectured to account for the entropy of the Strominger-Vafa
black hole in a similar manner as the $\sS 2$-wrapping M2-branes in
the  M-theory frame did for the current setup
\cite{Raeymaekers:2007ga}. Performing an S-duality, one obtains
another  interesting duality frame where only NS sector fields are
excited and which could be the starting point for a sigma-model
description. In another duality frame of interest, our system
describes the backreaction of D7-branes wrapped on  $\sS 3 \times
\calm_4$. In this frame, the field $\T$ is the standard
axion-dilaton of type IIB. Our configurations can then be viewed
as nontrivial solutions  of F-Theory.  For an overview, see \tableref{interestingframes}.
\begin{table}
\begin{center}
\begin{tabular}{|c|c|c|}\hline
theory & background branes & source branes\\ \hline
$M$ on  $\sS 2 \times T^6$ & $M5$'s & $M2$ on $\sS 2$ \\
$IIB$ on  $\sS 3 \times T^4$ & $D3-D3$ & $D3$ on $\sS 3$ \\
$IIB$ on  $\sS 3 \times T^4$ & $D1-D5$ & $D5$ on $\sS 3 \times T^2$ \\
$IIB$ on  $\sS 3 \times T^4$ & $F1-NS5$ & $NS5$ on $\sS 3 \times T^2$ \\
$F$ on  $\sS 3 \times T^6$ & $D3-D3$ & $D7$ on $\sS 3\times T^4$ \\ \hline
\end{tabular}
\end{center}
\caption[U-dual embeddings of 3d axion-dilaton system.]{Embeddings of the three-dimensional axion-dilaton system in various higher dimensional theories related by U-duality.
The axion-dilaton $\T$ is in each case a modulus of the toroidal compactification.
The background branes produce a near horizon $\AdS 3$ flux compactification, and the source branes couple to $\T$. For details of the embedding into different higher-dimensional theories, we refer to the appendix of \cite{Levi:2009az}, where the explicit U-duality chains are given.}\label{interestingframes}
\end{table}

\item{Instead of compactifying to three dimensions, one can instead consider the five-dimensional theory obtained by reduction on the Calabi-Yau alone. This reduction gives us the action of $\mathcal{N}=1$ supergravity in five dimensions. The complex scalar $\tau$ is then part of the universal hypermultiplet. To account for the charge and backreaction of probe branes wrapped on $\sS 2$  we must seek solutions with non-trivial \textit{hyperscalars} turned on. There is an extensive literature on the general framework of finding solutions in this situation \cite{Bellorin:2006yr,Huebscher:2006mr,Strominger:1997eb,Behrndt:2000zh}, though few specific examples are known. Our solution is a new example with non-trivial hyperscalars.
}

\end{itemize}

\section{Three-dimensional system and Ansatz}\label{s:GB-EOM}
In this section we discuss the equations of motion for supersymmetric solutions to the three-dimensional action
\be \label{3daction_bis}
{S_{3d}}=\frac{1}{16\pi G_3}\int dx^3\,\sqrt{-g}\left(R+\frac{2}{l^2}-(\m -1)\frac{\partial_\m \tau\partial^\m \bar \tau}{\tau_2^2}\right)\,.
\ee%
This system describes three-dimensional gravity with a negative cosmological constant $\Lambda = -1/l^2$, coupled to a complex scalar $\tau = \tau_1 + i\tau_2$, which has a typical axion-dilaton type kinetic term, describing a sigma model for the $\SU(1,1)/U(1) \simeq \SL(2,\Real)/\SO(2)$ symmetric space. There is an $\SL(2,\Real)$ global symmetry and a $U(1)$ local (gauge) symmetry, carried by the one-form $A$ (Hodge dual of $\tau_1$, see eq.~\eqref{Fvstau1}). In the $l\rightarrow\infty$ limit this action is that of the seminal paper on cosmic strings by \cite{Greene:1989ya}. \textit{Note}: we have introduced a coupling constant $\mu$ in front of the axion-dilaton action. Positive energy requires $\m >1$ and from the previous section we know that the M-theory reduction fixes $\m = 3/2$, which is the value we are interested in (the eleven-dimensional origin of the scalar $\tau_1$ is as the Hodge dual of the electric field sourced by the M2-branes, while $\tau_2$ is the dynamic Calabi-Yau volume). Nevertheless, we are  able to construct solutions for more general $\mu$, although it is not clear whether these allow for a supersymmetric embedding.


\subsection{Equations of motion and \ansatz{}\label{ss:GB-EOM}}

The equations of motion one obtains from the three-dimensional action \eqref{3daction} are%
\bea
R_{\a\b} &+&\frac{2}{l^2}g_{\a\b}=(\m - 1)\frac{\partial_{(\a}\tau\partial_{\beta)}\,\bar\tau}{\tau_2^2} \, ,\label{einsteineqns}\\
\partial_\alpha\left(\sqrt{-g}g^{\a\b}\partial_\b\tau\right) &+& i\sqrt{-g}g^{\a\b}\frac{\partial_\a\tau\partial_\b\tau}{\tau_2}=0 .\label{scalareqns}
\eea
It is important to note that, due to the specific `non-standard' kinetic term for the scalar $\tau$, equation \eqref{scalareqns} --- from varying with respect to $\bar \T$ --- only features derivatives of $\tau$ and not of $\bar \tau$. Furthermore,the probe branes we started from were moving with a constant velocity. By a linear coordinate transformation, one can make them static w.r.t.~the new time direction $t$ , i.e.~at a fixed time-independent position. The metric in these new coordinates is then stationary (we have shifted the rotation of the probe to a rotation of \spacetime{}).

Therefore, we seek \textit{stationary solutions} to the backreacted system (\ref{einsteineqns},\ref{scalareqns}) as well: $\partial_t\tau=0$, for a specific time coordinate $t$. As was shown in \cite{Greene:1989ya} for the flat space analog of our system ($l=\infty$), these two facts combined imply that the scalar equation of motion can be solved by choosing $\tau$ to be (anti)-holomorphic in the complex coordinate naturally made up of the remaining two spatial coordinates.  Moreover, this implies the solution is supersymmetric. One can see from \eqref{scalareqns} that this remains true even if the full three-dimensional metric is not flat but when the spatial part of $\sqrt{g}g^{\a\b}$ consists of constants.

\paragraph{Metric \ansatz:} Any stationary metric in three dimensions for which $\sqrt{g} g^{\a \b}=cst$ along  the spatial directions,  is equivalent to the form
\bea
ds^2_{3}&=&\frac{l^2}{4}\left(-(dt+\chi)^2+e^{2\phi(z,\bar z)}dzd\bar z\right)\label{stationaryLiouville} \,.
\eea
This metric describes a timelike fibre over two-dimensional flat space, which we parameterize in complex coordinates $z$.


\paragraph{Equations of motion bis:}

With the metric \ansatz{} (\ref{stationaryLiouville}), the $\T$ equation of motion (\ref{scalareqns})  reduces to
\be
\pa \bar \pa \T + i \frac{\pa  \T \bar \pa \T}{ \T_2} = 0 \label{taueqn}.
\ee
It can be solved by taking $\T$ to be an arbitrary holomorphic or antiholomorphic function. As we have shown in \cite{Levi:2009az} this gives a supersymmetric solution.

We further assume that $\tau$ is a holomorphic function
\begin{equation}
 \tau = \tau (z)\,.
\end{equation}
Using the metric ansatz (\ref{stationaryLiouville}), the Einstein equations \eqref{einsteineqns} are written as
\bea
\de \chi&=&\frac{i}{2}e^{2\phi}\,\de z\wedge \de \bar z\label{chieqn} \, ,\\
\partial\bar\partial\phi-\frac{e^{2\phi}}{4}&=&- (\m - 1)\frac{\partial\tau\bar\partial\bar\tau}{4 \T_2^2} .\label{liouvilleeqn}
\eea
The equation for  $\phi$ is the Liouville equation with a source term provided by $\T$.

\subsection{Solutions -- a short discussion} 

We have seen that choosing a holomorphic function $\tau(z)$ in principle determines the solution, specifying the warp factor $\phi$ and the one-form $\chi$ of the metric through the Einstein equations \eqref{chieqn},\eqref{liouvilleeqn}.  Note that the complex scalar $\tau$ lives on the upper half plane, as its imaginary part $\tau_2$, is positive (think about the interpretation of $\tau_2$ as the Calabi-Yau volume). In conclusion, \textit{the holomorphic maps from the spatial base with coordinate $z$ to the upper half plane with coordinate $\tau(z)$ classify the solutions.} However, unexpected difficulties appear. These are most easily discussed by starting with the flat space analog ($l=\infty$).

\begin{itemize}
\item[\ra] \underline{\emph{Flat space ($\Lambda=0$, or $l=\infty$).}} In this case, the three-dimensional system is that of \cite{Greene:1989ya,Gibbons:1995vg}. It describes the three-dimensional part of the geometry and the relevant fields for either cosmic strings in four dimensions or D7-branes in ten-dimensional type IIB supergravity ($\tau$ is then the axion-dilaton of type IIB string theory). The spatial base with coordinate $z$ is the complex plane. It is argued that the $\SL(2,\Real)$ symmetry associated with $\tau$, is broken to $\SL(2,\mathbb{Z})$ by quantum effects. One should then consider solutions that are $\SL(2,\mathbb{Z})$-invariant. This is achieved by considering holomorphic maps from two-dimensional space (the complex plane with coordinate $z$), to the fundamental domain $F$, where $F$ is the $\SL(2,\mathbb{Z})$-quotient of the $\tau$-complex upper half plane $H$ $F = H/\SL(2,\mathbb{Z})$. There is a function that provides a one-to-one map from the complex plane (coord.~$z$) to the fundamental domain (coord.~$\tau(z)$), namely the $j$-function. Solutions are constructed using this map.
\item[\ra] \underline{\emph{Negative cosmological constant, ($\Lambda<0$, or $l$ finite).}} This is our three\--di\-men\-sio\-nal system. Again, holomorphic maps $\tau(z)$ determine the solution. However, the spatial part of the geometry, with coordinate $z$ is no longer the entire complex plane. Instead, one should consider the spatial part of the geometry to be described by a manifold with a boundary. This can be seen for instance by considering the solution with constant $\tau$, namely $\AdS 3$ space. $\AdS 3$ has a different topology than flat space: the spatial part of the geometry has a boundary, and has the topology of a disk, namely a two-dimensional surface bounded by a line. Alternatively, we can describe the spatial part of the geometry by letting $z$ run over the upper half plane, as this has the same topology. Now things get more complicated. In order to find finite energy solutions, one would prefer to follow the prescription of \cite{Greene:1989ya}, and restrict to $\SL(2,\mathbb{Z})$-invariant solutions, by constructing a holomorphic map from the spatial part of the geometry (the complex upper half plane with coordinate $z$) to the fundamental domain $F$. However, such a holomorphic map cannot be found, because the domain and the range have different topology. We thus do not have a natural way of constructing $\SL(2,\mathbb{Z})$-invariant solutions.
\end{itemize}
In light of these difficulties, we ignore the $\SL(2,\mathbb{Z})$ invariance and study the identity map
\be
\T( z) = z\label{eq:tauw}
\ee
from the spatial base, considered as a complex upper half plane to the entire $\tau$-space $H$ (not restricting to a fundamental domain). The above map is one-to-one and has a first-order pole on the boundary at $z = i \infty$. More general multiple-to-one maps can be locally brought into this form by a conformal transformation. The metric for this choice of $\tau$ is derived in appendix \ref{app:GB-Solutions}. We discuss the form of the solution for the specific choice $\tau(z) = z$ in the next section. Note that because of the pole at the boundary, this solution has infinite energy (alternatively, infinite energy is seen from the fact that the map $\tau(z) = z$ covers an infinite amount of fundamental domains.). We will also discuss this issue in the next section.

\section{Our solution: G\"odel space}\label{s:GB-Soln}
We discuss a solution to the three-dimensional system given by the action
\be 
{S_{3d}}=\frac{1}{16\pi G_3}\int dx^3\,\sqrt{-g}\left(R+\frac{2}{l^2}-(\m -1)\frac{\partial_\m \tau\partial^\m \bar \tau}{\tau_2^2}\right)\,.
\ee%
The three-dimensional metric is  \eqref{stationaryLiouville}, where the spatial base with coordinate $z$ is taken to be the complex upper half plane. We consider the metric solution when the complex scalar $\tau(z)$ is the identity map on the upper half plane
\begin{equation}
 \tau(z) = z\,.
\end{equation}
We have proven that solutions with holomorphic axion-dilaton $\tau(z)$ are supersymmetric in \cite{Levi:2009az}. We do not give the details, only the resulting geometry.

\subsubsection{Metric: three-dimensional G\"odel space}\index{G\"odel space!in 3d|defn} 
{}Defining $z = x+ i y$, the solution for the metric is (see appendix \ref{app:GB-Solutions} for details)
\beq%
ds^2 = {\mu l^2 \over 4} \left[ - \mu(dt+ {\frac {dx} y})^2 +  {d x^2 + d y^2 \over y^2} \right] \, \label{godelUHPmain}\,,
\eeq%
after a rescaling of the time coordinate $t \to \mu t$. This is the metric of \textit{timelike warped Anti-de Sitter space}\index{Anti-de Sitter!timelike warped --} (see for instance \cite{Bengtsson:2005zj}). It describes a timelike fibre over Euclidean $\AdS 2$, also known as the hyperbolic plane. The latter is the upper half plane $(x,y)$ endowed with the metric $(\de x^2 + \de y^2)/y^2$.
  
We distinguish three cases:
\begin{itemize}
\item $\boxed{\mu=1}$ In this case, this is just the metric of global $\AdS 3$. This is no surprise, because in this case the complex scalar $\tau$ decouples from the metric (its coupling constant $\mu-1$ in the action becomes zero).
\item $\boxed{\mu>1}$ For $\m >1$, including the case of interest $\m = 3/2$, the timelike fiber is stretched, and the space is known to be the G\"{o}del geometry \cite{Rooman:1998xf}.  G\"{o}del's original solution \cite{Godel:1949ga} was four-dimensional: it is a direct product of the three-dimensional space above, with $\mu = 2$, and a line. This is one of the first examples of a three-dimensional supersymmetric G\"odel space in the literature, see also \cite{Israel:2003cx}. We discuss below that the stretching of the timelike fibre induces closed timelike curves.
\item $\boxed{\mu<1}$ Formally one could also take $\m <1$, in which case the timelike fibre is squashed  with respect to pure $\AdS 3$.
This space has no closed timelike curves \cite{Bengtsson:2005zj} and also appears as a solution to topologically massive gravity \cite{Anninos:2008fx}. However it  arises from an unphysical matter source:  as we can see from  (\ref{3daction}), it requires a `ghost' axion-dilaton with a wrong sign kinetic term. Alternatively, one can see it as coming from a perfect fluid source with negative energy density (see (\ref{enden}) below).
\end{itemize}
 
\subsubsection{Properties of G\"odel space}
We  restrict attention to  $\mu>1$ in what follows.

\noindent \emph{\underline{Link to original G\"odel solution.}} G\"odel's original solution was four-dimensional. The metric of four-dimensional G\"odel space has the form $\de s_4^2 = \de s_3^2 + \de z^2$, where $\de s_3^2$ represents the solution \eqref{godelUHPmain} above with $\mu=2$ and $z$ is a decoupled fourth direction. From now on, we  call the three-dimensional part of the geometry, with $\mu>1$, G\"odel space and denote it G\"odel$_3$.  G\"{o}del's solution was originally obtained as a solution of gravity with negative cosmological constant $\L = -1/l^2$ in the presence of \textit{a pressureless fluid source}. It is instructive to check that the energy-momentum tensor of our scalar field solution $\T = x + i y$ behaves exactly as a pressureless fluid:
\bea
T_{\m\n} &\equiv &  - { 1\over 8\pi G_3}\frac1 {\sqrt{-g}}{\d S_\T \over \d g^{\m\n}} = \frac{(\mu -1) }{8\pi G_3} \frac 1 {\T_2^2}\left[ \pa_{(\m}\T \pa_{\n)}\bar \T - \half g_{\m\n} \pa_\r \T \pa^\r \bar \T \right]
 \nonu
&=& \rho u_\m u_\n \, ,
\eea
where the unit vector is $u^\m = {2\over  l} \d^\m_0$ and the energy density of the fluid is
\be
\rho = \frac1{2\pi G_3 l^2} \frac{\mu-1}\mu\label{enden} \, .
\ee
Setting $\m = 2$ we again find the expression in \cite{Godel:1949ga}. The fluid flow is rotational since $ \star_3 ( u \wedge d u)$ is a nonzero constant, indicating that  G\"{o}del space rotates around every point.

\noindent \underline{\emph{Appearance of CTCs.}} It is well known that  G\"{o}del space suffers from causal pathologies in the form of \textit{closed timelike curves} (CTCs). These are most apparent in the coordinate system which has the Poincar\'e disk as the  spatial base. The following coordinate transformation takes us to this frame:
\bea%
z &=&i\frac{1+w}{1-w}\,,\nonu\
t&=& \tilde t+2 \arg(1 - w) \label{uhptodisk} .
\eea%
We define  $w= r e^{i \varphi}$ and drop the tilde on $\tilde t$ to get%
\be ds^2=\frac{ \m l^2}{4}\left[-\m(d
t+ \frac{2 r^2}{1-
r^2}d\varphi)^2 +4  \frac{d r^2+
r^2d\varphi^2}{(1- r^2)^2}\right]\,.\label{Godelcoords} \ee
In the above form of the metric, it is easy to see that the vector field $\pa_{\varphi}$ becomes timelike for $r > \frac{1}{\sqrt{\m}}$, so that  $ \varphi$-circles become closed timelike curves for these values of the radius.

\subsubsection{Problems with the solution}
We have already encountered one unwanted feature of our solution: the appearance of closed timelike curves. But there is more. The solution for $\tau$ reveals that it carries an infinite amount of $U(1)$ charge (remember that $\tau_1$ is Hodge dual to a gauge field for the local  $U(1)$-symmetry. In terms of eleven-dimensional fields, this $U(1)$-charge 
is just M2-brane charge). We comment in some detail.

\begin{wrapfigure}{R}{.5\textwidth}
{\centering \includegraphics[width=.47\textwidth]{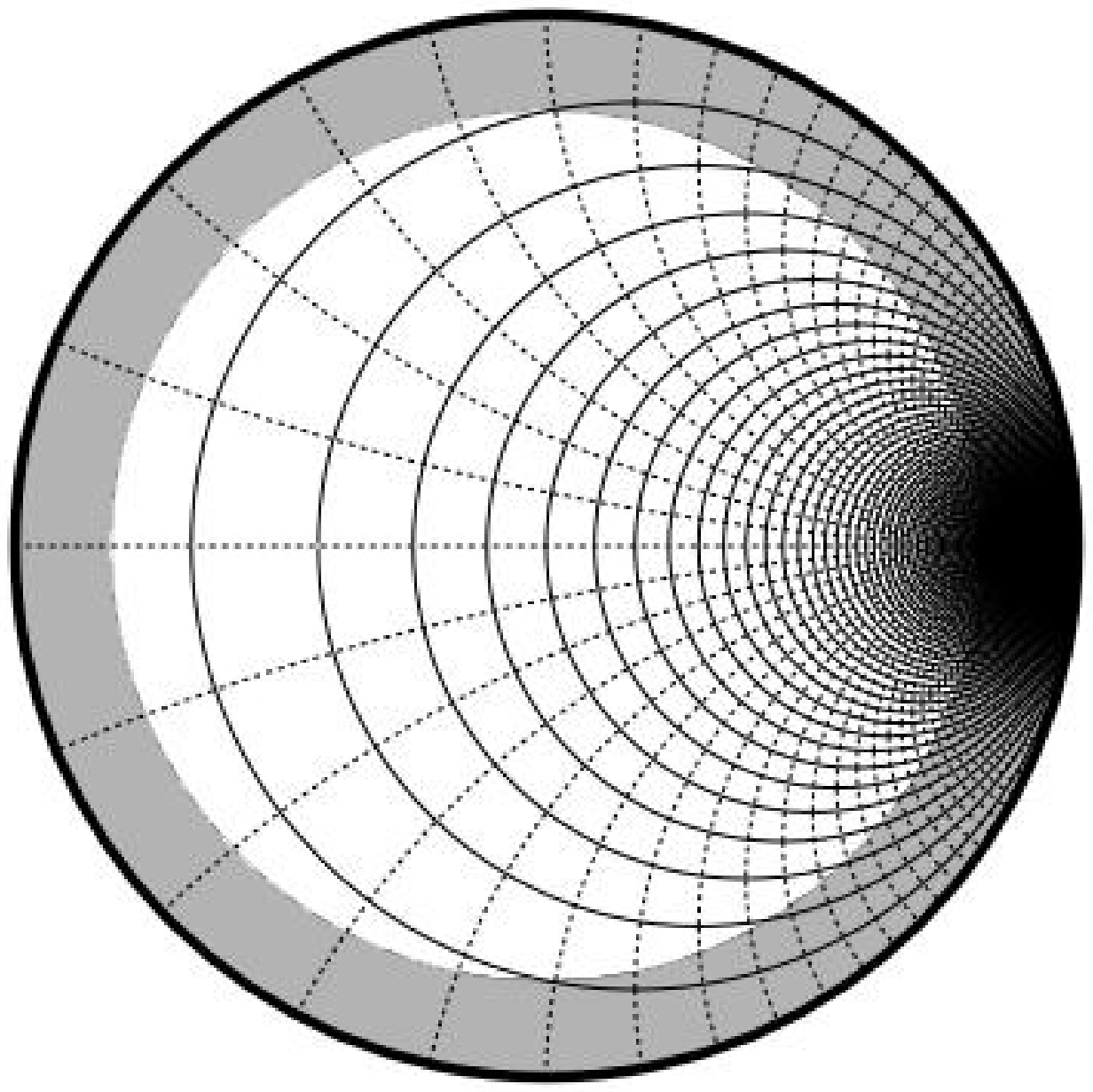}}
\vspace{-.015\textheight}
\caption[The 3d G\"{o}del solution in disk coordinates.]{The  G\"{o}del solution in disk coordinates. Circles within the gray zone and centered at the origin
are closed timelike curves.  The solid lines are equipotential surfaces (constant $\T_2$), the dotted lines
are electric field lines (constant $\T_1$). The brane source is at $w=1$. 
\vspace{-.015\textheight}
}
\label{fieldlines}
\end{wrapfigure}

In the upper half plane coordinates, the solution  $\tau(z)=z$ has a pole at infinity, so we expect to have some source there. We continue the discussion in the disk coordinate frame $(t,w,\bar w)$, as it is useful to visualize the axion-dilaton solution $\tau$. Using the mapping to the disk coordinates \eqref{uhptodisk}, we find
\begin{equation}
 \tau(w) =i\frac{1+w}{1-w}\,.
\end{equation}
The form of $\tau(w)$ reveals a source is located at the point $w = 1$ on the boundary. We argue this graphically. First, one can check the lines of constant dilaton $\T_2$ are circles tangent to $w=1$. Using the relation of $\tau_2$ to the gauge  field (\ref{Fvstau1}) 

and the metric (\ref{godelUHPmain}), one can furthermore show that  $\T_2$ also plays the role of the scalar  potential for the electric field, so these circles are also equipotential surfaces. The electric field lines are the lines of constant $\T_1$ and are orthogonal to the equipotential circles. These properties are illustrated in figure \ref{fieldlines}.

\newpage
{}From the form of the electric field lines, it is clear that there is a source at the boundary at $w=1$. Since the interpretation of the gauge field in eleven dimensions couples to M2 branes, the source reveals the presence of M2 brane charge. In our work \cite{Levi:2009az}, we found that the amount of charge is  \textit{infinite} (I omit the details here), leading to an infinite energy solution. The fact that the global G\"odel space carries an infinite amount of charge is probably due to the fact that the axion-dilaton solution covers an infinite number of fundamental SL$(2,\Z )$ domains. It seems plausible that one can obtain finite charge solutions by taking appropriate quotients of the global G\"odel space, but we do not have a clear view on how to achieve this at the moment.

\section{Joining G\"odel space to AdS space}\label{s:GB-Godel_AdS}

\begin{figure}[ht!]
\centering
\includegraphics[width=.8\textwidth]{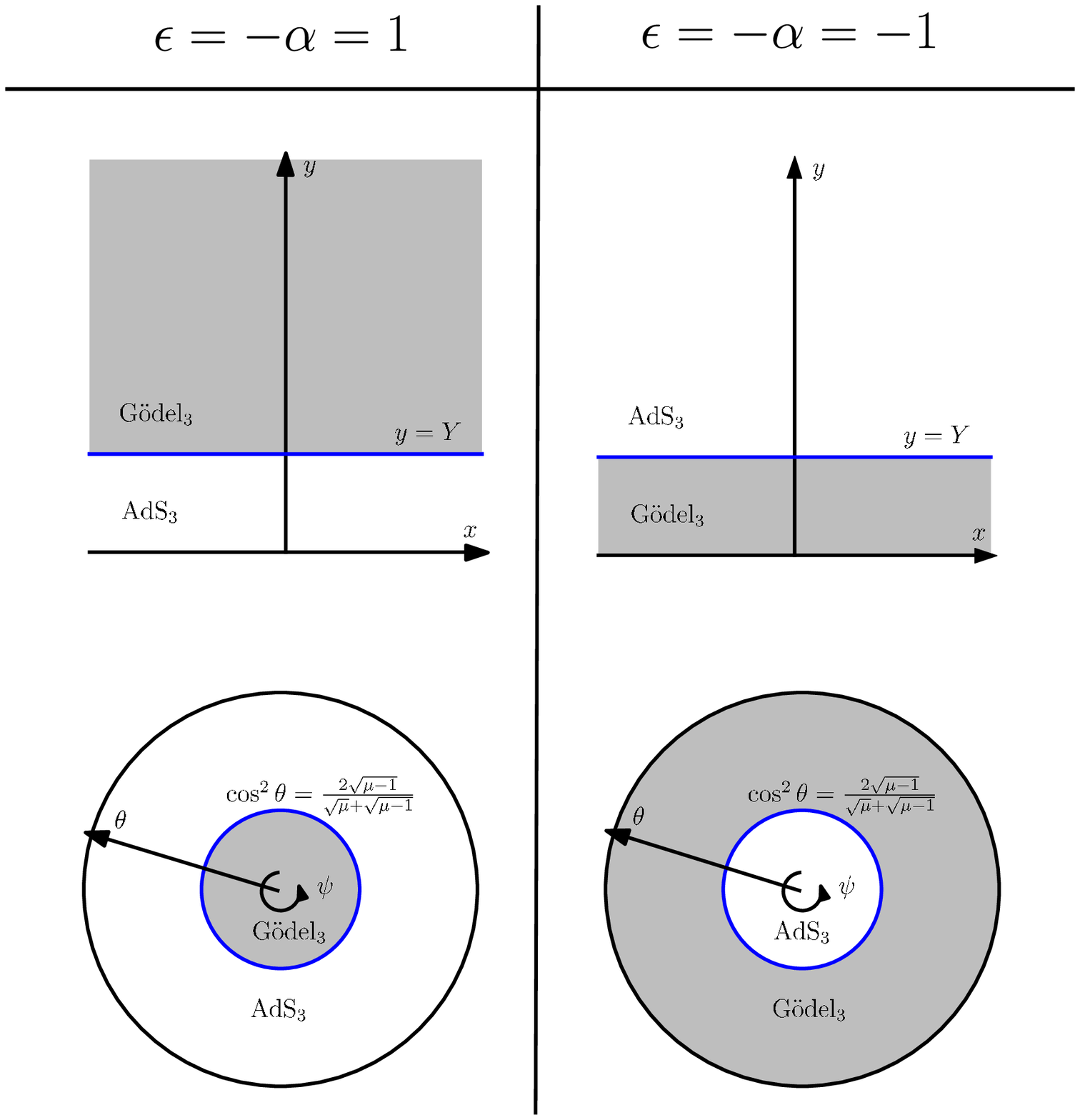}
\caption[Overview of the different gluings of G\"odel$_3$ to $\AdS 3$.]{Overview of the different gluings. On top we have presented the picture in the upper half plane coordinates $x, y$, most natural to G\"odel space at $t=cst$.  On the bottom the glued \spacetime{} is presented in global AdS coordinates. The disk shown here is a $\sigma=$cst slice of the cylinder that is $\AdS 3$. As discussed in our article \cite{Levi:2009az}, in these coordinates one still has to make an identification, inducing CTCs. Note that we have introduced the standard Penrose coordinate $\theta$, defined by $\tan\theta=\sinh \r$. The left hand side is that for the choice of  $\alpha=-1$. In this case the AdS part of space connects to the boundary, but the domain wall has negative tension. On the right hand side the situation is depicted for the opposite choice. Now the G\"odel part connects to the boundary and the domain wall has positive tension and an interpretation as smeared M2-particles. (the constant $\epsilon$ was introduced in \cite{Levi:2009az}, but can be neglected in the present discussion)}\label{gluedd}
\end{figure}

One of the original motivations in studying the three-dimensional system (\ref{3daction}), was to analyze solutions corresponding to branes wrapped around the $\sS 2$ of an $\AdS 3\times \sS 2$ geometry. However, we found a solution of the form G\"odel$_3\times \sS 2$, which no longer has AdS asymptotics. Both from the point of view of holography and from the black hole microstate motivation it would be interesting if there was some kind of `embedding' of these solutions into an asymptotic
AdS \spacetime{}. Probably the most straightforward way of realizing such a setup is by enclosing a G\"odel region carrying the M2-charge by a domain wall that cancels this charge. Then, as in three
dimensions all vacuum \spacetimes{} are locally $\AdS 3$, on the other side of the wall we are guaranteed to find a local $\AdS 3$ \spacetime{}. In \cite{Levi:2009az}, we realized exactly this idea, although it turns out that, under our assumptions, demanding that the AdS-side of the domain wall is connected to the boundary is equivalent to having a negative tension domain wall. In the case we have the G\"odel part of \spacetime{} on the outside then the domain wall is made up of more familiar positive tension, smeared out M2-branes. For an overview see \figref{gluedd}.
 
Another motivation to consider such a domain wall construction is the analogy to \cite{Boyda:2002ba,Drukker:2003sc,Gimon:2004if}. In these references the authors show that one can remove the closed timelike curves of G\"odel-like \spacetimes{} by introducing a domain wall that connects it to an AdS-like spacetime. Naively one would hope the same effect to take place in the current setup. However, for our setup, this seems not to be the case, as closed timelike curves are present in the \spacetime{} even after introducing the domain wall, as we show below. We were not able to find a more general construction that eliminates the closed timelike curves, see below. This however remains an interesting goal for future research. 

%

\begin{figure}[ht!]
\centering
\begin{picture}(200,100)
 \put(0,85){\epsfig{figure=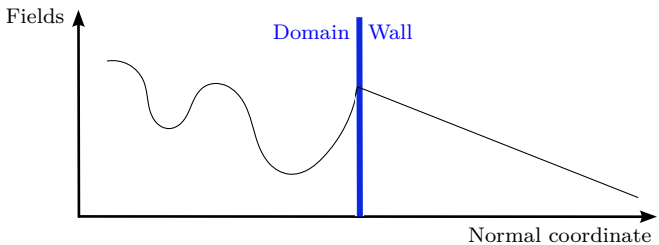,scale=.5,angle=-90}}
\put(-25,80){\footnotesize Fields}
\put(150,-3){\footnotesize Normal coordinate}
\put(76,74){\footnotesize \textcolor{blue}{Domain}}
\put(112,74){\footnotesize \textcolor{blue}{Wall}}
\end{picture}
\caption[Profile of fields across a domain wall (cartoon).]{Fields are continuous across a domain wall, but need not be differentiable. This means the domain wall acts as a source for the field equations.}
\label{fig:GB-DomainWall}
\end{figure}

\paragraph{Placing a domain wall:}
The idea is to look for the possibility of having a domain wall (a two-dimensional hypersurface, dividing three-dimensional \spacetime{} in two regions), such that on one side of the domain wall we have G\"odel space, and on the other side we have a locally AdS space. We ask the metric and the complex scalar to be continuous as we go across the domain wall, to have a good solution. However, the fields do not have to be differentiable as we go across the wall: there can be a jump in derivatives of the fields, inducing diverging second derivatives (delta-functions). For a cartoon, see figure \ref{fig:GB-DomainWall}.
We conclude the field equations can have delta-function singularities at the location of the domain wall. To cancel these delta-functions, we need to place source terms at the location of the domain wall.  We demand these source terms carry M2 brane charge, in order to cancel the M2 brane charge carried by the G\"odel region. Therefore, we propose a domain wall made up out of M2 branes. Since M2 branes couple to the Calabi-Yau volume $\tau_2$, we also propose to put the source branes on constant $\tau_2$. In the upper half plane coordinates of G\"odel space, we put the domain wall at $y=\tau_2 = Y$, with $Y$ a constant.

\paragraph{The glued solution:} In \cite{Levi:2009az}, we performed the calculation for a domain wall with the properties above (with M2 brane charge, at $y=Y=cst$). We do not present the details of the calculation, we summarize the results. Denote the tension of the domain wall as $\alpha$. We take units such that $\alpha$ is normalized to have norm one, but we allow negative tension $(\alpha=\pm1)$. We find two possibilities, 
summarized in figure \ref{gluedd}.

The case we would be most interested in is that of negative $\alpha$, as then the glued \spacetime{} is asymptotically AdS and we know how to do holography on such spaces. However, $\alpha=-1$ implies negative tension for the domain wall. Even though negative tension domain walls are not unheard of, either in supersymmetric theories (see e.g. \cite{Shifman:1999ri}), or as orientifold-type objects in string/M theory (see e.g. \cite{Dai:1989ua,Horava:1995qa}), clearly it is harder to interpret them in terms of fundamental M-theory branes. It might still be interesting to understand these glued spaces in more detail through holography. Moreover, we found that the solution with G\"odel glued to AdS, involves making a certain identification on the coordinates, that induces more CTCs. Instead of losing the bad properties of G\"odel space, we run into more trouble and do not have a well-behaved solution. Finding a well-behaved gluing with the desired properties (asymptotic AdS, no CTCs) remains a goal for future research.

\outlook{Note:}{as an aside, we attempted a similar gluing of G\"odel space to AdS space, without the presence of the scalar $\tau$. In this case, there is no gauge field and we do not need to require the domain wall to carry any charge. We were able to successfully perform the gluing in this case. Even though the interpretation in terms of the eleven-dimensional system above is lost, it is still a nice result. It is given in short in the next chapter.}

\section{Conclusions and future directions}\label{sec-con}

In this chapter, we have constructed supersymmetric solutions to three-dimensional axion-dilaton gravity with negative cosmological constant
which describe the backreaction of $\sS 2$-wrapped M2-branes in M-theory. We found a class of solutions where the axion-dilaton is holomorphic and where the local geometry is that of the three-dimensional G\"{o}del universe. These solutions preserve four supersymmetries in agreement with the analysis in the probe approximation.  We have also shown that our solutions can be glued, in a supersymmetric manner, into asymptotically $\AdS 3$ geometries by including a charged domain wall.

Let us comment on some aspects which deserve a better understanding and some interesting directions for future research. A first puzzle is that our backreacted solutions have M2-brane sources only on the boundary, whereas in the probe approximation discussed in the  section \ref{s:Motive}, it appeared as if the M2-branes could be placed anywhere. This could point to the existence of more general solutions with sources in the interior, but it could also be due to the fact that these are codimension-two objects producing long-range fields; hence the probe picture might be unreliable. Another feature of our solutions is that the brane charge residing on the boundary is actually infinite. This can be seen as a consequence of the fact that $\T$ takes values in the entire upper half plane. One way to obtain a `stringy' finite charge solution, would be to identify values of $\T$ related by $SL(2,\Z )$ transformations and make a similar identification on the coordinate $z$ of the base manifold. The M2 brane charge  would then be finite. A similar procedure would work for an arithmetic subgroup of $SL(2,\Z )$. One should note that such constructions in general involve identifications generated by timelike vectors and will produce  more closed timelike curves. Nevertheless, such configurations are finite-energy, finite charge BPS solutions, and one would expect them to contribute to the path integral. It would be interesting to  understand their role  better.

Our original motivation for studying this system was the black hole microstate or deconstruction proposal \cite{Gaiotto:2004pc,Gaiotto:2004ij,Denef:2007yt,Gimon:2007mha}, where it was argued that $\sS 2$ wrapped M2-brane probes have a large quantum mechanical degeneracy (coming from lowest Landau levels on the internal Calabi-Yau) that can account for the black hole entropy. An interesting question is whether this degeneracy can also be understood after including the backreaction. This might furthermore clarify the relation between these deconstruction states and other BPS solutions carrying the black hole charges that are more closely related to the original fuzzball proposal \cite{Mathur:2005zp, Balasubramanian:2008da,Mathur:2008nj}. Although various BPS solutions were explicitly constructed, see e.g. \cite{Lunin:2001jy,Giusto:2004kj,Giusto:2004ip, Giusto:2004id,Bena:2004de,Berglund:2005vb,Bena:2005va,Balasubramanian:2006gi,Rychkov:2005ji}, it was argued recently that these might only account for a subleading fraction of the black hole ensemble \cite{deBoer:2008zn,deBoer:2009un}. It would in particular be interesting to understand the deconstruction microstates in a dual CFT and see if and how they evade the bound of \cite{deBoer:2009un}. Our gluing procedure in the previous section is a first attempt in this direction.  It would be interesting to explore more general gluings into $\AdS 3$ and their holographic interpretation.

\cleardoublepage
\chapter{
\mbox{Aside: Relating chronology and} unitarity through holography}\label{c:GB}\label{app:HolUnit}

\punchline{We consider a ball of homogeneous, rotating dust in global $\AdS 3$  whose back\-reaction produces a region of G\"odel space inside the ball. We find the geometry outside of the dust ball and compute its quantum numbers in the dual CFT. When the radius of the dust ball exceeds a certain critical value, the \spacetime{}  contains closed timelike curves. Our main observation is that precisely when  this critical radius is exceeded, a unitarity bound in the dual CFT is violated, leading to a holographic argument for chronology protection.}

\paragraph{Relation to chapter \ref{c:TL}.}
In the previous chapter, we discussed a three-dimensional system of gravity, with a negative cosmological constant, coupled to a complex scalar $\tau$. We found a solution, three-dimensional G\"odel space, that has closed timelike curves (CTCs). A first try to eliminate the CTCs by gluing G\"odel space to a locally anti-de Sitter space failed, as discussed in the previous chapter. In this chapter, we discuss a similar idea, that gets rid of the CTCs and leads to an interesting observation (see below). 

We study the toy model of a three-dimensional combined gravity and matter system with negative cosmological constant, where we do not specify the microscopic matter content in detail. We assume that the matter sector can effectively produce a source of pressureless dust, but we do not consider any more dynamical equations it could come from. (In terms of the discussions of the previous chapter, we do not consider the scalar equation of motion for $\tau$, but only consider its effect as a source to the Einstein's equations.)

\section{Introduction}

Kurt G\"odel  was the first to emphasize that Einstein's equation, in the presence of seemingly innocuous matter sources, can lead to causality violating geometries containing closed timelike curves (CTCs) \cite{Godel:1949ga}. Since then, classical solutions with CTCs have popped up ubiquitously, including supersymmetric versions of G\"odel space in supergravity theories, both in 3+1 dimensions as well as in their higher-dimensional parent theories \cite{Gauntlett:2002nw}. Such \spacetimes{} lead to a variety of pathologies, both within classical general relativity as well as for interacting quantum fields propagating on them (see \cite{Friedman:2008dh} for a review and further references). This led Hawking to propose the Chronology Protection Conjecture, stating that regions containing CTCs cannot be formed in any physical process \cite{Hawking:1991nk}. It is expected \cite{Kay:1996hj} that a fully consistent treatment of such a dynamical argument behind chronology protection requires the issue to be addressed in a quantum theory where both matter and gravity itself are quantized.

The AdS/CFT correspondence \cite{Maldacena:1997re} proposes that combined quantum gravity and matter systems on anti-de-Sitter (AdS) spaces have a holographic dual description in terms of a unitary conformal field theory (CFT) in one lower dimension. It is therefore ideally suited to study the issue of chronology protection in asymptotically AdS spaces. Indeed, several examples are known \cite{Herdeiro:2000ap} where the appearance of CTCs in a BPS sector of the bulk theory is quantum mechanically forbidden as it would correspond to the violation of a unitarity bound in the dual CFT. In this chapter, we show that a similar conclusion holds for 2+1 dimensional G\"odel space  and give a simple argument that creating a patch of G\"odel space large enough  to contain CTCs would require violating unitarity in the dual CFT. The important novel feature is that the results do not rely on supersymmetry, but only on the  general properties of gravity theories on $\AdS 3$ that were established in \cite{Brown:1986nw,Strominger:1997eq}. 

\section{\ansatz{} for a G\"odel-AdS solution with dust}
We consider Einstein's
equation with negative cosmological constant $\L = -1/l^2$: 
\be
R_{ab} - \half R g_{ab}- \frac{1}{l^2}g_{ab}= 8 \p G_3 T_{ab} \,
,\label{einsteineqns2} 
\ee
 where $G_3$ is the (2+1)-dimensional Newton constant, $l$ is the $\AdS 3$ radius and we take 
\be 
T_{ab} =\frac{\rho}{2\pi G l^2} u_a u_b\, , \label{dustsource} 
\ee 
with $u$ a unit timelike vector and $\rho$ a dimensionless
number parameterizing the energy density. 

We solve Einstein's equations (\ref{einsteineqns2}) for a homogeneous ball of rotating dust, where we take the energy density $\rho$ to be nonzero and constant inside the ball and zero outside. Inside the ball, the metric is that of G\"odel space, while outside  we expect a metric of generalized  BTZ type characterized by a mass $M$ and angular momentum $J$.

\begin{figure}[p!]
\centering
\subfigure[]{
\begin{picture}(80,150)
\put(-40,-20){\includegraphics[width=.33\textwidth,angle=30]{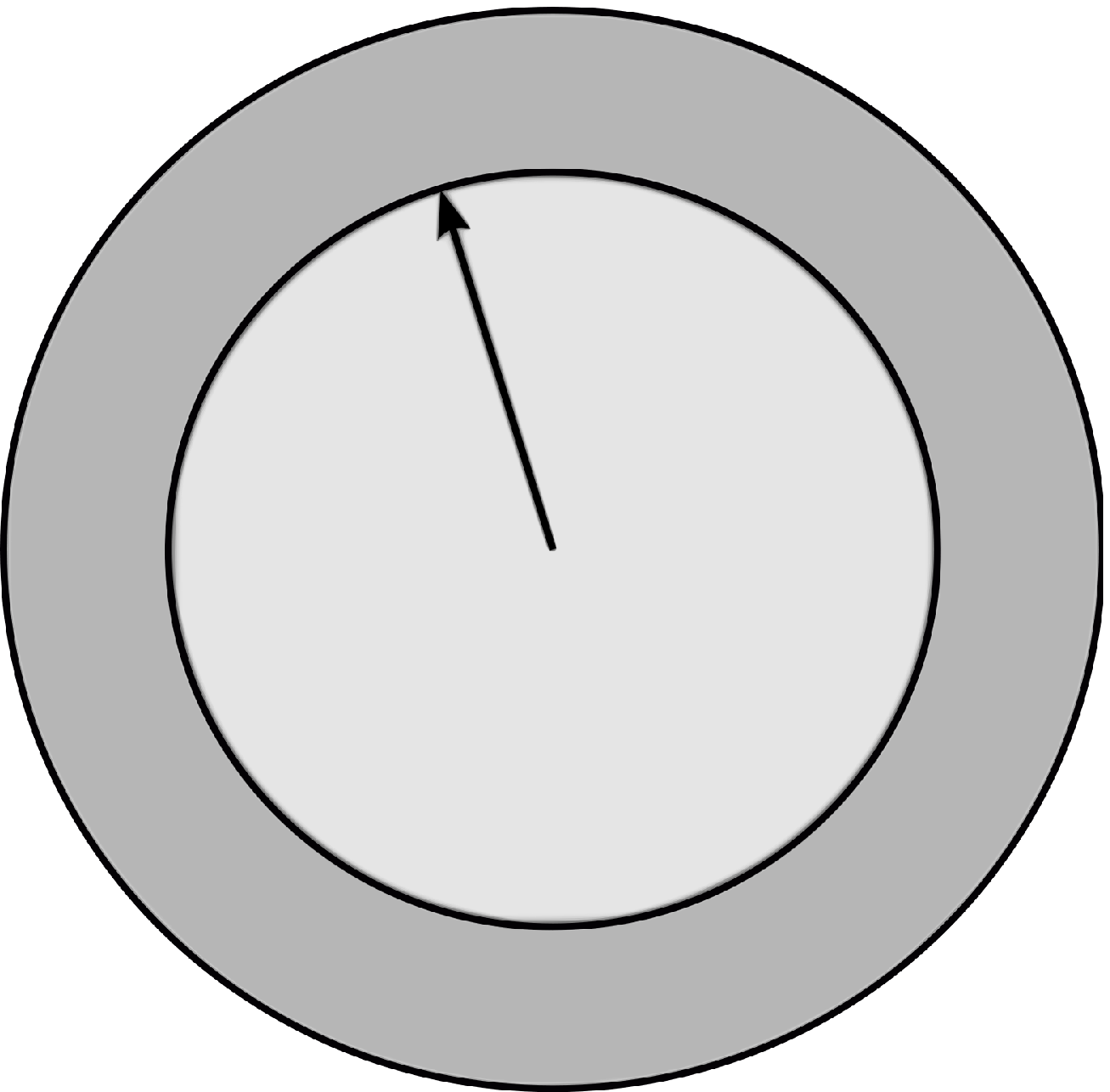}}
\put(20,80){\footnotesize$r_\star =1/\sqrt{1-\rho}$}
\end{picture}
}
\hspace{.2\textwidth}
\subfigure[]{
\begin{picture}(110,100)
\put(-10,0){\includegraphics[width=.33\textwidth]{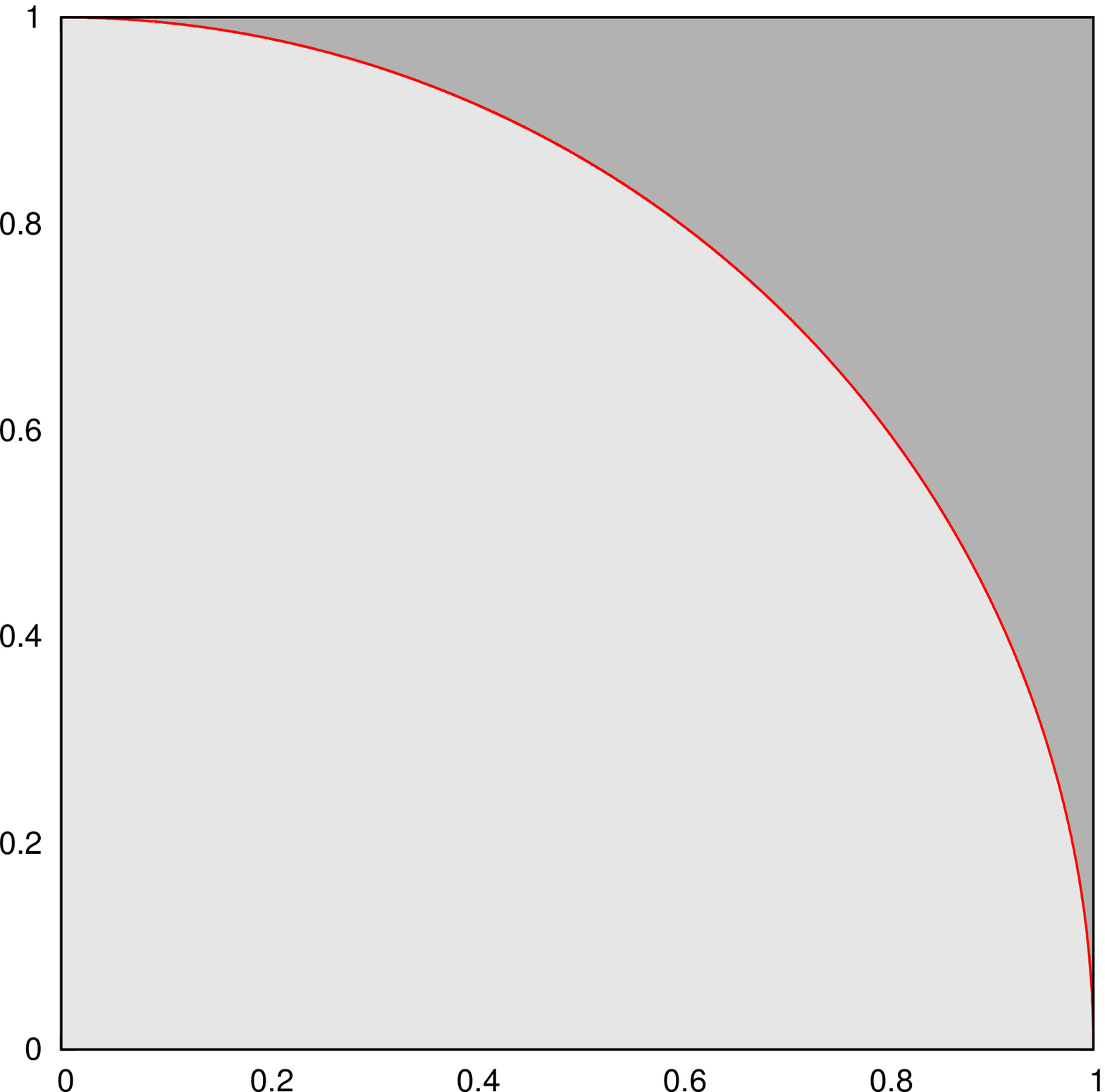}}
\put(-15,110){$\rho$}
\put(100,-7){$r_0$}
\end{picture}
}
\caption[The appearance of CTCs in the G\"odel region.]{The appearance of CTCs in the G\"odel region in terms of the physical parameters $r_0$ and $\rho$. On the left: when $r_0> 1/\sqrt{\mu}$, or equivalently $r_0>\sqrt{1-\rho}$, there are CTCs in the shaded region of the geometry. Figure on the right: the $(r_0,\rho)$ parameter space. Again, the (darkest) shaded region corresponds to having CTCs in the G\"odel space.}
\label{fig:GB-Goedel_CTCs}
\end{figure}

\paragraph{Inner (G\"odel) region}
On the inside, we have the G\"odel  space metric \be
ds_-^2=l^2\left[-(d t+ \m\frac{ r^2}{(1- r^2)}d\f)^2 + \m\frac{d
r^2+ r^2d\f^2}{(1- r^2)^2}\right]\,,\label{Godelcoords2} \ee where
$r$ runs between 0 and $r_0 \leq 1$, the radius where the dust
region ends. The angular coordinate $\phi$ is identified with
period $2 \pi$. The Einstein equations determine $\mu$ in terms of
the energy density $\rho$ of the dust as \be \mu ={ 1\over 1-\r
}\label{enden2} \, . \ee The physical values are $\rho \geq 0$, for
positive energy, and $\rho
<1$, for a Minkowski signature of the resulting metric. Note that for $\m = 1$, $(\rho=0)$, the metric describes
global $\AdS 3$.  When $r_0$ exceeds the critical value $1/\sqrt{\m}$, CTCs appear since
$\pa_\phi$ becomes a timelike vector in the region $r>
1/\sqrt{\m}$. We depict the regions with CTCs in figure \ref{fig:GB-Goedel_CTCs}, both for the geometry and in $(r_0,\r)$ parameter space.

\paragraph{Outer (locally AdS) region}
Outside of the dust ball, we take a metric \ansatz{} which is  a
vacuum solution to (\ref{einsteineqns2})  and  which generalizes
the BTZ metric: 
\bea ds_+^2 &=& l^2\left[ -  (u - M) d \tilde t ^2
+ J d \tilde t d \tilde \f + u d \tilde \f^2 + { d u^2 \over 4 f(u)}
\right]\,, \nonu f(u) &=&  u^2 - M u + { J^2 \over 4}. \label{genBTZ}
\eea 
The angle $\tilde \phi$ is identified with period $2\p$ and
the real parameters $M,\ J$ are the ADM mass and angular momentum
(in convenient units) respectively. This metric is taken to hold for values $u\geq u_0$. Below, we relate $u_0$ to $r_0$.

\begin{figure}[p!]
\centering
\setlength{\unitlength}{1.1pt}
\subfigure[]{
\begin{picture}(153,130)
\put(0,0){\includegraphics[width=.44\textwidth]{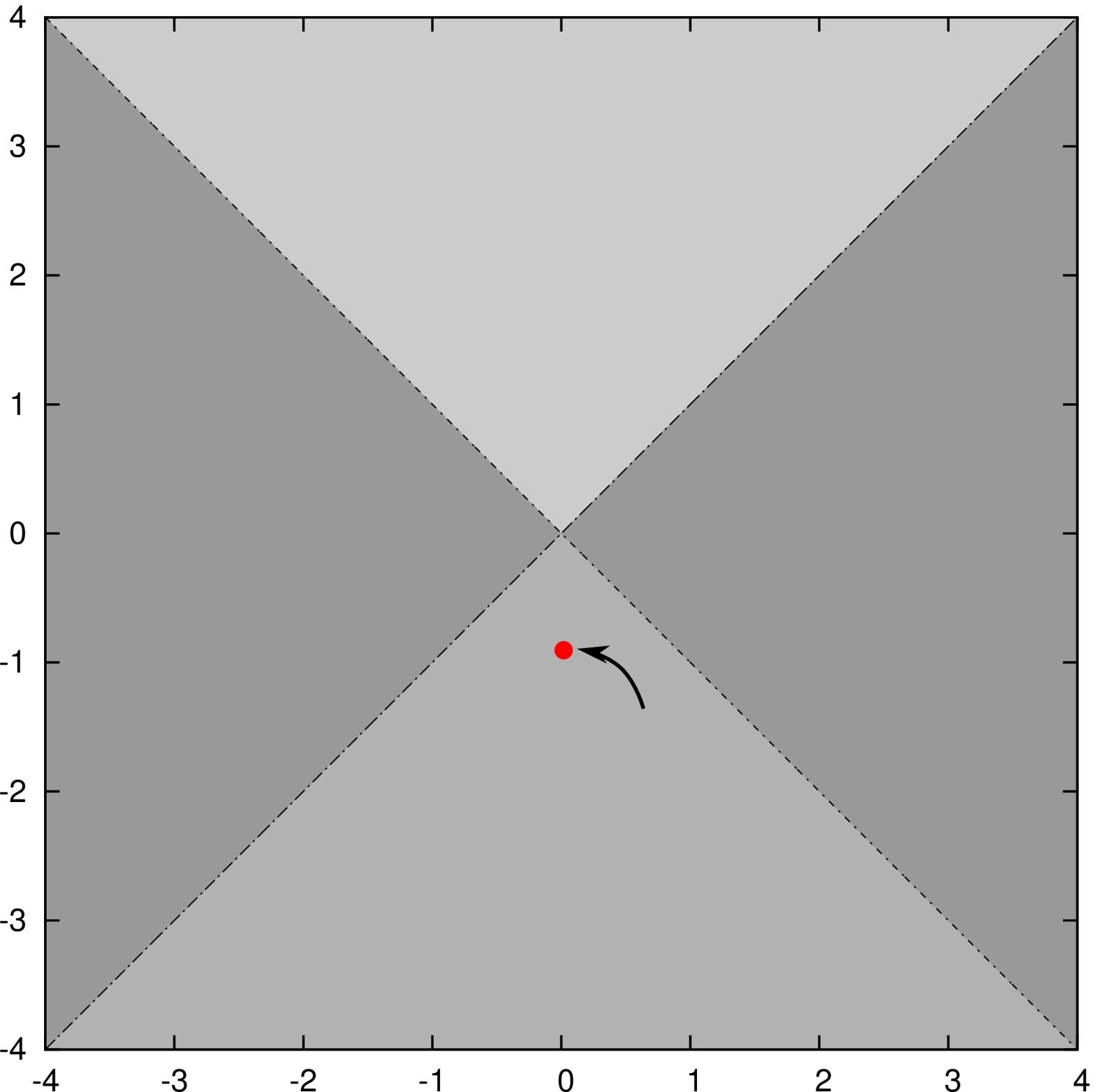}}
\put(130,-10){$J$}
\put(-10,130){$M$}
\put(70,40){$\AdS 3$}
\put(53,120){Region I}
\put(53,20){Region II}
\put(10,70){Region III}
\put(90,70){Region III}
\end{picture}
}
\subfigure[]{
\begin{picture}(153,100)
\put(0,0){\includegraphics[width=.44\textwidth]{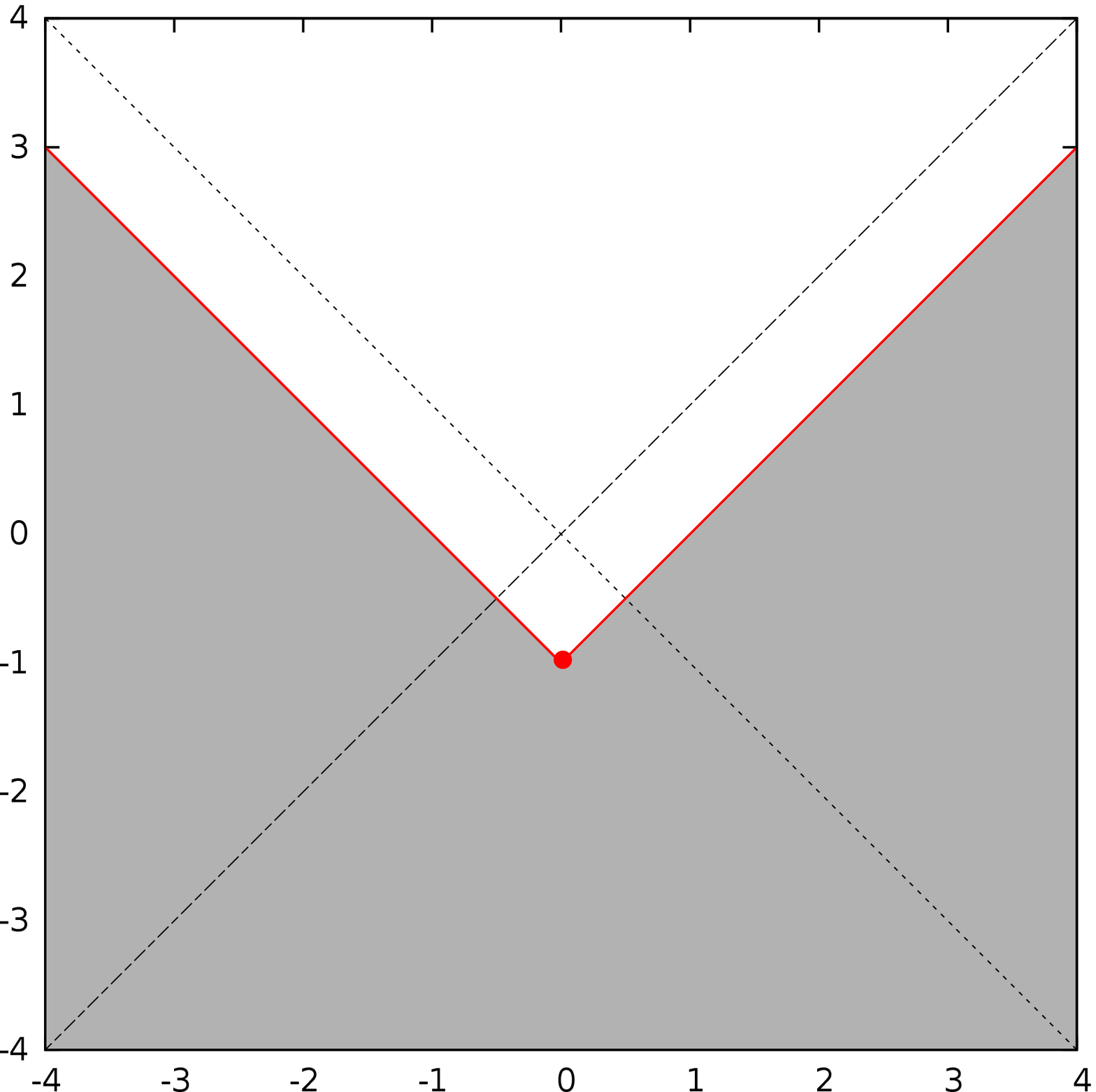}}
\put(130,-10){$J$}
\put(-10,130){$M$}
\end{picture}
}
\caption[The three types of locally $\AdS 3$ geometries.]{Left: the three different types of geometries described by the locally AdS metric \eqref{genBTZ}, in terms of the two parameters $(J,M)$. Note that global AdS corresponds to $(J=0,M=-1)$. Right: the same $(J,M)$-parameter space. The geometries in the shaded region correspond to a non-unitary dual CFT.} 
\label{fig:GB-AdS_Regions}
\setlength{\unitlength}{1pt}
\end{figure}

We discuss the properties of this  class of metrics  in $(J,M)$ parameter space. We have three regions: the locally AdS metric describes either the BTZ black hole, a spinning conical defect, or an overspinning object, see also figure \ref{fig:GB-AdS_Regions}.

\begin{itemize}
\item[\ra] {\bf Region I:} $M>|J|\geq0$. The metrics (\ref{genBTZ}) describe BTZ black holes
\cite{Banados:1992wn}. The coordinate $u$ ranges over $\Real$. The function $f$ has two positive real
zeroes, which correspond to the inner and outer horizons. 
(In the region $u \geq 0$,  the standard
radial BTZ coordinate is related to $u$ as $u = r^2_{\rm BTZ}$.)
\item[\ra] {\bf Region II:}  For $-M<|J|\leq0$, henceforth referred to as region II,
the metric describes a spinning conical defect. The function $f$
has two negative real zeroes, between which the signature of the
metric becomes Euclidean. One can verify that at the largest zero
$u_+$, the metric has a conical singularity arising from a
pointlike source. The range of the
$u$-coordinate is $u \geq u_+$ and, as before, there are CTCs in
the region where $u_+ \leq u<0$. 
 There is one exceptional point,
namely $M = -1,\ J=0$, for which the geometry becomes smooth global $\AdS 3$.
\item[\ra] {\bf Region III:} $-|J|<M<|J|$. The metric
describes an overspinning object. The function $f$ has no real
zeroes and the metric is free of curvature singularities, while $u$ ranges over the real line. The space contains a `naked' CTC
region for negative values of $u$.
\end{itemize}

Let us also briefly discuss part of the  AdS/CFT dictionary.  One finds that the outside spaces (\ref{genBTZ}) correspond to states with conformal weights\footnote{The Virasoro quantum numbers of the spaces (\ref{genBTZ}) can be extracted following the standard procedure of computing the renormalized boundary stress tensor and extracting its Fourier coefficients \cite{Balasubramanian:1999re}.}
\bea L_0 ={c \over 24} ( M+ J + 1 )\,,\qquad \bar L_0  ={c \over 24} ( M-J +1)\,. \eea The central charge of the CFT is given by  $ c = { 3 l \over 2 G}\label{BH} $ \cite{Brown:1986nw}.
Unitarity implies that conformal weights in the CFT are positive,
leading to the bound $L_0 \geq 0, \bar L_0 \geq 0$. In terms of
$M$ and $J$, this is equivalent to 
\be M+1 \geq
|J|\,.\label{unitbound} \ee 
States violating this bound are forbidden
by unitarity and, according to the AdS/CFT conjecture, cannot be
part of the spectrum in a consistent quantum gravity theory on
$\AdS 3$. Figure \ref{fig:GB-AdS_Regions} shows the region where this bound is violated in $(J,M)$ parameter space.

\paragraph{Matching the regions.}
We match the two geometries at $r=r_0$ to $u = u_0(r_0)$ by solving the Einstein equations across the shell $r=r_0$.  This matching gives $(M,J,u_0)$ in terms of the physical parameters of the solution, $r_0$ (radius of the G\"odel patch) and $\rho$ (the energy density). After some algebra (see \cite{Raeymaekers:2009ij}) one can check that there are 2 solutions, by choosing the angular momentum $J$ to be either positive or negative.
\bea
J &=& \pm \frac{2\,\rho\, r_0^4 }{(1-r_0^2)^2 (1-\rho )^2}\label{J}\,,\\
M &=& -\frac{(1-\rho )^2-2 r_0^2 \left(1-\rho ^2\right)+r_0^4 \left(1+\rho ^2\right)}{(1-r_0^2)^2 (1-\rho )^2}\label{M}\,,\\
u_0 &=&  \frac{r_0^2 (1-\r-r_0^2)}{(1-r_0^2)^2 (1-\rho )^2}\label{u0}\,.
\eea 
We fix this freedom by taking $J$ to be positive (the positive sign in the equations).

\begin{figure}[ht!]
\centering
\begin{picture}(330,140)
\put(0,5){\includegraphics[width=.35\textwidth]{RhoR0Red.eps}}
\put(-10,125){$\rho$}
\put(123,-5){$r_0$}
\put(147,70){\vector(1,0){30}}
\put(190,0){\includegraphics[width=.4\textwidth]{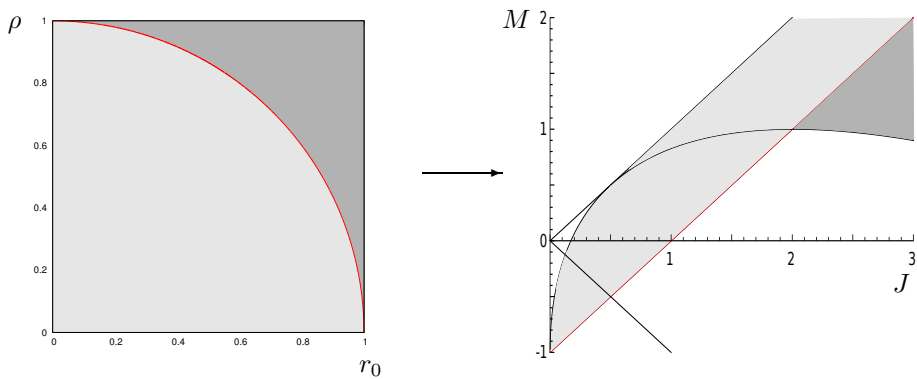}}
\put(177,125){$M$}
\put(325,25){$J$}
\end{picture} 
\caption[Solutions in $(M,J)$-space and causality/unitarity correspondence.]{The range of solutions in $(M,J)$-space that is traced out by all allowed possibilities for $(r_0,\rho)$ through eqs.~\eqref{J} and \eqref{M}. The darkest shaded region contains CTCs. We only give a specific range in $(J,M)$-space, the region with $J<0$ is the mirror image around the vertical $M$-axis.}
\label{fig:GB-MJ_intermsof_r0r}
\label{jmfig}
\end{figure}

As the range of the physical parameters is $0\leq r_0,\rho \leq 1$, one finds that the solutions above only give rise to spinning conical defects and overspinning objects. This can be seen by considering the map $(r_0,\rho)\to(J,M)$. Details can be found in \cite{Raeymaekers:2009ij}, we just give the pictorial representation in figure \ref{fig:GB-MJ_intermsof_r0r}.

\section{Chronology protection in gravity from unitarity in CFT}

The key observation is that, as can be seen from figure \ref{jmfig}, the region with CTCs in the geometry corresponds exactly to a violation of the unitarity bound in the dual CFT. We discuss this a little bit more.

\begin{itemize}
 \item\emph{\underline{CTCs in our matched solutions.}}
 Remember that a priori we have two regions  in which closed timelike curves can appear in the solution. The inside part of the metric (G\"odel space) has CTCs when $r_0>\frac{1}{\sqrt{\m}}=\sqrt{1-\rho}$, the outside metric when $u_0<0$. But observe that by (\ref{u0}) these two conditions are equivalent, hence either no closed timelike curves appear at all, or they appear both in the inside and outside parts of the metric. In parameter space these closed timelike curves can only appear in the darkest shaded region in figure \ref{jmfig}. 
\item\emph{\underline{The unitarity bound}} $M+1>|J|$ (eq.~(\ref{unitbound})). This gives an extra constraint on which outside metrics are physically acceptable. Taking a look at the relation between $M,J$ and $r_0,\rho$, we see that the bound for the absence of CTCs $r_0^2 \leq 1-\rho$  precisely coincides with the unitarity bound $M+1\geq |J|$. This can be seen directly as by (\ref{J}), (\ref{M}) \be M+1-|J|=\frac{4 \rho\,r_0^2 (1-\rho -r_0^2 )}{(1-r_0^2)^2 (1-\rho )^2}\,.\ee 
\end{itemize}
Hence our main observation: \textit{the condition of unitarity is equivalent to that of the absence of
closed timelike curves. }

\section{Outlook}

In this chapter, we have discussed an example where the appearance of CTCs in a G\"odel region within  $\AdS 3$ was shown to precisely coincide with the violation of a unitarity bound in the dual CFT. It was important that the G\"odel region was supported by pressureless dust such that we only needed to solve the Einstein equations (there are no other fields, e.g.~no scalar field  that supports the dust as in the previous chapter). Moreover, our observation does not rely on supersymmetry, contrary to similar discussions in the literature.

Based on our result and other examples in the literature \cite{Herdeiro:2000ap}, it would be natural to propose an AdS version of the Chronology Protection Conjecture, stating that regions with CTCs in AdS spaces cannot be formed as a result of any unitary process. The AdS/CFT correspondence could  in principle be used to address whether this proposal is true in general. If so, it would be very interesting to gain insight into the deeper dynamical mechanism that prevents the formation of regions with CTCs, see e.g. \cite{Drukker:2003sc} for some proposals in the context of string theory.

\emph{Note added:} In a recent work, we were able to find such a dynamical mechanism, explaining the removal of CTCs in the three-dimensional system studied here \cite{Raeymaekers:2010re}. The removal follows from an embedding of the three-dimensional system, studied in this chapter, in ten-dimensional type IIB string theory. The responsible mechanism is the condensation of extra light degrees of freedeom that exist in string theory, but do not reveal themselves in the (super)gravity approximation. We found that those degrees of freedom are 7-branes, wrapped on an internal 7-manifold. Those 7-branes are particles from the three-dimensional point of view. They condense on the $r_0$-circle, where they become massless and excise the problematic region of G\"odel space, much llike above. In this way, the results above relating unitarity in a dual field theory to absence of CTCs in bulk spacetime apply in that more intricate setup as well.

\index{AdS!see Anti-De Sitter}
\index{Anti-De Sitter!see AdS}

%
\cleardoublepage
\part{Conclusions and Appendix} 
\label{pt:Conc}
\cleardoublepage
\chapter{
Conclusions}\label{c:Concl}

\punchline{This chapter gives concluding remarks on the topics discussed in this thesis. These topics are the two main research fields mentioned at the introduction and a third one, which unexpectedly followed from the original motivations. For detailed conclusions with outlook, we refer to the conclusions to each chapter. Here, we only give the main idea of the research topics dealt with in this work, we highlight which were the new contributions and we give the most important pointers for future work.}

\section*{First-order formalism for black holes and beyond}
Due to the constraints by supersymmetry, the equations of motion for the many scalar fields that arise in (super)gravity for spherically symmetric black hole and $p$-brane solutions can be formulated as first-order equations in terms of a superpotential $\cw(q)$:
\begin{equation}
 G_{ij} \dot q^j = \pf {\cw(q)}{q^i}\,.\label{eq:CC-1stOrderEq}
\end{equation}
The $q^i$ are the scalar fields $\phi^a$ of the supergravity theory and a function $U$ in the metric (redshift factor). 
We investigated when eqs.~of the form \eqref{eq:CC-1stOrderEq} arise for non-supersymmetric solutions and what the necessary conditions are. New results:
\begin{itemize}
\item[$\rhd$] For gravity, coupled to a gauge field and one scalar field (a dilaton) in arbitrary number of \spacetime{} dimensions $D$, all black hole and $p$-brane solutions are found from first-order equations of the form \eqref{eq:CC-1stOrderEq}. 
\item[$\rhd$] Before, the assumption was made that the superpotential should take on the split form  $\cw(U,\phi^a) = \e^U W(\phi^a)$.\footnote{This intuition comes from the structure of the superpotential for supersymmetric black holes $\cw \sim \e ^U |Z|$, with $|Z|$ the central charge function.} We have shown that for general non-supersymmetric black holes, this factorization property does not hold.
\item[$\rhd$] The necessary condition: the description for the scalar fields must be integrable in the sense of Liouville (for $n$ scalar fields $q^i$, their must be $n$ constants of motion that commute under the standard Poisson brackets).
\item[$\rhd$] A method of principle to construct $\cw$ for scalar manifolds that are symmetric spaces illustrated in an explicit construction of $\cw$ for the non-extremal, non-supersymmetric dilatonic black hole.
\end{itemize}
There are new results \cite{Chemissany:2010zp} that clearly prove that exactly for scalar target spaces that are symmetric spaces, the system is Liouville integrable. This covers all supersymmetric and non-supersymmetric solutions to supergravity theories in four dimensions with $\cN>2$ and some with $\cN \leq 2$. What remains to be done is to give a construction of $\cw$ in all these cases and investigate it properties. For non-extremal black holes, this is still largely unexplored. A related question is: what about multi-center configurations? For supersymmetric solutions, these can be obtained from a set of first-order equations \cite{Denef:2000nb}. Maybe some of these results are applicable to more general multi-center solutions. A hint is given in \cite{Galli:2009bj}.

\section*{Black hole microstate in (super)gravity regime}

Ultimately, the question of what a black hole microstate is can only be answered in a full quantum theory of gravity. It is in this light that the fuzzball proposal of Mathur should be understood. Luckily, there is a large class of these `microstates' that are solutions to the classical equations of motion of the theory, i.e.~they are well described by classical supergravity. However, these classical microstates should be properly quantized to be able to explain the black hole entropy.  Nevertheless, much can be learned from the classical description. We focus on the interpretation as multi-center configurations in four dimensions. New results are:
\begin{itemize}
\item[$\rhd$] An exact relation between fuzzball state of five-dimensional black rings and four-dimensional multi-center configurations. The latter configurations are microstates of a two-centered solutions in four dimensions, with total D4-D2(-D0) charge.
\item[$\rhd$]  M2 probe states in an $\AdS 3 \times \sS 2 \times \CY 3$ background of eleven-dimensional supergravity play a role in a successful entropy counting for the D0-D4 black hole. We found the backreaction of this system is of the form $\text{ G\"odel}_3\times \sS 2 \times \CY 3$, where $\text{G\"odel}_3$ is the three-dimensional G\"odel universe.
\item[$\rhd$] The latter leads to a supersymmetric embedding of three-dimensional G\"odel space in M-theory, which was not done before.
\end{itemize}
An entropy counting for the D0-D4 black hole was not obtained. The reason is that we could not link G\"odel$_3$ to an asymptotically AdS \spacetime{}. Towards future research, it would be of importance to better understand the backreaction of these states, to shed light on the important question whether or not the entropy of the D0-D4 black hole can be understood from supergravity solutions alone, or one needs the full (quantum) string theory. This would shed light on the validity of using mainly classical geometries in the fuzzball and related proposals.

\section*{Causality in gravity and unitarity in quantum mechanics}

There have been many examples constructed of seemingly well-behaved matter sources that give rise to \spacetime{} geometries with closed timelike curves (CTCs). However, the presence of CTCs leads to problems with causality. This has led Hawking to state the Chronology Protection Conjecture: regions of \spacetimes{} with CTCs should not be formed in physical processes. This conjecture is similar to the Cosmic Censorship Conjecture of Penrose, which says that naked singularities should not be formed in nature. Both are physical requirements that, on the level of general relativity, \textit{one has to put in by hand}, to exclude unwanted features.
The extension of general relativity to supersymmetric theories and string theory, has shown that often, these requirements are automatically included by consistency. For instance, cosmic censorship often follows naturally in supersymmetric theories through the BPS bound for the mass of black hole solutions. Similarly, chronology protection seems to have a deeper meaning in supersymmetric theories. For many supersymmetric solutions to supergravity and string theory with CTCs, the presence of a causality violating regions in the \spacetime{}, goes hand in hand with unitarity violation in a dual quantum mechanics (by using the AdS/CFT correspondence).  We have found a three-dimensional example, that gives further evidence for a deeper meaning of causality violation:
\begin{itemize}
\item[$\rhd$] Through AdS/CFT, the condition for CTCs in the geometry is \textit{exactly the same} as the one for unitarity violation in the dual CFT
\item[$\rhd$] The observations \textit{do not rely on supersymmetry}.
\end{itemize}
A priori causality (a property of a classical geometry) and unitarity violation (in a dual quantum theory) are two totally unrelated concepts. The new example and earlier ones suggests that there is a \textit{general connection between the two}. It would be interesting to investigate this in more detail. A first idea would be to use aspects of the AdS/CFT, for example to compute the stress-energy tensor on the gravity side from the dual CFT and investigate its properties. Of related interest would be to extend these ideas beyond the AdS/CFT correspondence.

\label{c:Conclusions}

\appendix
\cleardoublepage
%
\chapter{
Some technical details}\label{app:SS}

\section{The \texorpdfstring{$\frac{\SL(3,\Real)}{\SO(2,1)}$}{} sigma model (details for chapter \ref{c:JP-Gradient_Flow})}\label{SL3}

We define the $\SL(3,\Real)/\SO(2,1)$ coset element in the Borel gauge
\begin{equation} L=\exp(\chi^1 E_{12})\exp(\chi^0 E_{23})
\exp(\chi^2 E_{13})
\exp(\tfrac{1}{2}\phi^1H_0+\tfrac{1}{2}\phi^2H_2)\,,
\end{equation}
where $H_{1}$ and $H_{2}$ are the Cartan generators of
$\mathfrak{sl}(3)$ and the $E_{\alpha}$ are the three positive root
generators. We use the fundamental representation of
$\mathfrak{sl}(3)$ and choose the following basis for the generators
\begin{equation}
H_0 = \frac{1}{\sqrt{3}} \begin{pmatrix} -1 & 0 & 0 \\
0 & 2 & 0 \\
0 & 0 & -1
\end{pmatrix},
\qquad H_1 = \begin{pmatrix} -1 & 0 & 0 \\
0 & 0 & 0 \\
0 & 0 & 1
\end{pmatrix},\end{equation}
and the three positive step operators
\begin{equation}\label{Es}
E_{12} = \begin{pmatrix} 0 & 1 & 0 \\
0 & 0 & 0 \\
0 & 0 & 0
\end{pmatrix},
\qquad E_{23} = \begin{pmatrix} 0 & 0 & 0 \\
0 & 0 & 1 \\
0 & 0 & 0
\end{pmatrix},
\qquad E_{13} = \begin{pmatrix} 0 & 0 & 1 \\
0 & 0 & 0 \\
0 & 0 & 0
\end{pmatrix}.
\end{equation}
The generators $T_{\Lambda}$, $\Lambda=1,\ldots, 8$, of $\SL(3,\Real)$
are given by
\begin{equation}\label{eq:T}
T_{\Lambda}=\{H_0, H_1, E_{12}, E_{23}, E_{13}, E_{12}^\mathrm{T},
E_{23}^\mathrm{T},E_{13}^\mathrm{T} \}\,.
\end{equation}

Then the coset element is explicitly given by
\begin{equation}\label{L}
L=\begin{pmatrix} \e^{-\tfrac{1}{2\sqrt{3}}\phi^1-\tfrac{1}{2}\phi^2}
& \e^{\tfrac{\phi^1}{\sqrt{3}}}\chi^1 &
\e^{-\tfrac{\phi^1}{2\sqrt{3}}+\tfrac{\phi^2}{2}}(\chi^0\chi^1
+\chi^2 )\\
0 & \e^{\tfrac{\phi^1}{\sqrt{3}}} &
\e^{-\tfrac{\phi^1}{2\sqrt{3}}+\tfrac{\phi^2}{2}}\chi^0 \\
0 & 0 & \e^{-\tfrac{\phi^1}{2\sqrt{3}}+\tfrac{\phi^2}{2}}
\end{pmatrix}.
\end{equation}
To find the metric on the coset we define the symmetric coset matrix
$\Mg$ via $\Mg=L\eta L^\mathrm{T}$ where $\eta$ is the matrix whose
stabiliser defines the specific isotropy group $\SO(2,1)$ of the coset.
To reproduce the sigma model (\ref{sigma}) we choose
\begin{equation}
\eta=\operatorname{diag}(+1,-1,+1)\,,
\end{equation}
whereas the other sigma model (\ref{sigma2}) has another $\SO(2,1)$
defined by
\begin{equation}
\eta=\operatorname{diag}(-1,+1,+1)\,.
\end{equation}
The metric is $\de s^2=-\tfrac{1}{2}\tr(\de \Mg\de
\Mg^{-1})$, the Cartan involution for
$A\in\mathfrak{sl}(3,\Real)$ is
\begin{equation}
\theta(A) = -\eta A^\mathrm{T}\eta\,.
\end{equation}

\section{Technical details for chapter \ref{c:TL}}\label{app:GB-Solutions}


We solve the equations of motion \eqref{chieqn}, \eqref{liouvilleeqn}.

It is instructive to note the differences with the flat space limit ($l\to\infty$) of \cite{Greene:1989ya}. In that scenario, the equation \eqref{liouvilleeqn} for the conformal factor $\phi$ is a Poisson equation with source, while we found a sourced Liouville equation (extra exponential term). Another difference is the topology of the spatial base manifold. In the presence of a cosmological constant, the spatial base  has the conformal structure and topology of the disk, as opposed to the flat case where the topology and conformal structure are those of the Riemann sphere. In our case the equations (\ref{chieqn}),(\ref{liouvilleeqn}) still have an elegant solution, but it is not straightforward to construct `stringy' solutions where $\T$ has nontrivial $\SL(2,\Z )$ monodromies as in \cite{Greene:1989ya}.

\subsection{Constant axion-dilaton: \texorpdfstring{$\AdS 3$}{} }
When $\T$ is constant, we are in the pure gravity case and the metric (\ref{stationaryLiouville}) describes local $\AdS 3$, written as a timelike fibration over Euclidean $\AdS 2$.  We illustrate how $\AdS 3$ can be written in this form and  introduce
two usful coordinate systems.

The general solution to the Liouville equation (\ref{liouvilleeqn}) without source and to the equation for the one-form $\chi$  (\ref{chieqn}) is\footnote{For notational simplicity we are a bit sloppy in distinguishing between the holomorphic partial derivative and the corresponding Dolbeault operator, denoting both with $\partial$. We trust the reader to distinguish between them by checking if the result is a scalar or differential form.}
\bea
e^{2 \phi} = {4 \partial g \bar \partial \bar g \over (1- g \bar g)^2}\label{Liouvillesolvac} \, , \qquad
\chi  =  2 {\rm Im}\, \pa \phi  + d f \label{chisolvac} \, ,
\eea
where $g (z)$ is an  arbitrary holomorphic function and $f (z, \bar z)$ an arbitrary real function. These arbitrary functions reflect the conformal invariance and shift symmetry of our
\ansatz{}. The resulting metric is locally $\AdS 3$. This is made explicit through the coordinate transformation to global $\AdS 3$ coordinates $(\s, \rho, \psi)$:
\bea
\s &=& {t + f\over 2} , \nonu
\tanh (\rho) \,e ^{i (\sigma - \psi)}&=& g(z),
\eea
in terms of which one obtains the standard global $\AdS 3$ metric (\ref{globalmetr}). $\AdS 3$ is what we expect, since $\tau = cst$ implies that the three-dimensional action just contains three-dimensional gravity and a negative cosmological constant. Moreover, this then gives the eleven-dimensional background solution $\AdS 3 \times \sS 2 \times \CY 3$.

\subsection{Holomorphic axion-dilaton solutions: G\"{o}del space}

We explore non-trivial (supersymmetric) solutions to the equations (\ref{chieqn},\ref{liouvilleeqn}),  with non-constant axion-dilaton field, which should be seen as the backreacted geometries due to wrapped M2-brane sources.  The class of solutions we find, has brane sources on the boundary, its local geometry is that of the three-dimensional G\"{o}del universe. In fact, the simplest solution is global G\"{o}del space, providing a new supersymmetric embedding of the G\"{o}del universe in string/M-theory. 

\paragraph{Solving the equations:}
We can solve the $\T$ equation (\ref{taueqn}) by taking  $\T$ to be a holomorphic function: \be \T = \T(z) \label{tausol}. \ee This leads to 1/2-BPS solutions (see appendix of \cite{Levi:2009az}).\footnote{We could take $\T$ to be antiholomorphic, which replaces brane sources with antibranes and preserves different supersymmetries.} Next we turn to the equation for the one-form $\chi$  \eqref{chieqn}.  It is solved by a simple modification of (\ref{chisolvac}):
\be
\chi = 2 \mathrm{Im} \left( \pa \phi + ( 1 - \m) \pa \ln \T_2 \right) + d f \label{chisol}.
\ee
Again, the arbitrary real function $f$ reflects a shift symmetry of the metric \ansatz{}.

It remains  to solve the Liouville equation (\ref{liouvilleeqn}) for the conformal factor $e^{2 \phi}$ in the presence of the source term. We write the equation as
\be
\cald e^{2 \f} \equiv \left( \pa \bar \pa \ln - \half  \right) e^{2 \f} = - (\m - 1) {\pa \T \bar \pa \bar \T \over 2 \T_2^2 } \, .%
\label{eq:Liouville}
\ee The source term on the right-hand side is quite special in that it is an eigenfunction
of the the non-linear differential operator $\cald =\pa \bar \pa
\ln - \half $. Indeed, we can write it as \be
- (\m -1 ) {\pa \T \bar \pa \bar \T \over 2 \T_2^2} = \cald \left( \m {\pa \T \bar \pa \bar \T \over  \T_2^2 }\right) .
\ee
Therefore a solution to the equation \eqref{eq:Liouville} is given by
\be e^{2 \phi } = \m {\pa \T \bar \pa \bar \T \over  \T_2^2 }\label{phisol} . \ee
This solution is a special case of those obtained in \cite{Semenov:2008}. Let us discuss the uniqueness of our solution. As far as we know, there is no proof of uniqueness of the solution (\ref{phisol}) in the literature.
Nevertheless, since we work in three-dimensional gravity, we know that a given energy-momentum tensor completely determines the local geometry, so that other
solutions to (\ref{liouvilleeqn}) (if any) must give a locally equivalent metric. Boundary conditions then provide the global structure. However, there seems to be a unique simply connected
and geodesically complete solution. In this solution  the spatial base, parameterized by $(z, \bar z)$, has  the conformal structure and topology of the disk, as in the case of global $\AdS 3$.\footnote{We have not considered the interesting generalization of taking the base to be a quotient of this disk  and $\T$ to have nontrivial $SL(2,\Z )$ monodromies. We hope to return to this in the future.}

\paragraph{The G\"odel solution:}
As discussed above, we can take $\T$ to be any single-valued (possibly multiple-to-one) meromorphic function from the upper half plane to itself. The simplest case, which we study in the remainder of
this work, is to take
\be
\T(z) = z\,,\label{eq:tauw2}
\ee
which is one-to-one and has a first order pole on the boundary at $z = i \infty$. More general multiple-to-one maps can be locally brought into this form by a conformal transformation.
With $f=0$ in (\ref{chisol}) and defining $z = x + i y$ the metric is%
\beq%
ds^2 = { l^2 \over 4} \left[ - (dt+ \mu {dx \over y})^2 + \mu {d x^2 + d y^2 \over y^2} \right] \, \label{godelUHPmain2} .
\eeq%
Rescaling $t \to \mu t$, one recognizes the metric of timelike warped AdS (see e.g. \cite{Bengtsson:2005zj}).

%

\cleardoublepage

{
\markboth{{Acknowledgments}}{{Acknowledgments}}
\def\rightmark{ACKNOWLEDGMENTS}
\def\leftmark{ACKNOWLEDGMENTS}

\chapter*{Acknowledgments}
\addcontentsline{toc}{chapter}{Acknowledgments}
\setlength{\parindent}{5mm}

First of all, thank you Toine, my supervisor, for your advice, your guiding hand, the chances of growing in the field. Thanks also to the members of the jury, Walter Troost, Okki Indekeu, D\'esir\'e Boll\'e at K.U.Leuven and thank you Jos\'e Figueroa-O'Farrill and Iosif Bena, for finding the time to read and assess my doctoral work and come to Leuven for the defense. 

Thank you, my collaborators, for many discussions and for shaping this work: thank you 
Bert, Jan, Paul, Thomas (times three: Thomas, Tokke and Tommy),  Dieter, Joris,  Walter, Jos\'e and Toine
(and also Frederik, Bram G., Fran\c cois, Jan D., Eric, Oscar, Andr\`es and Wieland for numerous discussions and work ``behind the scenes'' and my brother Wide Vercnocke for the cover illustration). \footnote{\it Note: for those looking for a thank you to all who contributed outside the work environment, see the printed version at \href{http://itf.fys.kuleuven.be/hep/phd/ThesisBertVercnocke_Cover.pdf}{\textcolor{blue}{http://itf.fys.kuleuven.be/hep/phd/ThesisBertVercnocke\_Cover.pdf}}.}

Finally, this work was supported by an Aspirant fellowship of the FWO-Vlaanderen during the 4 PhD years, and in part by the European Community's Human Potential Programme under contract MRTN-CT-2004-005104 `Constituents, fundamental forces and symmetries of the universe', by the Federal Office for Scientific, Technical and Cultural Affairs through the `Interuniversity Attraction Poles Programme -- Belgian Science Policy' P6/11- P, and by the FWO-Vlaanderen through project G.0235.05.

{}\raggedleft Bert Vercnocke,\\
\raggedleft August 2010
}

\selectlanguage{english}
\cleardoublepage
\phantomsection
\markboth{{Bibliography}}{{Bibliography}}
\addcontentsline{toc}{chapter}{Bibliography}
\bibliographystyle{toine}
\renewcommand{\bibname}{Bibliography}
		
\bibliography{ThesisBib}

%
%
 
%
\cleardoublepage
%

{
\markboth{{Publication List}}{{Publication List}}
\def\rightmark{PUBLICATION LIST}
\def\leftmark{PUBLICATION LIST}
\renewcommand\bibsection%
{
 \subsection*{\bibname
}
}

\chapter*{
Publication list}
\addcontentsline{toc}{chapter}{Publication List}
\makeatother
\vspace{-.3cm}
The original research presented in this thesis is based on several publications, which are referred to as refs.~\cite{VanProeyen:2007pe}--\cite{Raeymaekers:2009ij}. In particular, part \ref{pt:FirstOrder} considers the research of  \cite{Janssen:2007rc} and\cite{Perz:2008kh}  and part \ref{pt:Entropy} is based on \cite{Raeymaekers:2008gk}, \cite{Levi:2009az} and \cite{Raeymaekers:2009ij}.\footnote{The work of \cite{VanProeyen:2007pe} is briefly touched upon in part \ref{pt:FirstOrder}. The work \cite{FigueroaO'Farrill:2007rt} was performed during a three-month stay at the University of Edinburgh and is not discussed, as it is only remotely related to the other works. Finally, \cite{Perz:2009ag} is a proceedings contribution, summarizing 
\cite{Perz:2008kh}. }

\renewcommand{\bibname}{\vspace{-1.1cm}
}

\vspace{-1cm }


%
%

\newpage
\setcounter{section}{0} 

}

%
%
%
%

\end{document}